# One-dimensional Titanium Dioxide Nanomaterials: Nanotubes

*Kiyoung Lee,*[†,‡] *Anca Mazare,*[†,‡] *and Patrik Schmuki* [†,*]

[†] Department of Materials Science WW4-LKO, University of Erlangen-Nuremberg, Martensstrasse 7, 91058 Erlangen, Germany

‡These authors contributed equally.

* corresponding author: P. Schmuki – schmuki@ww.uni-erlangen.de

CONTENTS

## Introduction

In 1991, Iijima reported on the formation of carbon nanotubes[1] using a simple arc discharge reaction that led to an arrangement of the material in some µm long tubes with a diameter of only a few nm - this turned out to be a milestone in materials science and technology[2-7]. Only a few years later, a range of transition metal oxides[8-17] were reported to form nanotubular arrangements when oxide powder was heated in hot alkaline solutions, either by simple refluxing or under mild pressure in an autoclave[8-17]. Over the past decades, nanotubular geometries stimulated immerse research activity which is evident from the over 100'000 papers published to "nanotubes" up to 2013. The main reason for this interest is, except for scientific curiosity, the anticipated economic impact in the form of applications that are based on specific physical and chemical features of these 1D (or extremely high aspect ratio) structures. Among the transition metal oxides, particularly the ability to produce the classic wide band gap semiconducting $TiO_2$ in form of nanotubes found immediate interest, mainly due to the perspective of using the structures to enhance the properties in Grätzel-type solar cells[18] and photocatalytic materials[19]. These two photoelectrochemical applications made $TiO_2$ the most studied functional oxide over the past 30 years. Nevertheless, other features of $TiO_2$ (besides its classic use as a pigment), such as its excellent biocompatibility as well as ion intercalation properties, contribute largely to the high interest in this material.

The first hydrothermal $TiO_2$ nanotubes and carbon nanotubes have in common that they essentially consist of a rolled atomic or molecular plane with (in the ideal case) monomolecular layer thickness[20].

These nanotube geometries may be considered (in a physical sense) as “true” nanotubes, as quantum size and dimensionality may fully be effective on physical and chemical properties, such as electron mobility, optical band-gap, and surface reactivity[21-36].

In contrast to hydrothermal (titanate) tubes that consist only of one or a few atom layer thick wall, a considerable range of synthesis techniques leads to high aspect ratio $TiO_2$ tubular shapes that may be hundreds of micrometers long and some 10-1000 nm in diameter, but with a wall thickness that is typically in a range of ≈ 10-100 nm. Although in this case no considerable quantum size effects occur, these nanotubes attract tremendous interest due to other specific advantages or features, including geometric factors such as surface area, size exclusion effects, defined diffusion behavior, biological interactions, or directional charge and ion transport properties.

In this context, a most relevant geometric quality is provided by nanotubes that are produced as arrays, i.e. in an aligned form perpendicular to the substrate. In this case, the directionality of the ensemble provides inherent advantages, for example as large scale oriented electrode in photoelectrochemistry (solar cells, photocatalysts) or as highly size-defined bioactive coating. These aligned $TiO_2$ nanotube arrays that can be grown by self-organizing anodization (or template filling) have created enormous interest as reflected in a vigorous publication output over the past few years. Figure 1 gives a comparison of the publication activities broken down according to tube type over the last 20 years. From this compilation not only it is apparent that overall an almost exponential increase in work in this field can be observed, but also the fact that currently a vast majority of work deals with self-organized anodic $TiO_2$ nanotube arrays. Except for work towards improved synthesis conditions to tailor geometry, structure, organization or modification (doping, band-gap engineering, decoration), virtually every application of $TiO_2$ that has been based on nanoparticulate forms of titania is being examined using nanotubular geometries. Scientifically even more exciting are, of course, new aspects that arise from the specific geometry and its fine tuning.

Figure 2 gives schematically an overview over the most important (realized and anticipated) beneficial features of using $TiO_2$ nanotubes or nanotube arrays. Classic 1D quantum size effects on electronic properties may lead to reduced electron scattering (or in an extreme case to ballistic transport). Also extreme surface curvature may result in modification of chemical and physical properties. These effects may be exploited in virtually all electrical or photoelectrochemical arrangements (sensors, solar cells, photoreactors). The fact that diffusion length for minority carriers (holes) lies within the range of the tube wall thickness, and the comparably long electron life-time in $TiO_2$, allow orthogonal carrier separation (hole to the wall, electron to the back contact). Tube arrays enable core-shell structures (carrier separation) or interdigitated electrode assemblies, as well as decoration of the walls with e.g. light harvesting or sensing elements, while keeping well defined diffusion pathways for charge carriers or ion intercalation (ion insertion batteries). The definition of tight compartments, "nano-test-tubes", combined with a high observation length provides platforms for low volume / high sensitivity sensing (e.g., high optical contrast in light absorption or fluorescence emission).

Due to its high biocompatibility, $TiO_2$ nanotubular structures are explored in various biomedical applications, such as nanosize defined biocompatible coatings ($TiO_2$ is the prime coating material for biomedical implants) or drug-delivery devices.

A number of excellent reviews have been written mainly to specific aspects of $TiO_2$ nanotubes, such as dealing with hydrothermal tubes,[16,17,37] anodic tube synthesis, applications and self-organization,[38-44] or specific applications such as in solar cells,[45,46] sensing,[47] photocatalysis,[48,49] or biomedical use[50,51]. In the present review we try to give a comprehensive and most up to date view to the field, with an emphasis on the currently most investigated anodic $TiO_2$ nanotube arrays. We will first give an overview of different synthesis approaches to produce $TiO_2$ nanotubes and $TiO_2$ nanotube arrays, and then deal with physical and chemical properties of $TiO_2$ nanotubes and techniques to modify them. Finally, we will provide an overview of the most explored and prospective applications of nanotubular $TiO_2$.

# 1. Growth techniques for $TiO_2$ nanotubes

## 1.1 Overview

Over the past 20 years a considerable number of different strategies to synthesize $TiO_2$ nanotubes or $TiO_2$ nanotube arrays have been elaborated (Figure 3 provides an overview of typical morphologies and characteristic features of tubes synthesized by different approaches). Roughly one may divide the main routes into templating, hydrothermal, and anodic self-organization approaches. Templates to form tubular structures may be single high aspect-ratio molecules (such as cellulose), molecular rod-like assemblies (e.g. micelles), or defined organized nanostructures (such as ordered porous alumina or track-etch membranes)[20,52-62]. These templates then are coated or decorated with various deposition approaches (such as sol-gel or atomic layer deposition (ALD)[63-65]) to form $TiO_2$ in a nanotubular form. Such composite structures may be used while the $TiO_2$ is in/on the template, but most frequently the template is removed (selectively dissolved, evaporated, decomposed) to form “free” nanotubes, nanotube assemblies or tube-powder. Template-free approaches are based mainly on either hydro/solvothermal methods (where typically titanium oxide particles are autoclaved in NaOH to delaminate to titanate units and finally reassemble in the shape of tubes),[16,17,20,53] or nanotube arrays that form by self-organized electrochemical anodization of metallic titanium, typically in dilute fluoride-electrolytes[40,41,66-69].

One should note that Ti precursor/molecular template based processes and hydrothermal approaches result in single tubes or loose agglomerates of tubes or bundles that are dispersed in a solution where often a wide distribution of tube lengths is obtained. In order to make use of the structures in electrically contacted devices, the tubes are usually compacted to layers (similar to powders) on an electrode surface. However, this leads to an arbitrary orientation of the nanotubes on the electrode and this, in turn, eliminates many advantages of the one-dimensional nature of the structure (e.g., providing a 1D direct electron path to the electrode). Using aligned templates or self-organizing electrochemical anodization leads to an array of oxide nanotubes oriented perpendicular to the substrate surface (such as in Figure 3.2.b or 3.3.b). The tubes in the template can relatively easily be contacted by metal deposition. In the case of anodic tubes, the tube layers are directly attached to the metal surface and thus are already electrically connected. Additionally, electrochemical anodization allows to coat virtually any shape of Ti (and other metal) surfaces with a dense and defined nanotube layer. In the case of templates, the form of the template (molecule or aligned structure) determines to a large extent if electrodes with a back-contact perpendicular to the tubes can be obtained. We will discuss the main techniques to obtain the main types of $TiO_2$ nanotubes in more detail below.

## 1.2 Deposition into/onto templates

*Anodic aluminium oxide (AAO)*

Historically, the first effort to produce titania nanotubes was probably the work by Hoyer et al.[70] who used an electrochemical deposition method in an ordered alumina template. His electrodeposition approach was based on a $TiCl_3$ solution that was hydrolyzed and electrodeposited as a polymerized oxyhydroxide.

$$nTiOH^{2+} + mH_2O \rightarrow [Ti^{IV}O_x(OH)_{4-2x}]_n + 3nH^+ + ne^- \qquad (1)$$

Nowadays, to create aligned nanostructures (such as nanowires, nanotubes) by templating, more frequently than electrodeposition, other filling approaches are employed such as sol-gel techniques,[20,52-62,71] or more recently atomic layer deposition (ALD)[63-65]. The former approach is a common path of $TiO_2$ synthesis, which is based on the hydrolysis reactions of Ti-alkoxide, $TiCl_4$, $TiF_4$ precursors followed by condensation reactions (i.e., a gel-type polymeric Ti-O-Ti chain is developed, which further hydrolyzes and thus results in $TiO_2$ precipitates). For example, $TiO_2$ sol can be sucked into the pores of an alumina template, and after an appropriate heat treatment the alumina template can selectively be dissolved.[58] Various modifications of this process have been reported (for example, see [52,71-73]). In ALD, surfaces of templates (such as porous alumina) can be coated conformably with one atomic layer after the other by using alternating cycles of exposure to a titania precursor (such as $TiCl_4$, $Ti(OiPr)_4$), followed by purging and hydrolysis[63-65].

Most typical templates are porous alumina, ion track-etch channels, or occasionally ordered nanowires or rods, such as for instance ZnO nanowires[70,74-78]. The classic template for the synthesis of a variety of aligned nanomaterials is porous anodic aluminium oxide (AAO) which can be produced with a hexagonal pattern of nanopores in a long-range, virtually perfect order. Experimental details of the fabrication of highly ordered alumina pore arrays can be found, for example, in references [79-81]. To date, it is possible to fabricate well-defined self-ordered porous alumina with interpore distances between 10 and 500 nm,[79,82,83] with aspect ratios >1,000.

If electrodeposition is used for filling, one needs to consider that alumina is an insulating material and thus the thin barrier oxide (the pore bottom) has a high resistance. Therefore, prior to electrodeposition, the pore bottom is usually thinned or removed. For this, wet chemical etching of the anodic alumina film using a diluted phosphoric acid solution (also used as a pore widening treatment), or in situ thinning of the pore bottom using step-wise lowering of the anodic voltage at the end of anodization is typically employed[84,85]. For electrodeposition, usually pulsed current is used in order to overcome the remaining resistance of the barrier layer at the pore bottom, and to take into account diffusion processes during deposition. Here, in pulse breaks, the cation concentration gradient established during a deposition pulse is balanced by diffusion from the electrolyte to the pore bottom.[86]

Alternatively, the barrier layer can be entirely removed by chemical means. The resulting through-hole porous layer (membrane) may then be PVD coated with a metal, such as Pt, Au, Ag, to establish a contact for conventional electrodeposition approaches[87]. I.e., DC electrodeposition can then be used to fill the porous channels starting from the bottom. For $TiO_2$ electrodeposition to form tubes, e.g., a $TiF_4$ precursor deposited into a AAO template can be used,[76,77,88] as shown in Figure 3.2.a.

Another approach of synthesizing NTs in AAO templates is based on polymer wetting[90]. It is based on the observation that a low surface energy polymer preferentially wets the walls of pores of a material that has a high surface energy, such as $Al_2O_3$. To form oxide nanotubes in an alumina template, suitable oxide precursor compounds are mixed with polymers, then this mixture is used in the wetting process and after template removal (and polymer dissolution or burning off), the ceramic structure remains in a tubular form. For instance, ferroelectric and piezoelectric oxide nanotubes such as lead zirconate titanate (PZT, $PbZr_{0.52}Ti_{0.48}O_3$) and barium titanate ($BaTiO_3$) have been produced in this manner.[91]

*Molecular or molecular assembly templating*

$TiO_2$ nanoparticles or nanorods are also prepared by using micelle-templates of appropriate surfactants above their critical micelle concentration (the surfactant molecules aggregate and disperse in liquid, to so-called spherical or rod-like micelles, which are used as template for $TiO_2$ preparation). In this approach, nanotube formation is mostly carried out using water containing reverse micelles with a cylindrical shape. The Ti precursor can then react at the micelle surface, and after removal of the surfactant (burn off), a nanotube structure is obtained[92,93]. Usually $TiCl_4$ or Ti-alkoxide solutions are employed as the Ti-precursor. Using certain $H_2O$:micelle ratios allows to vary the dimensions of the nanotubes.[92,93] Nevertheless, only low aspect ratio tubes can generally be obtained by this technique.

A range of other methods involve the use or synthesis of other fibrous or rod-like templates. For example, Kobayashi et al. reported the preparation of $TiO_2$ nanotubes using gelation of an organogelator to a template that is coated using titanium alkoxides and alcohols[94,95]. The organogelator used was a cyclohexane derivative that was specially synthesized for this purpose. The outer and inner diameters of the $TiO_2$ nanotubes obtained were 150-600 nm and 50-300 nm, respectively.

Other examples are $TiO_2$ hollow nanostructures that are formed using cotton fiber as a template.[96] Here, chemical deposition of a $TiO_2$-precursor onto the cellulose template is used and the cotton thereafter can easily be burnt-off, forming pure hollow $TiO_2$ nanostructures.[96]

### 1.3 Titanium oxide nanotubes by hydrothermal reaction

In 1998, Kasuga et al.[20] reported for the first time on the hydrothermal synthesis of $TiO_2$ nanotubes. In general, the approach is based on alkaline treatment of a titanium oxide precursor, which may be rutile,[53,97] anatase, commercial P-25,[97-99] or amorphous $TiO_2$[100]. The powders or crystallites are typically heated in a NaOH solution with a concentration between 4 mol/L and 20 mol/L in an autoclave at temperatures between 100°C and 180°C for several (1-2) days.[97,101] The formation of nanotubes is facilitated with an increase in NaOH concentration[102] and temperature[97,101]. At higher temperatures, nano fibers and ribbons can be formed[103]. NaOH can be replaced by KOH which allows to increase the temperature to 200 °C.[99] LiOH, however, forms more stable $LiTiO_2$ compounds rather than oxide sheets or tubes.[104] After the alkaline treatment, usually the resulting powders are washed with water and 0.1 mol/L HCl aqueous solution until the pH value of the washing solution is lower than 7, and subsequently powders are filtered and dried at various temperatures[53,97-101].

Figure 3.1.b shows an HRTEM image of $TiO_2$ nanotubes, produced by Bavykin et al.[105]. The inner diameter of hydrothermal nanotubes usually ranges from 2 to 20 nm. The tubes generally have a multi-walled morphology. The distance between the wall-layers is approximately 0.72 nm in the protonated form. They are generally open-ended and have a constant diameter along their lengths[37]. Sizes and shapes depend on the synthetic conditions and on the size and structure of the used titanium and titanium oxide raw materials[37,106]. Generally, for higher hydrothermal temperatures and larger substrate precursors, longer tubes of up to several micrometers can be obtained[107]. Size and shape can be also influenced using other experimental conditions. Final tubes tend to agglomerate, but can be dispersed into aqueous colloidal solutions.

Mechanistic details of hydrothermal titanium oxide nanotube formation are discussed somewhat controversially in the literature. Kasuga et al.[20,53] believed that the nanotubes were formed in the washing step containing the hydrochloric acid. In 1999 they proposed the following formation mechanism: By treating the raw material with aqueous NaOH solution, some Ti-O-Ti bonds are broken, and Ti-O-Na and Ti-OH bonds are formed. In the washing step, the $Na^+$ ions in the Ti–O–Na bonds, present in the alkali-treated specimen, are exchanged by $H^+$, exfoliating the material to a sheet-like structure. By treating the material with HCl solution, Ti-OH bonds react with the acid and water to form new Ti-O-Ti bonds and anatase is formed. In this step, the titanium oxide sheets convert to anatase nanotubes by folding[53]. Other authors later supported this conclusion[11].

On the other hand, the formation of nanotubes was found by Du et al.[54] (2001) to occur even without washing the as-formed product with hydrochloric acid. The authors concluded that the nanotubes were formed in the hydrothermal treatment step. Since then, most work reported that the nanotube shape is created during the hydrothermal alkali reaction[15,107-111]. This is also supported by recent publications[99,103]. Wang et al.[112] concluded that the as-synthesized nanotubes were anatase rather than titanate. They proposed a formation mechanism as shown in Figure 4a-e[112]. In this approach, NaOH initially disturbs the crystalline structure (Figure 4.a) of raw anatase $TiO_2$ crystals (Figure 4.b). The free octahedra reassemble to link together by sharing edges, with the formation of hydroxyl bridges between the Ti ions resulting in a zig-zag structure (Figure 4.c), leading to growth along the [100] direction of anatase. Lateral growth occurs in the [001] direction and results in the formation of two-dimensional crystalline sheets (Figure 4.d). To saturate dangling bonds and reduce the surface to volume ratio the crystalline sheets roll-up, lowering the total energy; the result, seen in Figure 4.e, is anatase $TiO_2$ nanotubes[112]. Other mechanisms involving scrolling single-layer nanosheets[13,113] or curving of conjoined nanosheets (Figure 4.f-h)[105] were also proposed in literature.

It may well be that different experimental conditions are responsible for different findings. A detailed analysis by Sekino et al.[114] for an alkali treatment using refluxing at 110 °C leads to findings as illustrated in Figure 5.a and b. The figures show a sequence of typical XRD spectra and TEM images for samples taken during chemical processing. After the alkaline treatment, the product mainly consists of amorphous and crystalline phase corresponding to sodium titanate ($Na_2TiO_3$, Figure 5.a.2), but no clear morphological features can be observed (Figure 5.b.1). After water and HCl treatment (Figure 5.a.3 and 5.b.2), sodium titanate disappears completely and a low crystallinity phase is observed. In this step, a nanometer-sized $TiO_2$ nanosheet-like morphology is obtained. Subsequent water washing leads to a clear nanotube morphology (with an open-end structure and with individual tubes, Figure 5.b.3 and 4). The outer and inner diameters of the tubes are around 8–10 and 5–7 nm, respectively, and the length is of several hundred to several micrometers. In this process, the size of the obtained nanotubes does not depend on the starting materials, or whether KOH is used as a reaction solution rather than NaOH.

If the hydrothermal synthesis is carried out in an autoclave with a higher pressure during the process[97], not only $TiO_2$ but also Ti metal can be used as the source material for oxide nanotubes[110]. I.e., titanium is chemically oxidized in the alkaline solution prior to tube formation. Mostly a higher degree of size control, especially thick nanotubes, can be synthesized at higher temperatures. In addition, natural mineral sources can also be used for nanotube synthesis[115]. Overall, while some dispute exists about mechanistic details and the composition of titanate tubes (see also section on structure and properties), the hydrothermal method is a versatile approach scalable to large synthesis batches, and it is the only approach based on forming $TiO_2$ nanotubes with a wall thickness in the range of atomic sheets.

### 1.4 Self-organizing anodic $TiO_2$ nanotube arrays

Another most simple, low cost, and straightforward approach to fabricate titania nanotubes is self-organizing anodization of Ti-metal substrates under specific electrochemical conditions[40]. Typical examples of such type of tubes are shown in Figure 3.3.b and Figure 6.d. Mostly such tubes are formed in dilute fluoride electrolytes under several 10 V of anodization potential. In their most elaborated way they form highly self-organized hexagonal arrangements as in Figure 6.d.

Interestingly, first reports by Assefpour-Dezfuly et al.[66] in 1984 and later by Zwilling et al.[67] in 1999 on the formation of self-organized porous/tubular $TiO_2$ structures using anodization of Ti and some alloys in fluoride based electrolytes were widely overlooked, and the finding was mainly ascribed to Grimes[68] in 2001. However, all these $TiO_2$ layers, including early follow-up work[69], were far from perfect, i.e. they showed a considerable degree of inhomogeneity and were limited to tubes with length of about 500 nm. Later work showed significantly improved control over length, diameter, ordering and composition by the use of pH mediation[116], and particularly by the introduction of non-aqueous electrolytes[117,118,119]. It is noteworthy that fluoride based electrolytes were then also found to be an extremely versatile tool to grow ordered anodic oxide nanostructures on other metals, such as Hf,[120,121] Zr,[122-127] Fe,[128-130], Nb,[131,132] V,[133] W,[134-137] Ta,[138-142] Co,[143] and even Si[144-146].

Not only pure Ti, but a full range of alloys can also be used to form nanotubes – this turned out to become a unique way of direct doping the tube oxide by using alloys, with defined amounts of a secondary desired doping-metal, for anodization.

In the context of anodic treatments, it should be mentioned that except for self-organizing tube-formation in fluoride-electrolytes, also photoelectrochemical etch-channels,[147,148] or self-organized channel structures (obtained in hot glycerol electrolyte)[149-153] have been reported.

Furthermore, also some chloride or perchlorate containing electrolytes[134,154,155] can be used to grow anodic $TiO_2$ nanotubes – however, this type of tubes typically grows in the form of bundles on the Ti surface (see Figure 6.g)[156,157]. These tubes usually can reach few tens to few hundreds micrometer length within tens of seconds. In this case the anodization voltage must be sufficiently high to create a local breakdown of the oxide film that then represents the nucleus for the tube-bundle growth. This so-called "rapid breakdown anodization" can also be extended to other materials such as W and even to alloys such as Ti–Nb, Ti–Zr and Ti–Ta.[158] Although the process is very fast and thus is useful for producing large amounts of nanotubes in a short time, due to a lack of geometry control (uniform length, diameter control), these NTs received far less attention in comparison with self-organized $TiO_2$ nanotube arrays. The latter, in comparison, show a very high adjustability of geometry and as a result, these self-organizing tube layers have been over the last years the most widely investigated $TiO_2$ nanotube morphology. Due to this we will emphasize these self-organized tube arrays and provide more details on growth mechanisms and critical growth factors in section 2.

### 1.5 Electrospinning

A range of other approaches to form $TiO_2$ nanotubes have been reported and in particular electrospinning is certainly worth mentioning. In this process, a strong electric field is used to pull a thin jet out of a drop of polymer solution or melt through a suitable nozzle. The jet then is deposited in form of a nanofiber[159]. $TiO_2$ nanotubes are obtained e.g. by using titanate precursors to coat the fiber, and after polymerizing the precursor to $TiO_2$, applying a thermal treatment to remove (decompose) the organic fiber[160-162]. An example of such nanotubes is shown in Figure 3.4.b[163]. Most elegant is a simultaneous coating of the fiber while spinning. Li and Xia[164] reported the formation of $TiO_2$ hollow-nanofibers (nanotubes) by electrospinning of an ethanol solution (containing Ti tetraisopropoxide / polyvinylpyrrolidone) and of a heavy mineral oil or polymer through a coaxial, two-capillary spinneret, followed by selective removal of the cores and calcination in air. Under optimum conditions, $TiO_2$ nanotubes are formed with continuous and uniform structures (several cm range and well separated single tubes).

Another example is the work of Nakane et al.[165] in 2007 that used electrospinning to form precursor nanofibers of poly(vinyl alcohol) (PVA)-titanium compound hybrids, that then were calcinated to $TiO_2$ nanotubes. In the report, they examined the crystal structure and morphological change with different heat treatment processes. By heat treatment from 400 to 600 $^{o}C$ anatase phase was obtained, and above 600 $^{o}C$ anatase/rutile mixed phases were formed. The specific surface areas evaluated by BET decreased with increasing heat treatment temperature due to sintering of the tube structures.

In general, electrospinning allows producing nanotubes with extremely large aspect ratios. The nanofibers typically have diameter range from a few ten nanometers to a few micrometers. The ability to obtain such high aspect ratios rests on the fact that electrospinning is, like an extrusion processes, a continuous process. Moreover, by extension of the coaxial spinneret system multishell nanostructures can be formed. In addition, it is possible to use this method to obtain nanofibers/nanotubes, with specific surface topologies[166-168].

## 2. Ordered $TiO_2$ nanotube arrays

Electrochemical formation of ordered $TiO_2$ nanotube arrays such as shown in Figure 3.3 and Figure 6 is based on anodization of a metal in an electrolyte under conditions, where self-organization is established. The key to the "right" electrochemical conditions is an optimized steady-state situation of electrochemical oxide formation and chemical oxide dissolution. These conditions and key factors affecting them will be discussed in the following.

### 2.1 Electrochemical aspects of anodic growth of nanotube layers

Electrochemical anodization as such is a century old process mostly used in industry, to create "thick" compact or porous oxide layers on the surface of a metal substrate. It is carried out typically in an electrochemical cell, as illustrated in Figure 3.3.a, containing a suitable electrolyte, with the metal of interest as a working electrode (anode), and an inert counter electrode (usually platinum or carbon). Upon applying a sufficiently high anodic voltage to the metal of interest M, it is oxidized to $M^{z+}$ (eq. 1) that either forms a metal oxide, $MO_{z/2}$ (eq. 2a) or is solvatized and then dissolved in the electrolyte (eq. 2b). As a counter reaction, protons are reduced to produce hydrogen gas at the cathode (eq. 3).

$$M \rightarrow M^{z+} + ze^- \qquad (1)$$

$$M + \frac{z}{2} H_2O \rightarrow MO_{\frac{z}{2}} + zH^+ + ze^- \qquad (2a)$$

$$M^{z+} + solv \rightarrow M^{z+}_{solv} \qquad (2b)$$

$$zH^+ + ze^- \rightarrow \frac{z}{2} H_2 \qquad (3)$$

Regarding the question if an oxide layer is formed (eq. 2a) or dissolution (2b) dominates, thermodynamic aspects such as solubility products and oxide stability (that can be taken for example from Pourbaix-diagrams)[169] together with respective reaction rates need to be considered. In an electrolyte where the oxide is insoluble and no other side reactions occur, mainly reaction (2a) dominates, i.e. a high oxide formation efficiency is obtained. Anodic oxidation processes (thickening of the oxide) typically follow a high field law of the form:

$$I = Aexp(BE) = Aexp(B\Delta U/d) \quad (4)$$

where $I$ is the current, $\Delta U$ is the voltage across the oxide, $d$ is the layer thickness and $E$ is the electric field[170,171]. $A$ and $B$ are experimental constants. The key process is based on the field effect on ions migrating through the oxide layer as illustrated in Figure 6.b. As the transport of $M^+$ ions outward and of $O^{2-}$ ions inward are controlled by the applied field, with $E = \Delta U/d$, with an increasing film thickness the field (and thus the current, if a constant voltage is applied) drops exponentially. Finally the field is lowered to an extent that it is not able to significantly promote ion transport any longer and the film reaches a final thickness. If, however, a certain degree of solubility of the oxide is provided and an equilibrium of film formation and dissolution can be established, a considerable ion and electron flux is maintained in a steady state situation. For example, solvatization of $Ti^{4+}$ can be realized by the formation of fluoro-complexes such as in eqs. 5 and 6. One may consider pure chemical dissolution of the oxide (eq. 5) or direct complexation of high-field transported cations at the oxide electrolyte interface (eq. 6) – often this process is called ejection of transported cations to the electrolyte:

$$MO_2 + 6F^- \xrightarrow{H^+} [MF_6]^{2-} + 2H_2O \quad (5)$$

$$M^{4+} + 6F^- \rightarrow [MF_6]^{2-} \quad (6)$$

Figure 6.e schematically illustrates the observed electrochemistry represented in i/E and i/t curves for three cases of oxide "solubilities". First, if metal oxidation forms ions that are immediately and completely solvatized (eq. 2b), no oxide film is observed; this case is commonly described as active corrosion or electropolishing (EP). Secondly, one may obtain the formation of a stable (insoluble) compact metal oxide according to eq. 2a, and in accordance with the high field law (eq. 4). Thirdly, if there is a competition between solvatization and oxide formation, where the former reaction can be promoted by the addition of a suitable agent (such as $F^-$ in the $TiO_2$ case, Figure 6.b and 6.c), the established steady-state situation often leads to the formation of a porous oxide. If formation and dissolution are in an optimum range, highly self-organized oxide pore arrangement or nanotube formation is possible[44,172]. In this case, a typical current vs. time curve as shown in Figure 6.e for porous oxide (PO) is obtained and it can be divided into three different regions[38,40]. In region-I, the current is decreasing exponentially as a result of coverage of the anodized surface with an oxide film according to reaction 2a and eq. 4. Experimentally one observes that this layer is then partially penetrated by nanoscopic etch channels and a porous initiation layer is formed[40]. Porosification increases the surface area of the electrode and consequently in region-II a rise in current occurs. Underneath this initial layer, stable pore growth is initiated in region-III, the current then reaches a virtually constant value reflecting the establishment of a steady-state situation between dissolution and formation of oxide. In a typical self-ordering tube formation sequence, first the chemical dissolution is non-uniform (region-II) and the initiated pores grow progressively in a tree-like fashion. As a result, the individual pores start interfering with each other, and will be competing for the available current. This leads, under optimized conditions, to a situation where the pores equally share the available current, and self-ordering under steady state conditions is established (region III). Here,

oxide is continuously formed at the bottom under high field conditions, but part of the bottom layer is permanently dissolved, leading to a thinner oxide at the tube bottom than the corresponding final high field value. The steady state current density is typically in the order of some $mA/cm^2$.

In addition to affecting the $TiO_2$ dissolution according to eq. 5, fluoride ions, due to their small ion size, also migrate under the constant applied field through the growing porous oxide layer towards the metal, and form a fluoride-rich layer at the metal/oxide interface, as illustrated in Figure 6.f. The formation of this fluoride-rich layer is the most likely origin of the formation of a nanotube – rather than a nanopore morphology – as discussed in the next section, but it is also the cause for a reduced adherence of oxide nanotube layers on the metal substrate[173-176].

A large number of fluoride-based electrolyte compositions have, over the past ten years, been explored for tube formation such as mixtures of organic solvents (mainly EG, glycerol, DMSO) with $H_2O$ and fluoride sources such as HF, $NH_4F$, $BF_4^-$. Even ionic liquids containing $BF_4^-$ [177-180] were reported to successfully form tubes.

It is noteworthy that the formation of reaction products also increases the conductivity of the electrolyte and thereby increases the growth rate of the nanotubes (namely in the case of low conducting organic electrolytes).

Anodic growth of nanotubes takes place mainly by the competition of reaction (2a) viz. reactions (5) and (6). The purely chemical dissolution (eq. 5) leads to permanent thinning of the tube walls which is the strongest for the longest time exposed tube tops (Figure 6.a). After extended times, thinning to zero wall thickness prevents further overall growth of the nanotubes[40,116,181].

## 2.2 Why is a tubular shape formed?

In general, self-organizing anodization leads to hexagonal nanoporous cells (a honeycomb structure) due to the competitive space filling nature of pore growth. The role model is self-organized porous alumina formed in acidic solution[182-197] and in neutral fluoride containing solutions[198]. In contrast, for $TiO_2$ virtually under all self-organizing conditions a tubular shape is formed (rather than a porous morphology). This can be attributed mainly to the composition of cell boundaries; i.e., the main difference of porous alumina and self-ordered $TiO_2$ nanotube layers is that in the case of $TiO_2$ nanotubes, the cell boundaries can be etched (dissolved) under the applied electrochemical conditions. As mentioned above, a specific feature of fluoride electrolytes is that under high field oxide growth conditions, the small $F^-$ ions may migrate several times faster through an oxide layer than $O^{2-}$, thus a fluoride rich layer is formed at the metal/oxide interface (schematically shown in Figure 6.c). In addition to classical analytical techniques,[173,174] this fluoride rich layer can be observed as a haze at the bottom of the tubes such as shown in Figure 6.f. $H_2O$ easily dissolves the fluoride rich layers between $TiO_2$ "cells", and as a consequence, a tubular shape is formed. This occurs under a wide range of anodization conditions (Figure 7a)[40,174]. In case of porous alumina, even if formed in fluoride electrolytes, a porous morphology is obtained, as Al-fluorides are not easily soluble in $H_2O$[198]. Additionally, one may consider differences in the specific volume, structure and surface termination that contribute to separation into tubes.

For alumina it is, however, noteworthy that Chu et al.[199] reported cell boundary etching of porous alumina to form a tubular layer. Moreover, there are a few examples that produced alumina nanotubes by direct anodization[200-202]. An example is the work of Lee et al. [200,201] that used pulsed anodization process.

In contrast, *s*elf-organizing anodization of titanium leads only under very specific conditions to true porous layers, e.g. if the water content in the electrolyte is very low[203]. As illustrated in Figure 6.a and d, in many cases a transition from a porous to a tubular morphology is apparent along the self-organized oxide layer. That is, the bottom of the tubes has an ordered hexagonal nanoporous appearance, while towards the top, clearly a separated tube structure can be seen (due to different exposure times of top and bottom to the electrolyte). Not only the exact composition of the electrolyte, but also the anodization voltage affects the pore to tube transition[204]; this can be ascribed to a field effect on the fluoride ion mobility (faster or slower accumulation) and stress effects (electrostriction)[204], which affect the degree of fluoride accumulation at the metal/oxide interface.

### 2.3 Factors influencing the morphology of the anodic film

In practice, anodization of Ti in fluoride electrolytes can result in various morphologies. Formation of well-defined nanotubular structures depends on a number of factors such as the applied potential, the temperature of the electrolyte, the concentration of fluorides, etc.[38,40,44] Figure 7.a and 7.b show the regions of existence for nanotubes vs. other morphologies such as nanopores, nanosponge and compact oxide layers for a range of anodization parameters. Typically, high water contents[204] and high hydrodynamic flow in the electrolyte[181,206,207] favor a transition to a sponge structure rather than a nanotube structure. Under otherwise optimized tube formation conditions, the tube diameter can be controlled with the applied voltage.[38,40,181] The tube length can, over a considerable range, be adjusted by the anodization time until a steady-state between the growth of nanotubes at the bottom and the chemical etching at the nanotube top is established[38,116,181].

The control of tube length and diameter depends strongly on the electrolyte (see Figure 7.c-e). In aqueous electrolytes, typically tubes from ≈10 nm diameter to ≈ 100 nm diameter can be formed (by applying voltages between 1- 25 V in electrolytes with 0.1 - 0.5 wt.% $F^-$ [208,209] – as shown in Figure 7.c. In organic electrolytes, such as the most typical ethylene glycol with 0 - 13 wt.% of $H_2O$ and < 1 M of $F^-$, higher diameters can be achieved[203,211-217]; in optimized electrolytes diameters up to 800 nm have been reported[214]– see Figure 7.d.

In contrast to organic electrolytes, aqueous-fluoride-electrolytes exhibit a high chemical etching rate for the oxides. As a result, the steady-state between the growth of nanotubes and their dissolution is reached in a shorter anodization time. Thus, in aqueous electrolytes the upper limit of growth of the nanotubes is a few micrometers, while much longer nanotubes can be grown using organic electrolytes, where nanotubes of several 100 μm length have been reported. However, chemical etching thins out and finally penetrates tube walls, often some needle-like structures are present at the tube tops, in the end, these often collapse. This so-called "nanograss" (Figure 6.a) on the top can cover part of the nanotube openings[40] (some authors refer to this etched nanotube region also as nanowires or nanobelts)[218-220].

Various techniques have been explored to minimize the effect of chemical etching during anodization and hence to prevent the formation of nanograss, including approaches based on sacrificial coating layers such as thermal rutile layers, sacrificial tube layers, photoresist layers on the substrate, or alternatively supercritical drying of the nanotube layers[218,221-224].

Nevertheless, in virtually any electrolyte the nanotube wall thickness is constantly affected by chemical etching. As the etching is exposure-time dependent, the walls of the nanotubes are thinner at the upper part and the inner wall of the nanotubes has a "V" shape,[38,40,173] as illustrated in Figure 6.a and d.

Apart from the tube length, the amount of water in the electrolyte also influences other morphological features.[225] Ring-like structures, "ripples", on the wall of the nanotubes are formed in aqueous electrolytes (Figure 8.a and b), whereas organic electrolyte with almost no water produces very smooth nanotubes without ripples[38,40,117,181,226]. Whether or not sidewall corrugation (ripples as in Figure 8.b) occur, has been ascribed to a competition between tube growth rate and tube splitting speed.[225] That is, the rate of oxide tube bottoms eating into the metal substrate vs. the progress of the electrolyte between the nanotubes (chemical dissolution of the cell boundaries). Thus only for cases, where the water content is limited, i.e. in the case of organic electrolytes, the nanotubes can be grown to highly ordered arrays,[38,40,105] due to a much lower rate of cell boundary dissolution.

For $TiO_2$ nanotubes grown in many organic electrolytes, as shown in Figure 8.e, a multiwall structure is obtained[173,229,230]. For example, Figure 8.e shows a high resolution TEM image of tubes grown in an ethylene glycol (EG) based electrolyte where a carbon rich oxide layer at the inner part of the nanotubes is found. This inner layer consists of Ti-oxyhydroxide and carbon species from the EG electrolyte, these are mainly incorporated by EG decomposition, adsorption and overgrowth; the Ti oxyhydroxide species stem from precipitated Ti-ions ejected from the oxide. This inner layer is usually more prone to chemical etching by electrolyte fluorides; therefore the mentioned V-shape of the tubes is mainly defined by the dissolution of the inner layer[217].

The type of solvent used to grow nanotubes has a profound impact on the intrinsic chemical composition. In aqueous electrolytes, the inner tube layer is typically more hydroxide-rich than the outer layer, as is the case of porous anodized aluminum, in acidic aqueous electrolytes. In contrast to EG, in other organic solvents such as dimethyl sulfoxide (DMSO) or EG/DMSO mixtures, a single tube wall can be achieved (essentially only the pure outer part of the oxide is present), with significantly less carbon species incorporated,[214,229,230] as shown in Figure 8.f-h. These "single-wall" tubes provide considerably different properties from their double-wall counter part.

Another factor that significantly affects the thickness of the inner layer[119,204,231] is the electrolyte temperature – in fact, it is possible to virtually block the inner part of the tubes (and produce rod-like structures), if anodizing is carried out at sufficiently low temperatures[231].

To grow nanotubes rapidly, in addition to optimizing the typical growth conditions (fluoride content, $H_2O$ content, temperature), most recently various approaches using complexing agents such as EDTA[232] or lactic acid (LA)[233] were reported. The concept is that adding such complexing agents enables additional capturing of $Ti^{4+}$ arriving at the tubes' inner wall and thus allows to maintain a lower thickness bottom layer, i.e., a higher steady state field is present during tube growth. In fact, with additions of EDTA growth rates of 41 µm/h[232] could be reached, while with lactic acid growth rates of even up to 20 µm/min[233] were observed (as shown in Figure 7.e comparing thickness vs. anodization time for nanotubes grown with or without addition of such complexing agents).

Unusual morphologies such as in Figure 8.c – a tube in tube structure – can be obtained under specific anodizing conditions[234] or after annealing some tube layers under oxidizing conditions[173]. The tube in tube morphology occurs likely as a result of $Ti^{4+}$ precipitation to hydroxide for ions transferred across a double wall tube – in this case the carbon rich part is more prone to slow dissolution than the Ti-hydroxide precipitate layer and thus is semi-selectively etched out or decomposed (in the case of annealing), leaving behind an inner tube morphology.

Another interesting feature that may be observed is distinct patterns on nanotube bottom,[44,213] as shown in Figure 8.d. In literature, the origins of these patterns have not clearly been elucidated but the features resemble patterns that are observed in reaction-diffusion situations (Turing patterns[235]) or for spinoidal decomposition.

Recently, highly regular and organized short-aspect ratio tubes (see Figure 9.a and b), so-called $TiO_2$ nanotube stumps, were grown using concentrated $H_3PO_4$ / HF electrolytes under elevated temperatures[236-239]. In addition to a direct use, the short aspect ratio allows the use of such structures as highly defined templates for secondary material deposition (e.g., antireflection moth eye structures), as ideal UV photocatalyst, and possibly most importantly, as a patterned substrate to achieve maximum regular dewetting – see Figure 9.c. In this case, each nanotube stump is filled exactly with one metal particle (that for example can be used as co-catalyst loaded photocatalyst).

Another frequently asked question is if the metallurgical grain structure of the metallic titanium plays a role for the tube growth. Results in this respect seem at current somewhat ambiguous. While in the early initiation phase some influence of the orientation of an individual metal grain was found,[240] and some influence of the nature of the metallic substrate on tube growth was reported,[241] most frequently no significant influence of metallurgy is observed on the growth phase.

## 2.4 Advanced morphologies of anodic nanotubes

*Bamboo, Branched stacks, Multilayers*

Typically, anodic nanotubes are grown by applying a constant voltage or a constant current[39,40,242]. However, by using alternating voltages, a morphology that resembles bamboo can be grown,[227,228] as shown in Figure 10.a and b. The connecting rings between each section of the bamboo type $TiO_2$ nanotubes can resist chemical etching in fluoride containing electrolytes better than the actual tube wall. As a result, the tubular sections can selectively be etched out when extending the anodization for a sufficiently long time, and a two dimensional “nanolace” can be formed[227]. Similar bamboo structures can also be observed when optimizing the water content in the electrolyte[226] and thus optimizing a growth speed vs. splitting speed situation. Apart from the bamboo type, by voltage stepping, branching of the nanotubes can be triggered, or multiple nanotube layers with equal or two different tube diameters can be grown (Figure 10.c-e). Based on multiple anodization, stacks of nanotube layers can be produced – as shown in Figure 10.c and e. Depending on the detailed anodization conditions, the second tube layer can either be grown initiating between the original tubes layer, i.e., in the gaps, or through the bottom of the first layer[227,243,244].

Using the latter approach, amphiphilic nanotubes consisting of a bilayer of nanotube arrays can also be obtained (based on stacks of nanotubes as in Figure 10.e).[245] Here one layer acts as hydrophobic and the other one as hydrophilic entity. Such amphiphilic structures were proposed as a principle for drug delivery systems.

If tubes are grown in a branched manner by lowering the voltage, ideal branching (bifurcation) is achieved according to voltage lowering with a $1:\sqrt{n}$ ratio, where n = 2, 3, etc. I.e., branching into 2, 3, etc. nanotubes can be achieved,[246] an example for branched nanotubes is shown in Figure 10.d. Voltage stepping was also used to grow lower diameter nanotubes to considerable length.[247]

As not only pure Ti but also a range of other metals can be anodized to form nanotubes, it is possible to produce a self-ordered oxide nanotube "superlattice" structure by anodizing bi-metallic multilayer substrates (for example, Ti/Ta or Ti/Nb) under optimized conditions in a fluoride-containing electrolyte. Here the nanotube walls consist of alternating heterojunctions of two different metal oxides,[248,249] as outlined in Figure 10.f . Key to successful anodization is that the anodization recipe is adjusted to both metals in the stack. The formed self-organized one-dimensional nanotube superlattice structures can significantly modify the electrical, optical, or chemical properties of the nanotube systems.

A spectacular effect, two size-scale self-organization, was first observed when growing tubes on a complex biomedical Ti-based alloy[250,251]. In this case, under some anodization conditions, tubes of two distinct different diameters were formed highly ordered and arranged. This effect was later also reported for a range of more simple binary[252-255] and ternary[256,257] alloys. The origin of this highly unusual effect is still not entirely clear. While originally this effect was attributed to alloy composition, it seems however more related to geometry stabilization effects under certain anodization conditions[39,44]. This is most evident from very recent findings on two level size-scale stabilization on pure Ti[214,258].

*Free standing membranes*

The defined geometry (length and diameter) makes the nanotube layers also very interesting for membrane type of uses (filtration, microphotoreactors). A typical free-standing membrane bottom and a full membrane are shown in Figure 10.g and h. The strategy to produce such membranes typically consists of forming a tube layer, its separation from the substrate, followed by opening the tube bottoms. In the first paper on the fabrication of such free-standing flow through membranes,[118] nanotubes of 50-100 μm length were grown, then the underneath Ti metal was selectively dissolved in $Br_2/CH_3OH$, and subsequently the oxide bottom of the tubes was etched open by HF vapors. Later work reported several other approaches to form bottom opened $TiO_2$ nanotube membranes. The strategies to detach as-formed $TiO_2$ nanotubes from the metallic substrate can be based on mild ultrasonication in an alcohol (methanol or ethanol) solution,[259-262] exploiting the mechanical weakness of the oxide/metal interface. Nevertheless, it should be noted that tube layers can also be easily damaged by ultrasonication process[260]. Alternatively, some etchants (such as HCl, $H_2O_2$ solution) can be used to selectively separate the metal/nanotube interface[263-265]. The closed bottom of the freestanding $TiO_2$ nanotube layers can then be opened by an additional chemical etching process with HF or oxalic acid[263-265]. Another approach to produce freestanding bottom-open nanotubes is using a two-step process: First a tube layer is grown and partially crystallized using a heat treatment at 150-450 $^{o}$C then, an additional anodization step is performed to grow again an amorphous $TiO_2$ layer underneath the partially crystallized nanotubes.[38,45,265-269] The underneath amorphous nanotubes can then be removed by selective etching using chemical dissolution, or physically by dry sputter processes[38,45,270]. Other approaches to form bottom opened $TiO_2$ nanotube membranes are based on applying potential steps during anodization.[38,45,267-269,271] The first approach is the gradual reduction of the anodization potential at the end of the anodization process, e.g. from 100 V to 10 V in 0.5–1 min[272] (essentially following a strategy developed for porous alumina[84,85]). This leads to permanent thinning of the bottom oxide. The second approach is to increase the anodization potential at the end of the anodization process, for example, from 40 V to 100 V for 10 s.[273] The key effects of steps to higher voltages are believed to be local acidification and gas evolution that lead to a breakage/lift off of the membrane layer.[274]

Membranes formed by these approaches served first as flow-through photocatalytic reactors and were later explored for Grätzel type solar cells (see section 5.1). However, anodic $TiO_2$ nanotube membranes still face some challenges before they can be widely used in applications. For example, a large area lifting-off process of $TiO_2$ nanotube layers on substrate is not easy to control and the layers are prone to cracking (particularly in the case of oxide layer membranes).

In order to overcome such problems, a recent approach used a double metal layer, such as Al deposited on thin Ti foil, before anodization. In this case, the anodic oxide nanotube growth can be performed through a patterned Ti foil into an underlying Al metal layer[258]. Afterwards, selective dissolution of the Al and alumina layer leads to a very well defined both side open suspended $TiO_2$ nanotube layer[258]. Moreover, a photolithographic process to form a grid structure defines the etched area. The remaining Ti metal frame allows for a high mechanical flexibility and excellent electrical contact to the enclosed nanotube packs.[258]

On the other hand, anodic $TiO_2$ nanotube membranes are amorphous and thus have much less functional features than crystallized $TiO_2$ nanotubes; it remains a difficult task to thermally crystallize nanotube membranes. Most straightforward is first to crystallize $TiO_2$ nanotube membrane, and then detach them from the substrate[267,269]. Several applications of such $TiO_2$ membranes will be described in later sections (see section 5.1, 5.2, 5.4 ).

In this context, it is noteworthy that other anodic membrane formation approaches were reported[151] that led to ordered porous or channelar $TiO_2$ nanostructures with ~10 nm open pore diameter[151]. These layers can be formed by anodization of metallic Ti layer in hot phosphate/glycerol electrolyte and can be already partially crystalline. Such small open pore diameters are for example suitable for use in size-selective protein separation. Moreover, these membranes show photocatalytic activity that can be used for a photocatalytic opening of clogged membranes[151].

### 2.5 Maximizing ordering

As mentioned above, tube arrays (especially nanocavity types as in Figure 9.a, "stumps") can show a very high degree of order, but still show some flaws in perfect hexagonal long range order. For some applications such as photonic crystals[275-277] or high density magnetic storage media,[278] a very monodisperse pore diameter distribution, and a monodomain pore array of anodic oxide are necessary. Strategies to achieve an improved ordering for $TiO_2$ nanotube arrays follow, to a large extent, strategies that were developed for obtaining "perfectly" ordered porous alumina, i.e., involve the use of pure, large grain substrates to reduce the effects of contamination, inclusions, grain boundaries, or second phases. Moreover, there are a number of techniques that define initiation sites with ideal hexagonal spacing into the substrate surface, such as double anodization,[205,279-281] mold imprinting,[282] or ion beam dimpling[283-286]. As is the case of aluminium, key factors that were found to be crucial for the perfectness of the arrangement and the ideality of self-ordering process of $TiO_2$ are the anodization voltage for a given electrolyte and the purity of the material[205].

Most frequently used is multiple anodization – this was first demonstrated by Masuda et al. with anodic aluminum oxide[287]. The concept is based on the finding that the order in a growing self-organized oxide layer increases with time. In other words, if a first oxide layer is grown for some time and then removed from the substrate, dimples (footprints) of almost ideal order are present in the metal substrate. These ideally hexagonally imprinted surfaces then can serve as defined initiation sites for second anodization[205,279-281,288-297].

However, although this self-ordering technique leads to virtually ideal hexagonal order on the short range, on the long range still a poly-domain structure may be observed, e.g. due to the polycrystalline nature of the metal substrate. In order to overcome this limitation, Masuda[298-301] and others[302-304] introduced an approach, where the aluminium surface was pre-textured prior to anodization by nanoindentation, using an appropriate stamp (a mold consisting of tips of a hard material, e.g. Si-nitride, fabricated by e-beam lithography)[298] – examples are shown in Figure 11.a-c. In parallel, also the direct creation of FIB induced initiation sites[305] was successfully explored for Ti surfaces[306] – as presented in Figure 11.d-h. As for aluminium,[307] or silicon,[308] FIB initiation sites can guide anodization, allowing for unusual patterning of $TiO_2$ nanotubes in square, triangular, flower, and other tube arrangements[286,309].

Nevertheless, FIB defect writing is a sequential (slow) process, and in comparison, nanoindentation using predefined molds could offer a much higher throughput and low cost approach with a high resolution (see Figure 11.a-c)[310-314]. For titanium, the very high Young's modulus (116 GPa) hampers pre-patterning by imprinting, as molds of very hard materials would be required. At present, the most-used low-cost pre-patterning technique for Ti thus remains double anodization.

## 2.6 Theoretical considerations to self-organizing anodization

A key question for the reason of self-organization in porous or tubular oxide growth is the question why a non-smooth surface is stable during an anodization process, since self-scalloping/roughening is not energetically favored in terms of surface energy. Although considerable experimental data exist that identified crucial parameters for self-ordering of porous or tubular structures, no comprehensive model that directly converts into quantifiable experimental data is available yet[44]. One reason may be that the initiation and growth phase may be dictated by different effects. Figure 12.a-c gives an overview of simple considerations regarding self-ordering. From work on Al, Ti and TiAl, as well as other alloys, it is clear that once tube or pore formation is possible (experimental conditions established), the applied voltage determines linearly the length scale of self-ordering. I.e., if anodization starts at a specific point, it leads to a hemispherical oxide dome (Figure 12.a), where $r = f \times U$ (with f being the element oxide growth factor). Experimentally, a hexagonal arrangement of such domes allows a good estimate for the steady-state-growth situation and self-ordering length scale, once self-organization is established. For example, the different self-ordering length scale that is observed for different materials can be explained by the different high field growth factors f of the different oxides[252,315,316]. A rather crucial point in anodization is stress build-up in thin oxide films. As in case of other growing metal oxide films, also for anodic $TiO_2$ formation on a titanium substrate, Stoney curves (stress × thickness vs. thickness curves) as presented in Figure 12.b,[317] show during anodization and thus with growing film thickness a transition from compressive stress to tensile stress. One explanation given, in agreement with Nelson et al.[318], is that after the expected initial compressive stress the observed steady tensile stress is caused by free space generated at the metal/oxide interface by a metal to ion transition. This argument may be put forward to explain the need to form an initiation layer prior to steady-state pore or tube growth.

Another argument is space filling, usually when a sufficient number of oxide domes are initiated and grow sufficiently rapidly and compete for space (Figure 12.c)[44,243,319]. In some cases when the number of initiation sites is low, tubes may not compete for filling and may be expanding diameters to very high values, such as in the case of Albu et al.[214], where tubes of up to 800 nm diameter could be grown using a voltage of 100 V in DMSO electrolyte (I.e., this voltage is not sufficiently high to explain the large diameter D by classic $D = fU$ (with f = 2.5 nm/V)). However, in most cases, there are more initiation sites than growing tubes and thus the inverse is observed – that is extinction of small diameter in favor of larger tubes until an ordered structure of $D \approx fU$ tubes is obtained.

Regarding pore/tube initiation, most elaborate theoretical work has been performed for self-ordering of porous alumina. In order to explain self-ordering, perturbation methods, in particular, stability analyses were used, but only in a few cases also associated with specific physical phenomena[320,321] (in order to explain why a specific wavelength in perturbation analysis [corresponding to a certain tube or pore spacing] becomes stabilized [vs. a flat surface]). In the light of the contemporary experimental findings, approaches that base stable self-ordering either on the specific ion flux conditions (electromigration),[322] or on the stress generated during oxide growth,[323] seem most adequate. In stress based models a key question is if purely mechanical (volume expansion) or electrostrictive effects are dominant. While the length scale (wavelength) for pure Pilling-Bedworth volume expansion is far off to account for the observed self-ordering length (100 nm range), scales estimated for observed electrostrictive forces are very likely to be in the required range ($\lambda \sim 100$ nm)[322]. In other words, the compressive electrostrictive stresses that occur in the oxide film, and in particular at the film surface, could be minimized by an increase of the surface area, i.e. scalloping of the oxide film. However, the fact that optimized self-organization is typically observed for an optimized set of voltage combined with a specific current situation,[205] and some preliminary stress measurements,[323] support ion flux models (at least for initiation), which possibly need to be linked with approaches describing the final steady state (growth) ordering by saturation effects.

In general, one may assume that Ti anodization follows to a large extent approaches developed for Al[172,194,322-328]. Namely, Golovin's group[329,330] provided mathematical modeling of the initiation phase, i.e. the transition from compact oxide growth to pore formation. In their approach, the activation energies of the electrochemical interfacial reactions are assumed to be influenced by the Laplace pressure that is present due to surface energy and volume expansion of the oxide. Linear and weakly non-linear stability analyses were carried out for the situation when elastic stress is significant and when it is not[322]. In both cases, the instabilities in the oxide are generated by positive feedback between field-assisted dissolution and the perturbations at the two film interfaces. In the absence of stress, both interfaces of the oxide undergo long-wave instability, which can provide a wavelength selection mechanism for the initial pore geometry (diameter, inter-pore distance). For this case, weakly non-linear analysis showed that the dynamics of perturbations at oxide interfaces are governed by the Kuramoto–Sivashinsky equation[331] that can describe the formation and evolution of irregular pore cells with an average size distribution[332]. Linear stability analysis in the presence of significant elastic stress showed that long-wave perturbations are overwhelmed by two short-wave instabilities. These instabilities promote the formation of ideal ordered hexagonal pores. The authors found that the short-wave instability occurs in a limited range of the volume expansion, which is in agreement with experimental work on the formation of ideally organized pores for a narrow range of aluminum oxide expansion.[81] Later studies[323] considered electromigration of ions in the oxide layers and showed that short-wave instabilities, which lead to hexagonal self-organization of pores, can appear even in the absence of elastic stress. Only linear stability analysis was carried out for the updated model, and it was sufficient to demonstrate transition from the basic state of the oxide (uniform compact barrier layer) to a hexagonal ordered porous structure. It was concluded that ion migration through the oxide, coupled with non-linear reaction rates at the metal/oxide and oxide/solution interfaces, represents a sufficient condition for the generation of short-wave instability and subsequently formation and growth of hexagonally ordered pores.

On the other hand, remarkable modeling work has been carried out by Hebert's group by numerical computation based on the detailed pore geometry taken from the work of O'Sullivan and Wood[185] – Hebert analyzed two cases derived from the Poisson equation: In the first case, the potential distribution during steady-state growth of a pore is predicted from current continuity. A simplified model is needed for this case and it is assumed that the ion migration fluxes are much greater than the ion diffusion fluxes under the high electric field present in the barrier layer (~1 V $nm^{-1}$). In the second case, no space charge is assumed to be present in the oxide layer, and Poisson's equation becomes Laplace's equation.

The first and the second case are equivalent for planar 1D film geometry or for multidimensional geometries when the electric field is low. Anodic pores do not fulfill these conditions and consequently resulting equations differ.

Numerical simulations of the potential distribution for a real pore geometry showed that the current continuity equation is much more realistic than the case based on Laplace's equation. In the latter case, a huge difference (several orders of magnitude) between the current density at the metal/oxide and oxide/solution interfaces was obtained, i.e. violation of current continuity. On the other hand, the current continuity equation requires the presence of a space charge in the oxide. However, the equivalent of the charge is small (~0.007% of the charge density of $O^{2-}$ ions), and the first case becomes more realistic by adjusting the space charge to maintain the electric field distribution and continuity of the current. Moreover, in the model proposed by Houser and Hebert[328], both dissolution of the film and accumulation of oxide in the pore walls were taken into account, which is different from earlier models where only field-assisted dissolution was considered. The main drawback of this model is that conduction processes alone cannot explain ordering and the evolution of pores, because the metal oxidation rate at the pore cell border is enhanced by the convex shape of this region.

A key experiment by Thompson et al.[333-335] visualized, an up to that point unexpected behaviour, that mass transport in $Al_2O_3$ pore growth can be by plastic flow of the anodic oxide towards and up the wall. Aluminum films with incorporated tungsten markers showed that during the anodization the marker was partially flowing from the pore base up the pore wall.[333-335] The authors concluded that the barrier oxide exhibits plastic flow into the pore walls under the effect of stresses and field-assisted plasticity. As a result, the porous oxide obtained was thicker by a factor of 1.35 compared with that of compact-like oxide. Hebert's group took these experimental facts into consideration, and for simulation of porous film growth, not only the electric field but also the stress gradients were taken into account.[326] For the simulation of oxide flow, the oxide is assumed to act as a Newtonian-like fluid. The steady-state balance between pressure and viscous forces is determined by considering conservation of electrical charge, volume and momentum. The metal/oxide interface velocity is determined by the rate of metal consumption, i.e. directly proportional to the current density. The velocity of the oxide/solution interface is determined by the rate of oxide dissolution, i.e. ejection of $Al^{3+}$ in the electrolyte, migration, and deposition of freshly unbound $O^{2-}$ ions.

Figure 12.d shows results of the simulations that are the current density lines and potential distribution near the pore base. The current density is concentrated in the middle of the pore as it is highly dependent on the strength of the electric field that is decreasing from the reacting pore base to the inactive pore wall. Figure 12.e shows the dimensionless stress. The authors assumed that compressive stress is locally elevated by the lattice insertion rate of $O^{2-}$ at the oxide/solution interface. This leads to flow of the oxide shown by velocity vectors of the oxide flow. According to Figure 12.e, the oxide flows in the direction of the metal/oxide interface, which is continuously moving due to metal consumption, and then drifts laterally and finally pushes the pore walls to move upwards.[336] There is a very good agreement to the experimental data (Figure 12.f). The model supports that viscous flow occurs during the growth of porous anodic films. It also takes into account the specific geometry of the pores, a non-uniform distribution of electric field and stress. The authors consider that compressive stress is generated by the competition of strong anion adsorption with oxygen incorporation. Tensile stress is found to be located close to the walls and it is assumed to be the reason for void formation, especially at triple cell junctions in the hexagonal pore arrays[337]. This model of coupled electrical migration and stress-driven transport has been further developed and many other aspects are discussed in reference [194].

Common to these models is that they provide a mechanistic reasoning for the occurrence of self-organization and steady-state growth of ordered pores but remain semi-quantitative or qualitative models due to the multitude of experimental factors which influence the growth of self-ordered nanostructures. However, recently a quantitative model was proposed[172] which combines many parameters for self-organization to a simple "solubility" criterion. In this work, linear stability analysis was carried out and it was postulated that morphological instability is controlled by oxide dissolution and ion migration. The stress component in this analysis seemed to be small and it was considered not to play a significant role in perturbation of the system. Once the instability occurs, the oxide can change from its basic state of a compact uniform film to a modulated porous structure. Calculations revealed that the stability is highly sensitive to the oxide formation efficiency. Figure 12.g shows the dependence of the growth rate of perturbations on the wavenumber at different oxide growth efficiencies (ε). The dispersion curve for an oxide efficiency greater than 0.7 is in the negative range, i.e. the system is stable and only a compact anodic oxide is expected to form in this range of efficiencies. If the efficiency is below 0.66, perturbations persist and the oxide film is unstable. As a result, disordered porous structures or complete dissolution of the oxide film are expected. Only in a narrow range of efficiencies the disturbances exhibit a small-wavelength cutoff. For aluminum, this range is between 0.7 and 0.66, as shown in Figure 12.g.

From calculations, simple equations were derived for predicting the range of oxide growth efficiencies for formation of ordered structures. The limits of the oxide growth efficiencies are determined by a few parameters, namely the Pilling–Bedworth ratio and the ionic charge z.

Figure 12.h shows a plot of theoretical limits for Al (ε = 0.65–0.7) and Ti (ε= 0.5–0.58), as well as experimentally determined efficiencies for ordered porous structures, with a very good agreement between experimental and model growth efficiency measurements.

While these models provide increasingly improved understanding of self-organizing, they still do not account for the full range of effects such as point defects and incorporated electrolyte species, and further careful analysis of efficiency data is needed[44,338].

# 3. Properties of $TiO_2$ nanotubes

## 3.1 Crystal structure and compositional aspects

In nature $TiO_2$ exists mainly in three crystalline phases: anatase, rutile, and brookite. In addition, synthetic layered phases, such as $TiO_2$(B)[339] which can be produced hydrothermally,[340-343] and some high-pressure polymorphs have also been reported[344].

$TiO_2$ structures synthesized at low temperature, for instance $TiO_2$ from anodic or from sol-gel approaches, are typically amorphous. Generally, phase transformation to anatase occurs at around 300-400 °C, and from anatase to rutile at temperatures of 500-700 °C (as for example in Figure 13 and 14). The exact conversion temperatures depend upon several factors, including impurities, primary particle size, texture and strain in the structure[345-347]. Anatase and rutile are the most frequently used phases in practical applications.

Nevertheless, most recent literature seems to indicate that in $TiO_2$ nanotubes and similar geometries also unusual cubic $TiO_2$ – up to recently believed to exist only under high pressure conditions[348] – can be formed during Li cycling of $TiO_2$ nanotubes (see section 5.4 on Li battery applications and Figure 15[349]).

As bulk system (extended lattice), rutile is considered the thermodynamically stable phase[17,350]. For nanoscale materials, a large number of experimental and theoretical investigations conclude that at crystallite sizes smaller than approx. 10–30 nm, anatase becomes the most stable phase[344,351-355]. Some work showed that for anodic $TiO_2$ nanotube layers, i.e. grown and annealed on a titanium substrate, a tube diameter dependent phase stabilization in the nanotube walls is observed, where for small diameters (< 30 nm) rutile rather than anatase occurs upon annealing of amorphous tubes[347].

For hydrothermal tubes the situation is different. In this case, for the as-formed tubes several crystal structures based on sodium titanates, titanic acid, or anatase-like have been proposed in the literature.

Below, we will discuss the most important aspects in view of crystal structure for hydrothermal and anodic tubes in more detail.

*Hydrothermal tubes*

For hydrothermal nanotubes the detailed structure is still under debate, as it depends strongly on the preparation conditions. According to Morgado, Jr. et al.[103,107], the different crystal structures and compositions that have been proposed to describe the NT structure are $TiO_2$-anatase, $Na_xH_{2-x}Ti_2O_4(OH)_2$, trititanates $H_2Ti_3O_7$, $H_2Ti_3O_7{*}nH_2O$, $Na_xH_{2-x}Ti_3O_7$, tetratitanate $H_2Ti_4O_9{*}H_2O$, lepidocrocite titanate $H_xTi_{2-x/4-x/4}O_4$ and bititanate $H_2Ti_2O_5{*}H_2O$. Nevertherless, one may conclude that alkaline hydrothermal tubes, as formed, consist of $TiO_6$ octahedra arranged in corrugated layers[37]. After the alkaline treatment, as shown in Figure 5.a, the tube structure is close to sodium titanate. Typically, SAED patterns are anatase-like but some diffraction spots with belt-like spreading for a fibrous compound are found[114]. Interplanar spacing (d-spacing) of spots correspond to those of (101), (200), and (100) of anatase crystals[104]. I.e., the hydrothermal tube basically has a similar crystal structure as anatase, with the longitudinal direction of the nanotube corresponding to the a-axis [(100) direction], while the cross section is parallel to the b-plane [(010) plane] of the anatase crystal. On the other hand, also diffraction spots providing a d-spacing of 0.87 nm are observed, and correspond to the broad diffraction peak found at 2θ of around 9° in the XRD patterns of Figure 5.a.4, and also correspond to the spacing of 0.88 nm of the tube wall in Figure 5.b.4. This reflection has a considerably large deviation from the (001) of anatase structure (0.951 nm). This large interplanar distance is a typical characteristic of hydrothermal titanium oxide nanotubes.

Structure analyses have been carried out extensively. Chen et al.[109] investigated the structure of hydrothermally prepared nanotubes by using high resolution transmission electron microscopy and reported that the tube was titanate with a chemical composition of $H_2Ti_3O_7$ and proposed a structure model as shown in Figure 4e. On the other hand, Ma et al.[356,357] showed the tubes to be lepidocrocite, of the defect-containing titanate with a formula of $H_xTi_{2-x/4}\square_{x/4}O_4$ ($x$~0.7; □ represents a vacancy). Besides these structures, various compositions were reported, such as $Na_2Ti_2O_4(OH)_2$ or its protonated titanate of $H_2Ti_2O_4(OH)_2$[86] and $H_2Ti_4O_9$[110].

Upon annealing of hydrothermal tubes, the typical diffraction peak intensity 2θ at around 9° decreases with increasing temperature[114] (Figure 13.a), while the peaks that correspond to an anatase structure of $TiO_2$ become dominating, with an increasing crystallinity with increasing temperature. Thermogravimetry coupled with mass spectroscopic analysis for the as-synthesized nanotubes showed a weight loss that continued up to approximately 350 °C and detected the major species lost being $H_2O$. The specific surface area for pure nanotubes decreases with increasing annealing temperature[114] (see also Figure 13.b). Typically after formation and mild annealing up to 200 °C, tubes have BETs of ≈ 300 $m^2/g$, while at 400 °C it is ≈ 200 $m^2/g$ and at 500 °C ≈ 100 $m^2/g$.[114]

From TEM investigations it is evident that a nanotubular structure can be kept up to around 450 °C (as shown in Figure 13.a in the TEM images upon annealing at 400°C and 500°C). These facts imply that the as-synthesized nanotubes contain hydroxyl groups (–OH) and/or structural water ($H_2O$). By sufficient heat treatment (annealing) the nanotubes become stoichiometric $TiO_2$ with full anatase structure.

*Anodic tubes*

In contrast to hydrothermal tubes, anodic tubes after formation are amorphous. Typical XRD investigations on annealing of anodic $TiO_2$ nanotubes in air are shown in Figure 14.a. Conversion from amorphous to anatase starts at 280 °C. Due to heat transfer reasons, crystallization of the nanotubes in conventional furnaces starts from the Ti-substrate. With increasing the temperature, rutile starts to appear at 500 °C. This rutile originates mostly from thermal oxidation of the underlying Ti metal at this temperature. This is in line with the general observation that thermal oxide layers formed on metallic Ti by heating in presence of $O_2$ typically show this type of rutile formation[173,358,359]. At elevated temperatures often such rutile layers start growing underneath the nanotubes, when tubes are annealed while still on the metallic substrate. When annealing nanotubes that are detached from the substrate, plain anatase can be obtained up to 700 °C.[360,361] For both tubes on metal and tubes detached from the substrate, elevated temperatures of 700-800 °C typically lead to sintering and collapse of the nanotubular structures.[173,359] However, in some work using very short annealing times, e.g., flame annealing,[362,363] the tube structure can be maintained even at higher temperatures.

Nanocrystallites sizes can be evaluated from XRD patterns,[173] estimated by micro-Raman spectroscopy,[364] SEM[173] or TEM[173,229,361]. An overview of the nanocrystallite sizes as a function of heating rate is presented in Figure 14.b. It was observed that the nanocrystallite size increases from a few nm up to 200nm, and is influenced by the heating rate of the annealing treatment[173].

Upon annealing of $TiO_2$ nanotube layers, typically loss of $H_2O$, F and carbon compounds ($CO_2$) is observed, see Figure 8.h. Water is typically fully lost at 200 °C, while residual carbon (present from incorporation of organic electrolyte decomposition products) is strongly diminished at 600 °C[229].

The carbon stems mainly from the inner wall of the double wall of ‘classic’ tube layers that are grown in a typical EG–$H_2O$–$NH_4F$ electrolyte (Figure 8.e and g), and a large number of other organic electrolytes[214]. When tubes are formed under single walled growth conditions, namely in DMSO or EG/DMSO mixtures,[214,229,230] over the entire tube length a drastically reduced C-content is found (please note that the residual carbon found in EDX for single walled tubes is in the range of natural contamination)[214,229,230]. The strongly reduced carbon content is also evident from the thermal desorption measurements (TGA-MS) shown in Figure 8.h. That is, the thermal desorption profile for M=44($CO_2$), which is a typical thermal decomposition fragment observed for EG-prepared tubes, is of drastically lower magnitude for the single walled tubes[229]. This also means that for single walled tubes no carbon burn off takes place during thermal crystallization.

An appropriate annealing leads, for both tube types, to an anatase signature in XRD with a small amount of rutile. TEM for double walled tubes after annealing shows a convoluted image due to the overlap of inner and outer wall crystallites[229,230] – see Figure 14.c.

For the single-wall tubes defined crystallites are observed with a roughly ten times larger average crystallite size (≈100 nm vs. ≈10 nm) than for double walled tubes (as shown in Figure 14.e).

After annealing, the single walled tubes show considerably different physical properties from double walled tubes. A most impressive feature is that ‘single walled’ tubes after annealing exhibit a drastic improvement of their electrical conductivity by about 10–100 times[230].

*Low temperature annealing*

While freshly formed anodic nanotubes are amorphous, there are a number of reports on crystallization upon extended exposure time to water[365,366]. Generally, under hydrothermal conditions (> 100 °C, autoclave) amorphous oxides of different metals tend to crystallize[367-370]. Immersion of $TiO_2$ nanotubes in water under mild hydrothermal conditions, i.e. at temperatures of 80–90 °C, leads to some crystallization of the nanotubes. After several days of storage in deionized water at room temperature, nanotubes show an even stronger crystallization. Figure 14.f shows such morphology after 3 days storage in deionized water.

However, in water-annealed cases often XRD measurements reveal for $TiO_2$ nanotubes only a broad anatase peak[365,366] – see Figure 14.g. If the crystallite size is estimated using the Scherrer equation, it results as about 6 nm anatase. As a result, these "water-annealed" partially crystalline or nanocrystalline oxide tubes usually show a significantly inferior performance in photoelectrochemical applications, such as in solar cells, than material fully annealed to anatase (in furnaces at 400 – 500 °C).

*TEM artifacts*

Another point that should be emphasized regarding HRTEM investigations on crystallization of amorphous tubes is that amorphous $TiO_2$ is considerably prone to e-beam-induced crystallization,[371,372] as illustrated in Figure 16.a and b. These high-resolution TEM images show amorphous $TiO_2$ nanotubes, with images taken immediately and after e-beam exposure for several minutes at 200 kV.

Clearly, the formation of crystalline zones can be seen (lattice fringes appear) with increasing TEM observation time. The amount and size of these crystallites increase with time. Moreover, crystallization is accompanied by a shrinkage effect of the $TiO_2$ nanotubes as seen in Figure 16.c and d which show a tube bottom part immediately and after TEM exposure for some time. A similar shrinkage effect was also observed looking at walls[372]. This crystallization leads to considerable volume contraction and, therefore, to the deformation of the nanotube geometry. In other words, high-resolution TEM work that reports on crystallites being present in "as-formed" $TiO_2$ nanotubes (without additional proof) should be examined very carefully.

### 3.2 Electronic and optical properties

The specific electronic and ionic properties of $TiO_2$[373] strongly depend on their crystallographic form (amorphous, anatase, rutile, and brookite). As anatase shows the highest electron mobility,[346,374,375] this is in general the most desired crystal structure for many electron-conducting applications such as solar cells or photocatalytic electrodes. In view of single crystal material, rutile crystals are much easier to synthesize and produce defined surfaces, and thus they are much better characterized on an atomic level.

The optical band-gaps of anatase and rutile are reported as 3.2 and 3.0 eV, respectively. For $TiO_2$ nanotubes, amorphous and anatase samples show a similar band gap of 3.2 eV[358,376]. However, a very different magnitude and recombination kinetics is observed in photocurrent measurements,[35,377] as illustrated in Figure 17. The band gap can be obtained by measuring photocurrent spectra from $(i_{ph} \bullet h\nu)^{1/2}$ vs. $h\nu$ plots or $(\mathrm{IPCE} \bullet h\nu)^{1/2}$ vs. $h\nu$ plots of photoanodes[378]. Much less reliable are reflectivity measurements using a Kubelka-Munk approach,[379] bearing in mind that light absorption does not mean true electronic coupling of a state with a host lattice. To estimate doping concentration and flat-band potential, relatively frequently capacitance measurements in a Mott-Schottky type of approach are used[358]. Typically, doping densities $N_D$ ~$10^{20}$ $cm^{-3}$ for as-anodized $TiO_2$ layers and $N_D$ ~$2\times10^{19}$ for annealed are reported[358,380].

*Size confinement*

In nanotubes (nanotube walls), essentially a range of quantum size electronic effects could occur, such as ballistic electron transport or optical gap widening, due to quantum confinement.

Optical band gap confinement of nanosize semiconductors, for example of a 3D particle with a radius R, can be described using the classic Brus model[381]:

$$\Delta E_g = \frac{h^2}{8m_0R^2}\left(\frac{1}{m_e^*} - \frac{1}{m_h^*}\right) - \frac{1.8e^2}{4\pi\varepsilon_0\varepsilon_r R} \qquad (7)$$

where $h$ is Planck's constant, $R$ is the radius of the crystallites, $e$ is the charge of electron, $m_0$ is free electron mass, $m_e^*$ is the effective mass of the electron (for $TiO_2$ typically $m_e^*$ is between 5 to 30 $m_0$)[37], $m_h^*$ is the effective mass of the hole ( for $TiO_2$ typically $m_h^*$ is between 0.01 to 3.0 $m_0$) [37], $\varepsilon_0$ is the permittivity of vacuum, and $\varepsilon_r$ is the static dielectric constant (≈ 30-185).

Nevertheless, the wide spread of available data for the effective mass of the electron $m_e^*$ and the hole $m_h^*$ as well as the relative permittivity $\varepsilon_r$ allow only for very rough predictions of the expected size for onset of quantum confinement effects (exciton Bohr radius ≈1-10 nm).

Experimentally probably the best systematic data are the works of Anpo et al.[382], Kormann et al.[383] (for particle suspensions), and King et al.[384] for ALD layers, see Figure 18.a. These data show that experimentally a clear onset of quantum confinement for $TiO_2$ nanomaterials can only be expected, if size-scales are in the range <5 nm.

Thus it is not surprising that confinement effects strongly can be observed only for hydrothermal tubes. In fact, the band-gap energy for titanate nanotubes usually is similar to the value for a single “free”-nanotubular titania sheet, i.e., ≈3.84 eV[37]. This difference to the $E_g \approx 3.2$ or 3.0 eV for anatase or rutile is indeed generally ascribed to quantum confinement effects – in other words, as hydrothermal tubes are based on one atomic sheet, the observed 3.8 eV represent the practical limit of achievable optical band-gap confinement in a $TiO_2$ nanotube wall. The fact that hydrothermal $TiO_2$ nanotubes consisting of multilayers of rolled-up nanosheets still do not show a narrower band-gap is usually ascribed to only a weak electronic interaction between the stacked sheets. It should also be mentioned that except for quantum confinement there were also alternative explanations given for this observed band gap widening[37].

For hydrothermal tubes, namely of the type of a rolled-up nanosheet, the electronic properties may be conceived as shown in Figure 18.b-d[37]. If present as a sheet, the valence and conduction band can be described by parabolic functions. When the sheet is rolled up, the energy spectrum of the tube may be represented as shown in Figure 18.c: Here $k_I = 2n/d$, where d is the nanotube diameter and n is an integer, i.e., the separation of the sub-bands depends on the tube diameter.

The transition from a 2D geometry to a quasi 1D geometry has as a consequence the appearance of van Hove singularities. I.e., as shown in Figure 18.d, the resulting density of states shows a series of peaks. Also as a consequence of the rolling to a quasi 1D material, the band-gap of the tubes should be widened. Nevertheless, in practice, for both geometries values in the order of 3.8 eV are obtained.

Nanotubes prepared by other techniques in general all show band gaps corresponding to the bulk material, as their wall thickness typically is larger than ≈ 5nm.

In addition to confinement effects of the optical gap, phonon confinement can be observed for $TiO_2$ typically for particle size ranges < 20 nm. These phonon confinement effects are usually identified from Raman peak widening and peak shifts.[385,386,387] For example, for $TiO_2$ particles the peak position blue-shifts and the linewidth (FWHM) increases by decreasing the crystallite size (particularly for diameters less than 10 nm)[388], but also for $TiO_2$ nanotube walls or confined segments in $TiO_2/Ta_2O_5$ nanotube "superlattice" stacks[248].

For example, the $TiO_2$ /$Ta_2O_5$ superlattice nanotube structure in Figure 18.e contains $TiO_2$ size units that are confined to less than 12 nm in two dimensions. Figure 18.e shows the $E_g$ Raman mode from Raman spectra of these $TiO_2$ /$Ta_2O_5$ superlattice nanotube arrays, compared with pure $TiO_2$ nanotubes and also with large grain (≈200 nm) anatase ($TiO_2$) crystallites. The anatase peak position of the $TiO_2$ /$Ta_2O_5$ nanotubes is blue shifted compared with the reference materials and the full width at half-maximum (FWHM) of the peak also exhibits a significant broadening for the superlattice; this is in line with prediction from different theoretical models for phonon confinement in $TiO_2$[248] which are provided in the insets in Figure 18.e.

*Conductivity*

Several authors investigated the solid state electrical conductivity of anodic $TiO_2$ nanotube layers,[29-33] mainly as a function of annealing conditions or effects of dopants. The majority of the work is carried out with two-point measurements, such as in Figure 19.a, where a top contact metal (mainly Au, Pt, Al) is evaporated and the resistivity to the Ti-back contact is measured.

Only very little work is reported that measured single nanotubes in a four point geometry, as illustrated in Figure 19.b-d.

Depending on the techniques, considerably different conductivity values have been reported; for example, $10^4$ Ω cm (2 point top/bottom tube contacts as in Figure 19.a[29], $10^{-2}$–$10^{-3}$ Ωcm for 4-point,[32] or 1 Ωcm[34] (4 point on the tube as in Figure 19.b-d. This significant difference between 2 point and 4 point measurements may be attributed to the additional resistivity, if the tube is measured from a top contact to the underlying Ti substrate – i.e. namely the presence of a comparably resistive rutile layer at the bottom of the nanotubes affects these measurements.

The resistivity values from 4 point measurements are also low compared to values reported for polycrystalline bulk anatase ($10^2$ – $10^7$ Ωcm)[346,389,390], and even considering reported values from single crystalline bulk anatase (1.5 Ωcm)[391]. Another important point is that according to impedance measurements at a single tube,[32] not grain boundary transport is rate determining but bulk transport through the grains, see Figure 19.e. This is in line with the commonly reported high doping density[358,380] and the observed high defect density ($Ti^{3+}$, oxygen vacancies)[392] for anodic $TiO_2$ nanotubes. In line with this are findings by Docampo et al.[393] that analyzed the performance of solar cell devices and reported lower electron transport rates for $TiO_2$ nanotubes than for nanoparticles or nanowires.

From temperature dependent measurements[32] charge transport follows a Mott variable range hopping mechanism (i.e., is consistent with a model for a highly disordered system with high density of localized states). In this context one may also consider that the formation of oxygen vacancies is considerably easier in nanoscale material than in bulk material[394,395].

In term of carrier mobility, Figure 19.g shows a rough overview on electron mobility, data compiled from various sources ranging from single crystals to comparably loose particle agglomerates.[396] Data for $TiO_2$ nanotubes are comparably scarce. Docampo et al.[393] used a 2μm thick anodic $TiO_2$ nanotube layer in a solar cell configuration and obtained values similar or lower than nanoparticles. While this finding is in line with conclusions from above conductivity data and tetrahertz spectroscopy[397] that concludes there is a high defect and trapping density in anodic $TiO_2$ nanotubes, certainly further measurments for various annealing treatments are needed.

*Defects, $Ti^{3+}$, and oxygen vacancies*

As for bulk $TiO_2$, optical and electronic properties of $TiO_2$ nanotubes strongly depend on bulk or surface structural defects. Oxygen vacancies are quite common in $TiO_2$, and their presence and behavior can significantly affect the electric and optical properties of the materials. When $TiO_2$ is reduced, it forms $Ti^{3+}$ and an oxygen vacancy (OV)[373]. The two electrons coming from the removed oxygen are redistributed within the structure and thereby the electronic conductivity of $TiO_2$ structures is enhanced (self-doping). The resulting electronic $Ti^{3+}$ and OV states lie in the band gap of $TiO_2$ (typically 0.3 and 0.7 eV from the conduction band of anatase)[398] and are responsible for stronger changes in the electronic conductivity and optical properties. Generally, reduced samples appear dark or blue and show a light absorption above ≈ 2 eV. Except for native defects, $Ti^{3+}$ and OVs can be created by a reduction of $TiO_2$, which can be performed electrochemically,[38] by reducing gas annealing, or simple exposure to vacuum[48]. In the latter case, unsaturated Ti cations, such as, $Ti^{3+}$, $Ti^{2+}$, $Ti^{+}$ on a $TiO_2$ surface can be produced due to the splitting off of $O_2$ or $H_2O$ from terminal oxide or hydroxide groups and bridged oxide and $Ti^{3+}$states.[48]

The electronic properties of $TiO_2$ in nanotubular geometry are even more important because they determine the efficiency by which electrons can be transferred along the long path. In general, electrical conductivity of $TiO_2$ varies with temperature,[48] which is very characteristic for all $TiO_2$ nanotube layers. Using reducing annealing conditions, due to $Ti^{3+}$ formation, the conductivity strongly increases[29] (see Figure 19.a).

Titanate or sodium titanate nanotubes usually show no large contribution of oxygen vacancies or $Ti^{3+}$. Only once they are converted by annealing to anatase, particularly in the case of titanate tubes, e.g in ESR (Figure 19.f)[399] clear signatures ascribed to single electron trapped vacancies g = 2.003 and $Ti^{3+}$ g = 1.98 appear[37]. Sodium titanate tubes seem to be less prone to oxygen loss and reduction[37].

*Electrochemistry, photoelectroctrochemistry*

$TiO_2$ behaves in electrochemical I-V curves mostly as a typical n-type semiconductor with a current blocking characteristic in the anodic direction and a current passing behavior in the cathodic direction[29]. A general feature of highly-doped n-type semiconductors is that when a sufficiently high anodic bias is applied, valence band ionization and tunneling breakdown may occur[400]. As a result, valence band holes are generated, which can react with the environment, e.g. with $H_2O$ to form radical species ($H_2O \rightarrow OH^{\bullet}$). Thus, a reaction scheme similar to photocatalysis can be triggered in the absence of light on anodic anatase $TiO_2$, and $TiO_2$ nanotube surfaces,[400] including valence band holes and hydroxyl radical generation.

Recently, Lynch et al.[35] studied capacitance data and dynamic photoresponse of $TiO_2$ nanostructures in solution to investigate charge-carrier generation, transport, and recombination properties in different $TiO_2$ morphologies (anodic nanotubes, nanoparticle layers, compact layers). A typical behavior of photocurrent and capacitance vs. the applied bias for different $TiO_2$ nanostructure layers is shown in Figure 20.a. In each case, saturation of photocurrent and capacitance data close to the optical or capacitive flat-band potential $U_{fb}$ occur. In general, the capacitance follows a Mott-Schottky behavior sufficiently close to the flat-band potential. I. e., a space charge layer of the width $W$ is set up at the $TiO_2$ /electrolyte interface with:

$$w = \left|\frac{2\varepsilon\varepsilon_0}{qN_D}\left(U - U_{fb} - \frac{kT}{q}\right)\right|^{0.5} \qquad (8)$$

where $N_D$ denotes the donor concentration, $\varepsilon_0$ is the permittivity of the vacuum, $\varepsilon_s$ is the dielectric constant of the semiconductor, $q$ is the elementary charge, $k$ is the Boltzmann's constant and $T$ is the absolute temperature. $U_{fb}$ can be determined by plotting $1/C^2$ vs. $U$ (the so-called Mott–Schottky plots)[376,401,402] by measuring capacitance vs. $U$ curves and assuming $C = \varepsilon\varepsilon_0/w$. For $TiO_2$ nanotubes, a Mott-Schottky behavior is observed to the point where the space charge layer $W$ is approaching the thickness of the tube wall. For nanoparticle layers, photocurrent-saturation occurs at a lower applied potential due to the siginificantly lower doping concentration of commercial nanoparticles (~$10^{17}$ $cm^{-3}$)[403,404] than anodic nanotubes (~ $10^{19}$$cm^{-3}$)[376].

As can be seen from Figure 20.b, the incident photon to current conversion efficiency (IPCE) is much higher for the nanotube layer than for the nanoparticle layer of the same thickness,[35] although the particles have a 3 times higher surface area than the tube layers (120 and 29.8 $m^2$ $g^{-1}$, respectively). The higher IPCE can be ascribed to higher current collection efficiency for the tubes. Figure 20.c shows a plot of electron transport time constant ($\tau_c$) vs $\Phi^{-1/2}$ (where $\Phi$ is the photon flux) for four different tube lengths. While nanoparticle layers in the micrometer range yield a $\tau_c$ close to 20 ms, for a tube length of 3.6 μm a $\tau_c$ results as greater than 1 s. In other words, $\tau_c$ in nanotubes is much higher than in nanoparticles. In spite of the extremely large

electron transport time in nanotubes, the quantum efficiency is still remarkable. This, to a large extent, can be ascribed to much less surface-recombination occurring for tubes compared with particles.

This behavior becomes even more apparent if the hole transfer rate to the electrolyte is enhanced, i.e. the semiconductor intrinsic recombination can be reduced. For this purpose, a "hole scavenger" such as methanol is usually added to the electrolyte[405]. Significantly higher IPCEs for nanotube layers and nanoparticle layers are observed[35] after addition of 2 mol $dm^{-3}$ $CH_3OH$ to 0.1 mol $dm^{-3}$ aqueous electrolyte of $Na_2SO_4$. Additionally, for potentials sufficiently anodic to $U_{fb}$, the IPCE increases with decreasing $\Phi$.[35] These findings support the concept of a high density of trapping states which lead to a hopping transport in $TiO_2$ nanotubes. Lynch et al.[35] showed that apart from the tube layer thickness and the tube wall morphology, especially the wall surface roughness has influence on IPCE: compared to tubes obtained in water based electrolytes (rippled walls), tubes produced in organic electrolytes (smooth walls) exhibit a clearly higher IPCE.

Overall, a remarkable point is that in nanotubes an electron diffusion length of 24 μm is obtained, which is 30 times higher than for nanoparticle layers measured under the same conditions[35]. However, the high density of trap states present in the band gap of nanotubes makes the movement of the majority charge carriers extremely slow, i.e., in the order of seconds. Nevertheless, the high charge carrier diffusion length makes an application of such structures to electron transport devices, such as dye-sensitized solar cells, where the length of the nanotube layer defines the amount of the absorbed dye, very promising.

## 4. Modification of $TiO_2$ nanotubes

General modification strategies of the two main types of nanotubes, hydrothermal and anodic tubes, are quite different. Hydrothermal nanotubes are mostly modified by surface adsorption or ion exchange at their surfaces[406-410]. Nanotube arrays (and other type of nanotubes that consist of several nm thick $TiO_2$ walls) can be additionally modified by techniques used for nanopowders.

The main targets are usually to modify surface chemistry and physical properties (e.g., attach light harvesting or bioactive molecules), achieve electronic effects (such as doping or band-gap engineering), to induce electronic heterojunctions (secondary semiconductor particle decoration, core-shell type of wall cladding), or simply to increase the surface area. Therefore, these strategies are widely used, if nanotubes are applied in chemical sensing devices, solar cells, or photocatalytic electrodes.

Modification of anodic $TiO_2$ nanotube arrays is discussed in sections 4.1 and 4.3-4.6; hydrothermal tubes will be discussed in section 4.2-4.4.

### 4.1 Doping of anodic $TiO_2$ nanotubes

Strategies to alter optical and electric properties of anodic $TiO_2$ nanotubes resemble to a large extent approaches that are used for nanoparticles (e.g., thermal/hydrothermal treatments, ion implantation, etc.), but with the unique possibility to be able to dope the tubes with a species X using suitable Ti-X alloys as metal substrate for anodic growth. In general, active doping or band-gap engineering by introducing other elements into $TiO_2$ is widely explored to decrease the optical band-gap of $TiO_2$ (3-3.2 eV) and to enable a visible light photoresponse.

The electronic structure of $TiO_2$ can be altered by introducing intermediate state(s) in the band gap and / or by narrowing the gap itself. Figure 21 summarizes the relative positions (obtained by DOS calculations) for various 'doping' elements relative to the band edges of intrinsic $TiO_2$. First it should be mentioned that depending on the approach used in the calculation, the value for the $TiO_2$ band gap usually reported from DOS calculations is much lower (approx. 2.0 eV )[411,412] than the experimental value 3.0 or 3.2 eV[413,414]. Such a lower theoretical value is attributed to the shortcomings of plane wave calculations for $TiO_2$ using local density approximation (LDA) and generalized gradient approximation (GGA)[411]. However, there are some recent optimized approaches in closer agreement to the experimental values (3.05 eV for rutile, and 3.26 eV for anatase)[415,416]. Figure 21 provides values that were selected based on being close to experimental values, and preference is given to calculations that are based on higher number of atoms.

Preceeding any theoretical work, considerable experimental efforts have been undertaken to alter the band structure of $TiO_2$ by doping, and already early experiments indicated that doping can significantly enhance the photocatalytic properties of $TiO_2$[417,418,419]. Asahi et al.[420] reported nitrogen doping of $TiO_2$ and corresponding density of state (DOS) calculations. On this basis it was concluded that oxygen-substitutional nitrogen N(O, sub.) doping causes narrowing the band gap by introducing N2p states just above the $TiO_2$ valence band,[415,421] as shown in Figure 21. In addition, an intermediate state is formed almost in the middle of the band-gap due to interstitial nitrogen [N(int.)][422]. Similarly, oxygen-substituted carbon, sulphur and phosphorous form states near the valence band edge[415,416,423] – on the other hand, substitutional boron forms an intermediate state near the conduction band edge[424]. Moreover, Ti-substituted (Ti, subs.) nonmetals, sulphur and boron, also affect the band structure of $TiO_2$, resulting in the formation of an intermediate state in the case of sulphur, and lowering down the CBM in the case of

boron[423,425]. Various transition metals, such as vanadium, chromium, manganese, iron and nickel doped $TiO_2$ (i.e. Ti-substitutional doping) also show significant red-shift in their optical properties; DOS calculations indicated that these dopants form intermediate states in the $TiO_2$ band gap[411,426-429]. Early reports on comparably small amounts of Nb- and Ta- doping focused mainly on electrical properties,[430-433] later work showed their beneficial effect for solar cells in $TiO_2$ nanotubes[434-436]. Substituting an oxygen atom with fluorine in the lattice is considered to induce the formation of $Ti^{3+}$ species at the neighboring atoms, forming an intermediate state about 0.8eV below the CBM[437]. Similarly, W (VI) substitutional doping at Ti (IV) sites results in the formation of $WO_3$ doping in $TiO_2$ by taking an extra oxygen for each tungsten[438], which reduce the band gap by lowering the conduction band edge, as indicated in Figure 21. So far, the most studied and successful approach is nitrogen doping. Carbon doping is frequently explored but there is some well justified dispute over its effectiveness. [363,439] These C-doping attempts have to be distinguished from graphitization of the tubes[440] or conversion to oxy-carbides[441]. In the case of N-doping discussion exists on the mechanistic nature of the nitrogen in view of band gap engineering. In a typical non-metal doping process, ≈ 2% of nitrogen is present and it is problematic to assume that this low concentration is sufficient to rise the valence band level by > 0.5 eV. Therefore, the situation for most of the N-doped material may be best described as a high density of localized states.

However, it should be noted that, with respect to various methods used to achieve nitrogen doping, very different states of nitrogen are observed and the active species may be present in the bulk $TiO_2$ or on the surface. Proper ion-implantation of N and annealing shows a XPS peak at ~396 eV,[442] which is in accordance with the peak position found when sputtering $TiO_2$ in nitrogen environment,[420] and with the position obtained for titanium nitride[443]. Wet treatments in

amine based solutions lead typically to peaks at >400 eV, and are reported also to yield visible light response[444-446]. Such a peak in many cases can be interpreted as a surface sensitization (for e.g. with a N-C compound)[447]. Several groups claimed successful N-doping or N-F doping in $TiO_2$ while referring to a peak at 400 eV. However, this peak position at ~400 eV is also found for adsorbed molecular nitrogen on $TiO_2$. Most of these doping reports neither show visible photocurrent nor convincing photocatalytic activity. In these cases mainly absorption spectra were used as evidence – showing e.g. strong sub-band gap light absorption; however, a corresponding photocurrent spectrum may not show any significant response. Such effects may be obtained due to simple mixing effect of two light absorbing compounds.

Most unique to anodic $TiO_2$ nanotubes is that they can be doped by anodization of a homogeneous alloy of titanium with the dopant. Using this method, N-doped $TiO_2$ nanotubes can be obtained by anodizing N-containing Ti alloy substrates,[448,449] where the substrate is prepared e.g. by arc-melting of pure Ti and TiN powders. Similarly, W, Mo, Nb or Ta doped $TiO_2$ nanotubes can be obtained[450-452].

Additionally, doping of $TiO_2$ is reported to take place by ion pick up from the anodization electrolyte (e.g. for phosphorus anodization in a $PO_4$ electrolyte)[453], however, such attempts targeting nitrogen or N-F co-doping[454] mostly lead to XPS peaks at 400 eV (corresponding to adsorbed species, see e.g. [40]) and/or do not show convincingly electronic coupling of the doping species. For nanotubes prepared in organic electrolytes carbon-contamination can take place, leading to an enhanced visible absorption, due to the decomposition of the organic electrolyte under the applied voltage[173,400]. Additionally a number of reports show doping of tubes with Cr,[378] C,[455] and V,[456] with more or less beneficial effects to their properties.

## 4.2 Doping of hydrothermal tubes

There are three general approaches for doping of hydrothermal $TiO_2$ nanotubes: i) using predoped $TiO_2$ particles in the hydrothermal growth reaction,[457,458] ii) doping during the hydrothermal growth process due to a ”hydrothermal ion-intercalation” process,[459,460] or iii) post synthesis ion exchange[461-464].

Hydrothermally grown $TiO_2$ nanotubes can be doped with various elements including chromium, manganese, cobalt, niobium, vanadium, bismuth, boron, phosphor, gadolinium, platinum, iron, and neodymium[457,458,460-462,464-466]. Different co-doped[459,463] and tri-doped[467] $TiO_2$ nanotubes were also reported.

Typically, only small changes in morphology, surface area, and optical band gap of hydrothermal tubes are reported for doping with most metal cations. However, the doped hydrothermal tubes can exhibit a considerable increase in the electrical conductivity (for example, for Cr doping, a conductivity increase of 1–2 orders of magnitude can be obtained: $1.0\times10^{-4}$ S/cm for 0.08 mol.% Cr-doped hydrothermal tubes compared to $3.0\times10^{-6}$ S/cm for undoped)[114].

Cation doping of hydrothermal tubes also improves the thermal structural stability of the nanotube geometry, as shown in Figure 13.b. For non-doped hydrothermal tubes, structural degradation and accompanying decrease of surface area occur at around 400 °C. This critical temperature can be increased by approximately 50 °C (doping with $Mn^{3+}$, $Co^{2+}$, $Nb^{5+}$, $V^{5+}$) to 100 °C (doping with $Cr^{3+}$)[104].

Red shifts of absorption edge were observed in boron and nitrogen doped $TiO_2$ nanotubes, while the smallest band gap energy was obtained in (1% B, 1% N)-codoped nanotubes (2.98 eV),

as compared to undoped, 3% N doped (3.05 eV) and 1% B doped (3.08 eV).[459] N doping was confirmed by an XPS peak at ≈ 396.8 eV, ascribed to O-Ti-N bond and suggesting that partially O was substituted by N in the lattice of $H_2Ti_3O_7$; while, B doping (≈191.8 eV) indicated a Ti-O-B bond and that B could be localized at the interstitial position or act as a substitute for H in the lattice[459].

Phosphorus, neodymium and platinum doped $TiO_2$ nanotubes were also reported to decrease the band gap.[462,464,465]

Liu et al.[463] co-doped $TiO_2$ nanotubes with gadolinium and nitrogen, with 1.5 at.% and 2.3 at.%, respectively. Peaks for N were found at ≈ 399.9 eV and at ≈405 eV, the latter was attributed to oxidized nitrogen moieties.

For tri-doped $TiO_2$ nanotubes, Xiao et al.[467] used a post-treatment with thiourea to induce C, N, and S in the nanotubes - UV-vis absorption spectra showed an increase in the absorption edge to the visible light region. Authors[467] evaluated from XPS that: i) C was present in the form of O-C bonds (≈288.6 eV), possibly substituting some of the titanium atoms in the lattice and forming a Ti-O-C bond; ii) for N, a peak at ≈399.4 eV was ascribed to N interstitial doping, and iii) sulfur was present as $S^{4+}$ cation substituting $Ti^{4+}$ cation (≈168.5 eV).

Furthermore, it is possible to load various metals and/or compounds into the inside of the nanotubes and/or onto their surfaces. Figure 22 shows TEM images of hydrothermal tubes loaded with metals and sulfide compounds which were prepared by using various physicochemical processing methods[114].

### 4.3 Self-doping / Magneli phases / black titanium

*Self-doping*

Heating in vacuum or reduction with hydrogen at elevated temperatures are standard procedures to increase the conductivity of semiconducting oxides. In the case of $TiO_2$ self-doping occurs generally by $Ti^{3+}$ formation (see section 3.2)[468]. Namely, thermal hydrogen treatment of $TiO_2$ was found to lead to electron depletion from the surface, leading to less recombination and a higher photochemical quantum yields[469]. It was reported that photoactivity of $TiO_2$ can be enhanced,[470] and the treatment was used to improve the surface and photoelectrochemical properties of $TiO_2$[471-473]– the effect was ascribed to an extension of the hole life-time. Other reasoning for enhanced photocatalytic properties of reduced $TiO_2$ particles was that it had an altered Fermi level leading to an increased height of the Schottky barrier that repels electrons from the particle surface[469]. Electrical conductivity[474] and Fourier transform infrared spectroscopy[475] suggested OV and $Ti^{3+}$ to be present in the modified $TiO_2$. The presence of hydroxyl groups in the treated $TiO_2$ was confirmed by infrared spectroscopy by the appearance of OH absorption peaks, indicating that the hydrogen is bound to O atoms of the lattice,[470] however from those investigations it seemed not entirely clear if OH groups or $Ti^{3+}$/OV formation are the main reasons for the improved photoactivity of $TiO_2$ after a $H_2$ treatment.

The blue color that is associated with $Ti^{3+}$ and OV formation increases in intensity with the level of reduction, and is typically assigned to d-d transitions.[398] Most convincing proof for $Ti^{3+}$/OV formation is electron paramagnetic resonance (EPR). Hydrogen treated $TiO_2$ usually shows the presence of OV and $Ti^{3+}$ in the lattice as illustrated in Figure 19.f.[37,476] With increasing temperature, the signal intensity of $Ti^{3+}$ increases and reaches a maximum value at 600 °C, while the signal intensity of OV remains constant from 400-520 °C.[476] Moreover, experimental results

show that the EPR signal intensity of OV and $Ti^{3+}$ in $H_2$-treated $TiO_2$ after 10 months storage is still significantly higher than in the untreated $TiO_2$ catalyst.

Synthetic approaches to form 'self-doped' $TiO_2$ involve solvothermal treatments[477] or using imidazole to react with $O_2$ and also forming CO and NO as the reducing gas[478]. These works reported visible light absorption and improved photocatalytic activity of reduced $TiO_2$. Zuo et al.[478] reported that self-doped $TiO_2$ (mixed anatase/rutile) powder shows a strong EPR signal for $Ti^{3+}$. For water splitting experiments, reduced $TiO_2$ showed conversion in the visible light region (> 400 nm), while no $H_2$ evolution by commercial anatase $TiO_2$ could be observed under visible light[473,479]. Theoretical simulations support that the width of the band gap is related to the concentration of $Ti^{3+}$ or OV. It was further suggested that the high concentration of OV could break the selection rules for indirect transitions, resulting in an enhanced absorption for photon energy below the band gap[478].

Electrochemical reduction has also been used to reduce $TiO_2$ and fabricate self-doped $TiO_2$. Several studies investigated the effect of hydrogen loading by cathodic electrochemical treatment of various $TiO_2$ forms, such as single crystals,[468] sputtered layers,[480] thermal oxides,[481] and anodic nanotubes[482-484]. For single crystal $TiO_2$, hydrogen can be incorporated into the rutile lattice electrochemically. Depth profiling electron-stimulated desorption (ESD) shows a high density of hydrogen in a shallow surface layer[481]. With strong cathodic reduction,[481] hydrogen penetrates deeper into the $TiO_2$ electrode and an increased amount of hydroxy and/or oxyhydroxy groups were found by XPS. Moreover, hierarchical $TiO_2$ nanotubes were reduced electrochemically and were reported to show a remarkably improved and stable water splitting performance due to a higher electrical conductivity (evaluated from electrochemical impedance measurements)[482].

Theoretical calculations predict that the introduced localized OV states have energies of 0.75 to 1.18 eV below the conduction band minimum of $TiO_2$. i.e., they lie lower than the redox potential for hydrogen evolution, which, in combination with the low electron mobility in the bulk region due to localization, would make the photocatalytic activity of reduced $TiO_2$ negligible[485]. Other sources report that a high vacancy concentration can induce a vacancy band of electronic states just below the conduction band (~ 0.2 eV below the conduction band), and OV thus beneficially narrow the band gap and facilitate photocatalytic reactions.[486]

The mechanism of reduction using $H_2$ and the isothermal reduction kinetics of $TiO_2$ were investigated in detail in refs. [472,474].

Further techniques to produce self-doped $TiO_2$ include heating under vacuum[487] or reducing conditions (e.g. $H_2$,[474,488] CO[489]), chemical vapor deposition,[490] high energy particles (laser, electron, or $Ar^+$) bombardment,[491] and chemical reduction by $NaBH_4$ treatment[492]. For practical applications, the strategy enhances the performance in a number of applications, such as lithium batteries,[488] biosensors[489] and resistive switching devices[490].

*Magneli phases*

A special case of reduced $TiO_2$ are the so-called Magneli phases that are suboxide compounds of a defined stoichiometry such as $Ti_3O_5$, $Ti_4O_7$, $Ti_5O_9$, and $Ti_6O_{11}$. A key property of these phases is, as reported by Bartholomew and Frankl[493], that they possess a very high electrical conductivity. For example, $Ti_4O_7$ Magneli phases showed a conductivity 2 – 3 orders of magnitude higher than anatase. This effect is due to the crystallographic structure of $Ti_4O_7$: according to Goodenough's theory,[494] the amount of overlap between the d-electron wave functions to neighboring cations of titanium is a critical factor which determines if electrons are

localized or collective. In his view, $Ti^{3+}$ ions in shear layers move towards each other, leading d-electrons to be trapped in homopolar bonds between them and thereby forming a metallic-like phase. Generally, Magneli phases are formed by high temperature hydrogen treatment of $TiO_2$ but also acetylene treatments at more moderate temperatures were reported to convert $TiO_2$ nanotubes to such suboxides (or oxycarbides)[441]. Such Magneli-type anodic nanotubes show a semi-metallic behavior.

*Black Titania*

In 2011 Chen et al.[479] reported on the fabrication of 'black titania' that was obtained by a two-step synthesis process. First nanophase titania was formed by heating a precursor solution (consisting of titanium tetraisopropoxide, ethanol, hydrochloric acid, deionized water, and an organic template, Pluronic F127) at 40 ˚C for 24 h, followed by evaporation and drying at 110 ˚C for 24 h and final calcination at 500 ˚C for 6 h. Then the obtained highly crystalline anatase $TiO_2$ nanoparticles of approximately 8 nm diameter were exposed to $H_2$ atmosphere at 20 bar at 200 ˚C for 5 days. The resulting 'black powder' was identified as still mostly consisting of anatase, and was reported to show strong visible light absorption and exceptional photocatalytic properties. Under simulated sun illumination, 0.02 g black $TiO_2$ nanocrystals (decorated with 0.6 wt.% Pt) produced 0.2 mmol of $H_2$ per hour (i.e. 10 mmol $h^{-1}$ per g cat.) from a water/methanol mixture. This remarkable hydrogen production rate is about two orders of magnitude higher than for most other semiconductor photocatalysts[16]. Throughout testing cycles in 22 days, the high $H_2$ yield remained unchanged without catalyst regeneration, indicating an excellent stability for the black $TiO_2$. Under the same experimental conditions, no $H_2$ was produced from the unmodified

white $TiO_2$ nanocrystals loaded with Pt. Using only visible light illumination, the rate of photocatalytic $H_2$ production is however considerably lower.

From TEM images, the formation of a core-shell structure was observed where a highly disordered surface layer (approx. 1nm thick) with hydrogen dopants surrounded a crystalline (anatase) core[479]. XRD diffraction peaks indicated that the black $TiO_2$ was highly crystallized anatase. Raman spectroscopy used to examine structural changes in the $TiO_2$ nanocrystals showed that new bands emerged for the black $TiO_2$, in addition to the broadening of the anatase Raman peaks. From XPS results there was no detectable $Ti^{3+}$ found, and a broader peak of O 1s at 530.9 eV (for the $H_2$-treated samples) attributed to Ti-OH species. The onset of optical absorption of the black hydrogenated $TiO_2$ nanocrystals was found at about 1.0 eV (approx. 1200 nm), together with an abrupt change in both the reflectance and absorbance spectra at approximately 1.54 eV (806.8 nm). By valence band XPS, the density of states (DOS) of the valence band of $TiO_2$ nanocrystals was evaluated. For the black $TiO_2$ nanocrystals, the valence band maximum energy blue-shifts toward the vacuum level by approximately -0.92 eV. From FTIR reflectance spectra, the strength of the terminal O-H mode is reduced after hydrogenation of $TiO_2$[495]. By 1H NMR measurements, small and sharp resonances were observed for the black $TiO_2$, suggesting that the hydrogen concentration is low and there are dynamical exchanges between hydrogen in the different environments[495].

Unlike in the case of traditionally doped $TiO_2$, Chen et al. considered not $Ti^{3+}$/OV defects to be responsible for the long-wavelength absorption of black $TiO_2$, but assigned this effect to the formation of the disordered phase around the crystalline anatase nanoparticle core.[16,479,495] The dramatic color change was ascribed to the optical gap of the black $TiO_2$ nanocrystals that was substantially narrowed by intraband transitions. Additionally, the engineered disordered phase is

perceived to provide trapping sites for photogenerated carriers and prevent them from rapid recombination, thus promoting electron transfer and photocatalytic reactions. The authors compared DFT, without disorder, where defects yielded a gap state in $TiO_2$ nanocrystals, about 0.5 eV below the conduction band minimum. With DFT that considers lattice disorder, the presence of mid-gap electronic states leads to a band gap of ≈ 1.8 eV.

Follow up work used various reduction treatments, mainly without pressure, to achieve a visible response. For example, black $TiO_2$ nanoparticles obtained through a one-step reduction/crystallization process also exhibit a crystalline core/disordered shell morphology[496]. With valence band XPS, these $TiO_2$ nanoparticles exhibit a band gap of 1.85 eV, which well matches with visible light absorption. However, in this case, the presence of $Ti^{3+}$ was confirmed by EPR[496], i.e., visible light absorption may be attributed to the classic $Ti^{3+}$ formation.

A similar simple annealing treatment (without pressure)[497] in hydrogen was used to reduce $TiO_2$ nanowires and nanotubes. The results showed a fundamental improvement for photoelectrochemical (PEC) water splitting of rutile $TiO_2$ nanowires[497]. The hydrogen treatment was found to increase the donor density in $TiO_2$ nanowires by 3 orders of magnitude, via creating a high density of oxygen vacancies that serve as electron donors. In contrast, only a mild enhancement on PEC water splitting was found for hydrogen treated $TiO_2$ anatase nanotubes. On the other hand, it was also found that hydrogenated $TiO_2$ nanotubes show considerably enhanced capacitive properties for supercapacitors, which are attributed to the higher carrier density and an increased density of hydroxyl groups[498]. Most recent work by Liu et al.[473] show, however, a remarkable activation of $TiO_2$ nanotubes for noble metal free photocatalytic $H_2$ generation under open circuit conditions.

Hoang et al.[499] reported on a synergistic effect using a hydrogenation and nitration co-treatment of $TiO_2$ nanowire (NW) array that improves the water photooxidation performance. The two-step hydrogenation and nitration co-treated rutile $TiO_2$ wires show visible light (>420 nm) photocurrent that accounts for 41% of the total photocurrent under simulated AM 1.5 G illumination. From EPR spectroscopy, the concentration of $Ti^{3+}$ species is significantly higher than for samples treated solely with ammonia. It is believed that $Ti^{3+}$ enrichment by annealing in $H_2$ atmosphere also is the origin of higher N doping level observed for these tubes after traditional nitration[499]. At current, the treatment from Chen and Mao[16,479,495] and derivatives of it are widely explored for $TiO_2$ nanotubes and similar structures.

### 4.4 Conversion of tubes

$TiO_2$ nanotubes can comparably easily be converted to a perovskite oxide by hydrothermal treatments[500]. Perovskite materials, such as, lead titanate ($PbTiO_3$), barium titanate ($BaTiO_3$), strontium titanate ($SrTiO_3$), lead-zirconium titanate ($PbZrTiO_3$) show a variety of interesting piezoelectric or ferroelectric properties.[501-507] Particularly conversion to other photocatalytically active (semiconductive) materials such as $SrTiO_3$[500,504-507] or bismuth titanate[508,509] can extend the range of potential applications considerably, such as towards capacitors, actuators, electrochromics, gas-sensors, photocatalysts, bio-templates, and various electronic applications[501-509]. Furthermore, hydrothermal treatments can also be used to convert the ordered nanotubular layers into other geometries[510].

### 4.5 Particle decoration, heterojunctions, charge transfer catalysis

Decoration of $TiO_2$ nanotubes with nanoparticles (metals, semiconductors, polymers) is frequently used to achieve property improvements. Main effects include: i) hetero-junction formation that changes the surface band bending (metal clusters or other semiconductors), ii) suitable surface states are created for enhanced charge transfer with surroundings, iii) catalytic effects for chemical reactions (e.g. Pt for $H_2$ evolution, e.g. $RuO_2$, $IrO_2$ for $O_2$ evolution), iv) surface plasmon effects that lead to field enhancement in the vicinity of metal particles and thus allow for example for a more efficient charge harvesting.

If particle decoration is used to introduce locally on the $TiO_2$ surface variations in the band bending, a similar effect as by applying an external potential can be reached but under "open-circuit conditions'' (for example, metal particles can pin the Fermi level locally corresponding to their work function, see Figure 28 and 29.f). The geometric range of the effect depends mainly on the nature of the particle (i.e. its work function) and the doping concentration of the $TiO_2$.

For $TiO_2$ nanotubes, a range of approaches for decoration with foreign materials (metals or metal oxides) have been reported. Electrodeposition reactions into $TiO_2$ nanotubes essentially provide a very versatile tool to fill or decorate oxide nanotubes[25].

Complete filling of the empty tube space on the substrate is, however, not as straightforward as, for example, in the case of alumina,[86] because of the semiconductive nature of $TiO_2$[25], namely that for crystalline tubes under cathodic bias a forward biased Schottky function is established (i.e., almost metallic conductivity is established). Nevertheless, several filling-by-electrodeposition approaches have been reported. After a first approach of Cu electrodeposition in amorphous tubes[25] to establish a p-n junction, further attempts involved tube layers that were lifted off from the metal substrate, opened at the bottom, and the oxide tubes were filled from an

evaporated-noble-metal contact by electrodeposition[297] (in analogy to a treatment used for porous alumina)[287]. Of course, this treatment does not lead to an interdigitated structure. Cathodic metal deposition into intact crystalline tube layers in their most functional anatase or rutile form was only reported recently using more elaborate deposition techniques[26].

Complete filling of tubes with polymers is easier, as the deposition usually occurs under anodic conditions. In this case a reverse biased junction ($TiO_2$/electrolyte) is providing the insulating properties needed for easy bottom-to-top deposition. For example, electrodeposition of conductive organic polymers (polypyrrole, polyaniline, PEDOT, etc.) can even be tuned to selectively fill the intertube space or additionally the inner tube cavity dependent on the applied conditions[511-514].

Only partial decoration of $TiO_2$ nanotubes by noble metal nanoparticles (such as, Au, Ag, Pt, Pd, AuPd) is very frequently carried out in order to achieve co-catalyst effects[515-520]. Ag or Pt nanoparticles can be deposited on the tube wall by photocatalytically reducing Ag or Pt compounds on a $TiO_2$ surface by UV illumination.[515,521] Other metal nanoparticles are preferably deposited by UHV evaporation or chemical reduction techniques.[515,516,522] Ag/$TiO_2$ or Au/$TiO_2$ nanotubes show a significantly higher photocatalytic activity compared with plain nanotubes[516]. Ag decorated tubes were also found to enhance the performance of DSSCs significantly[523].

Oxide nanoparticle decoration of $TiO_2$ nanotubes by e.g., $WO_3$,[28] or tungstates,[524] $Cu_2O$,[525-527] $Fe_2O_3$,[528] $CuInS_2$,[529] ZnO,[530-532] $Bi_2O_3$,[533] ZnTe,[534] or $TiO_2$[529,535] has been obtained by slow hydrolysis of precursors electrochemically, or by CVD, PVD deposition. One of the most followed up schemes to establishing useful p-n heterojunctions ($Cu_2O$ – $TiO_2$) for solid-state solar energy devices is, however, the electrochemical deposition of $Cu_2O$[536]. Nevertheless, it should be noted that for many applied compounds, namely for II-VI type of materials or $Cu_2O$,

the long-time stability in photoelectrochemical applications must be questioned, not only due to corrosion or photocorrosion, but also due to instability of some of the co-catalysts under applied voltage. An elegant decoration approach for anodic nanotubes with noble metal particles is using low concentration Ti-X (X=Au,Pt) alloys[27,537] that can provide very uniform particle densities and defined particle diameters.

To increase the surface area in form of hierarchical structures, mainly hydrolysis of $TiCl_4$ is used that leads to layers of $TiO_2$ nanoparticles with 2-3 nm diameter that decorate the inside and outside of the tube walls.[535] In this case the beneficial effect is a surface area increase – if a similar treatment is used to deposit $WO_3$ nanoparticles, additionally junction formation between $TiO_2$ and the misaligned bands of $WO_3$ can be beneficially exploited[28].

More recent work deals with tube decoration using C60,[538] graphene,[539] Ag/AgCl or AgBr[540,541] to enhance mainly their photocatalytic activity. Decoration with nickel oxide nanoparticles has recently been shown to exhibit significant photoelectrochemical activity under visible light (possibly by charge injection from NiO states to the conduction band of $TiO_2$)[542]. A most simple but very successful approach for particle decoration is filling the $TiO_2$ nanotubes with a suspension of magnetic ($Fe_3O_4$) nanoparticles[543].

$TiO_2$ nanotubes can also be decorated by narrow band gap semiconductors, such as, CdS, CdSe, PbS quantum dots.[544-548] These quantum dots can be deposited on the nanotube wall electrochemically, by sequential chemical bath deposition methods, or by chemical treatment in presence of Cd-precursors. Such CdS/CdSe quantum dots have band gap values of 2-2.4 eV (i.e. they absorb visible light) and can inject the excited electron into the $TiO_2$ conduction band; i.e. essentially act as sensitizers.

There are also reports on sensitizing $TiO_2$ nanotubes with conducting (semiconducting) polymers[549] for solar energy conversion – however, it must be expected that such structures fail fairly quickly due to the photocatalytic degradation of the polymer.

**4.6 Self-assembled monolayers (SAMs)**

The properties of $TiO_2$ nanotubes can further be modified by decoration via defined monolayer coatings (SAMs), to tailor various properties of the surface, such as the wettability,[550-552] change the charge transfer properties, biological interactions,[553,554] to tailor morphology (e.g. when obtaining $TiO_2$ nanotubes by ALD),[555] or to trigger reactions (such as payload release)[245,543,556]. Attaching organic molecules is most straightforward by self-assembly of molecules from the gaseous or liquid phase. Typically attached molecules have a polar functional group and an organic tail. The attachment to the substrate can be based on covalent or non-covalent bonding.

$TiO_2$ as many other metal oxide surfaces are in ambient conditions at least partially terminated with hydroxyl groups.[557] This can be exploited to anchor monolayers by condensation reactions of a functional group. Various reactive groups can strongly interact with –OH terminated surfaces: carboxylic acids, esters, siloxanes and phosphonic acids can attach to the surface via condensation, chlorosilanes (and potentially also acyl chlorides) via elimination of HCl.[558] Amines can adsorb to a metal oxide surface via either formation of peptide-like bonds with the metal oxide or by interaction of the positively charged $NH_3^+$-group with the underlying substrate. Examples of the SAM adsorption process (reaction) are displayed in Figure 23a.

The initial adsorption of molecules occurs randomly with no systematic orientation of the organic chains. At low concentrations, submonolayers with a high degree of disorder and defect

density are produced. At higher concentrations a denser coverage with increased order of adsorbates and erected organic tails (e.g., hydrocarbon chains) will be obtained. According to Helmy et al., phosphonic acids and silanes (chlorosilanes, siloxanes) are especially suited to modify $TiO_2$ surfaces, with phosphonic acids adsorbing faster and forming more stable SAMs than silanes, even though a comparable final coverage is reached.[558] Silanes with reactive groups, i.e., chloro-, methoxy- and ethoxysilanes, are converted to hydroxysilanes in contact with water. While silanes show an insular growth pattern with cross-linking of neighboring molecules, phosphonic acids initially adsorb randomly, forming ordered monolayers with a higher surface concentration[558]. In this case, typically the maximum coverage is limited by the amount of available adsorption sites, i.e., the density of -OH groups on the surface determines the adsorbate density[559]. The mechanism of SAM formation is dependent on the interaction of the adsorbates with each other: in most cases adsorption data show that the molecules adsorb as monolayer without any interaction, the self-assembly follows Langmuir adsorption kinetics[560]. Nevertheless, also multilayer adsorption can be observed – in this case the coverage first approaches a constant value (monolayer coverage) and subsequently increases again. Multilayer adsorption is best described by the BET model.[561]

For $TiO_2$, most typical is the use of n-octadecylphosphonic acid under surface water split-off. Ethoxy- and methoxysilanes release ethanol or methanol upon condensation to $TiO_2$,[562] shifting the equilibrium to the covalently bound state at elevated reaction temperatures[563]. Carboxylic acids also react with surface –OH and typically represent the anchoring groups for the organic dyes in dye-sensitized $TiO_2$ solar cells[18,564-566]. For the latter applications it should be noted that even though the quality of the monolayer (packing density, attachment strength) is in the order phosphonate > silane > carboxylate, also the charge transfer reactions across the attached

functional group are important. For example, in DSSCs charge transfer from a dye molecule to the $TiO_2$ conduction band is significantly faster for $COO^-$ groups than for silanes.[40]

In order to build up a several stage functionalized surface, linker molecules that carry two terminal functional groups are commonly used. Prominent examples are amine terminated silanes. This linker SAMs find extensive application in a wide variety of both industrial and research-oriented applications, ranging from adhesion promotion of polymer films on glass,[567,568] fiberglass-epoxy composites,[569,570] and attachment of (noble) metal nanoparticles to silica substrates[571] to biomedical applications. For the latter, specifically 3-aminopropyltriethoxysilane (APTES) is used in lab-on-a-chip applications,[572,573] or as bioactive linker to promote protein adhesion to oxide surfaces relevant in implant technology[574]. For example, the enzyme horseradish peroxidase (HRP) was coupled to $TiO_2$ nanotubes via APTES and used for model drug release applications[245] as well as for the determination of the protein activity by ToF-SIMS.[575] Figure 23.b shows the attachment of proteins or other biomolecules to various typical linker SAMs: pure APTES (A) can bind the protein via free carboxylic acid groups (amino acids Asp, Glu); in combination with glutaraldehyde (AG)[562] or ascorbic acid (vitamin C, AV)[245,575] a free amino group of the protein (amino acids Arg, Asn, Gln, Lys) coordinates to the linker; carbonyldiimidazole (CDI, C)[576] also couples via free amine groups and is nearly completely replaced by the protein, i.e., the protein is adsorbed in close distance to the surface; HUPA, 11-hydroxyundecylphosphonic acid (H)[577] is a long chain linker molecule that provides a certain degree of steric freedom to the protein. The latter is adsorbed via free carboxylic acid groups. Proteins can also bind to the pristine oxide,[578,579] the interaction with the surface is strong enough to withstand surfactant rinsing,[580] i.e., the protein may even form a covalent bond with the oxide. It could be shown that the efficiency of the protein coatings immobilized on $TiO_2$ is strongly

dependent on the choice of bioactive linker SAM, with HUPA and CDI producing the most active protein coatings.[581] Gao et al.[582] modified $TiO_2$ nanotube arrays with APTES for the immobilization of an antibody to develop an ultrasensitive immunosensor system. Carboxyalkylphosphonic acid SAM modified $TiO_2$ nanotube surfaces constituted a highly sensitive fluorescence immunoassay for the detection of human cardiac troponin I as low as 0.1 $pg*ml^{-1}$ without the use of enzymatic amplification.[583]

Hydrothermal $TiO_2$ nanotubes have been used as adsorbent for organic dyes and organic vapors.[584] Modified with amines, they were found to be attractive adsorbents for $CO_2$ fixation that can be regenerated readily and energy-efficiently by temperature programmed desorption.[585]

It is noteworthy that bifunctional molecules, such as APTES with a silane and an opposing amino group, show different affinities of either functional group to amorphous, anatase, and rutile polycrystalline surfaces.[556]

Dependent on the strength of the interaction with the substrate, various drug and other payload release processes can be achieved, e.g., voltage induced,[400] by simple immersion in a solvent,[586,587] and by irradiation with UV light[245] and X-rays[588].

*Wettability*

Organic modification of nanotubes combined with their photocatalytic properties was further used to tune the wettability properties of $TiO_2$ nanotube surfaces[550,551]. Pristine nanotube layers (amorphous or crystalline) are superhydrophilic; only when treated with a suitable monolayer they become superhydrophobic, with the maximum achievable contact angle depending on the tube diameter[550,551]. The overall wettability behavior is in accord with the Cassie–Baxter model[589]. In typical photocatalytic reactions of monolayers on $TiO_2$ or $TiO_2$ nanotubes with UV

light, chain scission occurs, which makes the surface increasingly hydrophilic with the duration of illumination. Chain scission was observed to occur between the functional group of the SAM and the substrate for irradiation of siloxane or phosphonic acid SAMs on $TiO_2$, indicating strong, covalent bonding with the substrate[581,588]. Super-hydrophobic tubes are the basic material for the fabrication of amphiphilic nanotubes[245], organic solvents are needed to fill them with a liquid, e.g., an electrolyte. Of interest in this context is, however, the observation that on the microscopic level, all $TiO_2$ nanotube layers (non-modified and modified) show preferential wetting on the outer wall (the intertubular space) rather than on the inside (see Figure 23.c).[590] This observation is in line with those for dry anatase tubes: the inside of the tubes is not easily filled by aqueous electrolytes[591]. Another elegant way to adjust the wettability of nanotube layers is by applying mixed monolayers with a different degree of polarity or even actively switchable polarity. Such mixed monolayers of N-(3-triethoxysilyl)propylferrocenecarboxamide and perfluorotriethoxysilane were used to demonstrate electrical redox switching of attached ferrocene molecules and thus to induce alterations of the wettability on $TiO_2$ nanotube layers accordingly.[592]

## 5. Applications of $TiO_2$ nanotubes

### 5.1 Dye-sensitized solar cells

One of the most investigated applications of $TiO_2$ nanotubes is in Grätzel type dye-sensitized solar cells (DSSCs). This type of solar cell has a considerable history involving the observation of photoelectric effect on sensitized silver halide in the 1870's[593,594]. In the 1960's, the work of

Gerischer and Tributsch on organic dye sensitized semiconductive metal oxides showed a visible range photoresponse,[595-597] and the work of Spitler and Calvin reported that excited electrons from rose Bengal dye can be injected into the conduction band of ZnO (although only of a quantum efficiency of 4 x $10^{-3}$)[598,599]. In 1985, Grätzel et al. reported on an efficient photovoltaic system using $TiO_2$ nanoparticles and $Ru(bpy)_3^{2+}$ complex[600] that showed 80 % quantum efficiency under visible light irradiation,[601] and in 1991 Grätzel and O'Regan reported probably the most significant achievement that is a first fully functional solar cell device that they called dye-sensitized solar cell (DSSC) and operated at 11% of solar light conversion efficiency[18,602,603]. The photoelectrode is based on a 5-15 μm thick layer of compacted $TiO_2$ nanoparticles coated on a conductive glass electrode. The nanoparticles are modified with a monolayer of attached Ru−bipyridyl molecules that act as visible light absorber that inject light excited electrons from the dye's LUMO into the conduction band of the $TiO_2$ as illustrated in Figure 24. To refill the electrons of the dye, an iodine redox electrolyte is used that itself then is re-reduced at a platinized counter electrode. In these solar cells, the $TiO_2$ particle network plays only the role of an electron transport medium to the back contact[595-599,604]. Over the years, most of the efforts for enhancing conversion efficiency have targeted the optimization of suitable dyes[605-607] and optimizing metal oxide materials and structures.

Key to a high efficiency are the timescales of the individual processes, as illustrated in Figure 24. After light induces an electron excitation in the dye from the highest occupied molecular orbital (HOMO) to the lowest unoccupied molecular orbital (LUMO), the excited electrons are injected from the dye's LUMO to the conduction band of $TiO_2$ in a femto- to picosecond time scale. The oxidized dye molecules are reduced by the electrolyte redox reaction within nanoseconds. However, electron transport rates through the $TiO_2$ and the diffusion rates within

the electrolyte are comparably slow (micro- to milliseconds). This is the reason why the overall cell efficiency, is to a large extent, determined by the electron transport rate[608]. Electron transport competes with the recombination within the $TiO_2$, and with the dye and the electrolyte.

Often this is characterized by the definition of a charge collection efficiency ($\eta_{cc}$), which can be estimated from the electron transport constant ($\tau_c$) and recombination rate ($\tau_r$) constant according to:

$$\eta_{cc} = 1 - \frac{\tau_r}{\tau_c} \tag{9}$$

*Dye-sensitized solar cells with 1-D nanostructures*

Generally, the electron transport rate in $TiO_2$ nanoparticles is considered to be comparably slow due to surface states, intrinsic $TiO_2$ defects and grain boundaries which play a role as electron trapping and recombination sites[609-613]. In order to overcome the drawback of $TiO_2$ nanoparticles (mainly to provide direct and less defective electron pathways to the back contact), one-dimensional $TiO_2$ nanostructures such as nanorods, nanowires and nanotubes have been considered as substitutes in photoanodes in dye-sensitized solar cells[609-615].

Approaches involve the use of various nanotube geometries mainly produced by anodic self-organization and template assisted methods[57,70,74-78,96,616,617].

One of the earliest attempts was the use of nanotube powders that consisted of nanotubes formed by a surfactant template assisted technique.[57] The individual single-crystalline $TiO_2$ nanotube structures had a pore diameter of 5–10 nm and a length of approx. 30-500 nm. The use of such nanotube structures led to a higher short circuit current density than commercial Degussa P-25, not only due to higher dye loading but also due to a significant enhancement of the electron transport kinetics[57]. As a result, a solar cell efficiency of 4.88% with a 4 µm thick nanotube-

powder film layer was obtained. Such early results considerably stimulated further investigations of one-dimensional nanostructures in DSSCs.

More recent examples of using templating are $TiO_2$ hollow nanostructures that are formed on a cotton template. After burning off the template, the one-dimensional open morphology and high porosity provide a relatively high specific surface area (BET = 52 $m^2/g$) for dye-loading and good diffusional access of the electrolyte, resulting in a conversion efficiency as high as 7.15 %.[96]

Another typical approach to fabricate advanced DSSCs is based on the use of hydrothermal $TiO_2$ nanostructures[20,53]. Hydrothermally formed $TiO_2$ nanostructures generally have high specific surface area (with a BET over 100 $m^2/g$) that allows a high dye loading that finally leads to conversion efficiencies for DSSCs that range from 6.7 – 8.9 %.[618-623] However, the specific surface area of hydrothermally formed titanate nanostructures is drastically decreased by the required heat treatments[623]. Furthermore, the formed $TiO_2$ nanostructures are obtained as a powder, slurry or paste, and typically need to be deposited on a conductive glass substrate by doctor blading, screen-printing or electrophoretic deposition[619]. When using such deposition techniques, the one-dimensional nanostructure layers are oriented randomly, and due to this irregular arrangement the merit of one-dimensionality is to a large extent lost.

As outlined before a most directional charge transport is expected in an aligned arrangement of nanotubes perpendicular to the surface, i.e., to the back contact. Therefore, many aligned templates have been used for the fabrication of $TiO_2$ nanorods or tubes and used for DSSCs. Namely, porous alumina membranes[70,74-78,616] and ZnO nanorods/wires[617] have very frequently served for this purpose. Overall, DSSCs produced using such $TiO_2$ nanotubular structures typically show a 3-5 % conversion efficiency.[74-78,616,617] However, the fabrication process of such

a template assisted $TiO_2$ nanostructure is relatively complicated, and to reach the step to use the tubes in functional DSSCs takes a comparably long time.

*Dye-sensitized solar cells with self-organized $TiO_2$ nanotubes*

Due to the simple synthesis, anodically formed self-organized $TiO_2$ nanotube structures have been considered one of the most promising approaches to achieve vertically oriented fast electron pathways[614,615,624,625]. The first attempt of using anodic $TiO_2$ nanotubes in dye-sensitized solar cells was reported in 2005 by Macak et al.[564]. However, the aspect ratio and type of nanotubes in that report was not suitable for use in DSSCs (100 nm of diameter and 500 nm of tube length). It showed only 3.3% of incident photon to energy conversion efficiency (IPCE) in visible range and 0.036% of conversion efficiency in a fully fabricated DSSC[45]. Over time, anodic $TiO_2$ nanostructures have been improved, in particular, a smooth tube wall and high aspect ratio[116,117,211,626] led to a drastically enhanced solar cell performance[45,217,345,627]. Empirically, the conversion efficiencies of self-organized $TiO_2$ nanotubes are highly related to geometry, crystal structures, cell fabrication, etc. as shown in Figure 25.

However, there are some general important findings in $TiO_2$ nanoparticle layers and nanotubes:

i) Zhu et al.[615] investigated the electron mobility in DSSCs by measuring electron transport times and recombination rates. These authors found that the electron transport times in $TiO_2$ nanoparticle based and nanotube based DSSCs are similar, due to a similar average crystal size being present in tube walls as in nanoparticles[615]. Nevertheless, the recombination times in $TiO_2$ nanotubes were found to be 10 times slower than for $TiO_2$ nanoparticle layers - this results in a

25% higher charge collection efficiency for $TiO_2$ nanotube layers compared with $TiO_2$ nanoparticle layers.

ii) Jennings al.[624] reported the estimated electron diffusion length of $TiO_2$ nanotubes in DSSCs to be in the order of 100 µm, based on measurements of electron diffusion coefficients and lifetimes. These measurements were carried out on a 20 µm thick $TiO_2$ nanotube layer where a charge collection efficiency of close to 100% was obtained. The results were extrapolated to longer tubes using experimental data and numerical evaluation of electron transport and trapping properties in $TiO_2$ nanotube based DSSCs[624]. These findings indicated that nanotube layers considerably thicker than 20 µm could be used for optimized nanotube-based solar cells. The authors, however, observed that for higher layer thicknesses delamination of nanotube layers from the substrate occurred. It was only much more recent work that established anodization procedures to obtain considerably more robust (better adherent) nanotube layers[233].

In the following, we will discuss some key factors that strongly affect nanotube-based solar cells.

*Annealing effects*

As-formed anodic $TiO_2$ nanotubes are amorphous and need to be annealed (preferably to anatase) to show a sufficient electron conductivity for use in DSSCs. In general, by increasing annealing temperature the crystallinity of $TiO_2$ nanotubes is increased; this affects the final solar cell characteristics, mainly by an improved short circuit current and a higher open circuit potential which result in a higher overall conversion efficiency[345]. Figure 25.a shows a compilation of various literature data of cell efficiency, for solar cells fabricated with different $TiO_2$ nanotubes annealed at different temperatures. The increase in conversion efficiency with

higher annealing temperatures is generally explained by the formation of anatase with an increasing crystallinity that finally leads to improved electron diffusion coefficients and lifetimes[628,629].

On the other hand, as mentioned earlier, when $TiO_2$ nanotube layers are annealed on their metallic substrate due to direct thermal oxidation of the underlying metal, thermal rutile layers are formed at the metal/tube interface.[372] Generally, the higher the annealing temperature, the thicker are these thermal rutile layers.[362,372] As these rutile layers have lower electron mobility than anatase layers, the presence of thick layers considerably decreases the solar cell efficiency. Finally, at even higher temperatures (over 600 $^oC$) the tube structures sinter and collapse. An optimal annealing temperature range of anodic $TiO_2$ nanotubes on Ti metal substrates for DSSCs is generally found at 400-550 $^oC$[173,345]. However, Huang et al.[361] reported that $TiO_2$ nanotube membranes produced as described in section 2.4 (and thus Ti metal substrate free) could be annealed at temperatures higher than 700 $^oC$ without rutile phase formation (Figure 25.a). Such $TiO_2$ anatase nanotube layers were reported to show a 4 times faster electron transport than nanotubes annealed at 400 $^oC$. The fast electron transport for high temperature annealed tubes mainly contributes to an enhanced cell efficiency (50 % higher efficiency), even though the amount of absorbed dye was found to be 30 % lower than for tubes annealed at 400 $^oC$.

*Geometry effects (tube length, diameter, wall thickness and corrugation)*

As expected, the geometry of $TiO_2$ nanotube layers considerably affects the resulting solar cell efficiency. Diameter and length of tube layers influence the surface area and thus the specific dye loading but also influence light reflection, internal light management, and electrolyte diffusion effects – this makes a direct prediction of the resulting efficiency not always straightforward.

In the following we discuss most influential geometry factors using wherever possible data where only one parameter at the time was investigated.

Regarding tube length, in principle, by increasing thickness of anodic $TiO_2$ nanotube layers the specific surface area increases and the conversion efficiency should accordingly be improved until electron diffusion limits are evaluated. Nevertheless, thick oxide layers, namely >20 μm grown in the common EG electrolyte, often show only a weak adherence to Ti metal substrate and thus frequently a drop of efficiency is reported in literature as shown in Figure 25.b and c. From this data, an optimum length of $TiO_2$ nanotubes for DSSCs has, in early works, been considered as 15 – 20 μm[45,233,280,345,630]. However, recently So et al.[233] reported ultrafast anodic growth of $TiO_2$ nanotubes in a lactic acid additive containing electrolyte that can be grown to lengths >100 μm. These nanotubes show a considerably higher mechanical stability even for thick layers[233,631]. For these tubes, the optimal nanotube layer thickness for a maximum solar cell efficiency is ~40 μm. The solar cell efficiency is 20 % higher than for 15 μm thick nanotubes due to 2.6 times higher dye loading.

From Figure 25.d it is clear that also the diameter of $TiO_2$ nanotubes is an important factor influencing the final solar cell efficiency[345,632-635]. In direct comparison of different diameter nanotubes in DSSCs applications, small diameter nanotubes show generally a higher cell efficiency due to a higher specific surface area and accordingly a higher dye loading[345,632]. However, although small diameter aligned nanotube structures show clearly beneficial effects in DSSC electrodes, the growth of very small diameter nanotubes (such as 15 nm) with length > 2 μm is difficult[632]. It is, however, interesting to note that anodically grown one-dimensional $TiO_2$ nanoporous structures show much lower dye loading (62 %) and cell efficiency (70 %) than nanotube structures, even if the diameter and length of tube/pore layers are similar[632].

Recognizing the importance of surface area led to various tube geometry modifications. Additional gain has been reported for bamboo type nanotubes (as shown in Figure 10.a and b)[227, 228]. Such modulated $TiO_2$ bamboo nanotube structures can show higher cell efficiency due to a higher surface area and resulting higher dye loading (50 %)[228]. Additionally, Yip et al. revealed that such bamboo type $TiO_2$ nanotubes on conventional nanotube layers can be used as photonic crystal layers[636]. Such optical properties can also be used to enhance the overall solar cell efficiency of nanotube based DSSC[636].

The overall conversion efficiency of $TiO_2$ nanotube based solar cells is typically not fully matching the performance of classical nanoparticle based cells. A main reason is that $TiO_2$ nanotubes on Ti metal substrate show a considerably lower specific surface area than nanoparticles ($BET_{NT}$ = 20 – 30 $m^2$/g, $BET_{NP}$ = 50 – 150 $m^2$/g). The most straightforward approach to improve the specific surface area is surface modification with small $TiO_2$ nanoparticle layers.[535,637-642] Typically, a significant enhancement of surface area of the nanotubes can be achieved by nanotube wall decoration with a so-called $TiCl_4$ hydrolysis treatment.[280] By this treatment nanotube structures can be uniformly coated with 20 – 30 nm of $TiO_2$ nanoparticles with ~ 3 nm individual particle size (Figure 26). In comparison with bare $TiO_2$ nanotube electrodes, the overall efficiencies are enhanced by 20 – 30 %.[280,643,644] This reflects that the specific surface area is indeed the most important parameter for cell efficiency enhancement. So far the highest reported solar cell efficiency using classic anodic tubes under back-side illumination approaches with $TiCl_4$ treated nanotube structures is 7.6 %[643].

Another approach to enhance the efficiency of tubes is the use of single wall – instead of double wall – tube morphologies. As mentioned earlier (Figure 14.c-e), annealing of single walled tubes leads to comparably larger crystallite size with considerably higher electrical

conductivity compared with conventional tubes[259]. If such tubes are used together with an appropriate $TiCl_4$ treatment, efficiencies up to 8.14 % can be achieved.[217]

Another simple approach to improve geometry factors is growing nanotubes on already structured metal substrates such as Ti metal wire, mesh and bifacial ($TiO_2$ nanotubes grown on both metal surfaces) substrate[645-652]. Such structures may provide higher specific surface area than $TiO_2$ nanotubes on flat metal substrate. However, cell fabrication process is complicated for these 3D structures and reliable cell fabrication is difficult.

Frequently, it is also found that the solar cell performance is strongly affected by the morphology of the tube tops – namely an open tube top seems to be of a significant advantage.[222,614,643,653]

*Front-side DSSCs*

Anodic $TiO_2$ layers that are formed on a metal substrate have the drawback for optimized DSSCs fabrication that they can only be used directly (on the metal) in a back-side illumination configuration (Figure 24.b). This back-side illuminated cell configuration leads to loss of photons by light absorption in the electrolyte and by reflection at the Pt coated counter electrode.[654,655] The cell efficiency difference between front- and back-side illumination configurations with $TiO_2$ nanotube layers is estimated at 20 – 50 %.[656,657]

To fabricate front-side illuminated DSSCs with $TiO_2$ nanotubes, the most straightforward way is the growth of $TiO_2$ nanotubes on a transparent conductive oxide (TCO) glass such as FTO or ITO[658]. For this, thin Ti metal layer needs to be deposited first by sputtering or evaporation on the TCO.[338,656,658,659] The key parameter to achieve suitable $TiO_2$ nanotube layers on such substrates is the adherence between Ti metal and the TCO. An optimum thickness of $TiO_2$

nanotubes in front side illuminated DSSCs is considered to be 15 - 20 μm. In other words, to grow optimized $TiO_2$ nanotube layers on TCO glass, Ti metal layers need to be deposited to a thickness of approximately 5 – 8 μm (due to the volume expansion of metal to oxide of a factor of 2 – 3 during anodization)[338]. Using such a layer, the highest reported cell efficiency in a front-side illuminated configuration is 6.9 % (including a $TiCl_4$ treatment)[659]. Nevertheless, this efficiency is still considerably far from conventional nanoparticle based DSSCs, and lower than the reported values for back-side illuminated $TiO_2$ nanotube based DSSCs. This may indicate that the quality of the deposited Ti metal on TCO glass substrate governs the critical properties of $TiO_2$ nanotube layers in view of solar cell efficiency.

Due to these difficulties, indirect approaches have been considered to build reliable front-side configuration DSSCs. Most investigated are approaches that detach the $TiO_2$ nanotube layer from the Ti metal substrate and transfer the free standing nanotube layers on a TCO glass substrate. Means to detach $TiO_2$ nanotubes from Ti metal have been described in section 2.4. To fabricate DSSCs with such freestanding $TiO_2$ nanotube membranes, the membrane structures needs to be attached on the TCO substrate. For gluing the nanotube layer on the TCO substrate, $TiO_2$ nanoparticle paste or Ti alkoxide are usually used. A general finding from these investigations[38,262,660] is that a strong binding to the nanoparticle glue is needed, light reflection at the nanotube/nanoparticle interface needs to be minimized, and that a maximized redox electrolyte diffusion into nanotubes/nanoparticle structure should be achieved. To optionally satisfy these conditions, the bottom of the $TiO_2$ nanotubes should be opened[660]. Under such optimum conditions, the best conversion efficiency of front side illuminated $TiO_2$ nanotube based DSSCs is 8.0 % without additional particle decoration[361], 9.1% after a $TiCl_4$ treatment,[660] and 9.8% after using bottom opened membranes[627]. Nevertheless, it should be mentioned that

$TiO_2$ nanoparticle layer under nanotubes in NT/NP/FTO solar cell configuration also highly contributes to the overall conversion efficiency, and only few studies clearly separate the effects of tube layers from nanoparticle layers[266].

*Doped $TiO_2$ nanotubes for DSSCs*

Another direction to improve conversion efficiency is to increase the electric conductivity of $TiO_2$ nanotubes by introducing low concentration (less than 1 at.%) of doping elements such as niobium,[661] tantalum[630] and ruthenium[662]. By simple anodization of Ti alloys (Ti-Nb, Ti-Ta, and Ti-Ru), metal-doped $TiO_2$ nanotube structures can be obtained. Such low concentration of metal dopant in $TiO_2$ nanotubes mainly help to reduce the recombination rate by a faster electron transport[661]. If the concentration of the dopant in the structures is too high, usually the beneficial effects are lost. Under optimum doping condition, the cell efficiencies were reported to be enhanced by 15 - 35% compared with non-doped nanotubes.[630,661,662]

## 5.2 Photocatalysis

Ever since the groundbreaking work of Fujishima and Honda in 1972[19], $TiO_2$ is regarded as the key photocatalytic material. Here the semiconductive nature of $TiO_2$ is used to absorb UV light and thus create charge carriers (electrons and holes) that then individually react with their environment. Most photocatalytic investigations focus on: i) the conversion of sunlight directly into an energy carrier (namely $H_2$), ii) the degradation or conversion of unwanted environmental pollutants, and iii) to some extent, on photocatalytic organic synthesis reactions.

Among many candidates for photocatalysts, $TiO_2$ is almost the only material suitable for industrial use.[417] This is because $TiO_2$ combines a very high stability against photocorrosion with comparably low cost. Not only the electronic properties of a material, but also its structure and morphology can have a considerable influence on its photocatalytic performance. Therefore, in recent years, particularly 1D (or pseudo 1D) structures such as nanowires and nanotubes have received great attention, for a use namely as a photoelectrode.

In a general scheme of a photocatalytic reaction (as shown in Figure 27), $TiO_2$ absorbs light of a wavelength > $E_g$ and electron/hole pairs are generated in the conduction and valence band, respectively. These excited charge carriers can then have different fates:

i. They can separate, travel on their respective bands, and finally transfer to the surrounding and react with the red-ox species. This would be the desired photocatalytic pathway.
ii. They may recombine by a direct band-to-band transition or via trap (localized) states in the gap, either in the bulk or at the surface.
iii. If the holes reach the surface they may form an oxidized state of the semiconductor which can be detrimental (for many semiconductors such as CdS, Si, etc. hole accumulation can lead to full oxidation, e.g. $Si^0 \rightarrow Si^{4+}$, and this may finally lead to semiconductor dissolution [photocorrosion]). This problem hardly occurs for $TiO_2$ due to a favorable electronic structure.

The thermodynamic feasibility of a photocatalytic reaction is given by the positions of the valence and conduction band relative to the red-ox levels in the environment (as illustrated in Figure 27). From an application viewpoint the most important reactions are the transfer of valence band electrons to $H_2O$, $H^+$ or $O_2$ and the transfer of holes to $H_2O$, $OH^-$ or organic species.

If we consider Figure 27 and an aqueous environment, then the transfer of conduction band electrons may lead to the production of $H_2$. However, if $O_2$ is present in the electrolyte the conduction band electrons may "prefer" to react with $O_2$ (compare red-ox potentials in Figure 28 to form superoxide, hydrogen peroxide, or water). For the valence band holes, except for a reaction with $OH^-$ or $H_2O$ to form $O_2$, also OH• formation may occur and is often the desired reaction, namely for pollution degradation. In this case, formed OH• radicals are able to virtually decompose all organics to $CO_2$+$H_2O$. Nevertheless, if the $H_2O$ concentration is comparably small, valence band holes may also be transferred directly to the organics and lead to their decomposition. A maximum efficiency for the photocatalytic reaction (looking at it from the semiconductor side) is when all charge carriers react with the species from the surroundings rather than recombine.

However, hole and electron transfer may thermodynamically be possible but in many cases are sluggish and thus a slow photocatalytic kinetics is obtained. Therefore, frequently co-catalysts such as Pt, Pd for electron transfer or $RuO_2$, $IrO_2$ for hole transfer are used.

The thermodynamic feasibility of reactions is slightly different for anatase and rutile $TiO_2$. In the classic potential-pH diagram composed by Fujishima et al.[559] (Figure 31.a), the conduction band of anatase lies at more negative redox potentials than for rutile, and the valence band edges for both phases are at similar energies (redox potentials). However, a recent reevaluation of band alignment shows an inverse conduction band offset (Figure 31.b) and an according shift in the valence band positions.[670] Here both data are presented as still some ambiguity about exact band edge positions exist. These clearly need to be resolved to obtain a consistent picture on the thermodynamics of photocatalytic reactions on anatase and rutile $TiO_2$.

Most photocatalytic applications are carried out either with $TiO_2$ nanoparticle suspensions – that is under open circuit conditions (electron and hole transfer occur from the same particle) – or in photoelectrochemical two electrode configuration where $TiO_2$ is generally used as a photoanode together with an inert or catalytic cathode such as Pt, C, etc. In this latter case, classically compacted nanoparticle electrodes have been used. However, over the past years nanotube geometries and particularly anodic $TiO_2$ nanotube layers gained a lot of interest due to various potential advantages:

- for anodic self-organized tubes a key advantage is the fact that they are fabricated from the metal, i.e., no immobilization process is needed and the tubes are directly used as back contacted photoelectrodes.
- directionality for charge separation, i.e., as described in Figure 2 orthogonal separation of charge transport can be exploited.
- easy control of the photocatalytic size (diameter, length) is provided.
- controlled doping via substrate can be achieved.
- geometry for a defined chemical or electronic gradient or junction fabrication is provided.

*Some key factors for the photocatalytic efficiency of $TiO_2$ nanotubes*

The most important factors that influence the photocatalysis of $TiO_2$ nanotubes are the crystallinity, length, diameter of the tubes together with compositional effects. In early reports, less defined tubes were used just to show photocatalytic activity, but it could nevertheless be demonstrated that the nanotube layers can have a higher efficiency than comparable compacted nanoparticle layers[671]. In general, it is found that also for a photocatalytic use of nanotube layers, for low reactant concentrations a Langmuir-Hinshelwood kinetics holds,[48] and that sufficient

solution agitation (in most cases) prevents that reactant diffusion effects (onto the tube layers) play a significant role. As for particles, and as expected from a point of zero charge of $TiO_2$ of approximately 6-7, for acidic pH typically a better adsorption of e.g. $COO^-$ - containing molecules (for example dyes) is observed, and typically at least slightly increased photocatalytic kinetics is observed.[672] In the following some comparably well studied parameters are discussed for $TiO_2$ nanotubes.

*Annealing*

As formed (amorphous) $TiO_2$ nanotubes show a significantly lower photocatalytic activity than tubes annealed to anatase or rutile[29,358,359,626,673]. Figure 29.a shows comparison of the photocatalytic activity of $TiO_2$ nanotubes annealed at different temperatures and environments.[674] The photocatalytic activity increases with increasing the temperature (above 300 °C), first due to anatase formation at 300 °C and secondary due to crystallinity[345,358]. Above ≈ 500 °C, rutile phase starts forming with a highest photocatalytic activity for tubes annealed at ≈650 °C (Figure 29.a), i.e. when a mixed anatase/rutile structure is present. These results are, in this general form, manifold confirmed[671,675-677] – but it should be considered what light source is used for excitation - e.g. a broad spectral UV/vis lamp (such as a solar simulator) or pure UV (e.g. a laser). This is of special importance because a solar simulator spectrum possesses a strong intensity in the range of 3.0 to 3.2 eV. In other words, the small difference in band gap between rutile and anatase considerably influences the results – this is not the case if a deep UV source is used. In this case, explanations in terms of an anatase/rutile junction due to band offsets are more plausible than simple light absorption arguments[48,678]. As mentioned before, if the annealing temperature is higher than 650°C, the tubes start to collapse and the lower photocatalytic activity

is rather due to the destruction of the tubes than to a high rutile content[173]. If the annealing process takes place under slightly reducing atmospheres (Ar), a somewhat increased activity can be observed as a result of $Ti^{3+}$ formation. The effect of $Ti^{3+}$ formation has been attributed to a higher conductivity (better charge separation) or the formation of surface states that facilitate charge transfer[674].

Recently, so called ''water annealing'' was reported to convert amorphous $TiO_2$ nanotubes to crystalline material[366] and it is similar to some other low temperature approaches[674]. In these approaches conversion to anatase is only partial as shown in Figure 14.g, and the efficiency in photocatalytic or solar cell applications remains far below thermal annealing[674].

*$TiO_2$ tube length, diameter and type*

The photocatalytic activity of $TiO_2$ nanotubes, as of other $TiO_2$ morphologies, is commonly investigated by dye decomposition measurements, using dyes such as methylene blue or acid orange 7 (AO7). First experiments were carried out for tubes grown in aqueous electrolytes[671,679-681] and it was observed that they may be more efficient than comparable Degussa P25 layers[671]. An overview of more recent investigations on the photocatalytic activity for different lengths and types of $TiO_2$ nanotubes has recently been published[48]. Some typical results for two types of nanotubes (water based rough tubes and ethylene glycol based smooth tubes) are shown in Figure 29.b and c. In both cases, a strong increase in the degradation kinetics of AO7 can be observed with increasing tube length[48,671] up to a certain limit (this is expected as the open circuit decomposition to a large extent depends on the amount of absorbed light).

In the case of ethylene glycol based tubes, higher degradation kinetics of AO7 is observed for higher length of nanotubes until ≈16 μm (as shown in Figure 29.c). This is in line with several

other investigations by various researchers[505,681-684]. However, there are a number of investigations that either report a maximum in the photocatalytic activity for tube layer thicknesses around 3–7 μm,[151,681,684] or the absence of an influence of the tube length[685]. Some discrepancies exist also for the influence of tube diameter. Several reports find no significant influence,[48,683] but other works report a maximum efficiency at around ≈100 nm[682,684], or other trends[686]. These discrepancies can be attributed to the fact that it is very difficult to vary tube length independently from tube diameter (e.g. compare refs. [48,681]). Other very relevant morphological features of tubes seem to be their side wall morphology[35] or tube top features[279,293,687,688]. For example, when comparing Figures 29.b and c it can be seen that tubes grown in aqueous electrolytes are more active than tubes grown in ethylene glycol electrolytes - this has been ascribed to ripple formation on the tube walls for water grown tubes which may affect charge carrier trapping[35]. Another factor that is crucial is the top-geometry of the tubes[279,293,687,688], not only because it can affect the electronic properties of the tubes but can also strongly influence the reflectivity of a nanotube layer[48]. Besides self-organized nanotubes, there are also reports about other forms of self-organized structures such as self-organized mesoporous $TiO_2$[150]. These structures were termed ‘‘titania mesosponge’’ (TMS) or ‘‘nanochannelar’’ structures. These TMS layers, when formed, can contain significant crystallinity (anatase and anatase/rutile) and when annealed can show enhanced photocatalytic activity compared with P25 layers or TiNT layers, depending on layer thickness and annealing conditions[149].

*Applied voltage*

Using an applied voltage to carry out photocatalytic reactions on a $TiO_2$ based photoelectrode dates back to early 1990’s when Kamat et al.[689,690] reported electrochemically assisted

photocatalytic degradation of organic pollutant. By applying anodic potentials to the $TiO_2$ electrode, charge separation in the increased field of the Schottky barrier is accelerated and holes are driven more efficiently to the surface, enhancing the photocatalytic reactivity (Figure 29.e) [305,400,691]. A similar behavior was also shown for nanotube electrodes[400,692] and has been confirmed several times[672,674,693]. Additionally, at higher anodic voltages, Schottky barrier breakdown can occur and that leads to valence band ionization[694] and hole generation even in the absence of light. Such a "dark photocatalysis" approach may be particularly useful in environments where the use of UV light is hampered, for example in MEMS devices or lab on a chip that require a "photocatalytic" reaction or a self-cleaning step in the dark[400].

*Doping*

Several types of doped $TiO_2$ nanotubes have been explored for photocatalytic reactions[48,695]. A compilation of the photocatalytic activity in view of organic degradation for various mixed oxide tube layers is shown in Figure 29.d. In contrast to Al doping (one of the most efficient additives inducing carrier recombination),[452,695] $WO_3$ and $MoO_3$ mixed oxide tubes show a strongly enhanced photocatalytic activity compared with non-doped tubes for the degradation of AO7 dye. The highly beneficial effect for W and Mo cannot be explained by a better charge transport in the tubes but must be ascribed to modification of the band or surface state distribution of the doped nanotubes[696-698].

Graphene-$TiO_2$ nanotubes[539] showed higher photocatalytical activity than normal unmodified $TiO_2$ nanotubes.

Hydrothermal nanotubes can be easily doped (see section 4.2) during hydrothermal treatment in view of enhancing photocatalytic activity. Fe doped $TiO_2$ nanotubes showed increased activity

on the photodegradation of methyl orange[458]. Pt and N doped nanotubes[409] showed a higher activity than nanoparticles - as indicated in Figure 30. Gadolinium and nitrogen codoped $TiO_2$ nanotubes[463] have been shown to possess higher catalytic activity in the Rhodamine B degradation reaction (the presence of $Gd^{3+}$ leads to higher cystallinity,[699] can sensitize the surface of the nanotube,[463,700] and enhances the photocatalytic activity of $TiO_2$ in the visible light region). An increase in the photocatalytic activity was also observed for C,N,S-tridoped $TiO_2$ nanotubes[467].

Silica coated nanotubes[407] annealed at 650 $^0$C showed higher photoactivity than nanoparticles. Titania nanotubes modified with 4 wt.% $WO_3$[28,408] and annealed at 380 $^0$C also enhanced photocatalytic activity, compared with non-doped materials.

*Degradation of pollutants and undesired biological entities*

Several toxic organic compounds, such as organochlorine compounds,[697,701-704] aromatic pesticides[705], PCB, dioxins[706,707], DDT[697], azo dyes[528,708-717] and others[718] can be degraded relatively fast by photocatalysis, leaving small traces of intermediates. Several of these processes were also explored using $TiO_2$ nanotubes[693,704,719,720]. Similarly, $TiO_2$ nanotube layers were used for the destruction of gaseous pollutants as irritants,[684,709] or the photoreduction of $Cr^{6+}$[721]. A considerable advantage of tubes over powders is their easy applicability in static flow through reactors, possibly including the application of an aiding voltage. Nevertheless, this approach has hardly been explored up to now.

Early investigations on hydrothermal nanotubes by Liu et al.[722] showed that this type of $TiO_2$ nanotubes have a better photocatalytic activity than nanoparticles for the degradation of methylcyclohexane. On the other hand, Thennarasu et al.[723] showed that nanostructures obtained by hydrothermal treatment (titanate nanotubes and nanoribbons) have no significant

photocatalytic efficiency for rhodamine B degradation prior to calcination. Nevertheless, already calcination at 150 $^{0}C$ (24h or 72h) showed promising results. Hydrothermal $TiO_2$ nanotubes annealed at 400 $^{0}C$[724] were also used for degradation of brilliant red X-3B from aqueous solution. The influence of sodium on the photocatalytic properties has also been analyzed[725]. For example, at low annealing temperatures (< 500 $^{0}C$) the Na-TNT nanotubes did not show a significant activity in the degradation of formic acid, but with removal of Na, nanotubes annealed at 400$^{0}C$ presented an enhanced photocatalytic activity in the degradation of formic acid.

As most organic compounds are degraded on photoexcited $TiO_2$, similarly bacteria and cancer cells can be destroyed to some extent[726-729]. $TiO_2$ photocatalysis is considered to be effective in sterilization effects using bacteria such as *Escherichia coli* (*E.coli*), Methicillin-resistant *Staphylococcus aureus* (MRSA 11D 1677) and *Pseudomonas aaruguuinosa* (IFO 13736). For these bacteria $TiO_2$ substrates were reported to have a strong anti-bacterial effect,[730,731] even under very weak UV light[729]. This type of photocatalytic effect was also investigated in view of cancer cells[726,732].

*Water splitting*

The use of $TiO_2$ photoelectrodes (or $TiO_2$ suspensions) to produce hydrogen from water has been highly investigated over the past decades, as – in principle – using $TiO_2$ the photogenerated $e^-$ and $h^+$ can react with $H_2O$ to form $H_2$ and $O_2$ – i.e., direct splitting of water can basically be achieved. As shown in Figure 31, for $TiO_2$, at the conduction band the red-ox potentials for $O_2 \rightarrow O_2^-$ and $H^+ \rightarrow ½ H_2$ are very close, meaning that $H_2$ generation and $O_2^-$ formation are typically competing. At the valence band, $O_2$ can be formed from water via various pathways

including radicals that can react to $O_2$. The reaction rates of the photocatalytic processes on pure $TiO_2$ in water are typically limited by the kinetics of the charge transfer process to a suitable red-ox species. Therefore, at the conduction band often catalysts, such as Pt are used to promote $H_2$ evolution, and at the valence band, $O_2$ evolution catalysts as $IrO_2$ or $RuO_2$[733-735] and/or hole capture agents such as $CH_3OH$, are used to promote the overall reaction rate.

For water splitting, $TiO_2$ is most efficient as a photoanode in a photoelectrochemical arrangement[559], i.e. using $TiO_2$ as a photoanode, under an applied voltage (coupled with a suitable cathode). This is because the slow cathodic $H_2$ evolution reaction can be performed on a separate ideal electrode (such as Pt) and the reaction can be "aided" by an applied voltage (inducing band bending and thus efficient carrier separation).

Photoanodes based on $TiO_2$ nanotube layers have been reported to be more promising than nanoparticulate layers due to their well-defined geometry,[31,151,516,671,672,695,736-741] and the feasibility to easily incorporate co-catalysts and dopants[31,377,434,435,695,742].

Particularly promising results regarding the alterations of $TiO_2$ nanotubes have been reported regarding $RuO_2$ by in-situ doping (i.e. growth from Ru-containing alloys). Already very low concentrations can cause a significant increase in the photoelectrochemical water splitting efficiency. In this context it should be mentioned that an often neglected key point in photoelectrochemical arrangements is that majority carriers (electrons) have to travel through the $TiO_2$ layer to the back contact of the photoanode, i.e. electron life-time and in particular, conductivity within the $TiO_2$ structure becomes a very important factor for the overall efficiency. Therefore, doping nanotubular layers with appropriate elements such as Nb (in low concentrations) was also found very efficient to increase the water splitting efficiency in photoelectrochemical arrangements[31,434,435] (see Figure 32).

*$CO_2$ reduction*

Photocatalytic reactions on $TiO_2$ have also been examined for the reduction of undesired highly stable molecules such as carbon dioxide ($CO_2$). Photocatalytic reduction of $CO_2$ in the presence of $H_2O$ on suitable semiconductors can lead to the formation of desired products such as $CH_4$, $CH_3OH$, HCHO and other higher carbon chain molecules[743,744]. Therefore, photocatalytic conversion of $CO_2$ is considered in view of climate remediation by reducing the greenhouse gas stresses while producing useful chemicals as a product. Anpo et al.[745-747] studied the anchored $CO_2$ and $H_2O$ on $TiO_2$ using photoluminescence and proposed a possible mechanism based on ESR analysis.

The photoinduced electron transfer to the adsorbed $CO_2$ molecule splits the molecule to CO and O, followed by further cleavage of the CO bonds on the $TiO_2$ surface. In the same way, the photoinduced holes at the valence band of $TiO_2$ will interact with $H_2O$ molecules generating radicals that lead to the formation of mainly $CH_4$ and $CH_3OH$. A key issue is that $CO_2$ reduction may result in intermediate CO molecules formation in the product mixture,[747,748] which may affect the efficiency of additional co-catalysts.

In particular, it has been reported that the photocatalytic reduction of $CO_2$ can be accelerated by metals (Pd, Rh, Pt, Au, $Cu_2O$ etc.) deposited on $TiO_2$ catalysts that strongly enhance the photo-reduction to $CH_4$ (in decreasing order)[749]. It is also interesting that if $TiO_2$ is immobilized in various matrices, such as zeolite or multi-walled carbon nanotubes, it can exhibit a higher photo-catalytic activity and higher selectivity towards formation of $CH_3OH$ in case of zeolite,[750] while the selectivity changes towards $C_2H_5OH$ and HCOOH in the case of carbon nanotubes[751]. Using $TiO_2$ nanotube layers (and $Cu_2O$ decoration)[536] for $CO_2$ remediation, there are some

spectacular reports; however, more detailed results and analysis are needed to fully assess the potential impact.

*Membranes*

$TiO_2$ nanotube membranes are attractive to be used for size selective and flow through photocatalytic filtration due to their controllable dimension (diameter, thickness) combined with photocatalytic properties of $TiO_2$. Descriptions for the fabrication of $TiO_2$ nanotube membranes were given in section 2.4.

Early work on photocatalytic $TiO_2$ membranes was reported by Albu et al.[118]. Here, solutions penetrate through the membrane by diffusion and are photocatalytically treated at the same time. The membrane shows high pollutant removal by one flow-through cycle.

Another application of $TiO_2$ membranes is to exploit their photocatalytic properties for opening of clogged pores. This was demonstrated by Roy et al.[151] for a protein clogged membrane. The degradation rate in such membranes is highly related to the diffusion rate that is governed to some extent by the pore diameter in the membrane and the size of the particle, as the diffusion coefficient of a species is inversely proportional to its hydrodynamic radius[752].

## 5.3 Ion-intercalation (insertion) devices

As outlined in section 3.2, electrochemical reduction of $Ti^{4+}$ to $Ti^{3+}$ is possible in many electrolytes. The electrochemical reduction process ($Ti^{4+} \rightarrow Ti^{3+}$) is accompanied by small ion (such as $M = H^+$ and $Li^+$) insertion/extraction into the oxide structures to maintain overall charge neutrality according to $TiO_2 + xM^+ + xe^- \leftrightarrow M_xTiO_2$. This reaction is the key to using $TiO_2$ in ion insertion applications, namely i) ion intercalation batteries (e.g. Li ion and Li/air

batteries), and ii) electrochromic devices. In both cases, except for hydrothermal tubes,[753] mainly vertically aligned nanotube layers are currently strongly investigated[754]. This is due to the well connected electrode geometry that in the case of batteries provides shortest possible solvent diffusion distances (see Figure 33), and in the case of electrochromic devices allows defined light scattering geometries with a high observation length. A common characteristic of classic intercalation devices is their structural instability in repeated switching (e.g. pulverization of electrodes in Li-batteries due to volume expansion and reduction). If nanoscale host materials are used, strain during cycling can be much better accommodated and the intrinsic mechanical stability has been reported to drastically increase.

*Electrochromic devices*

In general the reduction of $TiO_2$ and formation of $Ti^{3+}$ leads to modifications of the electronic structure with a typical absorption apparent band gap of 2.2 ~ 2.5 eV, see section 3.2. This means the observed color of the oxide changes from visible/transparent to "blue/black". For many transition metal oxides and many electrolytes, the switching (oxidation and reduction of lattice ions) is, in principle, reversible[485,755]. Therefore, this effect (named electrochromism) is widely used for optical devices such as smart windows or displays.

The electrochromic ability of a metal oxide layer can be evaluated by the color efficiency η (cm²/C) that is defined by:

$$\eta(cm^2/C) = \Delta Od/Q \quad (10)$$

where $\Delta Od$ refers to the change in optical density and $Q$ is the specific charge density.

The optical density can be calculated from following equation:

$$\Delta Od = log(T_b/T_c) \quad (11)$$

where $T_b$ and $T_C$ refer to the transmittance of the electrochromic layer in its bleached and colored states, respectively.

In order to achieve an efficient electrochromic device, the materials should have an intrinsic high specific contrast (high absorption coefficient) as well as a high electrochromic capacitance. The rate determining step for color switching is typically the solid state diffusion/migration rate of intercalated ions into the host material. This penetration time of ion (τ) determines the switching time, and is proportional to the square of diffusion/migration depth (t), $\tau \propto t^2/D$ ($D$ - transport coefficient of an ion in a lattice)[756]. Typical diffusion depth of $H^+$ or $Li^+$ in $TiO_2$ is approximately 5 – 20 nm[757]. As a result, ideal electrochromic materials are structured to this size – this is also in the range of typical anodic nanotube layers that have wall thickness from 5 – 30 nm.

Early attempts that investigated the electrochromic properties of $TiO_2$ nanotube structures were performed first with hydrothermally formed titanate nanotubes,[753] then with anodic self-organized $TiO_2$ nanotubes[754].

Anodically formed $TiO_2$ nanotubes on Ti substrates show better electrochromic properties regarding $H^+$ and $Li^+$ intercalation from 0.1 M $HClO_4$ aqueous electrolytes or 1 M $LiClO_4$/acetonitrile electrolyte than compact oxide layers or nanoparticle layers.[754,758,759] The first report on the electrochromic properties of $TiO_2$ nanotubes showed that amorphous nanotubes present a higher switching capacity than anatase phase tubes, but anatase exhibits a higher reversibility and cyclability of the color reaction (due to a trapping of intercalated $Li^+$ in the non-stoichiometric amorphous structure)[758]. Generally, anodic $TiO_2$ nanotubes show a relatively high optical contrast, fast switching rates and good cycle stability. Nevertheless, modification of $TiO_2$ nanotubes can lead to an even higher optical contrast, a positive effect on

the threshold potentials for on and off switching, as well as a higher cycling stability.[28,249,450,451,456,760-762] Such modifications include: i) the formation of mixed oxide tubes such as grown from some alloys (e.g., Ti-Nb, Ti-V, Ti-W or Ti-Mo), or from multilayered metal substrates ($TiO_2$/$Nb_2O_5$ multilayer nanotubes), ii) the decoration of the nanotube layer with even more electrochromically active nanoparticles such as $WO_3$, $MoO_3$.

In particular, $TiO_2$ nanotubes, doped with a high concentration of Nb formed by anodization of Ti-45Nb alloys, was reported to show strongly enhanced electrochromic efficiency as well as a high increase in durability during repeated switching - these effects were ascribed to widening of the anatase lattice by substitutional Nb atoms[450]. Such lattice widened $TiO_2$ nanotubes even allowed insertion of larger ions such as $Na^+$.[450] In line with expectations regarding the mechanical stability, $TiO_2$/$Nb_2O_5$ multilayered nanotubes show, due to the small size of the segments (less than 10 nm), a much better mechanical stability than conventional $TiO_2$ nanotubes. Namely, structural degradation during repeated Li ion insertion/extraction[249] in such small 3D $TiO_2$ segments shows clearly less mechanical rupture by stress-induced burst.

In order to fabricate transparent $TiO_2$ nanotube based electrochromic devices (such as smart windows, display devices), $TiO_2$ nanotube layers need to be placed on a transparent conductive oxide glass e.g., ITO or FTO. One approach is that $TiO_2$ nanotube layers are lifted-off from metal substrates and transferred to the transparent conductive oxide glass[759]. On the other hand, a straightforward way to form $TiO_2$ nanotubes on transparent conductive oxide glass substrate is to sputter-deposit or evaporate Ti metal films on a transparent conductive oxide glass substrates and then completely anodize these layers[658,763]. Such nanotube layers on ITO show a higher transmittance than nanoparticle layers, combined with a higher optical contrast and a higher stability against disintegration (Figure 34).

*Li-ion batteries*

An even more important application of ion intercalation/storage is Li ion batteries (Figure 35).[764,765] Generally, carbonaceous electrodes (e.g. graphite and its lithiated form) are the most widely used material in rechargeable Li-ion batteries due to their high theoretical capacitance (372 mAh/g), low lithiation potential (~ 0.1 V vs. $Li/Li^+$), and good electric conductivity, which lead to a large potential difference between anode and cathode.[766,767] To have a relatively high cell potential, the most often used electrodes are graphite and $LiMO_2$ (M= Co, Mn, Fe). However, as the graphite-lithiate potential is close to the lithium plating potential, short circuiting and/or organic combustion (thermal runaway) are potential drawbacks of this system, with risks regarding safety and long-term stability.[768] $TiO_2$ is considered as one of the most promising substitute anode electrode material due to a combination of sufficient intercalation capacity with a higher lithiation potential (~1.6V vs. $Li/Li^+$), a good durability, a relatively high cell voltage, low production costs, light weight and environmental sustainability[769]. However, $TiO_2$ bulk materials show only a very limited $Li^+$ uptake and a poor electric conductivity that lead to a low capacity and a lower rate performance than theoretically possible[770-772]. In order to overcome these drawbacks, nanostructured $TiO_2$ materials have been considered for Li insertion/extraction; they mainly provide a larger specific surface area (increase the electrode/electrolyte contact) and an optimized $Li^+$ diffusion path[340-343,773-788] – these features allow an increase of the charge/discharge rate and an increase of the current density.

$TiO_2$ has a relatively high theoretical capacitance of 335 $mAhg^{-1}$, which corresponds to a structure of $Li_1TiO_2$ and a fully reduced lattice from $Ti^{4+}$ to $Ti^{3+}$.[789] In the schematic drawing in Figure 35.b an intercalation device with $TiO_2$ nanotubes (as shown in Figure 35.a) is arranged as

electrode in a Li-ion battery. Such a device consists of two electrodes, which can store $Li^+$, and a separator in a sealed container, which is filled with an organic solvent/ ionic liquid and a lithium ion source. The $Li^+$ ions are able to diffuse through the separator, from one electrode to the other, while the electrons move to the consumer.

Using a $TiO_2$ electrode in cyclic voltammetry as shown in Figure 35.c, in the potential range of 1 to 3 V, a pair of peaks at around 1.7 V and 2.3 V can be observed, these peaks correspond to Faradaic extraction and insertion reaction of $Li^+$ into the $TiO_2$ lattice. The peak position is dependent on the scan rate, the electrolyte, and on the $TiO_2$ polymorph. Most important are, however, capacitance measurements for battery testing which usually are done by galvanostatic charge and discharge experiments as shown in Figure 35.d. This provides direct information on how long a certain current density ($A/cm^2$ or A/g) can be sustained under constant load.

An early attempt of using $TiO_2$ nanotubes in Li ion batteries was reported by Zhou et al.[781] in 2003. In their study they used hydrothermally grown anatase phase $TiO_2$ nanotubes with 300 nm individual tube length and approximately 8 nm diameter. These nanotubes exhibited an overall capacitance of 182 $mAhg^{-1}$ at a charge/discharge rate of 80 $mAg^{-1}$.[781] Similar results were obtained later with hydrothermally grown $TiO_2$ anatase nanotubes with 10 nm of diameter and several hundreds (200 – 400) nm of tube length[790] that resulted in a capacity of 239 $mAhg^{-1}$ at a charge/discharge rate of 36 $mAg^{-1}$.[790] Up to now, hardly any further improvement of the specific capacity using hydrothermally formed anatase $TiO_2$ nanotubes can be found.

As an alternative, hydrothermally formed $TiO_2$(B) sheets have been considered as a promising material for Li-ion battery. The unit cell of $TiO_2$(B) lithanates contains 8 Ti sites and 10 $Li^+$ sites which are corresponding to a theoretically capacitance of 420 $mAhg^{-1}$.[340-343] The sheet structure leads to faster lithium ion transport than other crystal structures due to the low density compared

to other $TiO_2$ polymorphs, and an open channel structure along the b-axis (due to the perovskite like layered structure). Since Armstrong et al.[340] first reported hydrothermally formed $TiO_2$-B nanowire structures for Li-ion batteries in 2004, $TiO_2$(B) nanotube structures have also been formed with 10 – 20 nm outer diameters, 5 – 8 nm inner diameters and 1 μm single tube length[341-343]. Such nanotubes exhibited a capacity of 338 $mAhg^{-1}$ of at 50$mAg^{-1}$ for the first cycle.[342] In order to compare the geometry effect of $TiO_2$(B) nanostructures on Li-ion batteries, the same group evaluated the specific capacity with several types of $TiO_2$(B) nanostructures such as nanowires, nanoparticles, nanotubes and bulk materials (Figure 36).[779]. From the comparison, nanotube structures show relatively higher capacity than other structures but the enhancement is quite small (Figure 36). This finding can be ascribed to the fact that hydrothermally formed nanostructures are randomly oriented to the back contact, i.e. there is no potentially favored geometry for electron and ion transport. Moreover, to make stable electrodes with such $TiO_2$ nanotubes, proper binder and additional conductive materials should be used.

In order to overcome random orientation of nanotubes and eliminate binders, vertically aligned $TiO_2$ nanotubes have been suggested. The first anodic $TiO_2$ nanotubes for Li-ion battery were reported by Fang et al.[782] in 2008. The anodic anatase nanotubes were prepared in ammonium fluoride containing glycerol/water electrolyte. Such nanotubes (50 – 60 nm outer diameter, 10 – 15 nm wall thickness and approximately 1 μm tube length) showed an overall capacitance of 90 $mAhg^{-1}$. The capacity is relatively lower than hydrothermal anatase or $TiO_2$(B) nanostructures (Figure 36) due to the fact that the specific surface area of anodic $TiO_2$ nanotubes is considerably lower than for hydrothermal tubes.

However, the amorphous phase shows a twice higher capacity (229 $mAhg^{-1}$) than the anatase phase (108 $mAhg^{-1}$) at a current density of 1 $Ag^{-1}$.[783] This result is in line with later reports by

Ortiz et al.[784] using anodic $TiO_2$ nanotubes formed on sputter-deposited Ti layers on Si substrates. In 2012, Rajh et al.[349] reported in a detailed study on amorphous anodic $TiO_2$ nanotube structures for Li-ion battery that by intercalation/deintercalation cycles of Li ions, the nanotubes can be crystallized to a cubic closed packed crystal structure with high Li concentrations (>75 %). I.e., such ordered structures reach close to an ideal $Li_2Ti_2O_4$ stoichiometry that allows higher capacity, long-term stability and power density than $TiO_2$ anatase structures.[349]

Several groups have tried to enhance the capacity and long-term stability of nanotube based Li-ion batteries with modification or improvement of their geometry. Generally, the high surface area of $TiO_2$ nanotubes allows a high specific capacity or area capacity. Frank et al.[785] have explored different diameter of $TiO_2$ nanotubes for the use in Li-ion batteries. As expected, small diameter $TiO_2$ nanotubes (~ 21 nm inner diameter and ~ 40 nm outer diameter) formed by a low anodization potential (10 V) showed the highest normalized capacity at 10 $mVs^{-1}$ – the value is almost 2 times higher than $TiO_2$ nanotubes with the largest investigated diameter (~ 96 nm of inner diameter and ~ 130 nm of outer diameter).[785]

In order to increase surface area, long $TiO_2$ nanotubes have been considered, but an increased length of nanotubes not only increases the surface area but also the diffusion length for electrons and Li ions. Nevertheless, in general an increased areal capacity is observed with increasing tube length,[784,787,788,793] as shown in Figure 37.a ( it should be considered that the detailed anodization conditions for these data are different). Nevertheless, if the data is given as normalized areal capacities against tube length (areal capacity/tube length), similar values are obtained (see Figure 37.b).

In addition to tube geometry, top morphology and highly aligned $TiO_2$ nanotubes have also been considered for enhancing the capacity in Li-ion batteries. The main aim is to reach a higher diffusion rate of electrolyte. Recently, Wei et al.[788] investigated highly aligned anodic $TiO_2$ nanotube structures formed by two-step anodization, followed by a wall-thinning process using chemical dissolution. After 100 charge and discharge cycles, the modified anatase $TiO_2$ nanotubes show an areal capacity of 460 μAhcm$^{-2}$ at 0.05 mAcm$^{-2}$ that is more than twice higher than for conventional single step anodized nanotubes (200 μAhcm$^{-2}$ at 0.05 mAcm$^{-2}$)[296,788].

Another important approach to achieve better battery performance is to modify $TiO_2$ nanotubes with other active materials. Most frequently explored is decoration with noble metals, e.g., Ag,[782,794,795] or secondary transition metal oxides, e.g., Sn-oxide,[796-798] ZnO,[799,800] $Co_3O_4$,[801-803] NiO,[804-806] $Fe_2O_3$,[807-809] or non-metallic doping materials, e.g., C,[455] N,[420] S[810]. Such modifications increase Li-intercalation properties, increase the conductivity overall or locally,[778,811,812] or increase pseudo capacitive contributions[782,794]. From the compilation of the areal capacity for various modified anodic $TiO_2$ nanotubes (Figure 38), it is evident that different modification techniques lead to much higher areal capacity than bare $TiO_2$ anodic nanotubes.

The earliest attempts to enhance the nanotube performance with a noble metal coating were reported by Fang et al.[782] in 2008. Anodic $TiO_2$ nanotubes can be decorated with silver nanoparticles via simple dip-coating deposition in an $AgNO_3$ solution. This additional treatment increases the capacitance of the tubes from 90 to 110 mAhg$^{-1}$.[782] For example, Guan et al.[795] have attempted electrodeposition of Ag nanoparticles on bamboo type nanotubes. For an optimized electrodeposition, an impressive increase of the capacitance values of more than a factor of 2 (from 55 to 131 μAhcm$^{-2}$) could be observed[795].

In order to deposit secondary metal oxides on $TiO_2$ nanotubes, several techniques are used such as dip-coating, sputtering, electro-deposition, photodeposition, or atomic layer deposition. Yu et al.[815] reported 10 μm thick $TiO_2$ tubes that were modified with $Fe_2O_3$ particles by dip-coating using a $FeCl_3$ solution, thereby increasing the nominal capacitance from 300 $\mu Ahcm^{-2}$ to 600 $\mu Ahcm^{-2}$. For 1 μm long $TiO_2$ nanotubes with electro-deposited $Co_3O_4$ and NiO submicron particles, Keyeremateng et al.[816] reported an increase in the capacitance from 22 to 100 $\mu Ahcm^{-2}$ for $Co_3O_4$, and from 22 to 90 $\mu Ahcm^{-2}$ for NiO, respectively. Another approach by Fan et al.[793] used 5 μm thick $TiO_2$ nanotube array that was covered by $Co_3O_4$ via photo deposition from a $K_2Co(CO_3)_2$ solution - an improvement from 120 to 400 $\mu Ahcm^{-2}$ was obtained. Ortiz et al.[817] used a 2 μm thick nanotube array to grow several micrometers big Sn-oxide crystals on the top of the tube surface which led to a capacitance increase from 95 to 140 $\mu Ahcm^{-2}$. Alternatively, ALD coating of non-conductive amorphous $TiO_2$ nanotubes with ZnO led to a capacity enhancement from 74 to 170 $\mu Ahcm^{-2}$.[800]

### 5.4 Sensors

$TiO_2$ nanotubes have also been explored to a considerable extent for sensing applications, and in particular towards gas sensing. Many approaches target the use of $TiO_2$ in quantitative or qualitative analysis, with a maximum sensitivity, towards one specified type of gas[818]. The unique properties of titanium dioxide such as a high chemical stability, high temperature resistance combined with its semiconducting behavior make this material a very promising candidate for sensing devices. One of the first $TiO_2$ sensing layers was reported in 1983 by Logothetis and Kaiser[819] for high-temperature oxygen sensing to monitor and control the

combustion process of the air-fuel mixture of internal combustion engines. In 1989, the first report on hydrogen gas sensing showed compact crystalline $TiO_2$ to have a good sensitivity combined with a relatively fast response and recovery behavior.[820,821] In 1999, Dutta et al.[822] used $TiO_2$ nanoparticles instead of $SnO_2$, which was at that time the most extensively investigated semiconducting metal oxide for CO gas sensing[373] - the main advantage of $TiO_2$ being its stability in exhaust pipe environments. Meanwhile another important advantage of $TiO_2$ over $SnO_2$ is that defined nanostructures, such as nanotubes, nanoparticles, nanowires or other nanoporous assemblies can easily be obtained by a variety of fabrication methods[823]. From density functional theory (DFT) calculations the defect structure of the $TiO_2$ surface is found to be a crucial factor to sense e.g. $SO_2$ and $CO_2$ gas molecules.[824,825] According to theory, it is possible to estimate the number of oxygen vacancies by measuring the change of resistivity when the sample is brought into CO containing environment[825]. The high surface to volume ratio of nanomaterials results in a sensing response that is enhanced by several orders of magnitude compared to flat compact surfaces[826]. In addition, $TiO_2$ possesses suitable band edge positions to detect a number of gases. The reaction with a red-ox active gas species injects or consumes electrons from the $TiO_2$.

A variety of sensing strategies can be used that are based on field-effect transistors, electrochemical phenomena, fluorescence, acoustic wave speed and, most characteristic for metal oxide semiconductors, simply the electrical resistance. The response r of a sensor device can be defined as the relative change of the resistivity[827]:

$$r = \frac{(R-R_0)}{R_0} \tag{12}$$

where $R$ represents the resistance depending on the concentration of the tested gas and $R_0$ is the reference resistance, meaning the resistance for normalized conditions (e.g. in an atmosphere

of $N_2$, dry air or under ambient conditions). The sensitivity S of the sensor device is the dependence of the response to a change in the concentration M:

$$S = \frac{\partial r}{\partial M} = \frac{\partial}{\partial M}\frac{(R-R_0)}{R_0} = \frac{1}{R_0}\frac{\partial R}{\partial M} \propto \frac{\partial R}{\partial M} \qquad (13)$$

In general, the entire resistance change of the metal/metal oxide sensing devices can be attributed to changes of the Schottky barrier at the interface to the deposited metal contact due to changes in surface doping (or surface states) or due to changes in the neck-conductivity at a multigrain structure[828]. These effects are induced when atoms and molecules are physisorbed or chemisorbed to the surface of the sensitive material or are absorbed to the metal[826]. The degree of physisorption and chemisorption can be controlled by heating the sensor element. At temperatures higher than 100 °C, typically the influence of physisorption can be neglected. Since most of ceramic-based sensors are n-type semiconductors, their conductivity is increased if they are exposed to acidic gases (which inject electrons into the semiconductor surface) and decreased when exposed to basic gases. These changes of the electrical transport mechanism hold for single crystal materials. Anodic nanotube walls do not consist of a single crystal, but are built up from crystallites. In general, a change of the electron-depleted zones influences the conductivity at grain boundaries. A schematical picture of these three effects is shown in Figure 39.

The chemical reactions which are leading to the loss and the injection of electrons can be described by the following chemical equations:

reduction: $$\mathrm{R} + \mathrm{MO-O^-} \rightarrow \mathrm{MO} + \mathrm{RO} + \mathrm{e^-} \qquad (14)$$

causes the injection of an electron while the

oxidation: $$\mathrm{MO-O^-} + \mathrm{O} + \mathrm{e^-} \rightarrow \mathrm{MO-O_2^-} \qquad (15)$$

leads to consumption of an electron[826].

All reactions mentioned above take place at the surface of the semiconductor. As nanostructured sensors exhibit a high surface area, accordingly they have high fraction of electron-depleted zones. The relative change of the resistivity is therefore much higher when atoms are adsorbed to the surface, than in macroscopic sensors where the electron-depleted surface layer is only a small fraction of the material. A maximum of sensitivity can be achieved when the pattern size of the nanostructure (e.g. thickness of the tube wall) is in the range of the Debye length of the material.

Another benefit of nanostructures is the possibility to minimize the size of sensor devices by orders of magnitude, which is desirable for very large scale integration and the reduction of power consumption in view of heating.

*Response and sensitivity*

The main focus of most of the recent studies is to further decrease the lower limit of detection (LOD) of sensing devices. However, one of the fundamental challenges of sensors is to combine high sensitivity with extensive dynamic range. This is also the case when designing sensors from $TiO_2$ nanotubes as illustrated in Figure 40. It shows the relative response and the sensitivity of two different titanium oxide nanotube array-based gas sensors; the data is taken from work of Li et al.[47] and Lee et al.[835]. In general, the relative change of resistivity can be calculated according to equation (12) and the sensitivity is estimated by its slope. While the array in Figure 40.a is grown by anodic oxidation, the $TiO_2$ array in Figure 40.b is fabricated by atomic layer deposition (ALD) into a template of porous anodic aluminium oxide. The anodic nanotubes have an outer diameter of 70 nm and a length of 5 µm and are examined at 215 $^{o}C$ for $H_2$ sensing, while the ALD deposited nanopillars have a length of 750 nm and a diameter of 250 nm and are examined

at 100 $^{o}C$. When these sensors are compared, a considerable difference can be observed. While the sensing device in Figure 40.a exhibits a relatively linear sensing behavior, allowing for a quantitative sensing of $H_2$, the device in Figure 40.b shows a high sensitivity – in this case minimal concentrations of $H_2$ can be detected but only over a small concentration range and only with a semiquantitative accuracy. This illustrates that $TiO_2$ structure tuning in the nanometer range and sensor operation conditions can lead to drastic changes in LOD and the dynamic response that can be achieved.

*Preparation of sensing-devices from $TiO_2$ nanotube arrays*

The first anodic $TiO_2$ nanotube sensor for sensing $H_2$ was reported by Varghese et al.[829-831] in 2003. The nanotube layers were grown on a titanium foil in a water-based electrolyte. The lowest detection limit for $H_2$ was 100 ppm. Up to now studies report on the use of $TiO_2$ nanotubes-based sensors in detection of different gases e.g. oxygen (200 ppm $O_2$ at 100 $^{o}C$, 10 ppm $O_2$ in $CO_2$ at 600 $^{o}C$),[832,833] hydrogen (10 ppm $H_2$ at 25 °C in dry air),[47,834-841] carbon monoxide (100 ppm CO at 200 $^{o}C$),[822,835,842] ammonia (150 ppm $NH_3$ at room temperature),[843,844] ethanol (400 ppm $CH_3CH_2OH$ at room temperature)[842,845-847] and formaldehyde (10 ppm HCHO at room temperature)[847], nitric oxide (0.97-97 ppm NO at room temperature),[848] nitrogen dioxide (0.97–97 ppm $NO_2$ at room temperature),[848] sulfur dioxide (50 ppm $SO_2$ at 200 $^{o}C$),[824,849] thionyl fluoride (50 ppm $SOF_2$ at 200 $^{o}C$), sulfuryl fluoride (50 ppm $SO_2F_2$ at 200 $^{o}C$), hydrogen peroxide ($3\times10^{-5}$ mol $L^{-1}$ $H_2O_2$ in solution),[850] sulfur hexafluoride (99.999 % $SF_6$ at 200 $^{o}C$)[849] and humidity (11–95 %)[851].

Different techniques can be used to obtain $TiO_2$ nanotube electrodes for sensing. While tubes prepared by electrochemical anodization are already backcontacted by the Ti-substrate, top contacts are generally metals such as Pt,[851,852] Au ,[853,854] Ag,[849,853,854] Al,[855] stainless steel,[843] or Cu[856] that can be sputtered or evaporated in different configurations on the array of $TiO_2$ nanotubes[857]. In two terminal measurements, an overlap of dc-resistance and Schottky diode behavior will be seen (if no ohmic contact is used). Pure dc-conductivity requires four terminal measurements. Four and two point measurements essentially deliver a "resistivity" along the direction of the tubes. To determine the resistivity perpendicular to the growth direction of the tubes, the $TiO_2$ nanotube array has to be grown or transferred on an insulating substrate. Various geometries of anodic $TiO_2$ nanotube layers were examined by Perillo and Rodriguez[843] in 2012. They used a glycerol $NH_4F$ electrolyte and a 2 h 550°C heat treatment to produce tubes of 50, 90, 110, 150, 200 to 240 nm diameters. While the diameter is significantly increased, the tube wall thickness increased only slightly from 20 to 30 nm. The gas sensing behavior was tested with 400 ppm ethanol and 150 ppm ammonia gas at different relative humidity of 40 and 90 %, at room temperature. The authors found that the tube size does not have a significant impact on the sensing behavior. However, a low humidity leads for both gases to a drastically improved response. While in relative wet air a relative response for ammonia of 0.06 can be detected, in drier air a relative response of 0.4 is observed.

The use of amorphous $TiO_2$ nanotubes, grown in a $(NH_4)_2SO_4$ water based electrolyte with 0.5 wt.% $NH_4F$ for the use of $O_2$ sensing was reported by Lu et al.[832] in 2008. The 2.3 µm long amorphous tubes with a 40 nm wall thickness showed a relatively good linear $O_2$ sensing behavior between 200 ppm and 20 % at 100°C. This corresponds to a maximal sensitivity of 170. In contrast, crystalline anatase $TiO_2$ nanotubes with comparable morphology showed only a

maximal sensitivity of 20. The mechanism behind this phenomenon is the change of the charge carrier concentration on the nanotube surfaces. Amorphous nanotubes are more disordered and contain more sensing-active defects such as oxygen vacancies[373] but often show a decreased long term stability and reversibility.

A very interesting sensing material is $TiO_2$-B.[858] An important inherent advantage of this material is the crystal structure, which exhibits intrinsically abundant surface states and oxygen vacancies. If the material is used to synthesize nanowires (with a diameter of 20 to 50 nm), humidity in the range of 5% to 95% relative humidity could successfully be detected[858].

*Temperature-effects*

In most sensing studies, temperatures around 200 – 500 °C are used. This is due to the fact that at temperatures close to room temperature metal oxide sensors will show cross-sensitivity for humidity along with the sensing gas. Therefore, the very first step towards selective sensitivity is heating. While water from the atmosphere is physisorbed, oxygen is chemisorbed, i.e., is stronger adsorbed. When the sensor is heated to 100 – 200 $^{o}$C, the water (including surface adsorbed layers) is released so that the influence of humidity can be neglected. Nevertheless, there are strong efforts to make reliable sensor devices to operate at room temperature[859]. For example, Palacios-Padrós et al.[840] managed to detect $H_2$ with a concentration of 9 ppm at a temperature of only 80 °C using a $SnO_2$-nanotube array as a sensing device. Chen et al.[834] built a room temperature hydrogen sensor based on highly ordered $TiO_2$ nanotube arrays using a crystallized mixed anatase and rutile phase. For hydrogen sensing tests the arrays were contacted with Pt electrodes, and $H_2$ concentrations in dry ambient air between 10 ppm and 3000 ppm were used. Even at the lowest concentration the conductance shows a well-defined peak with a

response time of 53 s. Up to a concentration of 1500 ppm the change in resistivity is reversible, however at higher values a different behavior is observed. Not only is the resistivity not recovering to its original value, but also the induced change in resistivity for higher concentrations gets smaller. It is suggested that this behavior is related to the saturation of the nanotube surface with hydrogen. However, with this device it seems possible to analyze hydrogen gas in dry air up to concentrations of 1500 ppm without heating. Lee et al.[835] conducted a study of the sensing behavior at different temperatures of a device made from ALD-deposited $TiO_2$ nanotubes. The relative resistivity and response time at different temperatures are shown in Figure 41.a and b. The relative response shows a maximum at 100 $^oC$, while the response time decreases with increasing temperature.

*Improved sensing performance*

To enhance the sensing performance, in particular the sensitivity and response of $TiO_2$ nanotube arrays, considerable efforts have been made over the last two decades. Preferential doping, decoration of nanoparticles and optimizing sensing parameters were explored for different gases.[373,818,823,826,860]

The sensing properties of $TiO_2$ nanotubes can be drastically enhanced when the arrays are decorated with secondary materials such as nanoparticles of $Ni(OH)_2$ (glucose),[861] Au ($H_2O_2$),[850] Ag ($O_2$),[862] Pt (benzene, $H_2$),[838,863] Ir (benzene),[863] PtIr (benzene)[863] or Pd ($H_2$)[848].

The most important effects of such a decoration are a change in the Schottky barrier due to the different workfunctions of the used material as well as chemical catalysis effects. In presence of Pd or Pt nanoclusters on the surface of $TiO_2$, for example, hydrogen gas split-up into hydrogen

atoms can be facilitated. This leads to an increased interaction of adsorbed dissociated hydrogen with $TiO_2$, and results therefore in an enhanced hydrogen sensing performance[864].

Jiang et al.[865] showed in 2013 the decoration of $TiO_2$ nanotubes with Ag for $H_2O_2$ detection. 15 nm Ag particles were deposited on the walls of the nanotubes by electrophoretic deposition. Compared to plain $TiO_2$, the decorated nanotubes showed a three times enhanced sensitivity (up to 184.24 $mA/Mcm^2$) and a low detection limit (85.6 nM)[865].

Modified hydrothermal titania nanotubes with Pt-Ir by MOCVD were reported by Colindres et al.[863]. The nanotubes with a length of several hundred microns and a diameter of 7-10 nm were decorated with Pt and Ir particles. While Ir or Pt forms single particles during the deposition, a bimetallic particle was grown when both precursors were used. The combination of $TiO_2$ nanotubes with the Pt-Ir nanoparticle mixture showed the highest activity, when tested for the detection of cyclohexene.

Pd and Pt nanoparticle decoration for $H_2$ sensing was also used by Han et al.[866] in 2007. Mixing the precursors in the right concentrations led to the hydrothermal synthesis of Pd and Pt decorated titanate nanotubes with a diameter of 100 nm and with small, dispersed noble metal particles. The best sensing was obtained by a mixture of Pd and Pt on titanate nanotubes. For all temperatures, the mixed nanoparticles showed an enhanced performance compared to single metal decoration; the best sensing performance was achieved at 250°C. The rate of oxidation reactions on Pd/Pt decorated nanotube surfaces was found to be almost twice as high compared with other catalysts[866].

In the context of non-enzymatic glucose sensing, Gao et al.[861] reported a significant improvement of the anodic current density for glucose oxidation when carbonized $TiO_2$ nanotube arrays ($TiO_xC_y$ NTs) are decorated with $Ni(OH)_2$ nanoparticles without the aid of a polymer

binder[861]. Improved properties such as a wider linear range, a low detection limit, fast response and long-term stability were observed.

Another efficient way to modify the physical and chemical properties of $TiO_2$ is to dope or alloy the semiconductor system. The sensing sensitivity or the sensing stability can be improved by adding different dopants e.g. Ni,[855] Rh,[820] Nb,[867] $WO_3$,[868] ZnO,[869] $Ga_2O_3$,[870] $Al_2O_3$ and $V_2O_5$[871]. While in most cases thin films in form of nanoparticles are used as sensing layer, only few reports with doped or alloyed $TiO_2$ nanotube arrays are published.

Li et al.[872] used a NiTi alloy (50.8 at.% Ni) and further annealing treatment (425 $^oC$ for one hour) after forming the oxide nanotubes in a non-aqueous ethylene glycol/glycerol electrolyte. The grown nanotubes (diameter of 65 nm, length of 500 nm) were used to fabricate a sensor device that could detect 1000 ppm hydrogen at room and elevated temperatures. Alloying of the substrate led to $TiO_2$ mixed oxides and was perceived to result in a change of the semiconducting behavior; an amount of approximately 7 at.% of Ni was reported to show a p-type behavior, which resulted in a resistance increase during $H_2$ sensing.[872]

The same group showed the doping of $TiO_2$ nanotubes by anodizing a TiAl6V4 alloy.[871] Al and V doping were reported to lead to a reduction of the band gap, with a p-type sensing behavior for hydrogen compared to non-doped $TiO_2$ nanotube films.[871] The authors found the annealing temperature to have a significant influence on the sensing behavior of the anodic nanostructures. The best performance was observed for samples annealed at 450 °C, where the saturation response of the films was more than twice the value of the 550 °C annealed sample (sensing conditions: 200 °C, 1000 ppm) and a response at room temperature was obtained.

In 2010, Moon et al.[873] showed the beneficial effect of Pd decoration on $TiO_2$ nanofibers produced by electro spinning. The relative response of pure $TiO_2$ at 200 °C for 2.8 ppm $NO_2$ is

30%, while the Pd doped fibers have a response of 55%. Even more impressive is the sensing ability for very low concentrations of 0.16 ppm $NO_2$ at 180°C for the Pd doped sensor.[873]

*Free-standing $TiO_2$ nanotube arrays*

Another approach to enhance the intrinsic sensing properties of $TiO_2$ nanotube arrays is to remove them from their substrates as described in section 2.4 and open the bottom to build a flow-through sensor. Chen et al.[259] reported large-scale free-standing nanotube arrays of 7-50 µm thickness by ultrasonic splitting of as-anodized $TiO_2$ nanotube films. For such a nanotube array of a thickness of 25 µm the relative resistivity was found to double when the film was free-standing.[259] This significant increase in the relative resistivity may justify the comparably large effort of producing free-standing membranes.

*Single $TiO_2$ nanotubes as sensing devices*

The ultimate miniaturization of a sensing device is achieved when only a single nanotube (or a few nanotubes) are used as a sensing device. Such a single TiNT sensing device can be expected to have a very short response time, since there is no need for the gas to diffuse into a complex structure. Nevertheless, only very few reports on sensing with single $TiO_2$ nanotubes can be found. Techniques as electron beam induced deposition and electron beam lithography have to be used to fabricate electrical contacts to a nanotube of only a few micrometer length[32]. The high resistivity observed for a single $TiO_2$ nanotube in these reports (in contrast to section 3.2) requires an advanced system for electrical measurements. Lee et al.[874] report the fabrication of a single $TiO_2$ nanotube device produced by ALD into a AAO template for bio-chemical sensing, where contacts were fabricated with electron beam induced deposition and electron beam

lithography as mentioned above. The resistance of the $TiO_2$ nanotube is obtained in a two terminal measurement before and after the tube surface was functionalized with Glycine ($C_2H_5NO_2$), Lysine ($C_6H_{14}N_2O_2$) or gamma-Aminobutyric acid (GABA, $C_4H_9NO_2$). In all three cases the resistance reaches a final value after 12 hours of amino acid treatment and a relative change of resistivity of 2.88 (Lysine), 2.19 (GABA) and 0.065 (Glycine) is found.

### 5.5 Memristive behavior

In 2008, Strukov et al.[875] (revisiting some earlier electrochemical[876] and theoretical[877] work) demonstrated that a thin $TiO_2$ ($TiO_x$) film sandwiched between two platinum contacts shows a voltage dependent on/off resistive switching. In such a memristive device, the conductivity of the $TiO_2$ layer depends strongly on the history of previous voltages applied. The effect is illustrated in the I-V curve in Figure 42.a. If voltage is applied to a Pt/$TiO_2$/Pt sandwich, such as across a $TiO_2$ nanotube bottom (shown in Figure 42.b), in the positive direction, the current increases quite steeply; however, once a threshold voltage is passed, the current drops almost instantly and remains low in the reverse cycle until a lower threshold voltage is reached where the current suddenly increases again. If the voltages are kept below the upper and lower thresholds, one can cycle the voltage and notices either a very shallow or steep IV curve (inset in Figure 42.a). In other words, applying a sufficiently high voltage step (above threshold potential), depending on the sign of the voltage a higher or lower resistivity state is established in the oxide.

Works by Yang et al.[878] in 2008 and follow-up work[879-882] on thin $TiO_2$ layers demonstrate convincingly that the effect is associated with the presence of $TiO_x$ (suboxides) in the oxide layer. In the meantime, it is generally accepted that the memristive effect is based on mobility of

oxygen vacancies or $Ti^{3+}$, i.e., that some degree of defectiveness of the film is required[883] to cause such switching effects. The concept is that vacancies, originally present in a layer of $TiO_2$ can be moved across the oxide using a sufficiently high applied voltage (field). A negative electrode attracts vacancies; a positive electrode rejects vacancies. A sufficiently connected vacancy path (percolating) throughout the oxide can lead to a conductive path in the oxide (often described as a conductive filament) — i.e. as a result, the oxide as a whole shows high electron conductivity. By a sufficiently high reverse pulse, vacancies are repelled from the positive electrode and the oxide as a whole shows a high resistivity. In other words, by sufficiently high voltage pulses that open or cut conductive filaments one can switch the resistivity state forth and back reversibly,  and hence this effect can be used as a data storage element. Such memristor effects have attracted tremendous attention for use in non-volatile memories. For a rapid and high magnitude memristive switching with a fast field-aided transport, thin oxide layers (some nm-thickness) are preferred, as they create sufficiently high fields already at low voltages.

In 2011, Szot et al.[884] suggested an alternative to vacancy mobility, i.e., that memristive switching might take place in amorphous $TiO_2$ by formation of magneli-phases. Based on their XRD results, it was observed that under applied voltage crystalline filaments of $Ti_5O_9$ or other magneli-phases are formed. Also using this explanation it seems clear that the formation of conducting channels is localized and not uniform along the whole metal/$TiO_2$ interface.

Frequently used methods to fabricate thin film memristors involve atomic layer deposition or sputtering of $TiO_2$ (or derivatives of it), followed by an adequate heat treatment (to adjust the vacancy concentration), and finally establishing a top contact (mostly Pt)[878,885-889].

For $TiO_2$ nanotube layers, reports exist on the observation of memristic effects using entire layers of amorphous nanotubes[889] or highly defined nanotubes with conformally metal filled

bottoms[236,237] – as shown in Figure 47.b, where only the ≈ 30 nm thick tube bottom is used to create a memristive response. More recently, Liu et al.[26] showed that reliable switching can also be achieved using crystalline (anatase) $TiO_2$ nanotubes that are exposed to a reductive treatment in $Ar/H_2$ atmosphere. These findings are in line with explanations given by Yang[878], i.e., that in any case sub-oxide or vacancies in the film must be present to achieve a memristive response.

### 5.6 Supercapacitors

Nanostructures, such as nanotubes, are also of high interest in electrochemical capacitors, namely supercapacitors or ultracapacitors. These devices are a focus of large interest due to their higher energy density than conventional capacitors and a higher power density than batteries. Based on charge-storage mechanism, two type of supercapacitors can be distinguished. The first one is called electrochemical double layer capacitor (EDLC) that is based on the charge separation at electrode/electrolyte interface without a Faradaic reaction. In this case, electrodes are mostly made of high surface-area carbon materials (e.g. CNT, active carbon, carbon aerogel)[890-892]. The second type of electrodes involves redox reaction, namely of metal oxides such as $MnO_2$, $RuO_2$, $SnO_2$, NiO, etc., or redox charge within conductive polymers. In this case, Faradaic processes, i.e. redox-state switching in the oxide occur[893-897]. $TiO_2$ nanotube structures have been considered as promising materials for the second type of supercapacitors due to their combination of a high specific surface area and a defined ion and charge transport directionality, as well as for their semiconductive properties and chemical stability.

However, the reported capacitance of $TiO_2$ nanotubes is not as high as for other active materials such as $RuO_2$, NiO, or conductive polymer nanostructures. Nevertheless, some reports

on the capacitance of anodic $TiO_2$ nanotubes discuss considerably high values (911 $\mu Fcm^{-1}$ at 1 $mVs^{-1}$) which are significantly higher than reported for nanopowder based electrodes (181 $\mu F\ cm^{-1}$ at 1 $mVs^{-1}$)[898,899]. In order to enhance the capacitance, hydrogenated,[900] nitrided $TiO_2$ nanotubes[901,902] or annealed $TiO_2$ in Ar atmosphere[903] were used; such modifications of $TiO_2$ nanotubes lead mainly to an increase of the electric conductivity[900-904]. Nevertheless, the absolute capacitance value of $TiO_2$ nanotubes is still not comparable with conventional materials due to the relatively low electric conductivity of $TiO_2$.

Another approach to use self-organized $TiO_2$ nanotubes for supercapacitors is to use the nanotubes as template of active material. In other words, most active materials such as $RuO_2$ or NiO, and conductive polymers cannot be directly grown in self-organized nanotube structures. By coating or filling the active materials on/into $TiO_2$ nanotubes, several groups target the fabrication of electrodes with an enhanced specific capacitance[905-911].

### 5.7 Biomedical applications

The use of $TiO_2$ nanotubes towards biomedical applications is still in a very early stage but the inherent biotolerance of $TiO_2$,[912,913] combined with a nanotubular geometry, bear considerable potential[50,208,914]. Numerous possibilities are explored, namely towards advanced tissue engineering, novel drug delivery systems, coatings that are antibacterial or enhance osseointegration of a biomedical implant[50,51,226,915-918].

Titanium and titanium alloys are one of the most important biomaterials, due to the high biocompatibility and corrosion resistance of $TiO_2$ layers in biological environments.[912] Implant surface chemistry and morphology on the micro- and nanoscale were widely found to affect

biointeractions[919]. In the past decade, interest in terms of length scales shifted from micrometer to nanometer surface topographies[919-921]. Of all surface nanopatterning techniques, self-organized anodic oxidation is one of the most convenient methods to induce controllable topography and chemistry directly on biomaterials surfaces, i.e. coating of an implant with an ordered nanotube layer presents a viable option of achieving a well-defined and controllable nanotopography.

In this context, anodic 1-D nanotubular structures have been explored in the view of cell interactions, hydroxyapatite formation and even in some in vivo tests. Anodizing allows control over the dimension of the nanotubes and can be used to easily coat complex shapes[922], thus it can be directly applied to coat an implant material. Titania nanotubes obtained by the hydrothermal method have the disadvantage of being obtained finally in powder form[17] and thus cannot be directly obtained aligned perpendicular to the implant surface. Nevertheless, hydrothermal nanotubes have been molded to pellets or tested in composites for *in vitro* or *in vivo* experiments[98].

*$TiO_2$ nanotube interactions with cells*

The influence of $TiO_2$ nanotubes on living cells (their growth, proliferation and differentiation) has been intensely investigated in ongoing nanotube research[50,208,554,923-927]. A key finding is that for cells on anodic $TiO_2$ layers, the nanotube diameter significantly affects virtually any aspect in cell viability (Figure 43). In vitro/in vivo biocompatibility of 1-D titania nanostructures has been tested with mesenchymal stem cells, hematopoietic stem cells, endothelial cells, osteoblasts, and osteoclasts (eukaryotic cells), with tubes varying from 10 – 150 nm in diameter[209,923].

The effect of different nanotube diameter on cell adhesion was first time reported in 2007[208], showing that mesenchymal stem cells react in a very pronounced way to the nanotube diameter.

15 nm diameter nanotubes were shown to strongly promote cell adhesion, proliferation and differentiation, whereas 100 nm diameter nanotubes were found to be detrimental, inducing programmed cell death (apoptosis). Furthermore, osteogenic differentiation of mesenchymal stem cells was stimulated on 15 nm but impaired on 100 nm diameter nanotubes.

Nevertheless, it is worth mentioning that other studies yielded conflicting results (see e.g., Oh et al.[923]): i) small diameter nanotubes (30 nm) promoted adhesion without noticeable differentiation, and ii) larger diameter nanotubes (70-100 nm) induced osteogenic differentiation. In the course of reasoning such different findings, a number of critical factors were screened such as: the role of $TiO_2$ crystallinity, residual fluorides remaining in the nanotubes after formation, surface pretreatment, or the cell type used[554,924-927]. However, investigations of these factors further supported the beneficial effect of 15 nm diameter nanotubular layers and the universal nature of the cell stimulating influence (see Figure 43.a and b). Similar size-dependent responses (Figure 43.c) were present not only for mesenchymal stem cells, but also for hematopoietic stem cells, endothelial cells, osteoblasts and osteoclasts. It was shown that the size effect clearly dominates over nanotube crystal structure (amorphous, anatase or rutile), fluoride content, as well as to some extent over the wetting properties[554,924,925]. Even by changing the substrate material to other valve metals (e.g. $ZrO_2$), similar size effects as for $TiO_2$ nanotubes were observed[924].

Generally, models to explain these findings are based on integrin interactions with the nanotopography or the nature of the adsorbed proteins[928-931]. The main hypothesis for the clear size effect was related to integrin clustering in the cell membrane leading to a focal adhesion complex with a size of about 10 nm in diameter, thus being a perfect fit to the tube openings of about 15 nm – as depicted in Figure 43.d and e; whereas nanotubes larger than 70 nm in diameter

do not support focal contact formation and thus trigger apoptosis. An alternative approach to account for the adhesion of cells to nanotubular structures is based on modeling of the charge distribution at the nanotube tops[932,933]. The attraction between a negatively charged nanotubular surface and a negatively charged osteoblast is assumed to be mediated by charged proteins (proteins with a quadrupolar internal charge distribution such as fibronectin, vitronectin)[932]. Some authors concluded that smaller diameter nanotubes have on average more sharp convex edges per unit area than larger diameter tubes; therefore, a stronger binding affinity is present on smaller diameter nanotubular surfaces[933].

An interesting experiment[934] using micropatterns with defined areas of stimulating 15 nm diameter tubes within apoptotic 100nm tube environments showed that the micropatterned mixed 15 and 100 nm nanotube surfaces responded initially in line with tests on mono-diameter surfaces[208]; however, the extracellular matrix (ECM) produced by "active" cells on regions of small tube diameters led to a spreading of cells to neighboring "unfavorable" larger nanotube regions, enabling after some time settling of vital cells on the 100 nm nanotube patterns (Figure 44.a-c.). Similar effects[934] were observed for nanotubular layers with less well-defined long-range order (as for nanotubes obtained in water-HF electrolytes). Due to the ECM spreading effect, the defects present in nanotube layers act strongly as point of attachment and activation of cells – thus such defects are a crucial experimental factor.

*In vivo* experiments with anodic nanotubular surfaces in adult domestic pigs demonstrated that nanotubular surfaces can enhance collagen type I and BMP-2 expression[922] and that a higher bone contact can be established if implants are coated with nanotubes. Furthermore, recent studies[935] pointed out that nanotube diameter can be designed to support cellular functions of osteoblasts and osteoclasts *in vivo*, including differentiation and protein expression, and therefore

offer a powerful tool for the controlled formation of peri-implant bone around medical implant devices.

$TiO_2$ nanotubes showed potential applications as blood-contacting implant materials, presenting a good hemocompatibility.[936-938] As it is possible to obtain nanotubular structures also on titanium alloys,[40,939] it is a straightforward method of additionaly improving the *in vitro* adhesion, proliferation or osseointegration of nanostructured surfaces. Such alloys include Ti-6Al-4V,[257] Ti-6Al-7Nb,[940,941] Ti26Nb13Ta4.6Zr[250,940] or binary alloys as TiZr,[942,943] TiTa[944] etc.

A different approach using $TiO_2$ nanotube layers and cells interactions concerns macrophage cells in the expectation of controlling and optimizing the inflammatory response to a Ti implant surface.[945-947] A nanotube size dependence of macrophage adhesion and proliferation was suggested,[947] but possible optimization of the inflammatory response is hard to conclude without further *in vivo* tests.

### *$TiO_2$ nanotubes for improved cell interactions*

Efforts were also undertaken to combine the nanotopography provided by $TiO_2$ nanotube surfaces with surface functionalization, namely by the immobilization or attachment of bio-active molecules (proteins, peptides, enzymes)[948-951]. Of potential interest is the functionalization of surfaces with biomolecules which have been shown to be involved in bone development and regeneration during fracture healing, such as growth factors.

Some examples of significant growth factors that were successfully used for implant surfaces and have also been used for $TiO_2$ nanotube modification[948-950] include: EGF that influences the regulation of cellular proliferation or survival of osteoblasts, members of the TGFβ and BMP

family that induce osteoinduction and increase the activity of bone cells including collagen synthesis[952-955].

$TiO_2$ nanotubes were decorated with growth factors using epidermal growth factors (EGF),[948] vascular endothelial growth factors (VEGF),[951] or bone morphogenetic protein 2 (BMP2)[948-950].

Immobilization of EGF[948] was found to affect the response on 100 nm nanotubes, specifically enabling the cells seeded on the 100 nm nanotubes to attach and proliferate. Immobilization of VEGF[951] proved more beneficial when performed via heparin-VEGF interaction, presenting higher bioactivity and also inhibiting bacterial adhesion (*S. aureus*).

Immobilization of BMP-2 can be performed by polydopamine,[949] by an amino-functional organisilane (APTES),[956] or by carbonyldiimidazol[948,950]. BMP-2 functionalized nanotubes via polydopamine proved beneficial for cell proliferation and differentiation (higher osteocalcin and osteopontin levels),[949] while the uncoated nanotubes showed the clear size effects of small diameter nanotubes. BMP-2 functionalized nanotubes via covalent immobilization with carbonyldiimidazol (CDI)[948] (schematic is shown in Figure 44.d) did not induce beneficial conditions on cell adhesion and proliferation, but led to enhanced osteogenic differentiation on 15 nm nanotubes and to chondrogenic differentiation on 100 nm nanotubes (while rescuing MSCs from apoptosis generally occurring on uncoated 100 nm nanotubes)[950] – see Figure 44.d. Functionalization with CDI[948,950] maintained the surface structure of the nanotubes without evidence of blocking the nanotubes, whereas some authors[956] reported functionalization via APTES to alter the surface structure – namely, a blocking of the tube tops. Further approaches consist of loading different biocompatible materials inside the nanotubular structure to increase its bioactivity and osseointegration, e.g. gelatin-coated gold nanoparticles[957].

Recent work[98] using nanotubes obtained via sol-gel method and hydrothermally modified to calcium doped titanate nanotubes showed their beneficial use for bone regeneration; *in vivo* tests in a rat's femur put into evidence a good bioactivity in terms of bone regeneration speed.

After the development of the anti-osteoporosis drug strontium ranelate, strontium has received significant clinical interest and in this respect, Zhao et al.[958] have used a Sr loaded nanotubular structure produced by hydrothermally transforming the titania nanotubes into $SrTiO_3$, to demonstrate no cytotoxicity and good osteogenic activity.

Investigation into surface chemistry modifications in terms of surface wettability alterations[555] indicated no significant influence on the size-dependent cell behavior over large observation times. Nevertheless, drastic changes in the wetting behavior of $TiO_2$ nanotubes from super-hydrophilic to super-hydrophobic (i.e. by means of octadecylphosphonic acid – a self-assembled monolayer),[959] induced modifications in the adsorption of characteristic ECM proteins and improved the attachment of mesenchymal stem cells. Comparing super-hydrophilic conditions with super-hydrophobic conditions, in the latter case, adhesion became independent of the nanotube diameter; however, this effect was only of a temporary nature (3 days)[959]. Superhydrophobic nanotubular surfaces obtained with SAMs showed an improved blood compatibility[960].

Overall, in all cases studied and with all possible modifications of the nanotubular layers (of surface chemistry, substrate material and wettability) it seems evident that the observed size effects dominate over surface chemistry of the nanotubes as long as the material is not actively cytotoxic (e.g. Ag).

*Antibacterial behavior*

The size effects of $TiO_2$ nanotubes on interactions with cells also pose the question of using nanotopography to create bacteria-repellant surfaces. However, till now only few studies have addressed the influence of nanostructured titanium surfaces on bacterial interactions.

One key aspect related to implant materials consists of the fact that cells have to compete with bacteria for the surface of the implant (also known as a "race for the surface")[961]. Once bacteria attach to an implant surface, a biofilm will form with time,[962] and bacteria present in the biofilm cannot be replaced by cells and are difficult to eradicate[963]. Compared to interactions between cells and implant surfaces, interactions involving bacteria are not as well understood[931,964].

Existing research indicates a similar trend for bacteria, as for cells, i.e. that larger diameter nanotubes (60 or 80 nm) decrease the number of live bacteria (*S. aureus* and *S. epidermidis*) as compared to lower diameter (20 nm) nanotubes[965,966]. However, adhesion of bacteria (*S. aureus*, *S. epidermidis*, and *P. aeruginosa)* on nanotubes (60- 70 nm) suggested that the increase in bacterial attachement (compared to conventional or nanorough titanium) could be explained by a decreased number of living bacteria and a large number of adherent dead bacteria, the latter helping in the adhesion of subsequent live bacteria.[967]

Other approaches for decreasing bacterial interactions are based on using titanium alloys containing elements which could inhibit bacteria (eg. zirconium) or using the nanotubes for active coatings (to release preincorporated bactericidal agents as silver ions, growth factors/chemokines/peptides). Grigorescu et al.[968] suggest that small diameter nanotubes on Ti50Zr alloy have higher antibacterial effect against *E. coli* compared to larger diameter.

The active coating method has been extensively studied for $TiO_2$ nanotubes. An often used bactericidal agent is Ag, which can be easily incorporated in nanotubes[969-971]. Incorporation of silver in anodic nanotubes[969,970] and hydrothermally obtained hydrogen titanate nanotubes[971]

showed promising results in view of constructing bacteriostatic materials with long-term silver ion release capability, however showing some cytotoxicity. By further controlling the Ag release rate, it may be possible to achieve an optimum effect, i.e. to accomplish both long-term antibacterial ability and bio-integration (but no cytotoxicity).

Other less commonly investigated inorganic antimicrobial agents that may be suitable for active coatings on titanium implants are copper, fluorine, calcium, zinc, and nitrogen. For example, incorporation of low doses of Zn by hydrothermal treatment in anodized nanotubular layers led to good antibacterial effects and enhanced osseointegration.[972,973]

As mentioned before, immobilization of VEGF inhibited *S. aureus* bacteria adhesion (due to a highly hydrophilic and negatively charged surface)[951].

A preferred method for decreasing bacteria adhesion could be loading the $TiO_2$ nanotubes with antibiotics (such as gentamicin,[916] vancomycin, cephalotin, etc.); however, it is difficult to ensure a constant release rate – as will be discussed in the following section.

*Drug delivery and release of other payloads*

The geometry of $TiO_2$ nanotubes suggests that their surface may be used as a drug-delivery capsule by separating and stabilizing the nanotubular layers or as a drug-eluting coating on biomedical devices.

Nanoscale encapsulated ferromagnetic structures can be transported (or held) at pre-targeted locations within the human body using an external magnetic field. The goal is to perform a specific bioactive function with good precision regarding time and place.

Shrestha et al.[543] showed that $TiO_2$ nanotubes can be completely filled with magnetic $Fe_3O_4$ particles and thus became magnetically guidable. Using suitable linker molecules, the nanotubes

can be coated with drugs – a schematic of the release mechanism is shown in Figure 45.a. Apart from UV reactions, drug release can also be triggered electrically (voltage induce catalysis)[400] or by X-rays[543] allowing *in vivo* treatments through living tissue. $TiO_2$ nanotubes filled with magnetic particles can be used directly for photocatalytic reactions with cells or tissue, such as site-selective killing of cancer cells[543]. Kalbacova et al.[732] showed the possible use of nanotubes as photocatalyst for killing of cancer cells, the only drawback being that there is a need for direct access of UV light to the $TiO_2$ nanotubes.

With regard to drug eluting coatings based on titania nanotubes, reported data are conflicting. Some reports show that for loading with paclitaxel, sirolimus and BSA, there are elusion time constants of minutes,[974,975] while for similar morphologies and drugs the elusion time is of days or weeks[976]. Despite promising possibilities for local drug release on titanium based implant materials, the release kinetics from these drug loaded $TiO_2$ nanotubes are not highly controllable.

It seems that a facile approach for ensuring the use of $TiO_2$ for drug delivery is by simple physical adsorption or by deposition of such drugs from simulated body fluid (SBF), e.g., penicillin/streptomycin, dexamethasone, etc.[975] An increased drug elution was observed in the case of drug deposition from SBF;[975] however, in this case the nanotubular structure was not retained – and it is not clear if the noticed effect was due to the drug or to a possible calcium phosphate coating.

The most difficult aspect is to be able to create a drug delivery system which can allow a controlled delivery (release kinetics). The key point is the membrane (interface) which allows the drug elution. A more complex system using an amphiphilic $TiO_2$ nanotubular structure consisting of nanotubes that provide a hydrophobic cap (monolayer) that does not allow water (body fluid) to enter into the nanotubes unless the cap is opened by a photocatalytic interaction

was developed,[226,245] the concept is schematically represented in Figure 45.b. By this approach, the hydrophobic layer avoids leaching of the hydrophilic drug, and the opening of the hydrophobic layer (achieved by UV induced chain scission of attached organic monolayers) would lead to washing out by the body fluids of the hydrophilic drugs loaded into the nanotubes.

Recently, with the same goal of controlled delivery, the drug-loaded nanotubular layers were capped with biopolymer. For example, coating the structure with chitosan or other polymers, and based on the thickness, properties and degradability, a controllable and sustained drug release could be achieved[977]. Nanocarriers used for designing of nanoparticle drug delivery systems could also be integrated into the nanotubular layers[977].

*In vivo* tests of a drug delivery system consisting of N-acetyl cysteine (NAC)-loaded $TiO_2$ nanotubes used as dental implants in rats[978] indicated that NAC delivery from the nanotubular titania implant led to a higher degree of osseointegration.

*Hydroxyapatite formation*

A key factor for a successful osseointegration of biomedical implants (e.g. dental screws, hip-replacements) is hydroxyapatite formation leading to a bone-binding ability of biomaterials (hydroxyapatite exhibits bioactive behavior and integrates into living tissue resulting in a physicochemical bond between implant and bone). For Ti based biomaterials, the high bone-binding ability has been assigned to a spontaneous modification of the passive Ti surface by calcium and phosphate ions during exposure to a biological environment[979].

The mechanism of hydroxyapatite (HAp) formation is sequential: firstly, $Ca^{2+}$ is adsorbed around a surface OH group (or oxide ion) and then the $HPO_4^{2-}$ group is adsorbed to form the apatite layer[980]. In literature, it has been established that surface hydroxyl groups such as Ti-OH

are efficient inducers for apatite formation[981,982]. Consequently, as the as-grown amorphous nanotube layers contain a high amount of Ti-OH groups on their surface,[983] it would be expected that they are most efficient for apatite formation (particularly more efficient than the layers obtained by annealing at high temperatures with a drastically lower amount of surface hydroxide). However, it was shown that it takes more time to initiate apatite formation on an amorphous surface than on anatase or a mixture of anatase and rutile[984]. In other words, other more effective factors (e.g. crystal structure or surface morphology effects) must override the hydroxide effect.

Regarding $TiO_2$ nanotubularstructures, three important aspects have to be mentioned: i) the nanotubular structure clearly enhances the formation of apatite compared with a flat surface; ii) transforming the tubes from amorphous to crystalline structure facilitates the formation of apatite; iii) a mixed anatase/rutile nanotube structure is even more efficient than a plain anatase structure[984-988]. The formation and growth of apatite precipitates on nanotubular surfaces might be more homogeneous and faster due to a higher number of nuclei formed in the initiation stage, whereas the apatite growth on flat $TiO_2$ proceeds in a more heterogeneous, mushroom-like manner[986].

Further examinations reported that the key factors for hydroxyapatite are crystallographic structure, geometry, porosity, or presence of foreign elements in titania. Different geometry titania nanotube surfaces have also been investigated with respect to the growth kinetics of hydroxyapatite, reporting an acceleration of the hydroxyapatite growth[984-988]. In all studied cases, when crystalline forms of titania are used instead of amorphous titania, an improvement of hydroxyapatite growth was observed[984,986,989]. Furthermore, a nanotube size influence on

hydroxyapatite growth was also observed (indicating 100 nm diameter nanotubes are best for HAp formation)[986].

Using nano-lithographic approaches it was possible to construct well defined microstructures,[988] where the micro-patterned titania nanotubes are surrounded with compact oxide; based on the different hydroxyapatite formation rates, by immersion in simulated body fluid one can preferentially grow hydroxyapatite only on the nanotubular layers, thus being able to deposit hydroxyapatite in designed locations on biomedical devices[990,991].

Calcium phosphate coatings (leading to hydroxyapatite) can also be obtained by alternating immersion methods (AIM) (e.g. cycles of alternating immersions in $Ca(OH)_2$ and $(NH_4)_2HPO_4$)[915,987,992]. It was shown that $TiO_2$ nanotube layers are a highly suitable host for such synthetic hydroxyapatite coatings formed by AIM, leading to a uniform deposition of HAp and even loading the primer HAp inside the nanotubes[987]. Obviously, AIM enhances apatite formation in SBF environments for $TiO_2$ nanotubes. However, most striking are the acceleration effects on amorphous nanotubes when under the same SBF exposure conditions and in the absence of AIM primer no apatite forms, whereas in the presence of AIM primer, several μm thick hydroxyapatite layers can be obtained[987]. The apatite growth rate for amorphous AIM-treated nanotubes was higher than on AIM-treated or AIM-free anatase/rutile nanotube layers[987,992].

Compact and homogenous HAp coatings on the TNTs substrate can be easily prepared by electrochemical deposition[993-995]. When electrodeposition occurs in the presence of a magnetic field, the crystal orientation, shape, and size of HA particles were influenced by the intensities and directions of applied magnetic field that can accelerate the nucleation rate of HA crystals[995]. In order to achieve more uniform HA coating with higher bond strength, alkali treatments in

NaOH before deposition of HA can be used[994]. Following the alkali treatments, titanates are formed on the top of titania nanotubes enhancing the formation of HA during electrodeposition process,[994] or during immersion in simulated body fluids[996].

HAp crystals can be synthesized under hydrothermal conditions, e.g. using urea as precipitation agent to control the HAp growth and to mediate its morphology[997]. Under these conditions,[997] nanotubular surfaces favour prism-like hexagonal HAp crystallization.

$TiO_2$ nanotubes obtained by anodization can be useful also as template for deposition of mixed coating based on HAp and other materials, e.g. carbon nanotubes,[998] improving the bonding strength of the material. Furthermore, $TiO_2$ nanotubes obtained by hydrothermal methods could be used in obtained composite coatings with bioactive materials. An example of the potential of hydrothermally obtained nanotubes is by Ca-nanotubes, which are molded to a shape and immersed for only 1 day in simulated body fluid induced deposition of apatite crystals[98].

Indifferent of the method to obtain nanotubes on the implant surface, it was shown that the nanotubes contribute to increasing surface area necessary for the coating deposition, acting as anchor and enhancing hydroxyapatite nucleation and growth. Due to these aspects, $TiO_2$ nanotubes present an interesting approach to increase osseointegration, enhance bond strength and reducing interfacial failure of implants.

## 6. Summary

In the present article we have tried to give an overview on the state of the art of research on $TiO_2$ nanotubes, their formation, properties and applications. Currently, this is a rapidly expanding research area, where methods for the fabrication of increasingly defined $TiO_2$ nanotubes and aligned arrays produced by anodic and ordered template techniques make fast

progress. In the review, we deal with most common attempts to produce nanotubular $TiO_2$ - that is hydrothermal tubes as well as templated structures and, with a certain emphasis, anodic nanotube arrays. For anodic tubes, the understanding of the interactions of anodization parameters, self-ordering, tube morphology, composition and structure evolve daily and leads to ever refined morphologies and to a largely improved control over the growth of such structures. In parallel, theory and modeling of the anodic self-organizing process enable not only further understanding but also entirely new perspectives for property modification (e.g. models describing plastic flow of the oxide during formation).

In approaches based on templating, certainly the current wide use of ALD techniques provides not only a new platform to increasingly defined structures but also may be one key to further improve tube decoration (cladding) to form core shell structures, with unprecedented definition. Many fields of classic $TiO_2$ applications have been explored using tubular structures, such as their use in dye-sensitized solar cells, photocatalysis or as biomedical coatings, and many attempts have already led to very promising features and findings.

Still a large potential for property improvement exists, for example by reducing defects in anodic tubes, formation of singe crystalline tube walls, shrinkage of tube walls to achieve electronic size effects, or optimized doping and modification approaches. Further improvement of ordering may allow the use of $TiO_2$ nanotube arrays, and thus the high refractive index of $TiO_2$, in even more applications, such as in photonic guiding structures.

In addition to reviewing current knowledge and perspectives to these exciting nanostructures, we hope to have reflected to the readers some of the fascination involved in the research in this field.

**Acknowledgements**

The authors would like to acknowledge DFG and the Erlangen DFG cluster of excellence (EAM), and European Research Council (ERC) for financial support and Manuela Killian, Robin Kirchgeorg, Christopher Schneider, Ning Liu, Sabina Grigorescu, Alexei Tighineanu, Ole Pfoch, Nabeen Shrestha, Sergiu Albu and Chong Yong Lee for their valuable contributions.

**FIGURE CAPTIONS**

Figure 1. Research trend: Number of research article publications in the field of $TiO_2$ nanotubes separated by different fabrication approaches (self-organizing anodization, hydrothermal synthesis, and other approaches) in the period 2000 – 2012 according to "Scopus (Elsevier)".

Figure 2. Beneficial features of nanotubes: a) size confinement (directional or ballistic charge transport), b) atomically curved surface (modified chemical and physical properties), c) electron-hole separation (higher efficiency of charge separation devices), d) ion intercalation (low diffusion length in batteries, electrochromic devices), e) p-n junction and core shell structures (efficient charge separation), f) harvesting functionality (for light absorption of DSSCs, chemical sensors), g) interdigitated electrode structure (e.g. for memristive devices), h) small confined volumes with high observation length (e.g. high-sensitivity/low analyte volume sensing). Reproduced with permission from ref. 50. Copyright 2010 Wiley-VCH Verlag GmbH & Co.

Figure 3. Overview of differently synthesized $TiO_2$ nanotubes: a) Schematic synthesis path, b) typical morphology in TEM or SEM images, c) characteristic features of $TiO_2$ nanotubes formed by different synthetic approaches. $TiO_2$ nanotubes are fabricated by 3.1) hydrothermal method (Reproduced with permission from ref. 105. Copyright 2004 The Royal Society of Chemistry.), 3.2) template assisted formation (Reproduced with permission from ref. 71. Copyright 2002 American Chemical Society, and Reproduced with permission from ref. 89. Copyright 2013 IOP Publishing Ltd.), 3.3) anodic self-organization (Reproduced with permission from ref. 40. Copyright 2011 Wiley-VCH Verlag GmbH & Co.), and 3.4) electrospinning (Reproduced with permission from ref. 163. Copyright 2010 Koji Nakane

and Nobuo Ogata. Originally published in ref. 163 under CC BY-NC-SA 3.0 license. Available online at: http://dx.doi.org/10.5772/8155.).

Figure 4. Schematic illustration of hydrothermal $TiO_2$ nanotube formation: a)-e) Mechanism of hydrothermal formation of $TiO_2$ nanotubes involving delamination of starting crystal into sheets (a/d) and $TiO_6^{2-}$ units (a/b), reassembly (c/d) and scrolling (e). Reproduced with permission from ref. 112. Copyright 2004 Copyright 2004 Materials Research Society. f)-h) Three different types of loop closing f) snail, g) onion, and h) concentric. Reproduced with permission from ref. 105. Copyright 2004 The Royal Society of Chemistry.

Figure 5. Crystal structure a) and morphology evolution b) during hydrothermal $TiO_2$ nanotube formation: a) XRD spectra and b) TEM images. a)-(1.) anatase-type $TiO_2$ raw material, a)-(2.) and b)-(1.) after alkaline reflux (10 M NaOH, 110 °C, 24 h), b)-(2.), a)-(3.) and b)-(2.) after 0.1 M HCl treatment and washing, a)-(4.), b)-(3.) and b)-(4.) final product. Reproduced with permission from ref. 114. Copyright 2010 Springer-Verlag.

Figure 6. Anodic self-organized $TiO_2$ nanotube formation: a) Schematic of growth sequence of $TiO_2$ nanotube growth. b)-c) Oxide growth by field aided transport either to compact oxide or tubular structures in presence of fluorides (rapid fluoride migration leads to accumulation at the metal–oxide interface). d) SEM images of typical $TiO_2$ nanotubes taken at the top, from the fractures in the middle, and close to the bottom of a tube layer, illustrating the gradient in the tube-wall thickness, the inner opening diameter increase and the increasing intertube spacing from bottom to top. Reproduced with permission from ref. 205. Copyright 2007 Wiley-VCH Verlag GmbH & Co. e) Typical current–time (j–t) characteristics after a voltage step in the absence and presence of fluoride ions in the electrolyte. Either compact oxide, CO (fluoride free) or porous/tubular metal oxide, PO formation (fluoride containing) with different morphological stages (I–III). The inset shows typical linear sweep voltammograms (j–U curves) for different fluoride concentrations resulting in either electropolished metal

(EP, high fluoride concentration), compact oxide (CO, very low fluoride concentration), or tube formation (PO, intermediate fluoride concentration). f) SEM image of the fluoride rich layer at the bottom of nanotubes. Reproduced with permission from ref. 40. Copyright 2011 Wiley-VCH Verlag GmbH & Co. g) $TiO_2$ nanotubes prepared by rapid breakdown anodization in chloride containing electrolyte. Reproduced with permission from ref. 156 and 157. Copyright 2007 Wiley-VCH Verlag GmbH & Co. and Copyright 2011 Elsevier B.V.

Figure 7. Critical parameters for anodic $TiO_2$ nanotube formation: a)-b) Diagrams of different morphologies formed during anodization in fluoride containing electrolyte depending on a) applied voltage and water concentration in electrolyte (Reproduced with permission from ref. 204. Copyright 2010 Wiley-VCH Verlag GmbH & Co.) and b) applied voltage, hydrodynamic parameters and $F^-$ concentration in electrolyte (Reproduced with permission from ref. 207. Copyright 2012 Elsevier B.V.). c)-e) Influence of applied voltage on c)-d) nanotube diameter and e) growth rate shown for classic anodization in a fluoride containing ethylene glycol electrolyte and the effect of complexing additives EDTA and lactic acid. Data taken from various sources.

Figure 8. Different morphologies in $TiO_2$ nanotube formation: SEM images of a) smooth nanotube walls (Reproduced with permission from ref. 226. Copyright 2010 Elsevier B.V.), b) rippled $TiO_2$ nanotube wall (Reproduced with permission from ref. 226. Copyright 2010 Elsevier B.V.) c) tube in tube morphology, d) "brain" morphology on bottom of the $TiO_2$ nanotubes (Reproduced with permission from ref. 213. Copyright 2012 Elsevier B.V.), e) double and f) single wall nanotubes (with corresponding TEM image underneath) (Reproduced with permission from ref. 229. Copyright 2013 The Royal Society of Chemistry). g) Carbon content in the double wall and single wall nanotubes analyzed by EDX measurements. The data correspond to 3 different cross-sectional locations (red boxes in SEM image). Reproduced with permission from ref. 229. Copyright 2013 The Royal Society of Chemistry. h) Thermal desorption profile for $CO_2$ and $H_2O$ by TGA-MS analysis for the double

wall and single wall $TiO_2$ nanotubes. Reproduced with permission from ref. 229. Copyright 2013 The Royal Society of Chemistry.

Figure 9. Highly ordered $TiO_2$ nanotube: a)-c) SEM image of $TiO_2$ nanotube "stumps", a) top-view, b) cross-section. Reproduced with permission from ref. 238. Copyright 2013 Elsevier B.V. and c) dewetted Pt layer to from single particle per tube structures. Reproduced with permission from ref. 236. Copyright 2013 Wiley-VCH Verlag GmbH & Co.

Figure 10. Altered $TiO_2$ nanotube layers and membranes: SEM images of a)-b) bamboo-type tubes grown by altering voltage conditions during growth with a) a sequence of 5 min at 120 V and 5 min at 40 V for 4 h anodization and b) with a sequence of 1 min at 120 V and 5 min at 40 V for 12 h anodization in a $NH_4F$/ethylene glycol electrolyte Reproduced with permission from ref. 228. Copyright 2008 American Chemical Society. c)-d) SEM images of branching of nanotubes by voltage stepping. Reproduced with permission from ref. 246 and 227. Copyright 2011 American Chemical Society and Copyright 2008 Wiley-VCH Verlag GmbH & Co. e) SEM image of an example of a hydrophobic/hydrophilic stack of nanotubes. Reproduced with permission from ref. 245. Copyright 2009 American Chemical Society. f) HAADF-STEM images of $TiO_2/Ta_2O_5$ superlattice nanotubes. Reproduced with permission from ref. 248. Copyright 2010 Wiley-VCH Verlag GmbH & Co. g) SEM image of bottom opened $TiO_2$ nanotube membranes and h) optical image of full nanotube membrane lifted-off from substrate. Reproduced with permission from ref. 173. Copyright 2008 Wiley-VCH Verlag GmbH & Co.

Figure 11. Towards higher order: a) Schematics of the indentation of Ti under a pressure of 25 $kN/cm^2$ (I: imprint master stamp consisting of a hexagonal array of pyramids of $Si_3N_4$, S: mechanically polished Ti substrate, SEM image of b) the nanoindented surface of the Ti substrate and c) anodized Ti at 10 V in ethanolic 0.5M HF for 240 min. Reproduced with permission from ref. 282. Copyright 2004 Elsevier Ltd. d)-h) FIB induced initiation sites:

SEM image of d) FIB guiding pattern in hexagonal arrangement with 300 nm distances. The upper inset is the AFM image and the lower inset is the surface topology along the line in upper inset. SEM images of e) $TiO_2$ nanotube arrays from Ti with the hexagonal FIB pattern (Reproduced with permission from ref. 284. Copyright 2011 The Royal Society of Chemistry) and f) FIB patterned concaves with different pore depths on an electropolished Ti surface. The inset is the surface topology along the line in f). Reproduced with permission from ref. 309. Copyright 2011 American Chemical Society. g) Surface and h) cross-sectional SEM image of $TiO_2$ nanotubes formed by anodization of f). Reproduced with permission from ref. 309. Copyright 2011 American Chemical Society.

Figure 12. Overview of various factors influencing self-ordering: a) oxide formation initiation at a nucleation spot and resulting hemispherical oxide dome formation. b) Stress effects at growing thin films. Stress*thickness vs. thickness curves obtained upon anodization in $HNO_3$ at various current densities indicating transition from compressive to tensile stress with increasing oxide thickness. Reproduced with permission from ref. 317. Copyright 2006 The Electrochemical Society. c) (1-2) Volume filling by pore widening. Reproduced with permission from ref. 319. Copyright 2008 Wiley-VCH Verlag GmbH & Co. (3) Self-organization of cell arrangement at high-current-density (high electric field). Reproduced with permission from ref. 322. Copyright 2005 Elsevier Ltd. d)-e) Results of simulation of steady-state growth, and comparison of experimental and simulated tracer profiles.[336] d) Current lines (thin red lines) and potential distribution (color scale) for anodic film growth in oxalic acid at 36 V, e) velocity vectors (arrows) and mean stress (color scale) for the same conditions as in d). f) Comparison of W tracer profiles from TEM cross-sectional images with simulated profiles. Reproduced with permission from ref. 333. Copyright 2006 The Electrochemical Society. g) Effect of oxide formation efficiency on stability and morphology for growth of anodic $Al_2O_3$ at an electric field of 0.8 V nm−1. h) Comparison of measured oxide formation efficiencies with predicted limits of self-ordered growth of $Al_2O_3$ pores and $TiO_2$ nanotubes. Reproduced with permission from ref. 172. Copyright 2012 Macmillan Publishers Ltd.

Figure 13. Annealing of hydrothermal $TiO_2$ nanotubes: a) X-ray diffraction patterns for different heat treatments and corresponding morphology change. b) Temperature dependence of BET surface area of cation-doped $TiO_2$ nanotubes (cation concentration of ca. ~0.1 mol%) and corresponding morphology change. Reproduced with permission from ref. 114. Copyright 2010 Springer-Verlag.

Figure 14. Crystallization of anodic $TiO_2$ nanotubes: a) XRD spectra of nanotubes annealed at 500, 600, 700, 800 and 900 °C for 12 h. Reproduced with permission from ref. 173. Copyright 2008 Wiley-VCH Verlag GmbH & Co. b) Size of crystals in tube walls in dependence of the heating rate for nanotubes annealed at 500 °C (from different methods); the inset shows a comparison to the crystal for nanotubes annealed at 400 °C and 500 °C. Data taken from various sources. c)-d) TEM images of annealed c) double wall and d) single wall $TiO_2$ nanotubes. Reproduced with permission from ref. 229. Copyright 2013 The Royal Society of Chemistry. (e) Size distribution of grains for double wall (red) and single wall (black) $TiO_2$ nanotubes after annealing at 500 °C for 1h. Reproduced with permission from ref. 229. Copyright 2013 The Royal Society of Chemistry. f) SEM images of $TiO_2$ nanotubes after 3 days of storage in $H_2O$[336]. Reproduced with permission from ref. 365. Copyright 2012 Elsevier Ltd.. g) Corresponding XRD spectra for such "water annealed" nanotubes[336].

Figure 15. Crystallization induced by cycling of Li-insertion: Amorphous $TiO_2$ nanotubes are crystallized at discharge stages of cycling. a) Synchrotron XRD measurements at the stages of discharged to 1.25 V vs. $Li/Li^+$ (brown curve) and 0.9 V vs. $Li/Li^+$ (red), followed by charged to 2.5 V vs. $Li/Li^+$ (black). Inset shows pre-edge feature in the Ti K-edge XANES for the same samples. b) SEM and c) TEM images of charged $TiO_2$ nanotubes. d) HR-TEM image and e) selected area electron diffraction (SAED) at a region of the same tube (red square). Reprinted with permission from ref. 349. Copyright 2011 American Chemistry Society.

Figure 16. Crystallization of $TiO_2$ nanotubes induced by TEM observation: High-resolution TEM images of amorphous $TiO_2$ nanotubes before (a and b) and after (c and d) e-beam exposure for several minutes. Reproduced with permission from ref. 372. Copyright 2010 Wiley-VCH Verlag GmbH & Co.

Figure 17. Photocurrent transients: Photocurrent transients excited at a wavelength of 350 nm for 10 s with a) as-formed compact anodic $TiO_2$ layer, b) annealed compact $TiO_2$ layer, c) as-formed $TiO_2$ nanotubes, d) annealed $TiO_2$ nanotubes. Reprinted with permission from ref. 376. Copyright 2011 Elsevier Ltd.

Figure 18. Size confinement effects: a) Experimental observation of quantum confinement for $TiO_2$ nanoparticle (left) and ALD layer (right): Optical band gap for different particle sizes and bandgap shift for $TiO_2$ ALD film thicknesses. Also shown are expected quantum shifts calculated using the Brus model with reduced effective masses of 0.50, 0.10 and 0.05. Reproduced with permission from ref. 384. b)-c) Band diagrams of hydrothermal structure for: b) a 2-dimensional nanosheet and c) a quasi-1-dimensional nanotube. Reproduced with permission from ref. 37. Copyright 2009 The Royal Society of Chemistry. d) Energy vs. density of states for nanosheet ($G_2D$) and nanotubes ($G_1D$). $E_G^{1D}$ and $E_G^{2D}$ are the band gaps of 1-D and 2-D structures, respectively and kx and ky are the wave vectors. Reproduced with permission from ref. 37. Copyright 2009 The Royal Society of Chemistry. e) Phonon confinement in anodic $TiO_2$ nanotubes and $TiO_2/Ta_2O_5$ superlattice of Figure 10.f): Raman spectra of the $E_g$ mode for annealed $TiO_2$ /$Ta_2O_5$ superlattice nanotube arrays, $TiO_2$ nanotubes and large grain $TiO_2$ anatase crystal powders. The FWHM and peak shift of the Raman line in comparison with theoretical calculations as a function of $TiO_2$ feature size (solid lines) are also shown. Reproduced with permission from ref. 248. Copyright 2010 Wiley-VCH Verlag GmbH & Co..

Figure 19. Conductivity and defects of $TiO_2$ nanotubes: a) Conductivity of $TiO_2$ nanotube layers for different annealing temperatures performed with 2-point conductivity measurement (data taken from ref. 30). b)-d) procedure of 4-point conductivity measurement with a single nanotube. b) Single $TiO_2$ nanotube before fixing it on the substrate, c) The nanotube fixed with WCx, d) After producing the electrical contacts. Reproduced with permission from ref. 32. Copyright 2013 AIP Publishing LLC. e) Current-voltage characteristics for single $TiO_2$ nanotubes with different temperatures. Inset shows Nyquist plot of a single $TiO_2$ nanotube (the red line is the fit to the experimental data assuming the equivalent circuit shown in the inset). Reproduced with permission from ref. 32. Copyright 2013 AIP Publishing LLC. f) ESR spectra of calcinated titanate nanotubes under vacuum conditions. Reproduced with permission from ref. 399. Copyright 2007 Springer-Verlag. g) Electron mobility for various forms of $TiO_2$. Reproduced with permission from ref.396. Copyright 2013 Wiley-VCH Verlag GmbH & Co.

Figure 20. Photoelectrochemical and electrochemical capacitance properties of $TiO_2$ layers: a) Potential dependence of the space charge layer capacitance and photocurrent density and b) IPCE for a $TiO_2$ compact oxide (60 nm), a nanoparticle layer (2.4 µm thick), and a nanotube layer (2.4 µm thick). Capacity was measured at 1 Hz. The photocurrent and IPCE were measured in 0.1 M $Na_2SO_4$ under incident wavelength of 350 nm. c) Carrier collection time ($\tau_c$) from IMPS (intensity modulated photocurrent spectra) measurement of anodic nanotube layers with different thicknesses using an alternating diode light source of 360 nm. Reproduced with permission from ref. 35. Copyright 2010 The Electrochemical Society.

Figure 21. Schematic illustration of energy level positions for various dopants in $TiO_2$ relative to band-edges. Reprinted with permission from ref. 48. Copyright 2012 Wiley-VCH Verlag GmbH & Co.

Figure 22. Various hydrothermal $TiO_2$ nanotube-metal nanocomposites: (a) Pd nanoparticle loaded $TiO_2$ nanotubes, (b) Ag nanoparticle loaded $TiO_2$ nanotubes, (c) Ni nanoparticles loaded $TiO_2$ nanotubes, (d) ZnS quantum dot loaded $TiO_2$ nanotubes. Reprinted with permission from ref. 114. Copyright 2010 Springer-Verlag..

Figure 23. Self-assembled monolayer (SAM) modification of $TiO_2$ surfaces: a) Attachment of SAMs to –OH terminated surfaces [581]. b) Immobilization of proteins (R) via linker SAMs (protein not to scale)[581]. (c) SEM image of microscopic wetting of $TiO_2$ nanotube surfaces, showing that wetting takes place preferentially between tubes. Reprinted with permission from ref. 590. Copyright 2010 The Electrochemistry Society.

Figure 24. Schematic diagram of DSSCs: a) principle of a dye-sensitized solar cell; b) different configuration using nanoparticle and nanotube layers: front-side illuminated (left) and back-side illuminated (right) (⊖ Platinized FTO, ⊜ Iodine electrolyte, ⊛ $TiO_2$ layer, ☙ FTO substrate, and ❧ Ti metal).

Figure 25. Key factors affecting efficiency of $TiO_2$ nanotube based DSSCs: Comparison of solar cell efficiencies for a) different annealing temperatures, b)-c) different tube length for b) back side illuminated and c) front side illuminated cell configuration, and d) different tube diameter at same length for back-side illuminated cell configuration. Data are taken from various sources.

Figure 26. Effect of $TiCl_4$ treatment ($TiO_2$ nanoparticle decoration) on $TiO_2$ nanotubes: Top and cross-sectional SEM images of a), c) before and b), d) after $TiCl_4$ treatment of $TiO_2$ nanotubes. Reproduced with permission from ref. 535. Copyright 2009 Elsevier B.V. e) TEM image of $TiCl_4$ treated $TiO_2$ nanotubes. Reproduced with permission from ref. 45. Copyright

2010 The Royal Society of Chemistry. f) Comparison of solar cell efficiency before and after $TiCl_4$ treatment for back-side illuminated cell configuration.

Figure 27. Scheme of photo-induced processes at a $TiO_2$ semiconductor/electrolyte interface: Light (hν) excites valence band electron to conduction band. Electron and hole react with environment acceptor (A) and/or donor (D). Acceptor and donor species are reduced and oxidized (=photocatalytic reactions). Competing reactions are recombination and trapping of electrons and holes (=reducing photocatalytic efficiency). Grey boxes give typical reactants and reaction products in photocatalytic reactions on $TiO_2$. Reprinted with permission from ref. 48. Copyright 2012 Wiley-VCH Verlag GmbH & Co.

Figure 28. Redox potential and work function: Relative positions of various red-ox couples and work functions of various metals relative to the band-edges of $TiO_2$. Reprinted with permission from ref. 48. Copyright 2012 Wiley-VCH Verlag GmbH & Co.

Figure 29. Key factors affecting the activity of anodic $TiO_2$ nanotube based photocatalysts: Photocatalytic degradation of AO7 for different a) annealing conditions (temperature and atmosphere) using $TiO_2$ nanotubes of thickness ~ 1.5 μm. Reproduced with permission from ref. 674. Copyright 2012 Springer-Verlag. b)-c) tube formation and nanotube thickness, $TiO_2$ nanotubes grown in b) glycerol and c) ethylene glycerol based electrolyte. Reproduced with permission from ref. 452. Copyright 2010 Wiley-VCH Verlag GmbH & Co. d) Photocatalytic degradation of AO7 with doped nanotubes (mixed oxides) and particle decoration on $TiO_2$ nanotubes. Reproduced with permission from ref. 40. Copyright 2011 Wiley-VCH Verlag GmbH & Co. e) Energy diagram of n-type $TiO_2$ semiconductor for the case $E_{f,sc}>E_{f,redox}$ and applying an anodic bias (+ΔU) that leads to an increase in band bending and an increase in the space charge layer width(W). Reproduced with permission from ref. 48. Copyright 2012 Wiley-VCH Verlag GmbH & Co. f) Schematic of effect of noble metal particle on a n-type

semiconductor surface ($TiO_2$) in a band diagram Reproduced with permission from ref. 48. Copyright 2012 Wiley-VCH Verlag GmbH & Co.

Figure 30. Photocatalytic activity of hydrothermal $TiO_2$ nanotubes and comparison with P25 nanoparticles.: Effect of several modifications of $TiO_2$ on photocatalytic $H_2$ production in glycerol/water mixture solution under UV/vis (17% of UV and 83 % of visible) light illumination. (Vsolution = 400 mL, glycerol concentration = 10 vol.%, catalyst amount = 0.4 g) Reproduced with permission from ref. 409. Copyright 2012 John Wiley & Son, Ltd.

Figure 31. Band positions of $TiO_2$ anatase and rutile: a) Energy bands of $TiO_2$ to relative redox potentials of water as a function of pH according to Fujishima et al.[559] Reproduced with permission from ref. 559. Copyright 2008 Elsevier B.V. b) Alternative model for relative positions of valence and conduction band for the anatase and rutile interface according to David et al.[670]. Reproduced with permission from ref. 670. Copyright 2013 Macmillan Publishers Ltd.

Figure 32. Doped $TiO_2$ nanotubes for water splitting: Enhancement of photoelectrochemical water splitting current using Nb doped (Ti0.1Nb) $TiO_2$ tubes, shown as dark/light I–V curves for TiNb and plain $TiO_2$ nanotube layers. The measurements were carried out in 1M KOH under AM1.5 (100 $mW/cm^2$) conditions. Reproduced with permission from ref. 31. Copyright 2011 The Royal Society of Chemistry.

Figure 33. Ion intercalation: Schematic diagram of beneficial effects of nanotubular structure regarding diffusion length L for lithium ion ($L_{ion}$) lattice insertion and electron ($L_{electron}$) in $TiO_2$ nanotube structures.

Figure 34. Electrochromic device with $TiO_2$ nanotubes: a) Schematic diagram and optical images of a electrochromic device made from transparent nanotube electrodes on TCO glass. Shown are different coloration states upon anodic (bleached) and cathodic (colored) polarization. Reproduced with permission from ref. Error! Bookmark not defined.. Copyright 2009 American Chemistry Society. b) Optical images during in situ switching of a 1 µm NT layer (top), 10 µm NT layer (middle), and 10 µm NP layer (bottom). Reproduced with permission from ref. 759. Copyright 2008 Wiley-VCH Verlag GmbH & Co.

Figure 35. Li-ion battery using anodic $TiO_2$ layers: a) Top and cross-section SEM images of $TiO_2$ nanotubes that are used for electrode in Li-ion battery. b) Schematic diagram of Li-ion battery device. c)-d) Example of cyclic voltammetry and galvanostatic charge/discharge behavior during Li-ion battery test.

Figure 36. $TiO_2$ nanostructures for Li-ion batteries: Specific capacity vs. the current density for various $TiO_2$ nanostructures formed by different synthetic methods.

Figure 37. Anodic $TiO_2$ nanotubes for Li-ion batteries: a) Areal capacity and b) normalized areal capacity vs. the current density for different thicknesses of anodic self-organized $TiO_2$ nanotubes.

Figure 38. $TiO_2$ nanotube modifications for Li-ion batteries: Normalized areal capacity vs. the current density for different modifications of anodic $TiO_2$ nanotube electrodes.

Figure 39. Sensors: Schematic representation of sensing mechanisms for ceramic-based sensor devices. a) Change in the Schottky barrier between semiconductor and metal contact

when active gas arrives in that area. b) Gas triggers change in the conductivity of the grain surfaces. c) Gas triggers change in contact resistivity between two touching grains.

Figure 40. Comparison of response and sensitivity in $H_2$ sensing using different $TiO_2$ nanotube arrays: a) anodic $TiO_2$ nanotubes with a length of 5μm at a working temperature of 215 °C; b) $TiO_2$ nanotube produced by atomic layer deposition into anodic aluminium oxide nanopores, at sensing temperature of 100 °C. Data taken from ref. 47 and 835.

Figure 41. Temperature effect for $H_2$ sensing: a) Response at various hydrogen concentrations and b) response times of anodic $TiO_2$ nanotube sensors to 1000 ppm hydrogen as a function of sensing temperature. Reproduced with permission from ref. 835. Copyright 2011 Elsevier B.V.

Figure 42. Memristive effect with $TiO_2$ layer: a) I-V curve of a memristive $TiO_2$ nanotube/Pt layer as shown in the cross-sectional SEM image of figure b) formed at the bottom of a $TiO_2$ nanotube stump . Resistive switching in the I-V curves occurs at approximately $\pm$ 1 V. The inset shows I-V curves for conductive and resistive state (obtained after voltage pulses $\pm$ 1.5 V). Reproduced with permission from ref. 237. Copyright 2013 Elsevier B.V.

Figure 43. Size-effect of $TiO_2$ nanotubes diameter on cell interactions: a)-b) Fluorescence images of adherent rat mesenchymal stem cells (GFP-labeled) on $TiO_2$ nanotubes of 15 nm and 100 nm diameters. 3 days after plating an identical density of 5000 cells/$cm^{-2}$ on the two surfaces a strongly different cell density and cell spreading is obtained for 15nm (43.a) and 100nm (43.b). Reproduced with permission from ref. 948. Copyright 2011 The Royal Society of Chemistry. c) Comparison of cell activity for mesenchymal stem cells, primary human osteoblasts, osteoclasts and endothelial cells for different $TiO_2$ nanotube diameter (3 days after seeding). d)-e) Model illustrating how nanotubes of different diameter may affect

formation of focal contacts. Reproduced with permission from ref. 208. Copyright 2007 American Chemistry Society.

Figure 44. a)-c) Effect of neighboring 15 nm/100 nm diameter tubes on cell activity shown one fluorescence micrographs of GFP labeled MSC on concise 15 nm/100 nm patterns: a) 500 µm wide stripe pattern after 1 day in culture, b) after 3 days in culture, c) cell numbers on isolated tube surfaces (15 nm, 100 nm, respectively) and on 15 nm/100 nm stripe pattern. Reproduced with permission from ref. 934. Copyright 2012 Elsevier Ltd. d) Effect of immobilized growth factors with schematic illustration of BMP-2 immobilization by covalent reaction of an amino group of the protein with grafted CDI. Fluorescence micrographs show that differentiation of MSCs to osteoblasts is strongly supported on 15 nm BMP-2-coated nanotubes but much less on uncoated nanotubes as indicated by osteocalcin staining. No osteogenic differentiation occurred on 100 nm nanotubes. Reproduced with permission from ref. 950. Copyright 2012 Wiley-VCH Verlag GmbH & Co.

Figure 45. Drug delivery principles using $TiO_2$ nanotubes: a) Diagram showing the release principle of active molecules (model drug) from the functionalized magnetic $TiO_2$ nanotubes upon irradiation with UV light. A fluorescent dye (active molecule) was attached to the $TiO_2$ nanotubes with a siloxane linker. Reproduced with permission from ref. 543. Copyright 2009 Wiley-VCH Verlag GmbH & Co. b) Schematic of the amphiphilic $TiO_2$ nanotube arrays fabricated by a two-step anodization procedure combined with hydrophobic monolayer modification after the first step. The outer hydrophobic barrier provides an efficient cap to the drug inside the nanotube, providing also a controlled drug release after photocatalytic cap removal. Reproduced with permission from ref. 245. Copyright 2009 American Chemistry Society.

**Figure 1**

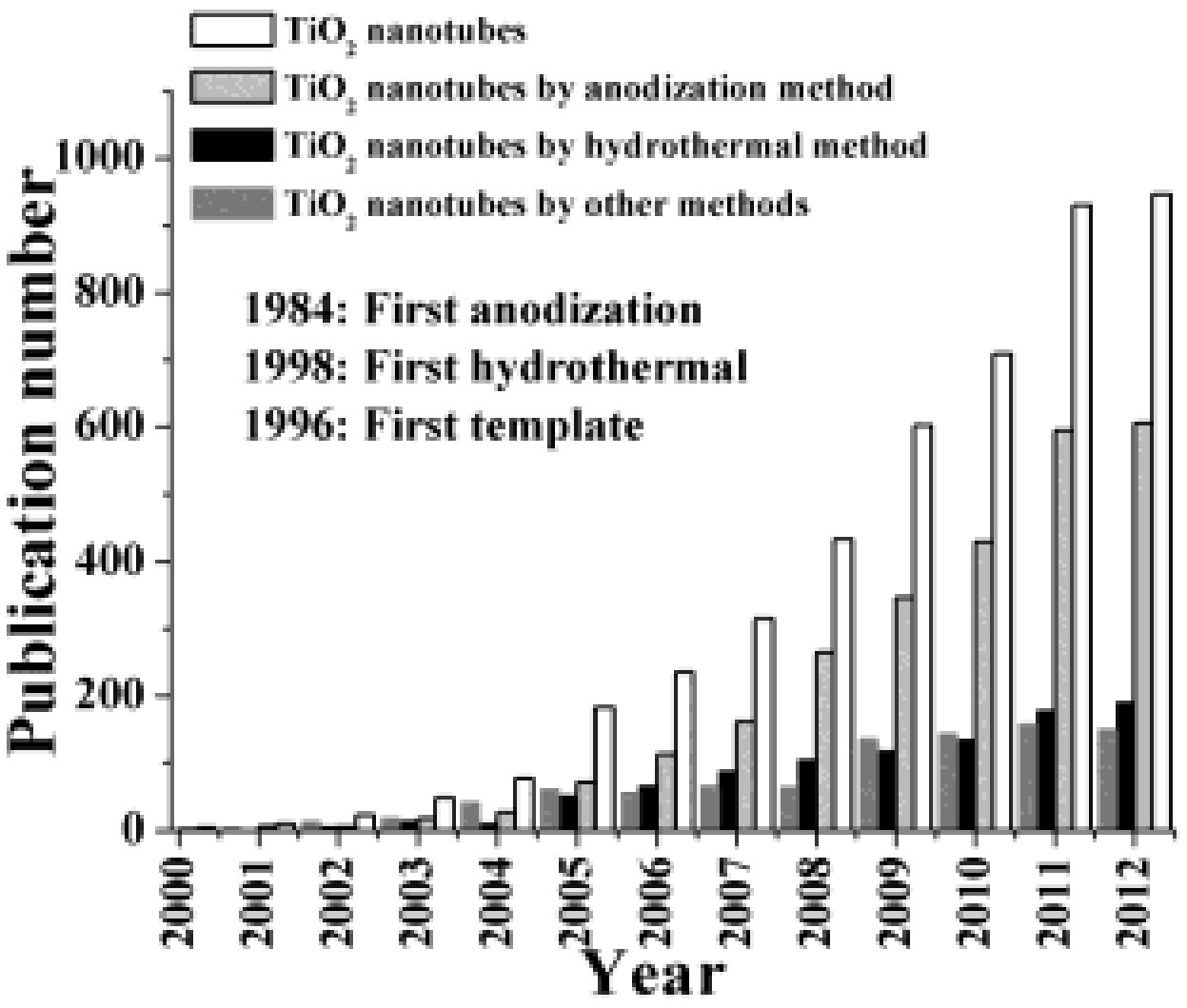

**Figure 2**

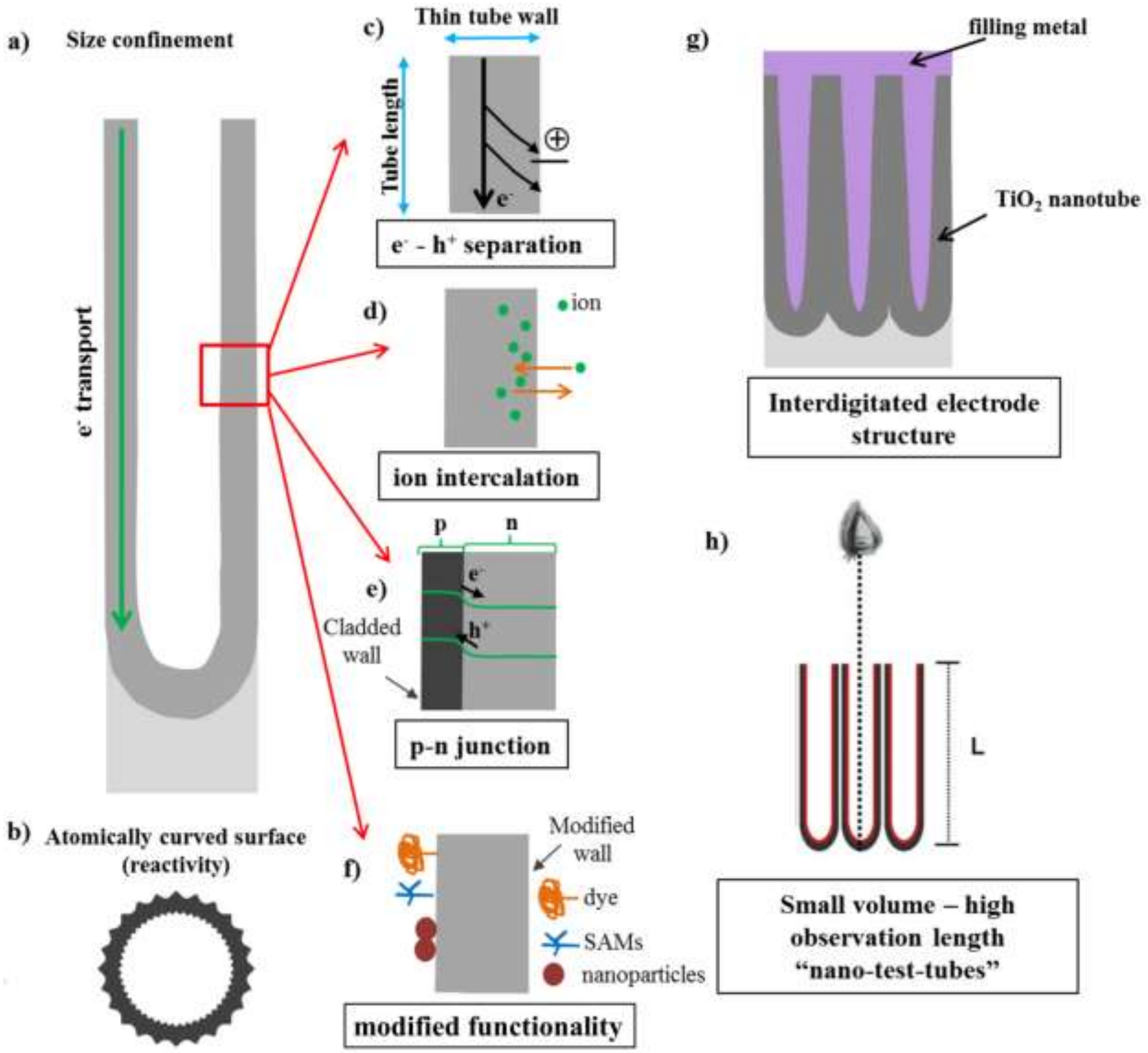

**Figure 3**

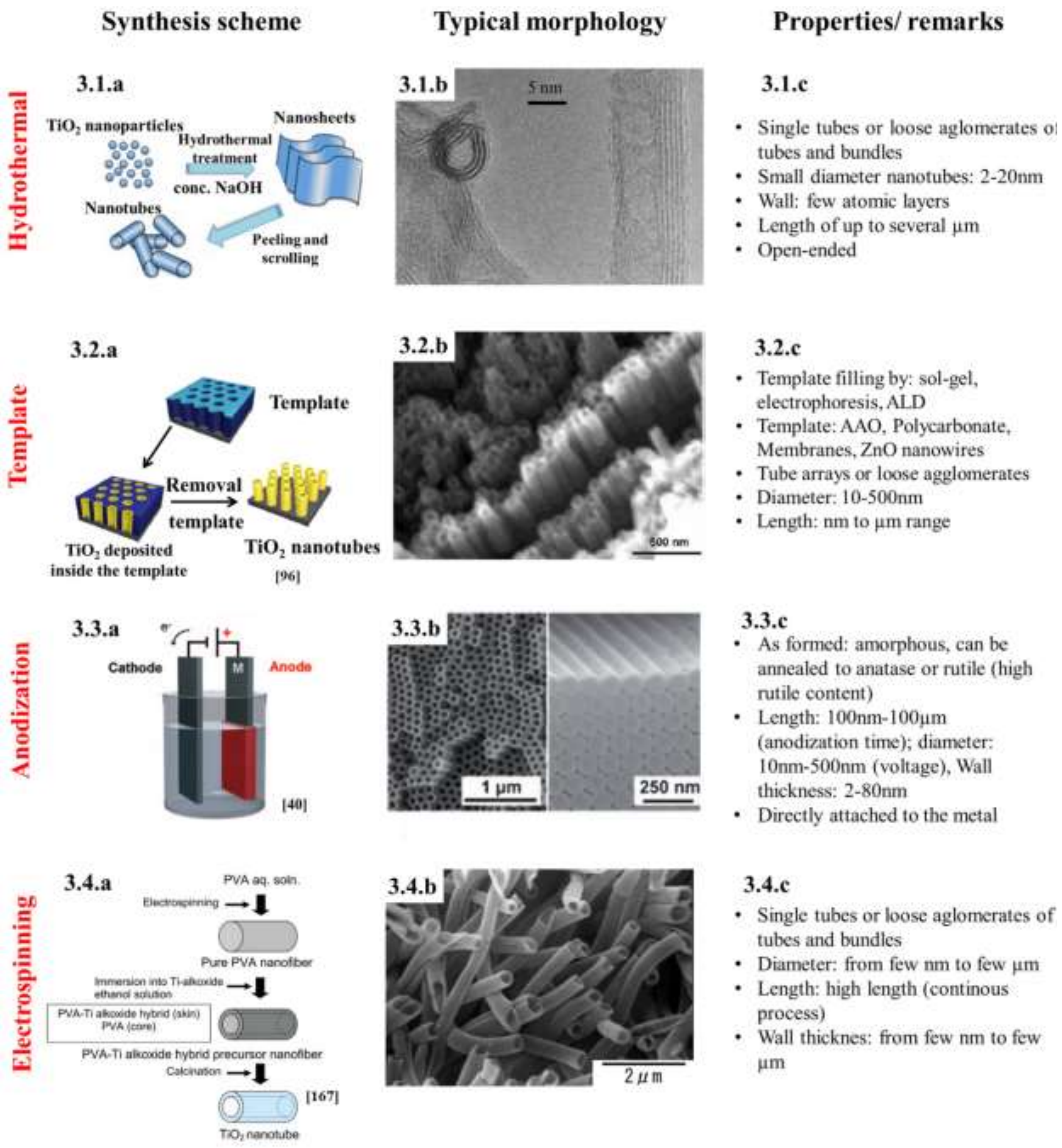
Synthesis scheme
Typical morphology
Properties/ remarks
Hydrothermal
3.1.a
TiO2 nanoparticles
Hydrothermal treatment
conc. NaOH
Nanosheets
Nanotubes
Peeling and scrolling
3.1.b
5 nm
3.1.c
• Single tubes or loose aglomerates of tubes and bundles
• Small diameter nanotubes: 2-20nm
• Wall: few atomic layers
• Length of up to several µm
• Open-ended
Template
3.2.a
Template
Removal template
TiO2 deposited inside the template
TiO2 nanotubes
[96]
3.2.b
500 nm
3.2.c
• Template filling by: sol-gel, electrophoresis, ALD
• Template: AAO, Polycarbonate, Membranes, ZnO nanowires
• Tube arrays or loose agglomerates
• Diameter: 10-500nm
• Length: nm to µm range
Anodization
3.3.a
Cathode
M
Anode
[40]
3.3.b
1 µm
250 nm
3.3.c
• As formed: amorphous, can be annealed to anatase or rutile (high rutile content)
• Length: 100nm-100µm (anodization time); diameter: 10nm-500nm (voltage), Wall thickness: 2-80nm
• Directly attached to the metal
Electrospinning
3.4.a
PVA aq. soln.
Electrospinning
Pure PVA nanofiber
Immersion into Ti-alkoxide ethanol solution
PVA-Ti alkoxide hybrid (skin)
PVA (core)
PVA-Ti alkoxide hybrid precursor nanofiber
Calcination
[167]
TiO2 nanotube
3.4.b
2 µm
3.4.c
• Single tubes or loose aglomerates of tubes and bundles
• Diameter: from few nm to few µm
• Length: high length (continous process)
• Wall thicknes: from few nm to few µm

**Figure 4**

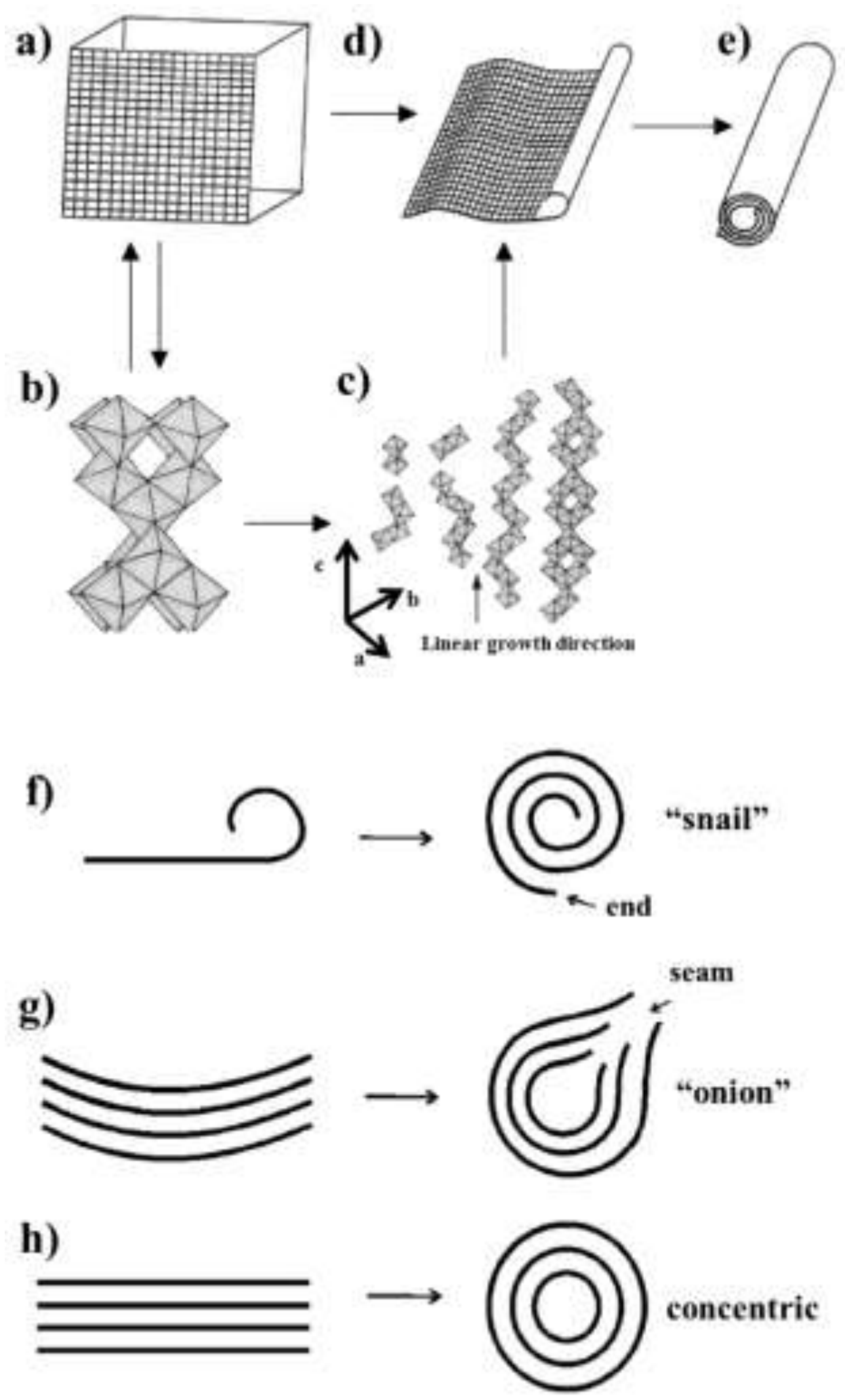

**Figure 5**

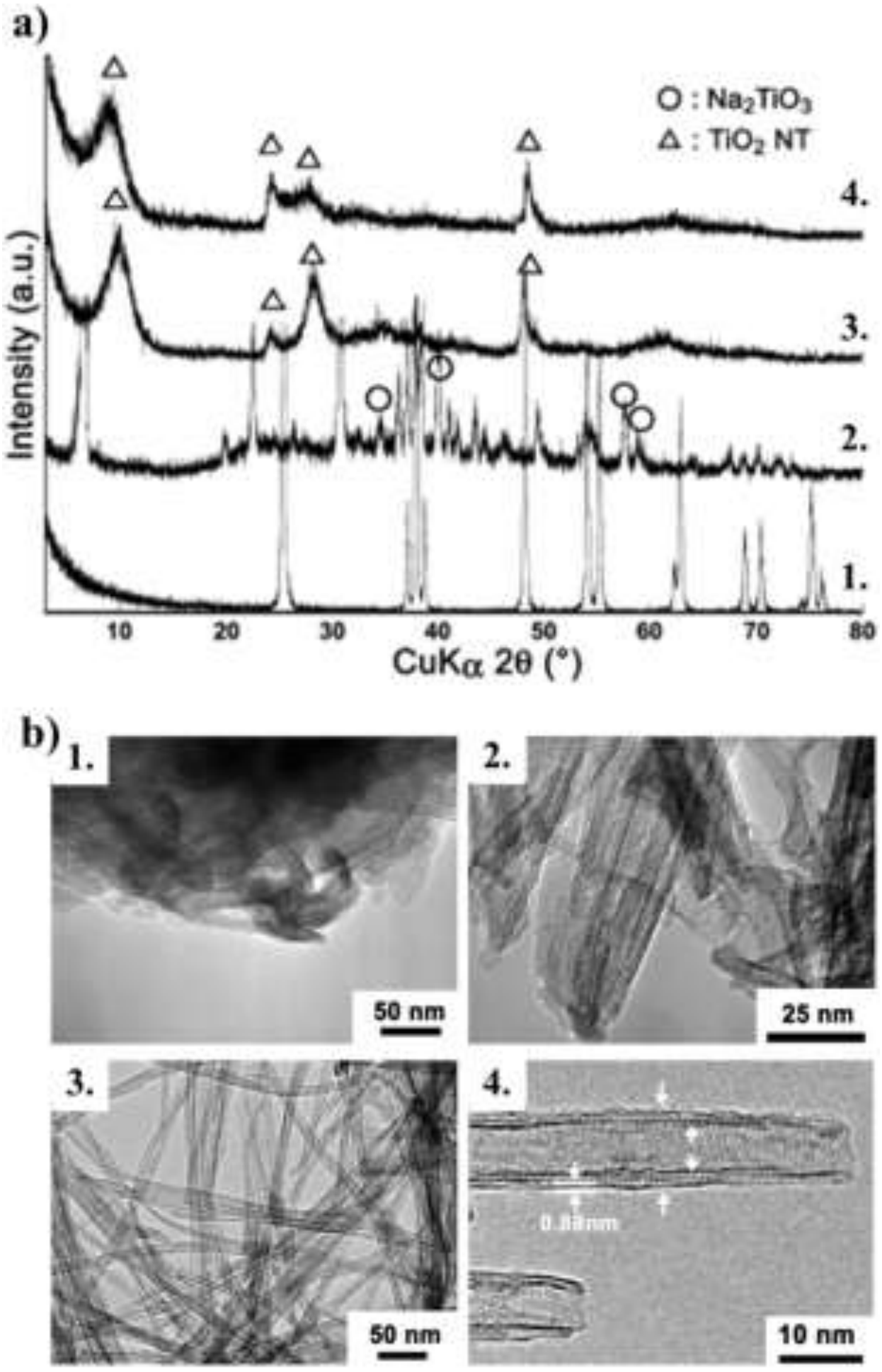

**Figure 6**

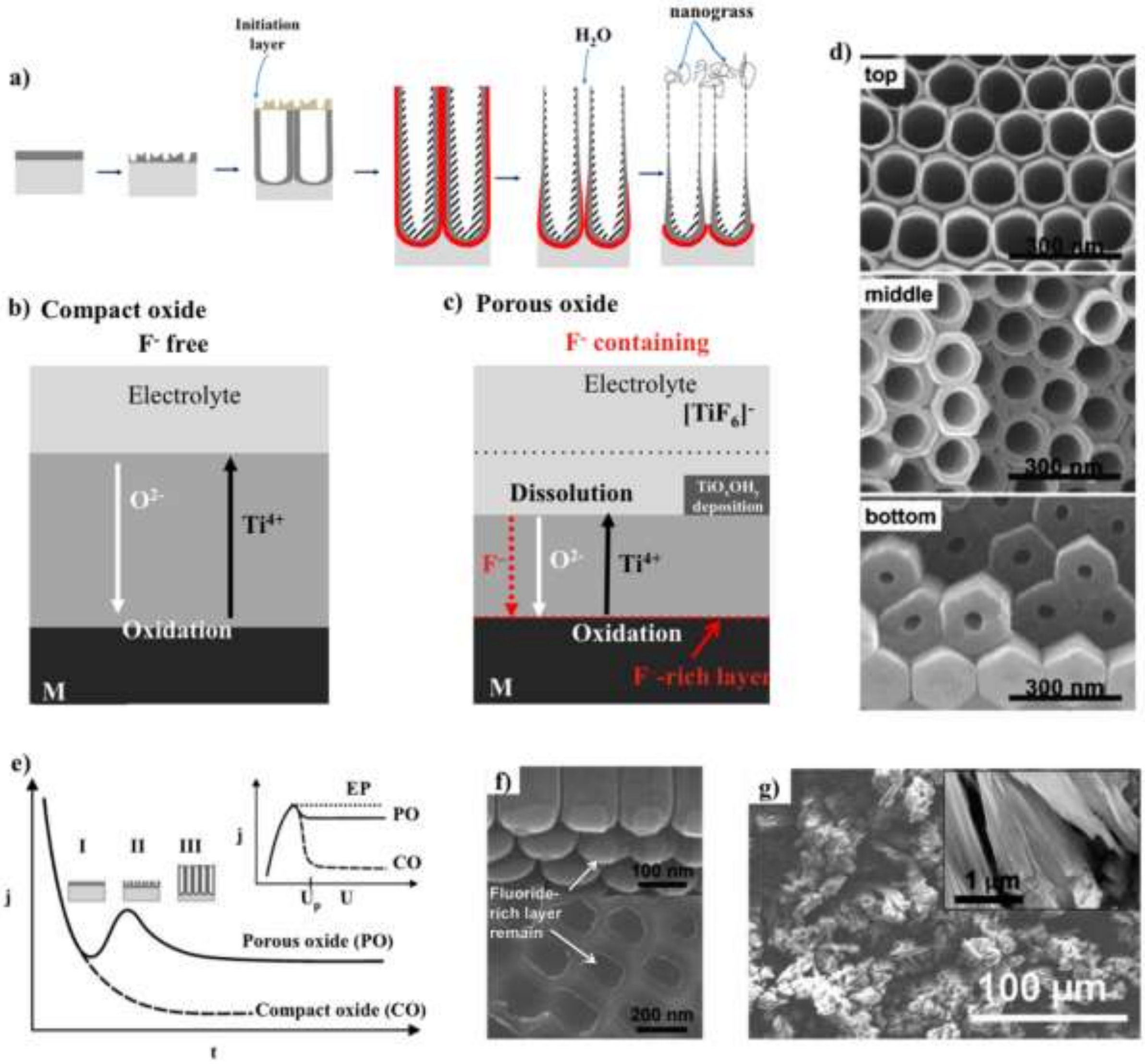

## Figure 7

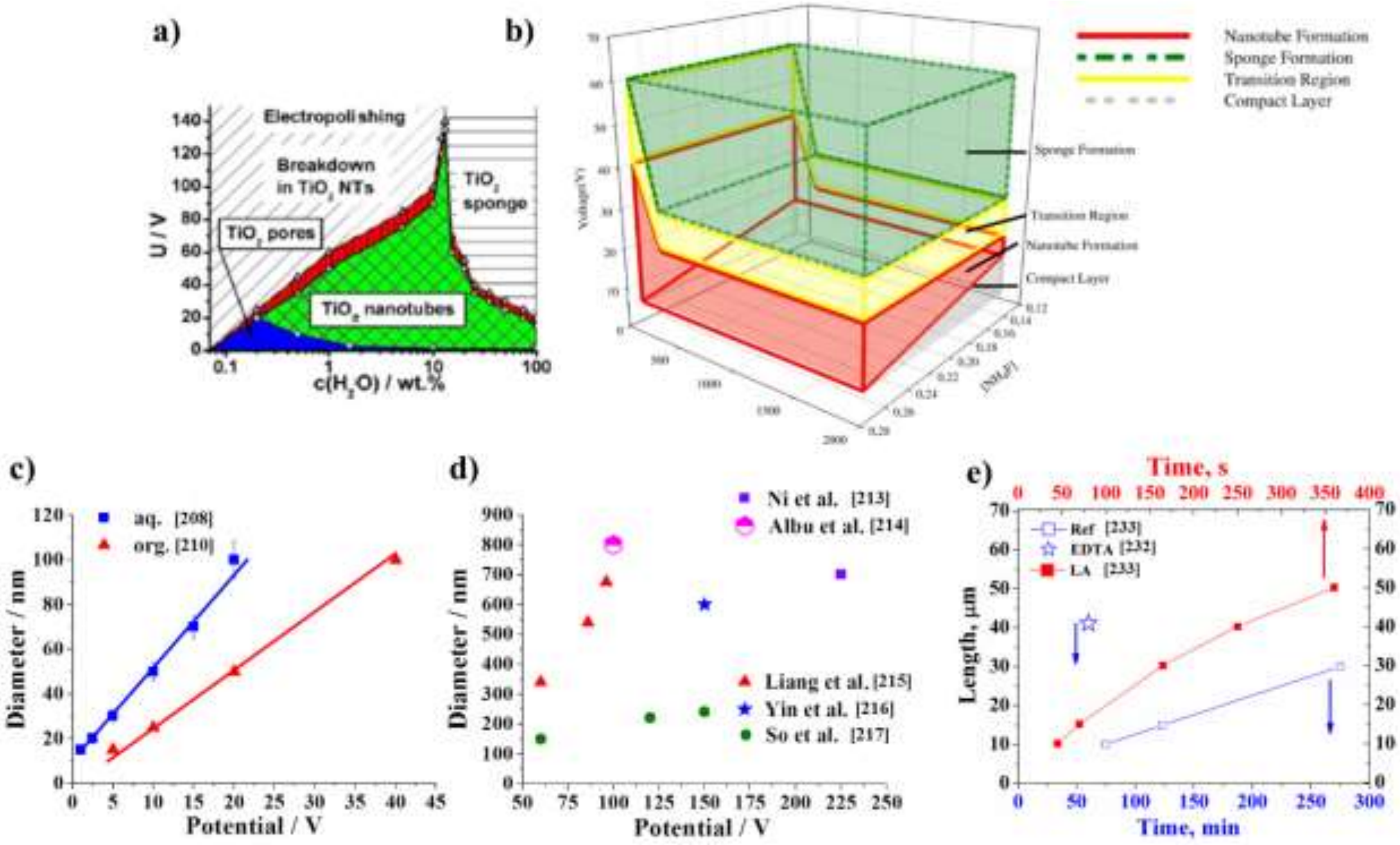

**Figure 8**

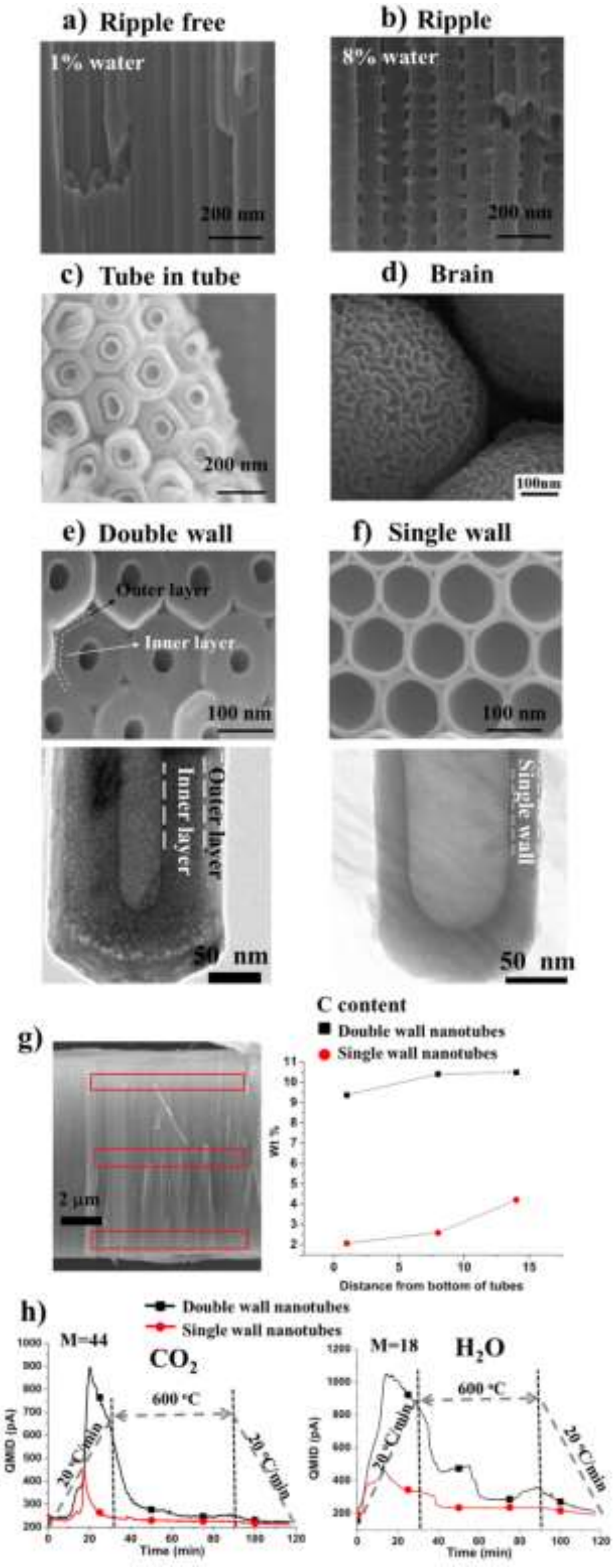

**Figure 9**

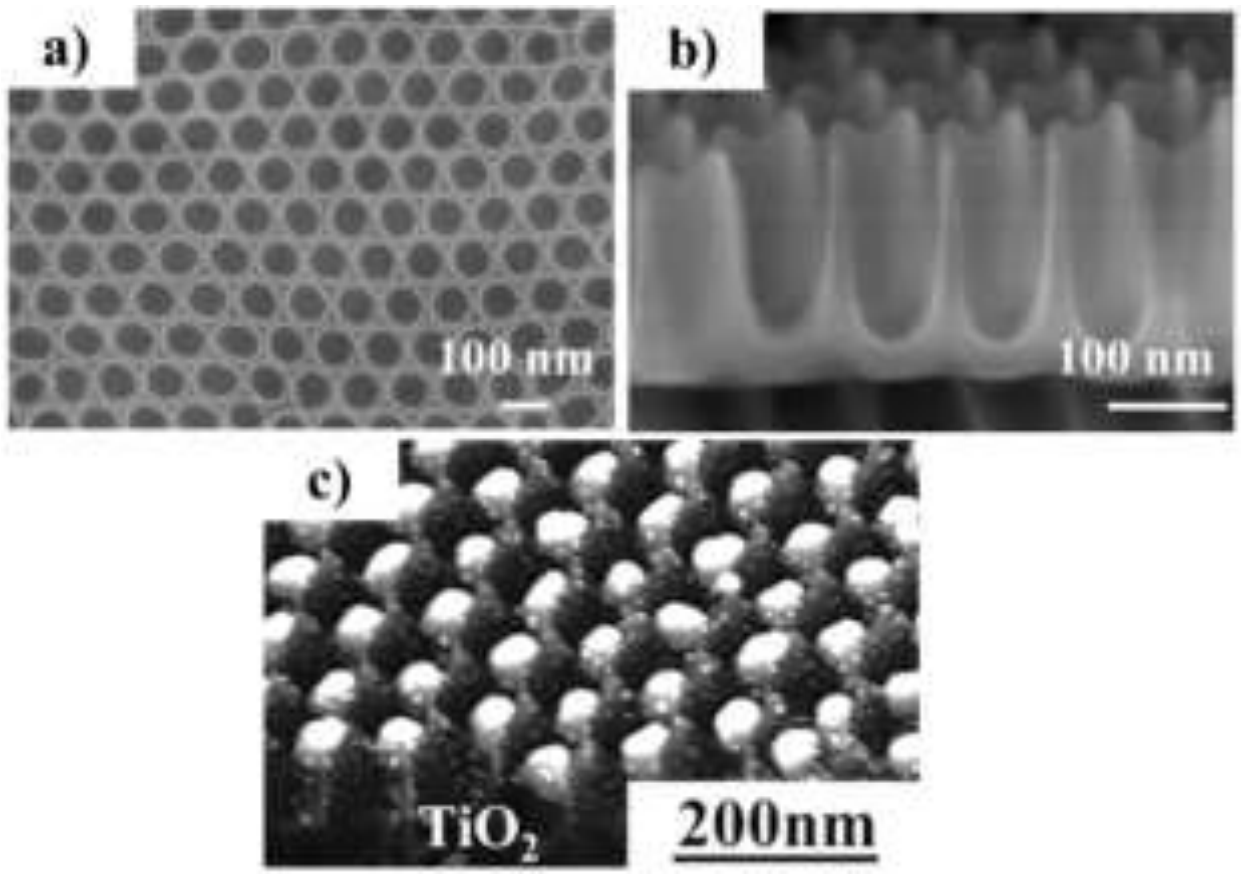

**Figure 10**

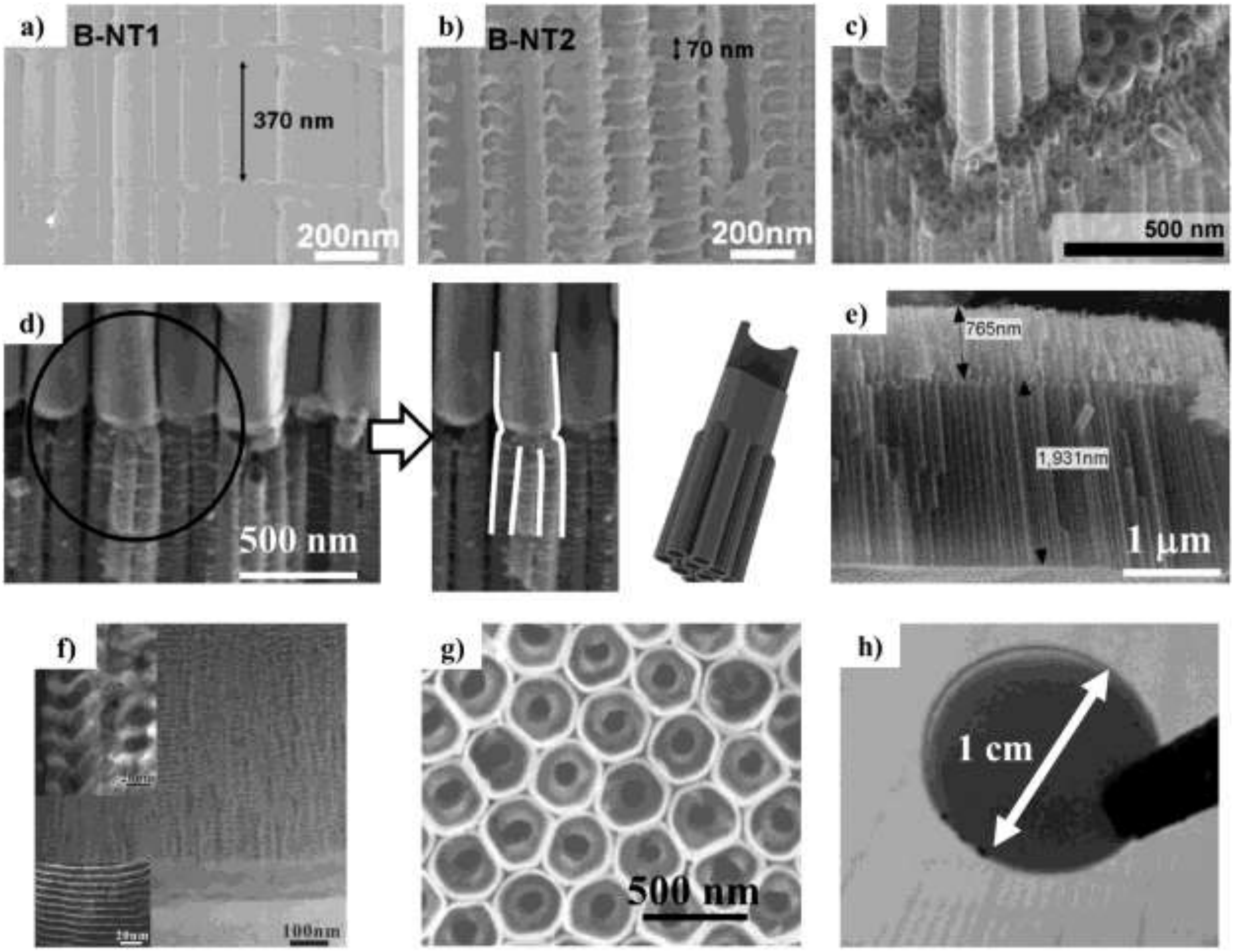

**Figure 11**

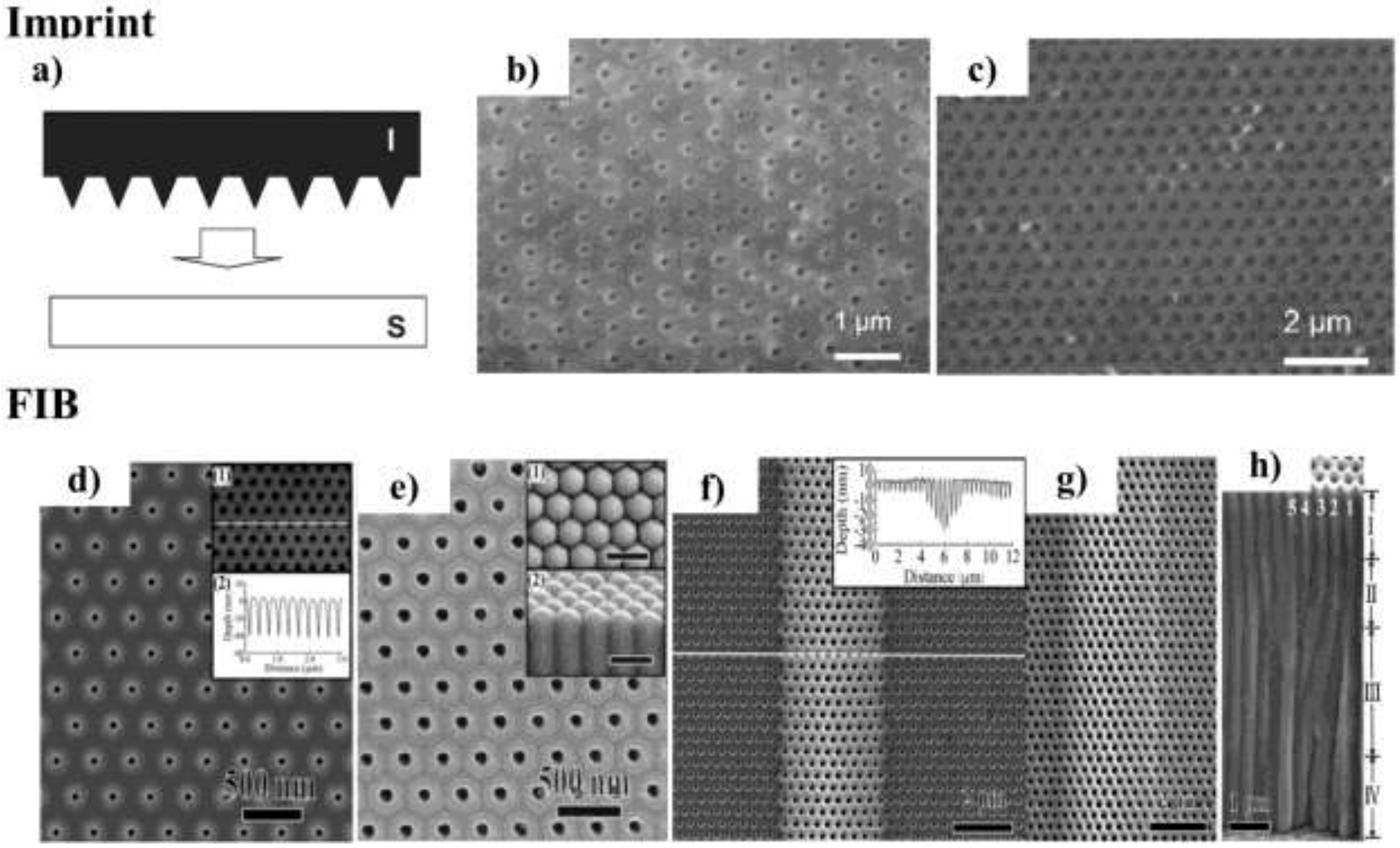

**Figure 12**

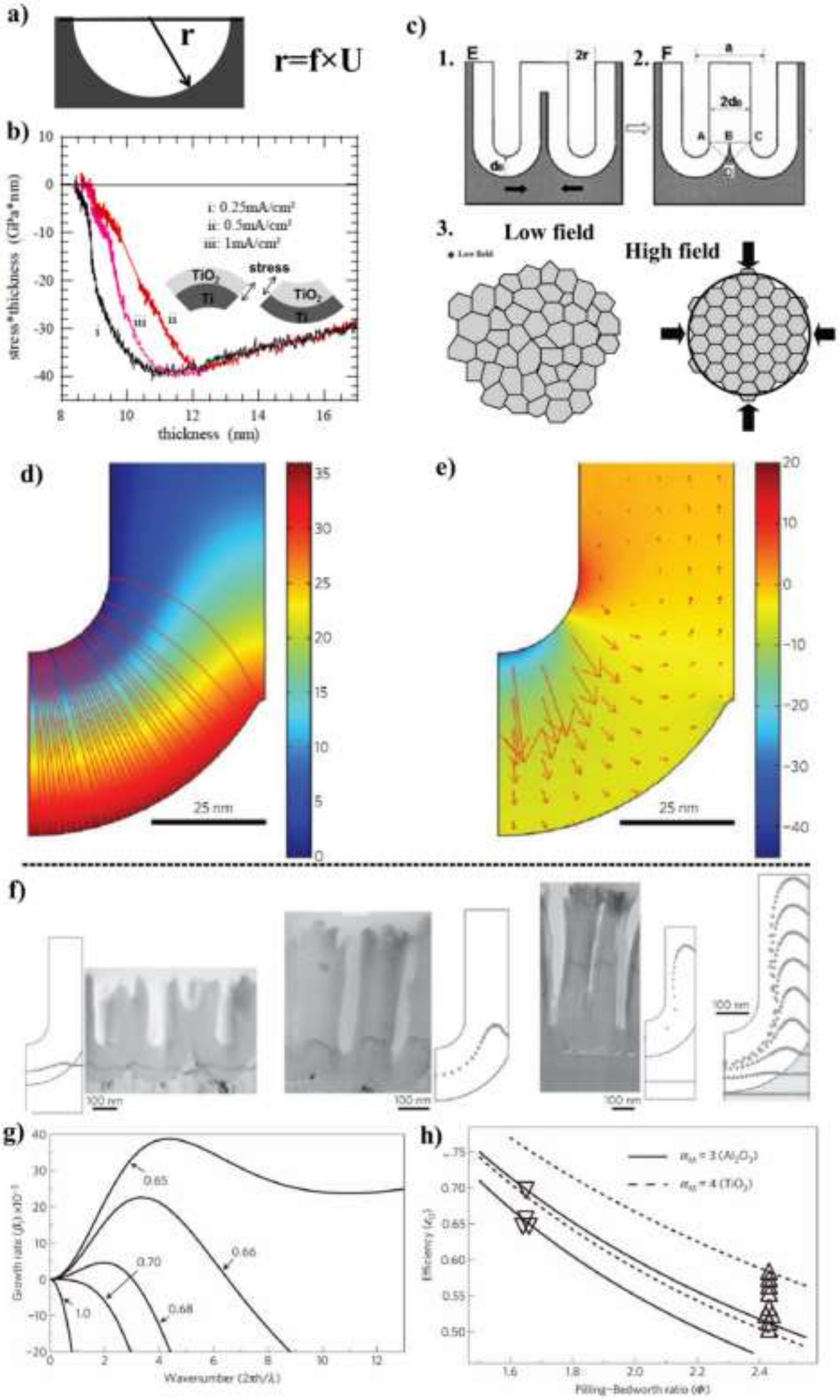

**Figure 13**

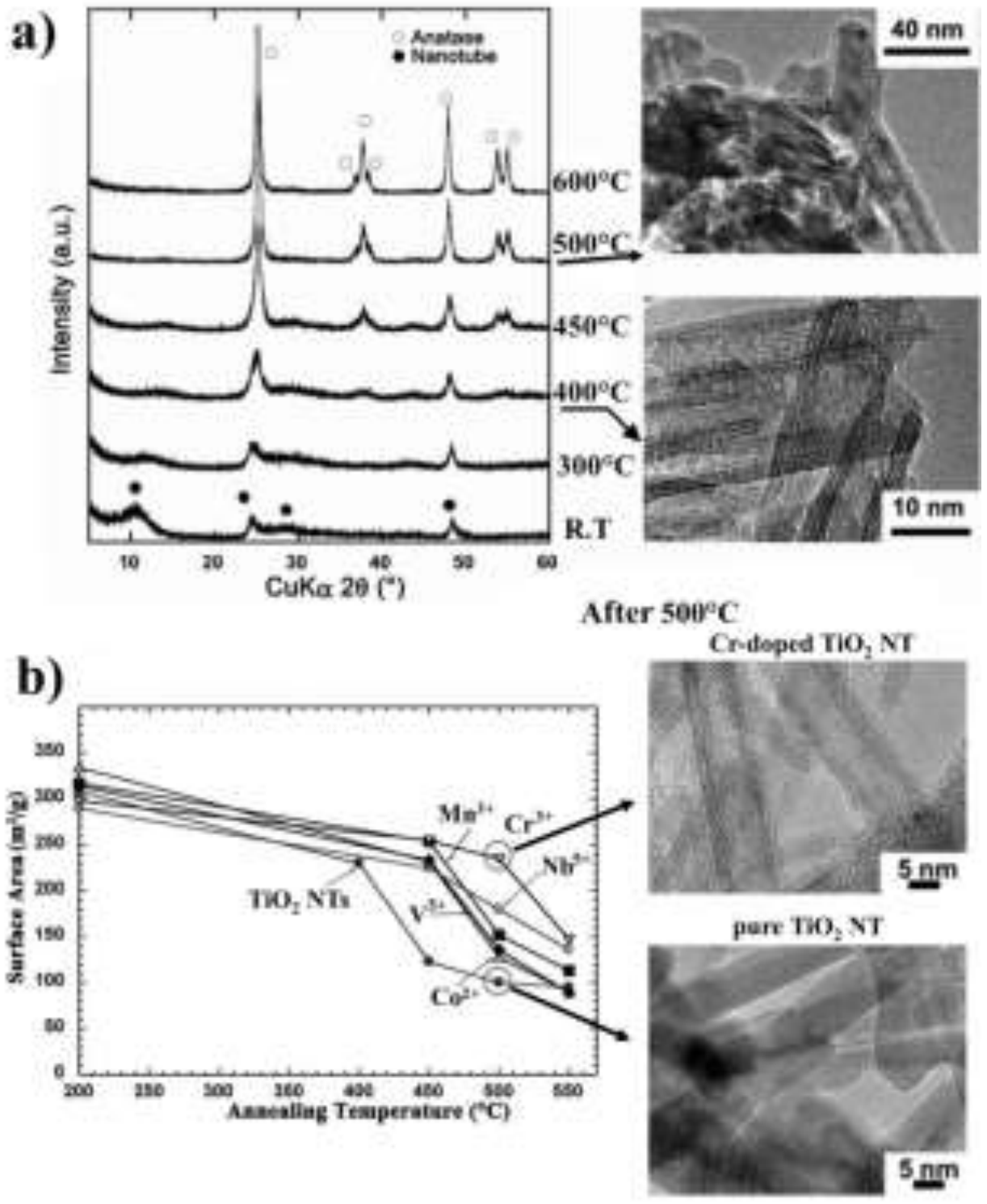

**Figure 14**

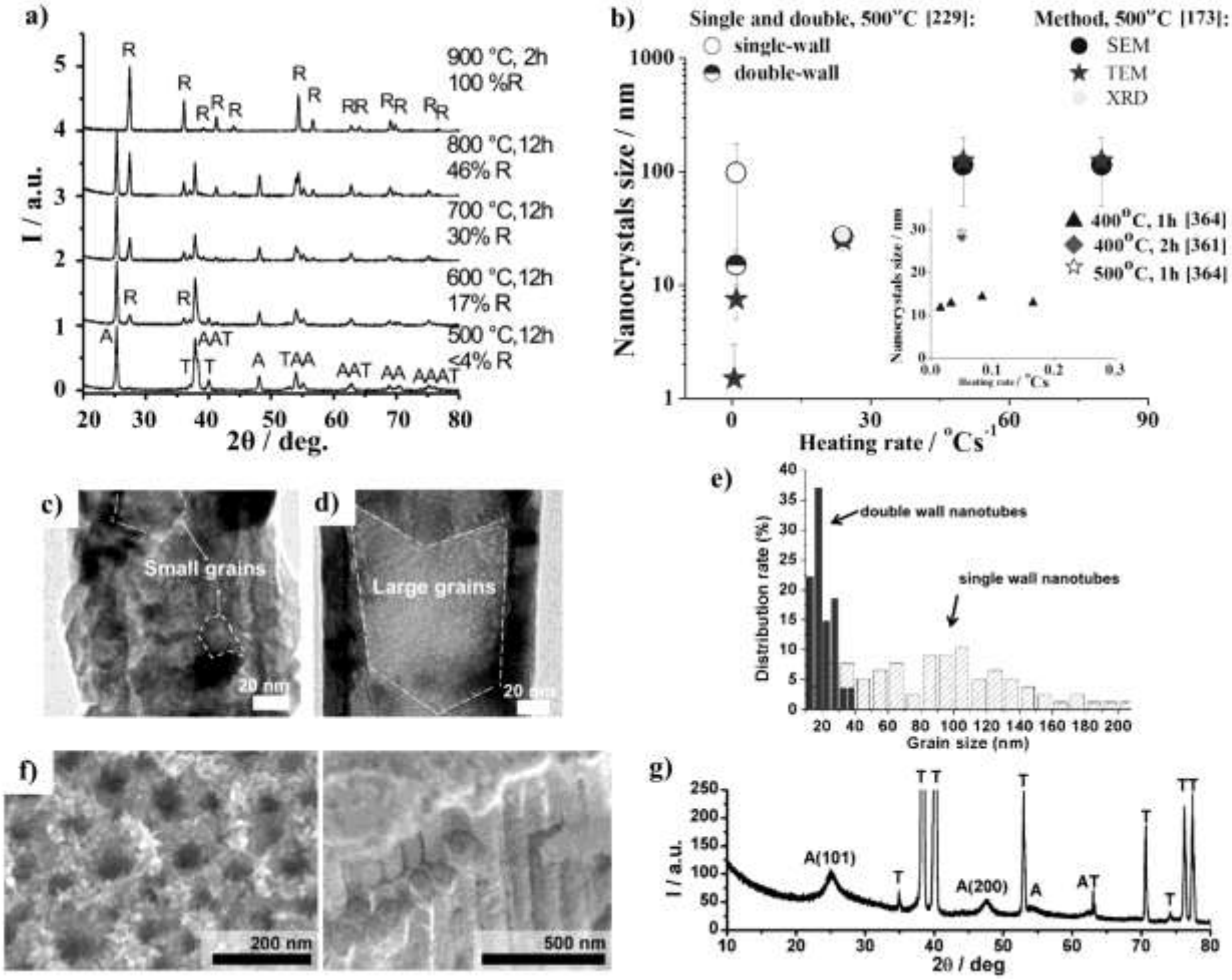

**Figure 15**

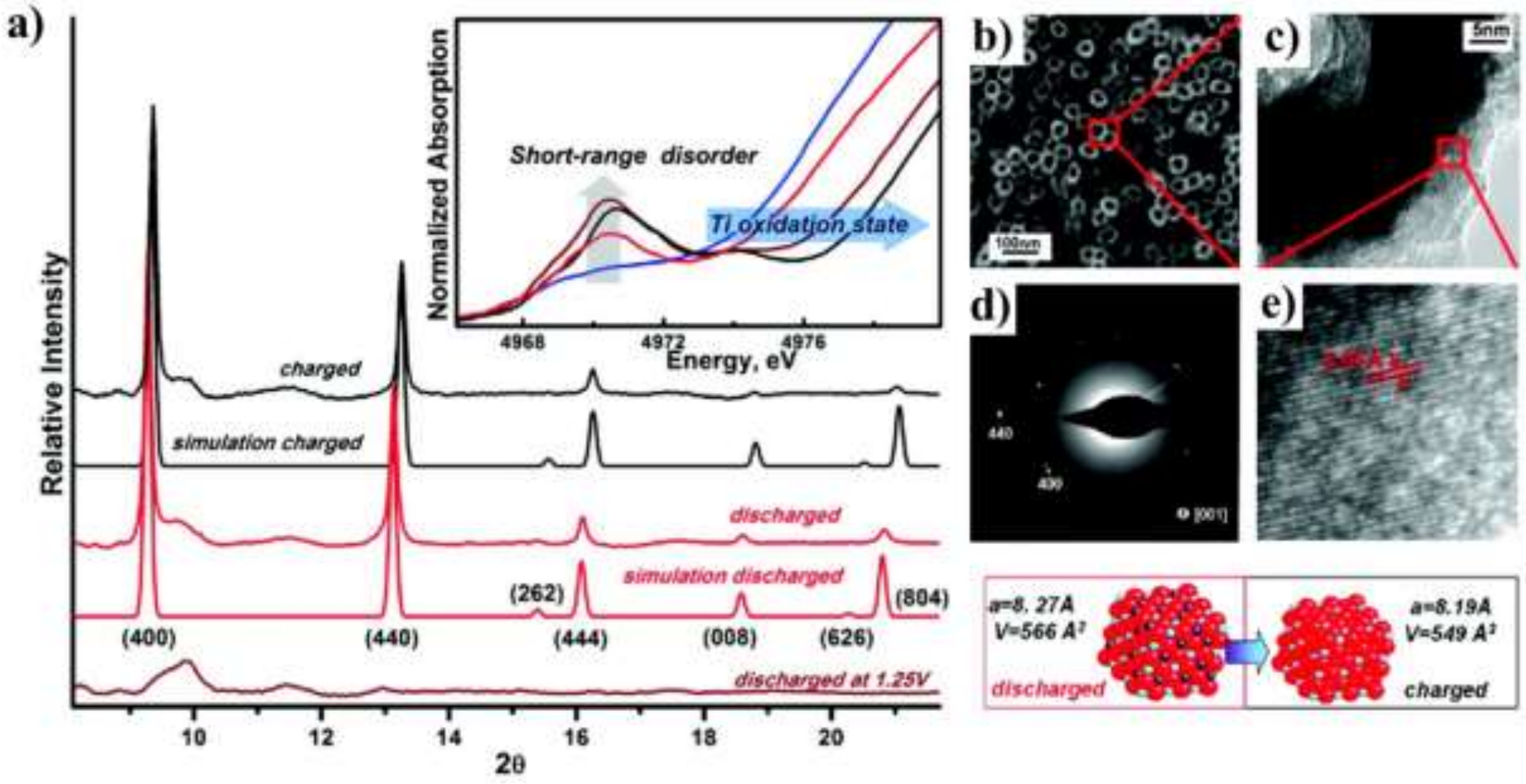

**Figure 16**

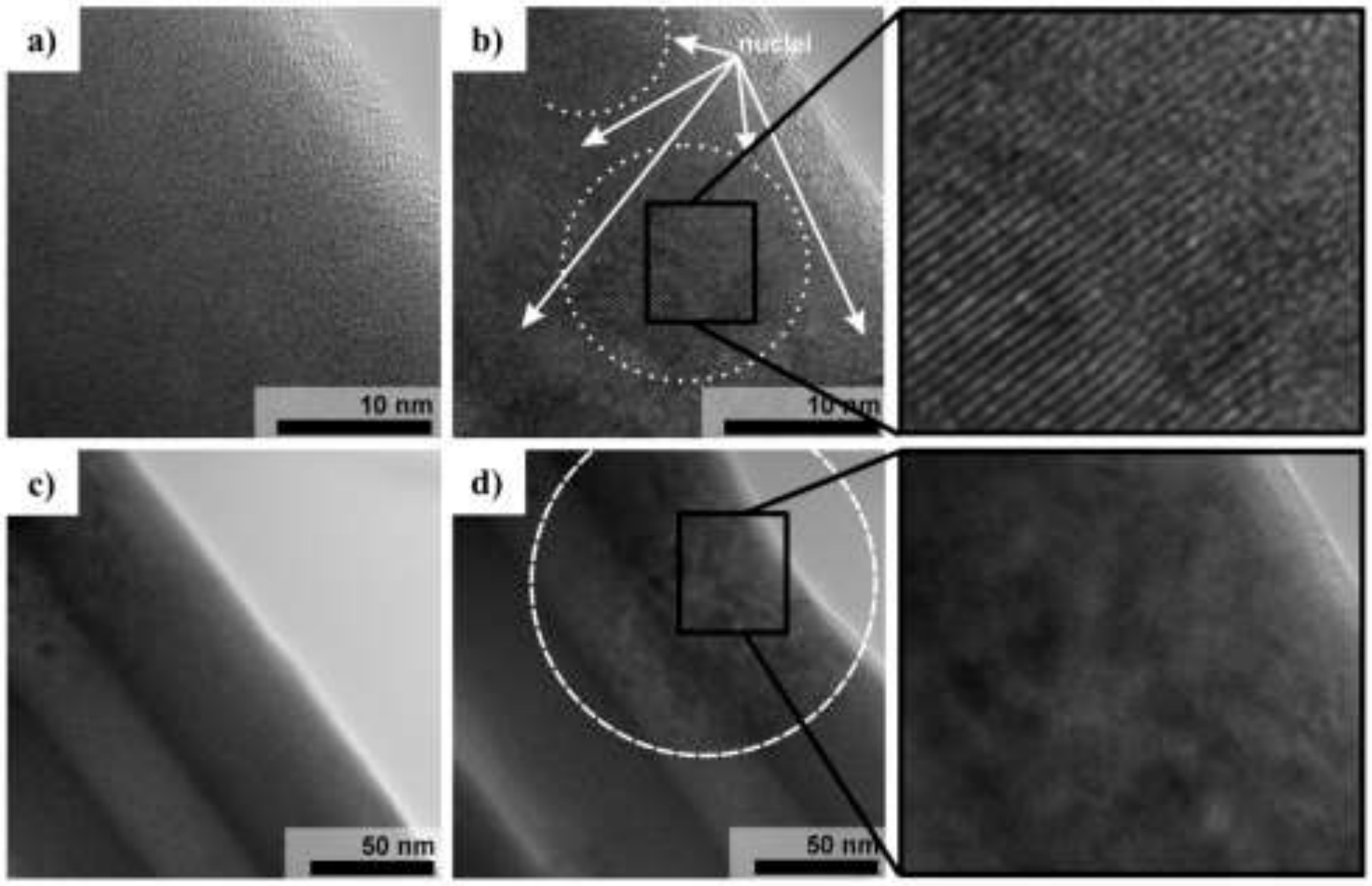

**Figure 17**

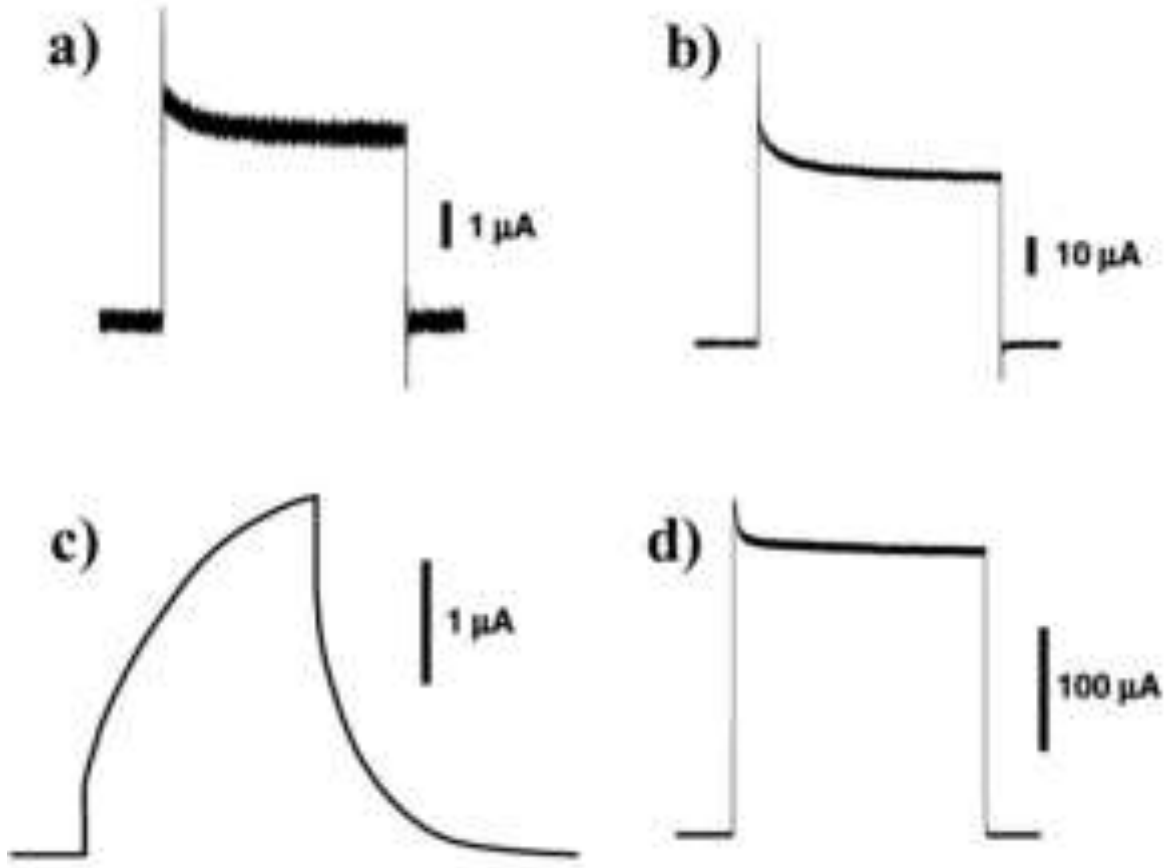

**Figure 18**

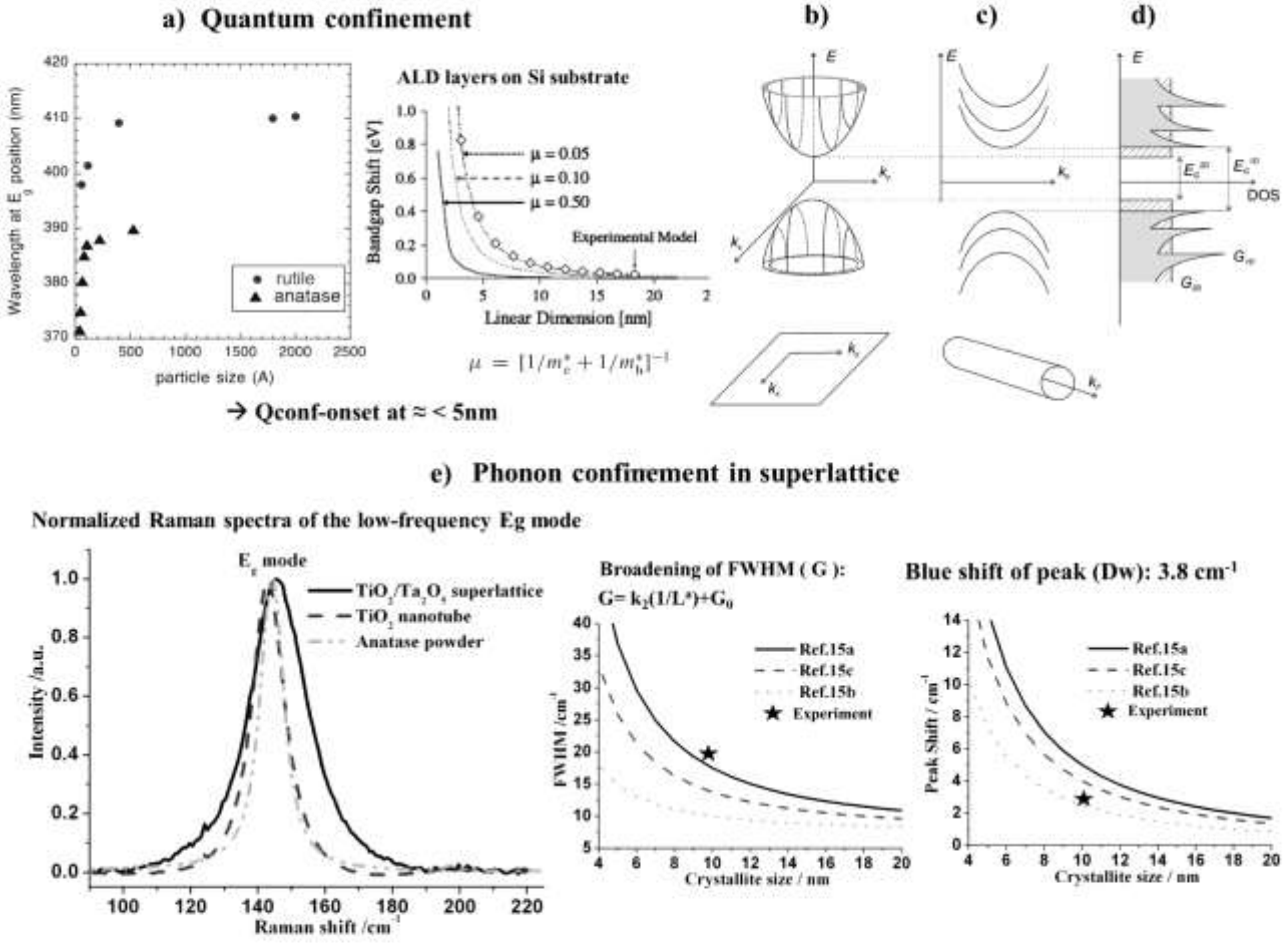
a) Quantum confinement
Wavelength at $E_g$ position (nm)
particle size (A)
rutile
anatase
ALD layers on Si substrate
Bandgap Shift [eV]
Linear Dimension [nm]
μ = 0.05
μ = 0.10
μ = 0.50
Experimental Model
$\mu = [1/m_e^* + 1/m_h^*]^{-1}$
→ Qconf-onset at ≈ < 5nm
b)
c)
d)
DOS
e) Phonon confinement in superlattice
Normalized Raman spectra of the low-frequency Eg mode
$E_g$ mode
$TiO_2/Ta_2O_5$ superlattice
$TiO_2$ nanotube
Anatase powder
Intensity /a.u.
Raman shift /cm$^{-1}$
Broadening of FWHM ( G ):
$G = k_2(1/L^a) + G_0$
Blue shift of peak (Dw): 3.8 cm$^{-1}$
FWHM /cm$^{-1}$
Peak Shift / cm$^{-1}$
Ref.15a
Ref.15c
Ref.15b
Experiment
Crystallite size / nm

**Figure 19**

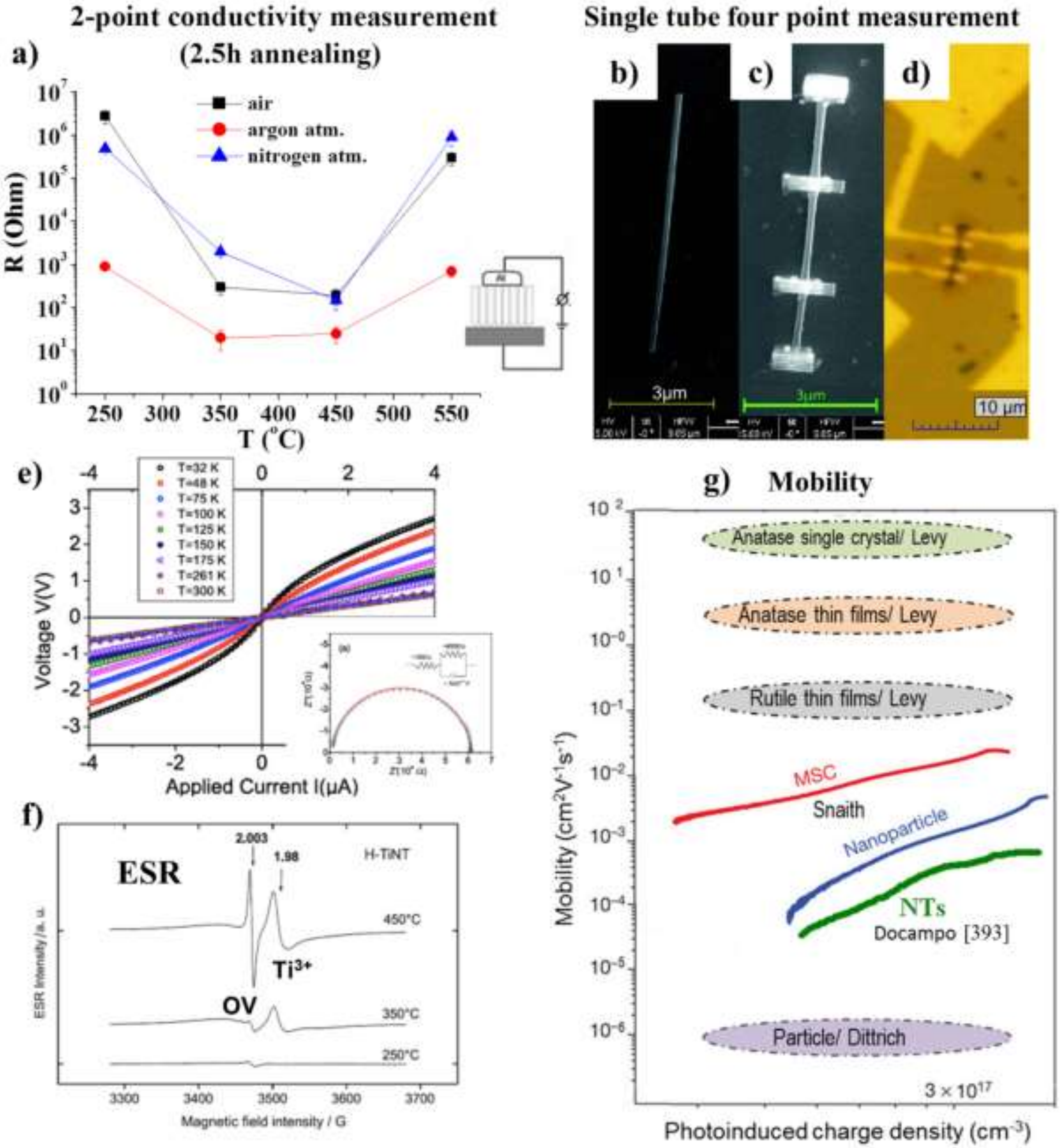

## Figure 20

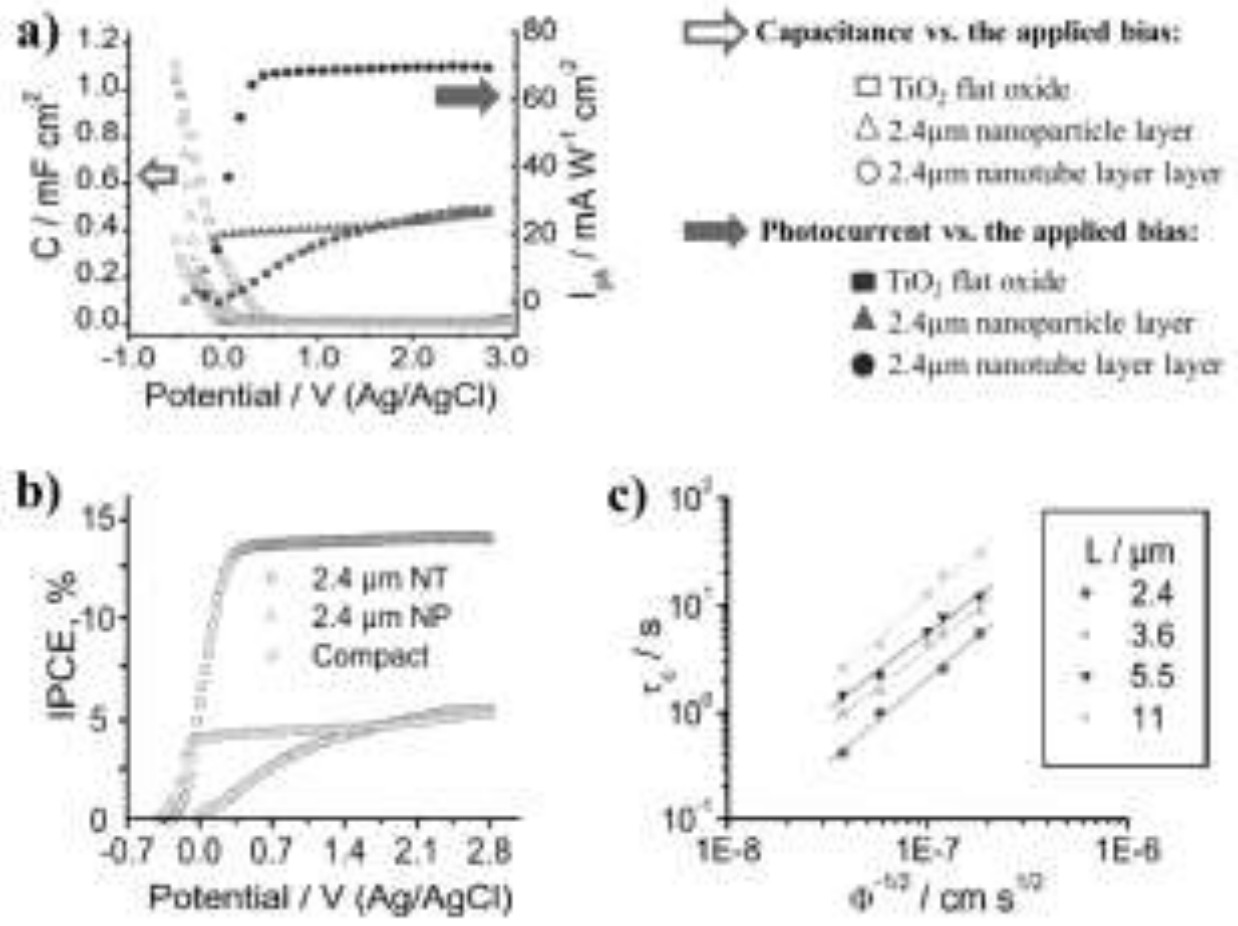

a)
C / mF cm$^{-2}$
$I_{ph}$ / mA W$^{-1}$ cm$^{2}$
Potential / V (Ag/AgCl)
Capacitance vs. the applied bias:
TiO$_2$ flat oxide
2.4μm nanoparticle layer
2.4μm nanotube layer layer
Photocurrent vs. the applied bias:
TiO$_2$ flat oxide
2.4μm nanoparticle layer
2.4μm nanotube layer layer
b)
IPCE, %
2.4 μm NT
2.4 μm NP
Compact
Potential / V (Ag/AgCl)
c)
$\tau_c$ / s
L / μm
2.4
3.6
5.5
11
$\Phi^{-1/2}$ / cm s$^{1/2}$

**Figure 21**

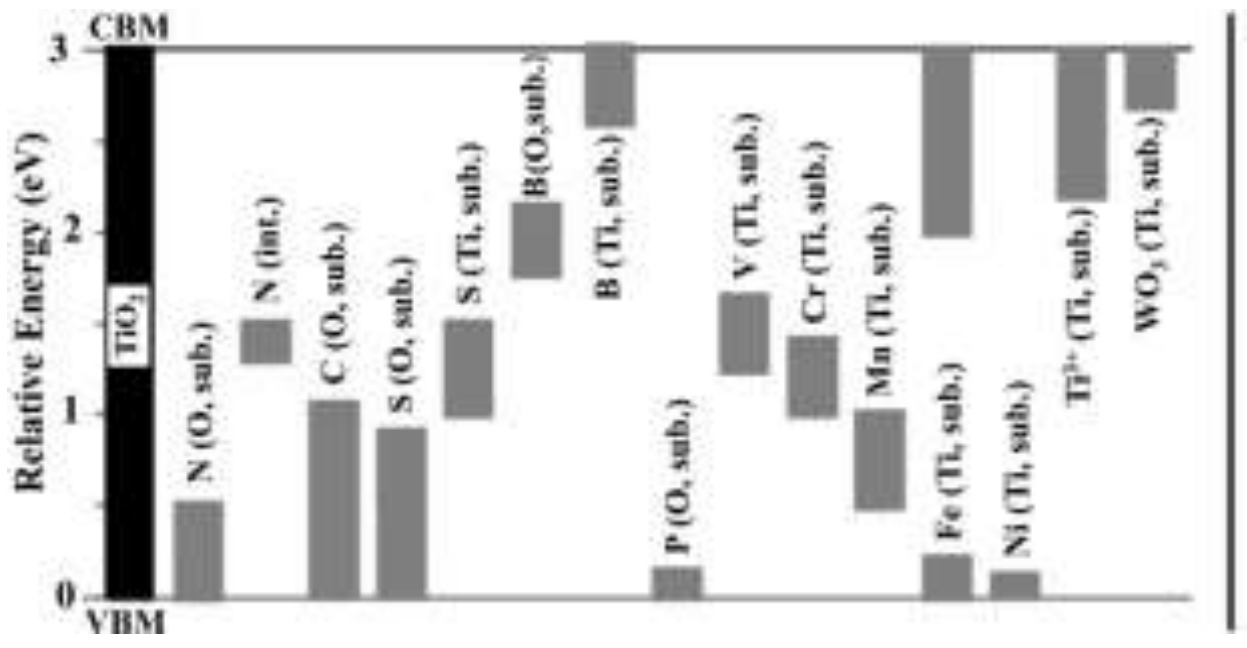

**Figure 22**

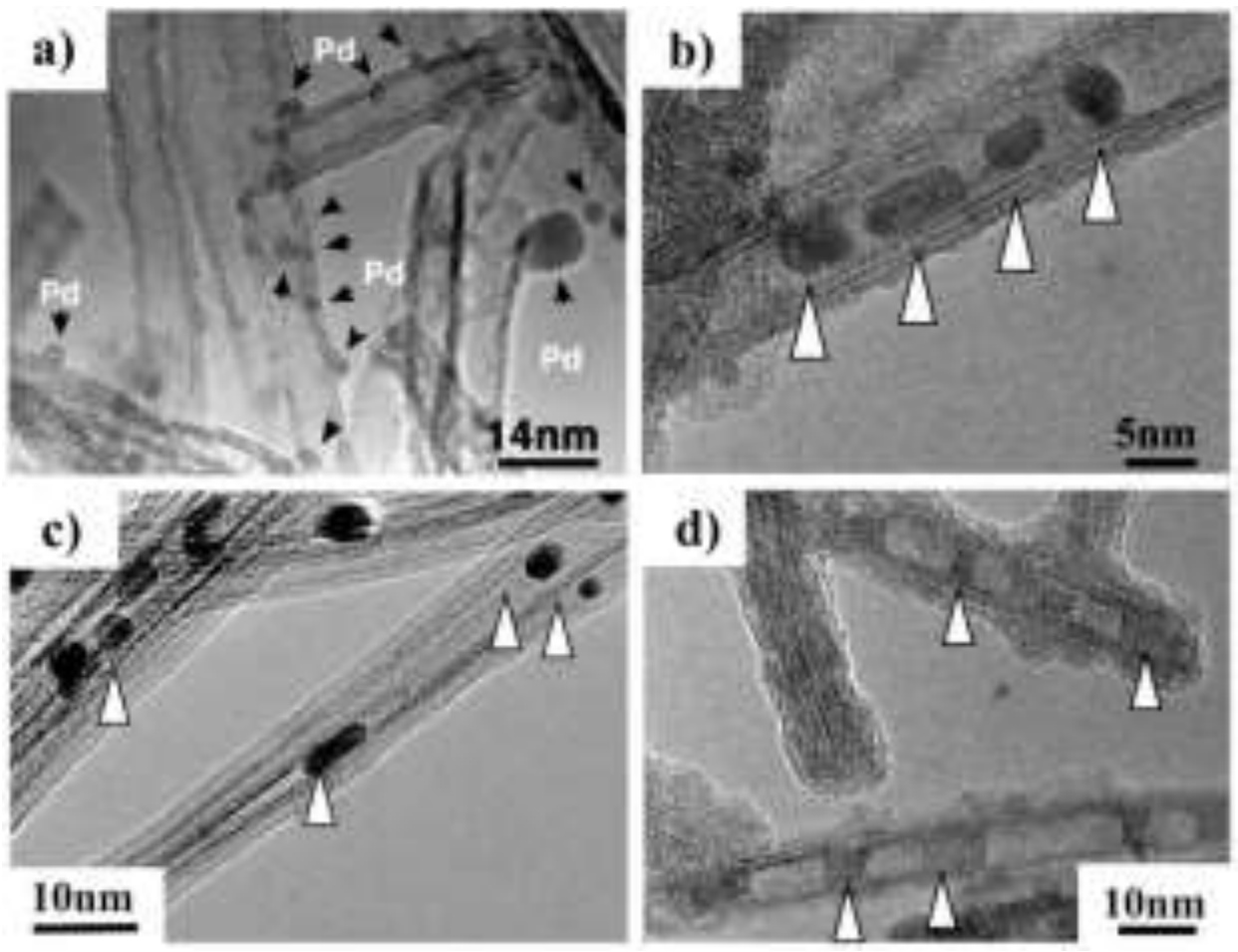

**Figure 23**

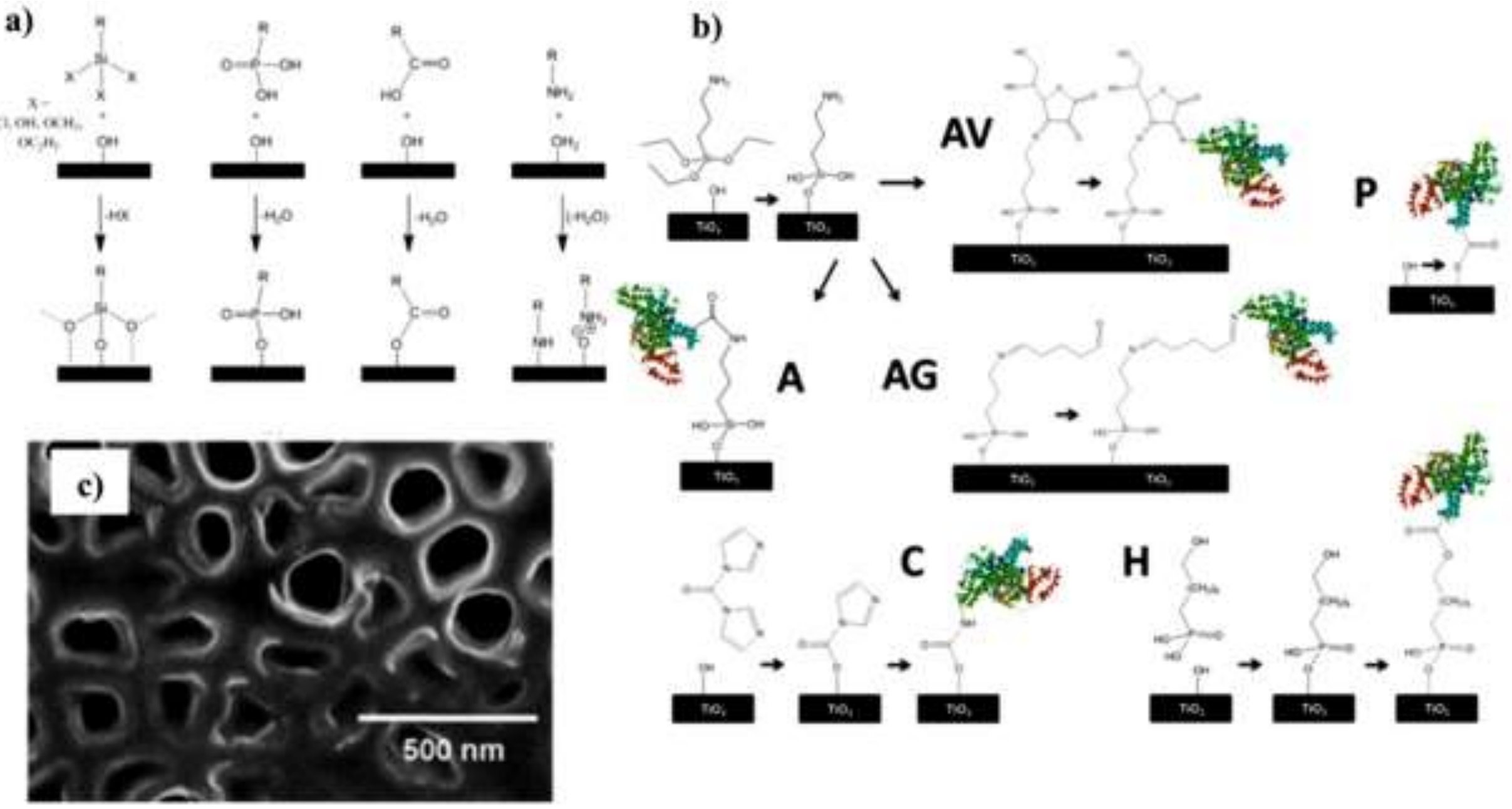

**Figure 24**

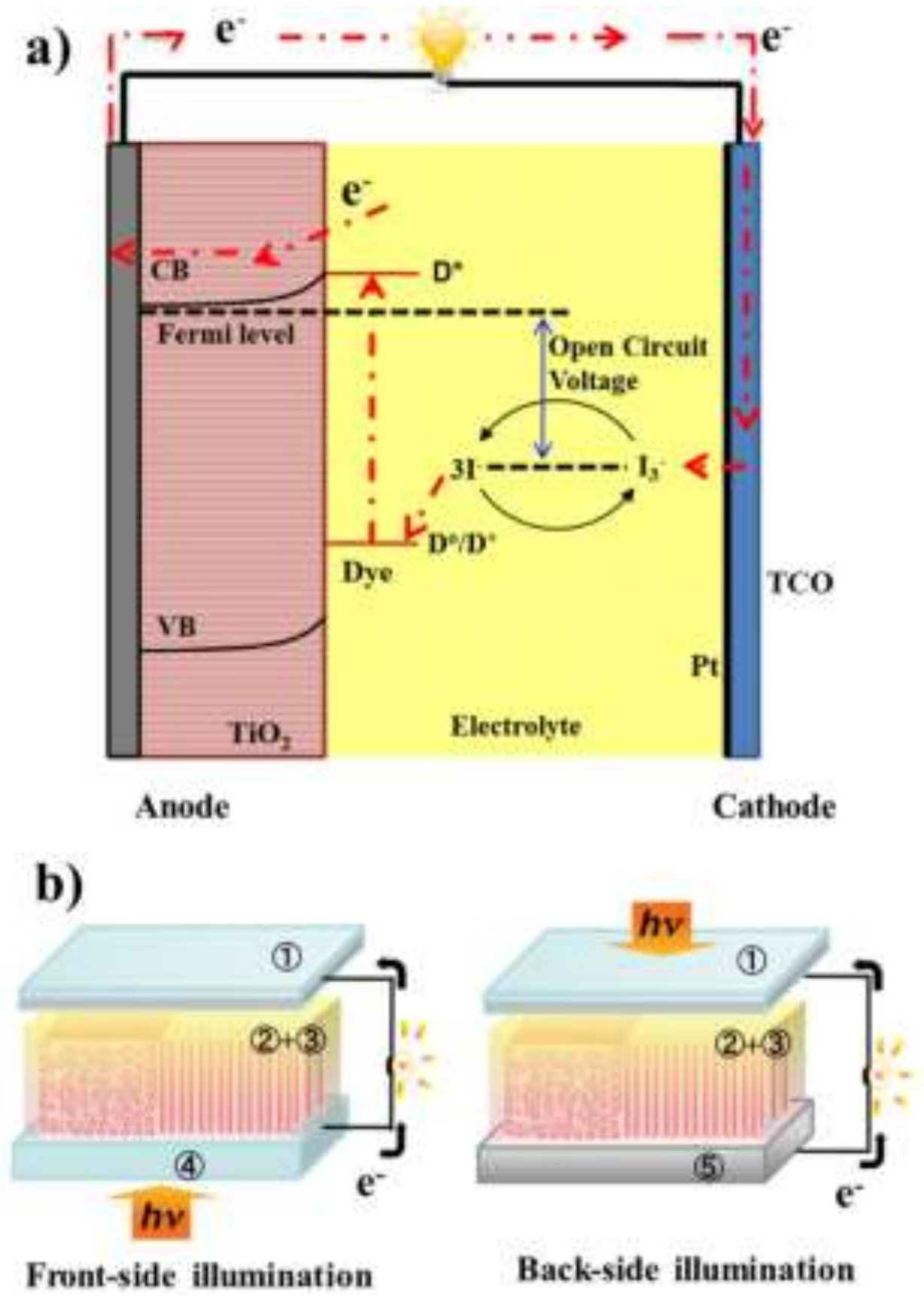

**Figure 25**

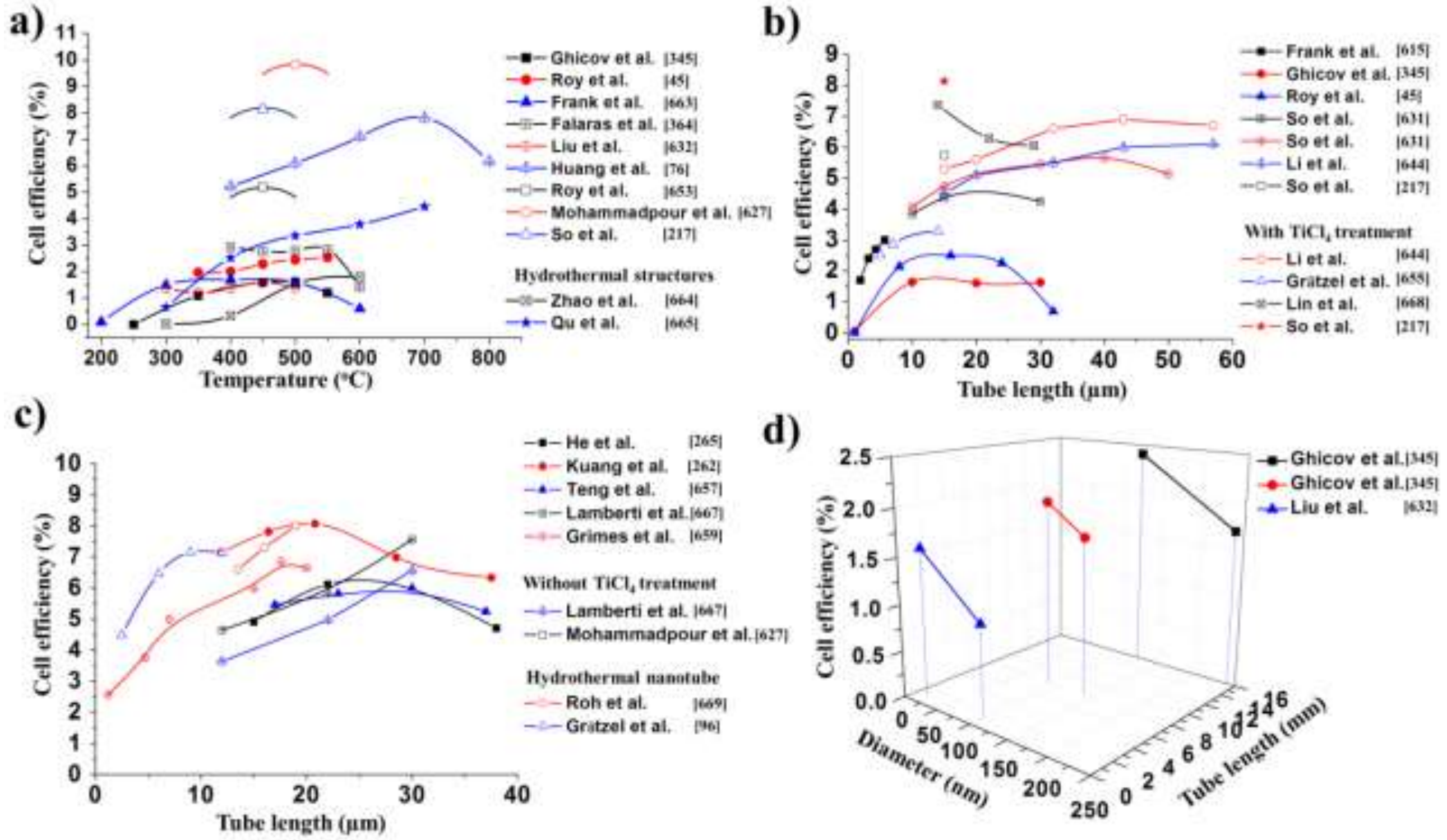
a)
Cell efficiency (%)
Temperature (°C)
Ghicov et al. [345]
Roy et al. [45]
Frank et al. [663]
Falaras et al. [364]
Liu et al. [632]
Huang et al. [76]
Roy et al. [653]
Mohammadpour et al. [627]
So et al. [217]
Hydrothermal structures
Zhao et al. [664]
Qu et al. [665]
b)
Cell efficiency (%)
Tube length (µm)
Frank et al. [615]
Ghicov et al. [345]
Roy et al. [45]
So et al. [631]
So et al. [631]
Li et al. [644]
So et al. [217]
With TiCl4 treatment
Li et al. [644]
Grätzel et al. [655]
Lin et al. [668]
So et al. [217]
c)
Cell efficiency (%)
Tube length (µm)
He et al. [265]
Kuang et al. [262]
Teng et al. [657]
Lamberti et al. [667]
Grimes et al. [659]
Without TiCl4 treatment
Lamberti et al. [667]
Mohammadpour et al. [627]
Hydrothermal nanotube
Roh et al. [669]
Grätzel et al. [96]
d)
Cell efficiency (%)
Diameter (nm)
Tube length (mm)
Ghicov et al. [345]
Ghicov et al. [345]
Liu et al. [632]

## Figure 26

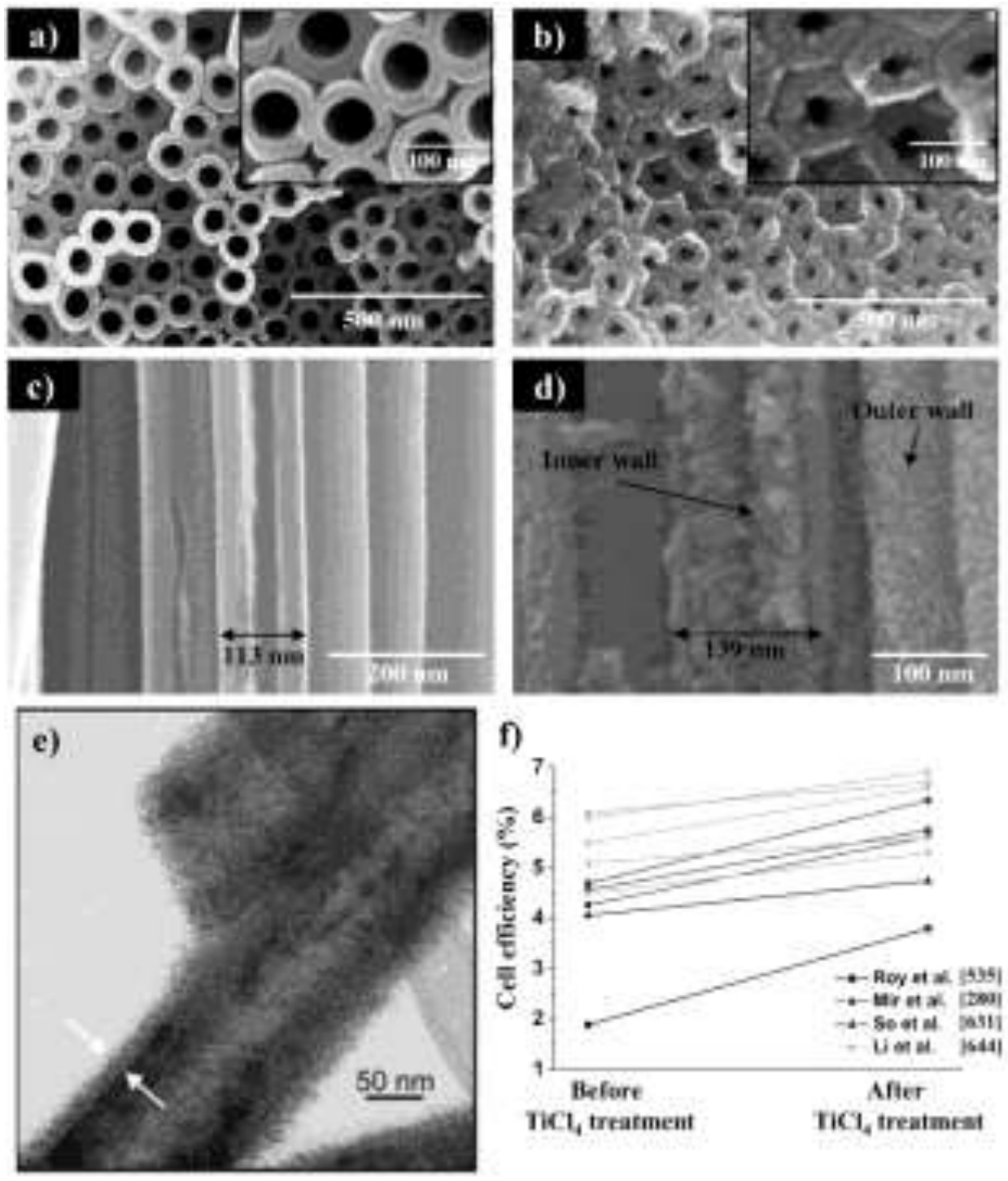

## Figure 27

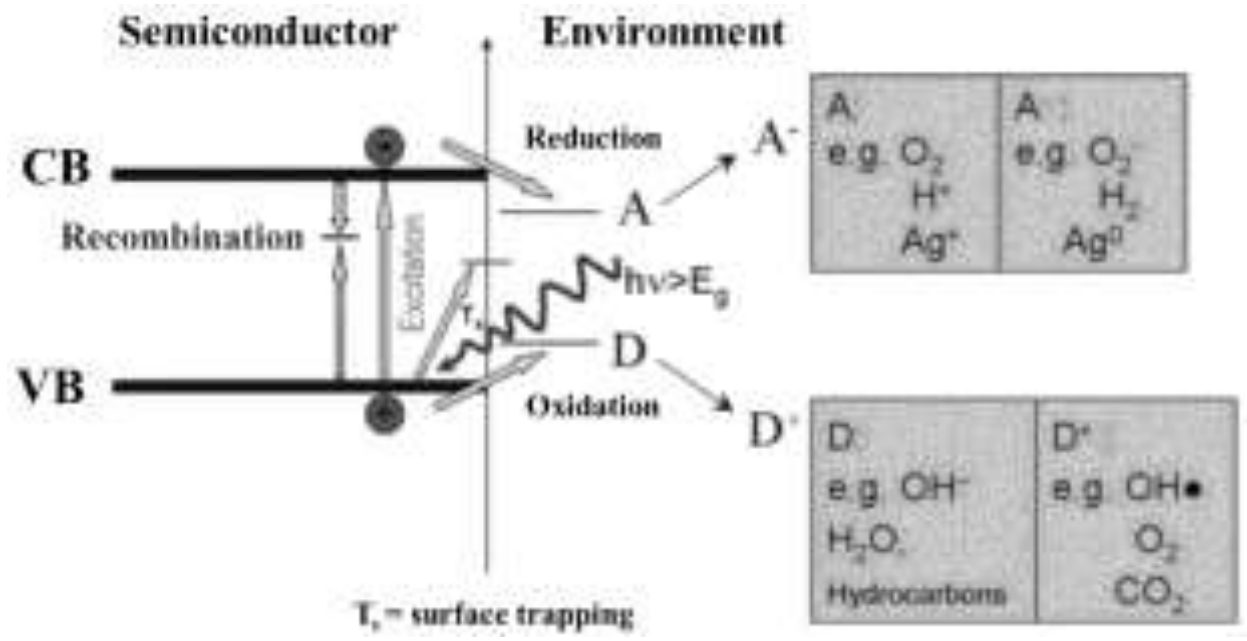

Semiconductor
Environment
CB
VB
Recombination
Excitation
Reduction
Oxidation
A
A⁻
D
D⁺
hν>E_g
A: e.g. O2, H+, Ag+
A⁻: e.g. O2⁻, H2, Ag0
D: e.g. OH⁻, H2O, Hydrocarbons
D⁺: e.g. OH•, O2, CO2
Ts = surface trapping

**Figure 28**

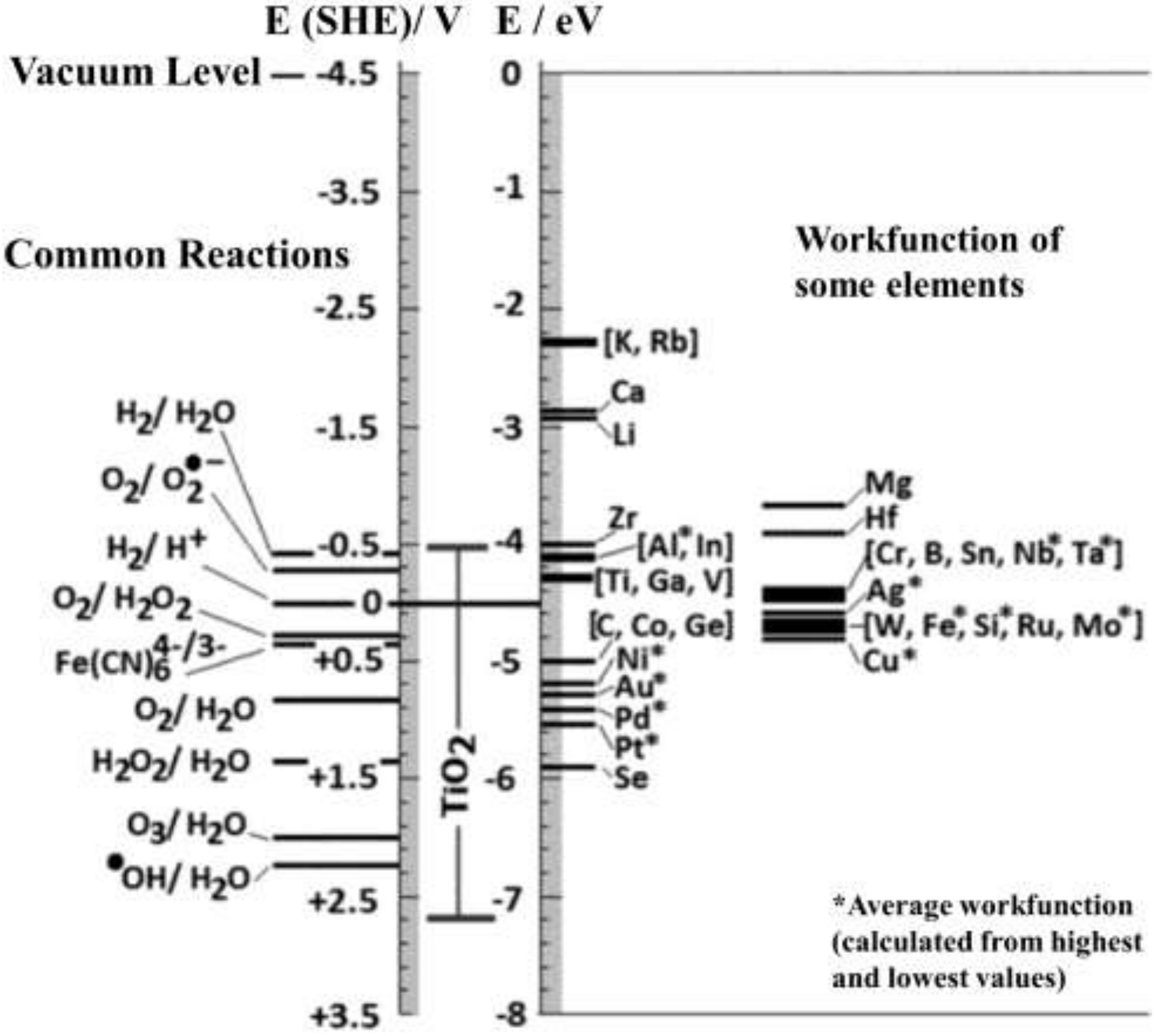
E (SHE)/ V
E / eV
Vacuum Level — -4.5
0
-3.5
-1
Common Reactions
Workfunction of some elements
-2.5
-2
[K, Rb]
Ca
Li
H2/ H2O
-1.5
-3
O2/ O2•−
Mg
Hf
H2/ H+
-0.5
-4
Zr
[Al*, In]
[Cr, B, Sn, Nb*, Ta*]
[Ti, Ga, V]
Ag*
O2/ H2O2
0
[W, Fe*, Si*, Ru, Mo*]
[C, Co, Ge]
Fe(CN)6 4-/3-
+0.5
-5
Ni*
Cu*
Au*
O2/ H2O
Pd*
Pt*
H2O2/ H2O
+1.5
-6
Se
TiO2
O3/ H2O
•OH/ H2O
+2.5
-7
*Average workfunction (calculated from highest and lowest values)
+3.5
-8

## Figure 29

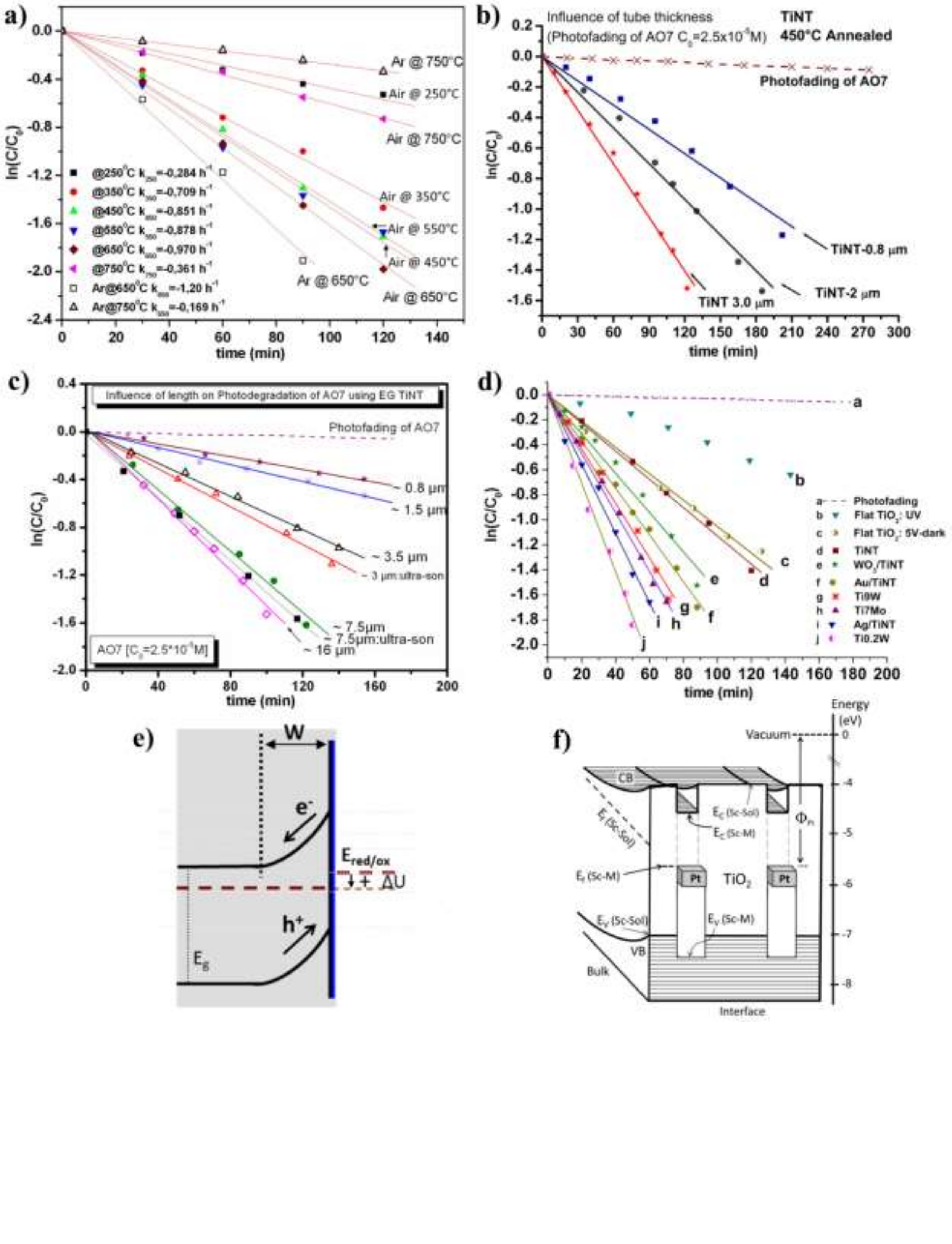
a)
Ar @ 750°C
Air @ 250°C
Air @ 750°C
Air @ 350°C
Air @ 550°C
Air @ 450°C
Ar @ 650°C
Air @ 650°C
ln(C/C0)
time (min)
b)
Influence of tube thickness
TiNT
450°C Annealed
Photofading of AO7
TiNT-0.8 μm
TiNT-2 μm
TiNT 3.0 μm
c)
Influence of length on Photodegradation of AO7 using EG TiNT
Photofading of AO7
~ 0.8 μm
~ 1.5 μm
~ 3.5 μm
~ 3 μm:ultra-son
~ 7.5μm
~ 7.5μm:ultra-son
~ 16 μm
d)
a Photofading
b Flat TiO2: UV
c Flat TiO2: 5V-dark
d TiNT
e WO3/TiNT
f Au/TiNT
g Ti8W
h Ti7Mo
i Ag/TiNT
j Ti0.2W
e)
W
e-
h+
Ered/ox
ΔU
Eg
f)
Energy (eV)
Vacuum
CB
TiO2
Pt
VB
Bulk
Interface

**Figure 30**

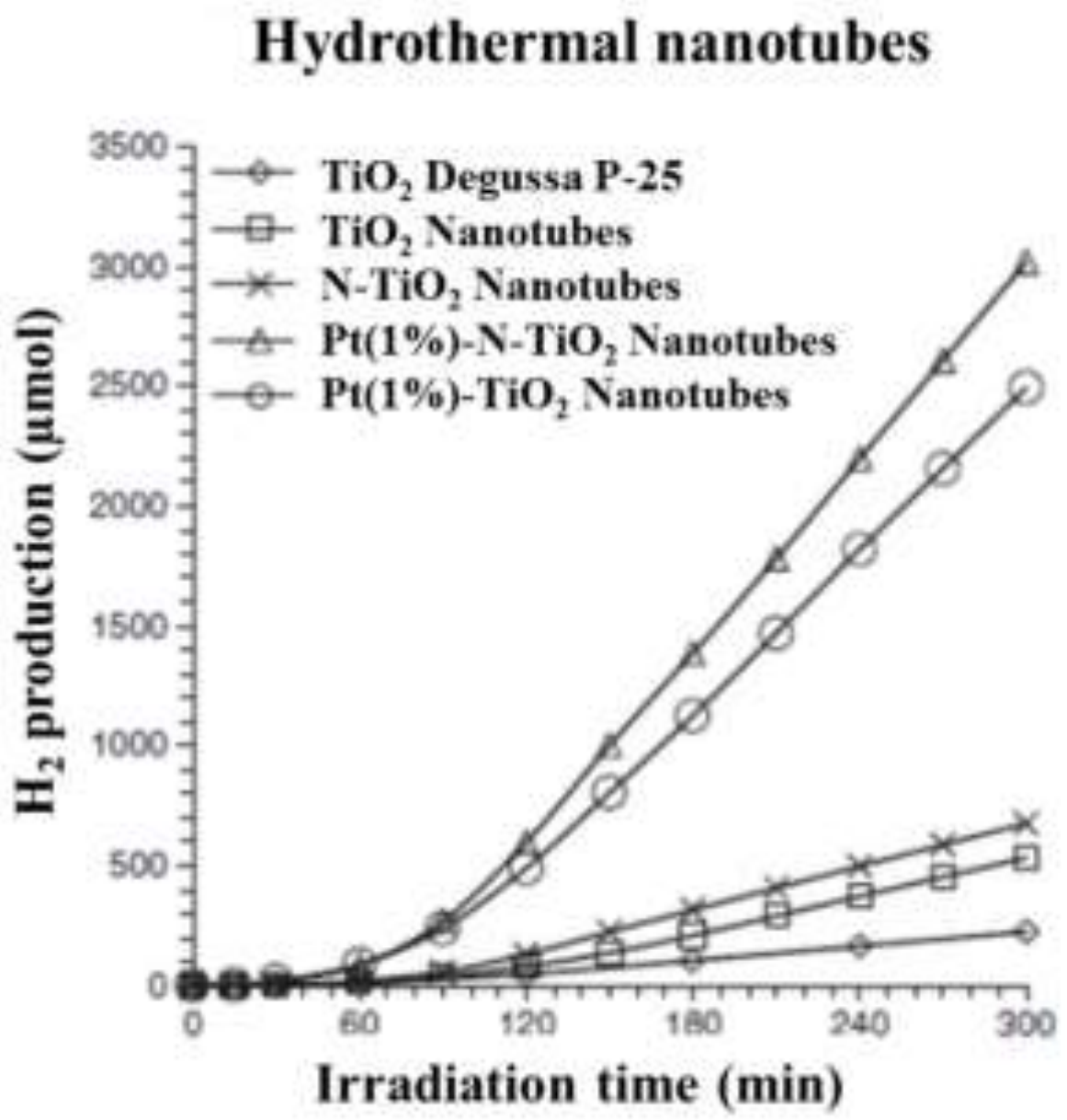

**Figure 31**

a)

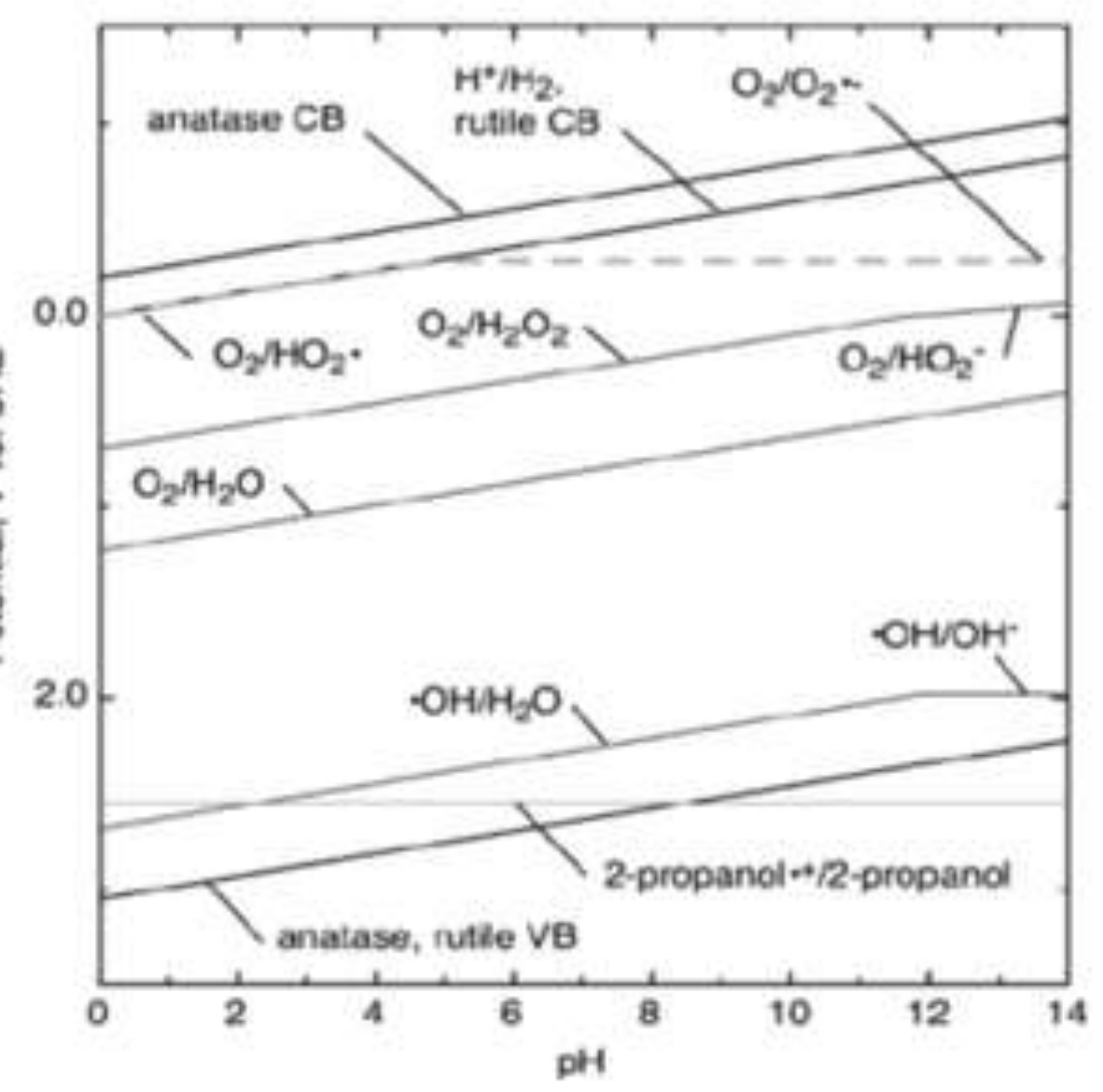

b)

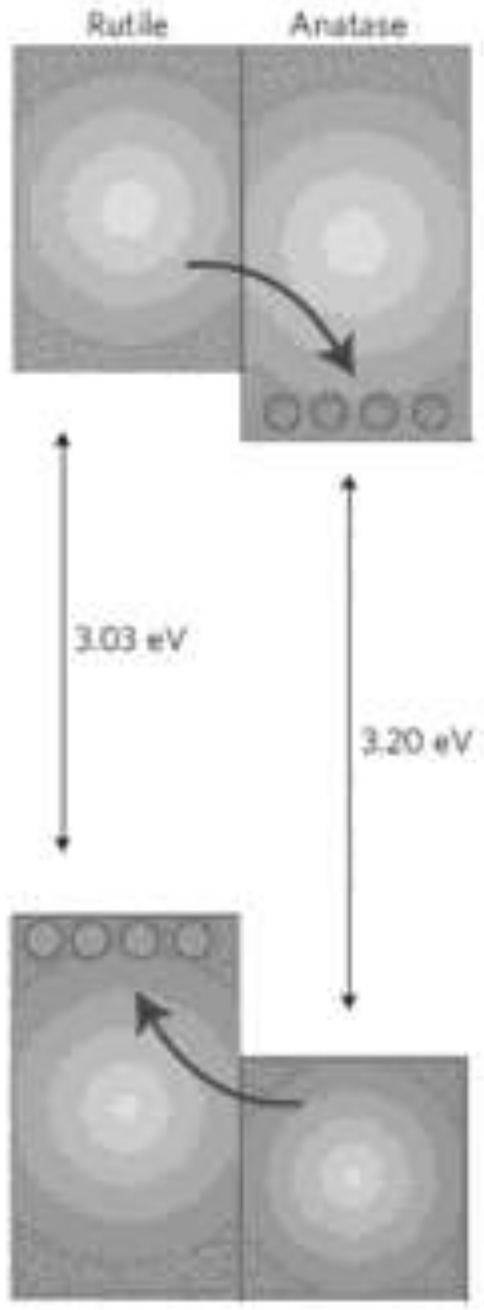
Rutile
Anatase
3.03 eV
3.20 eV

**Figure 32**

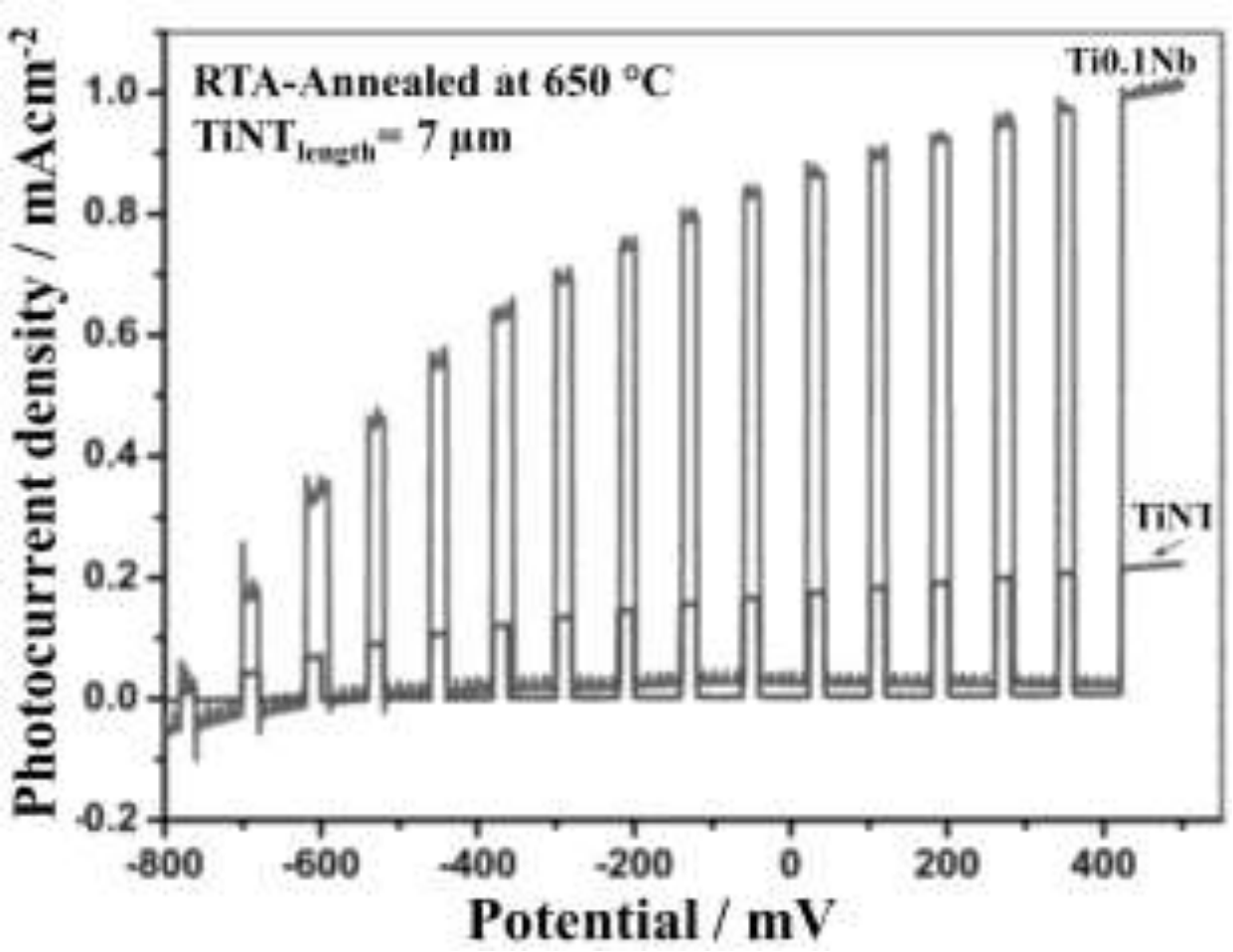

**Figure 33**

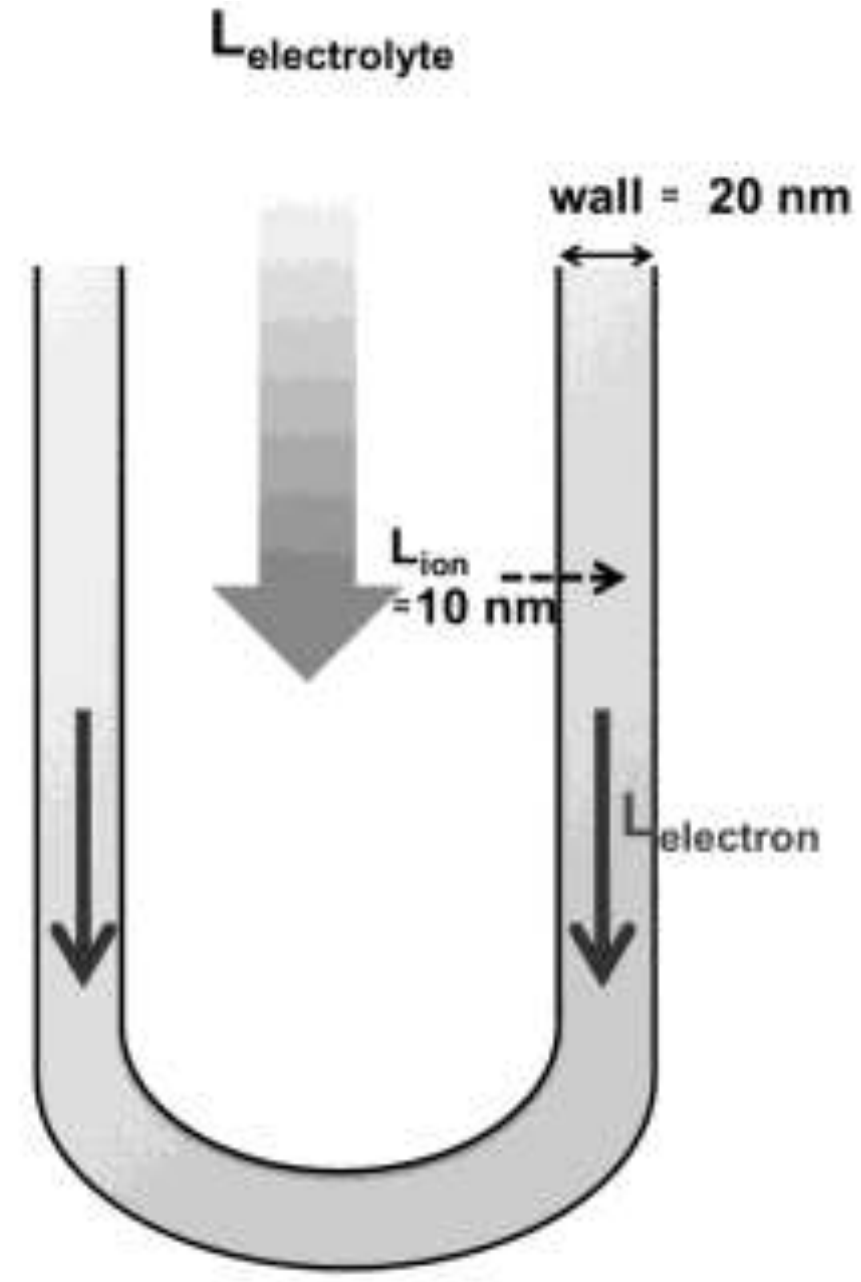


$$xLi^{+} + TiO_2 \leftrightarrow Li_xTiO_2$$

**Figure 34**

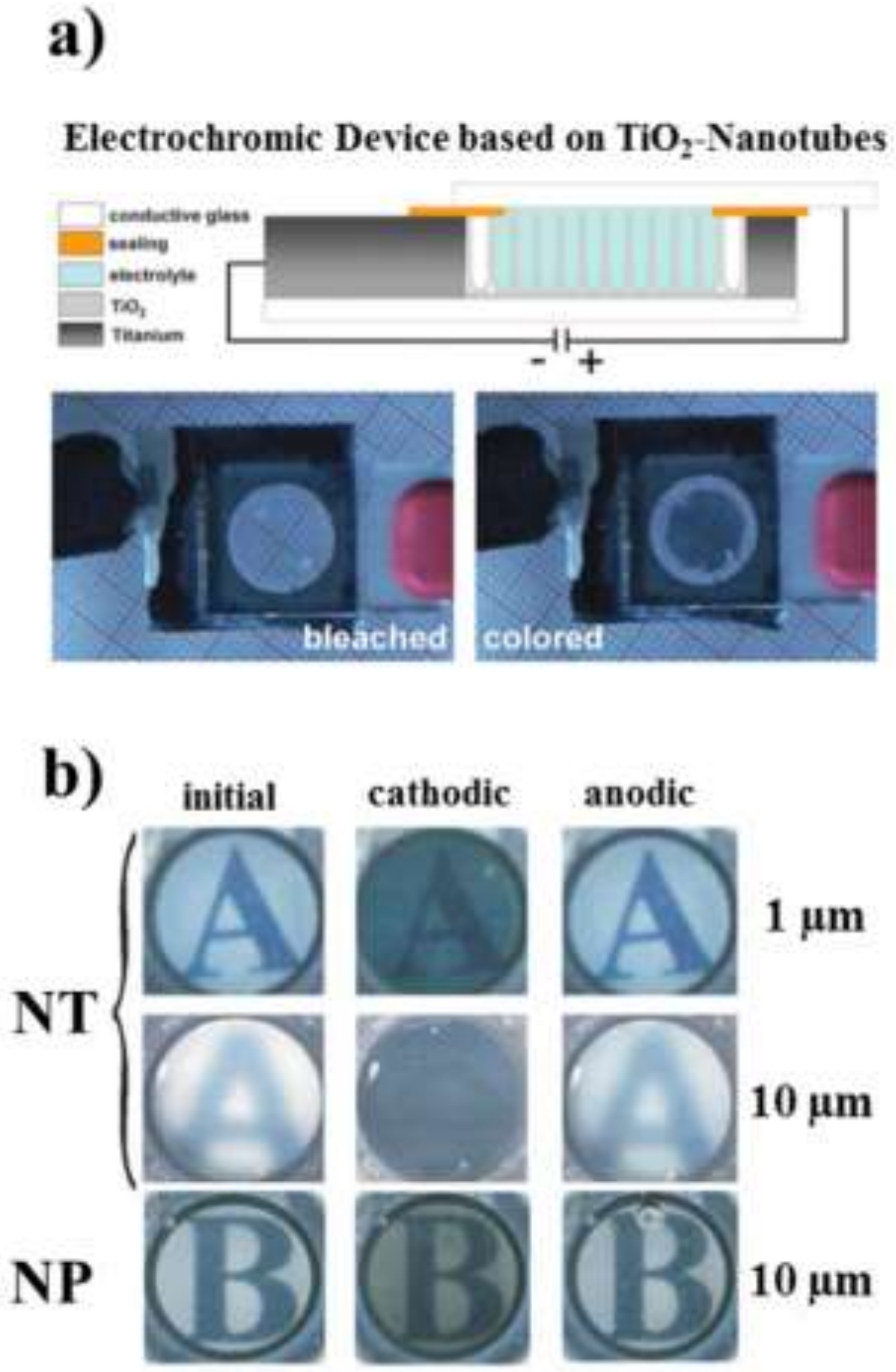

**Figure 35**

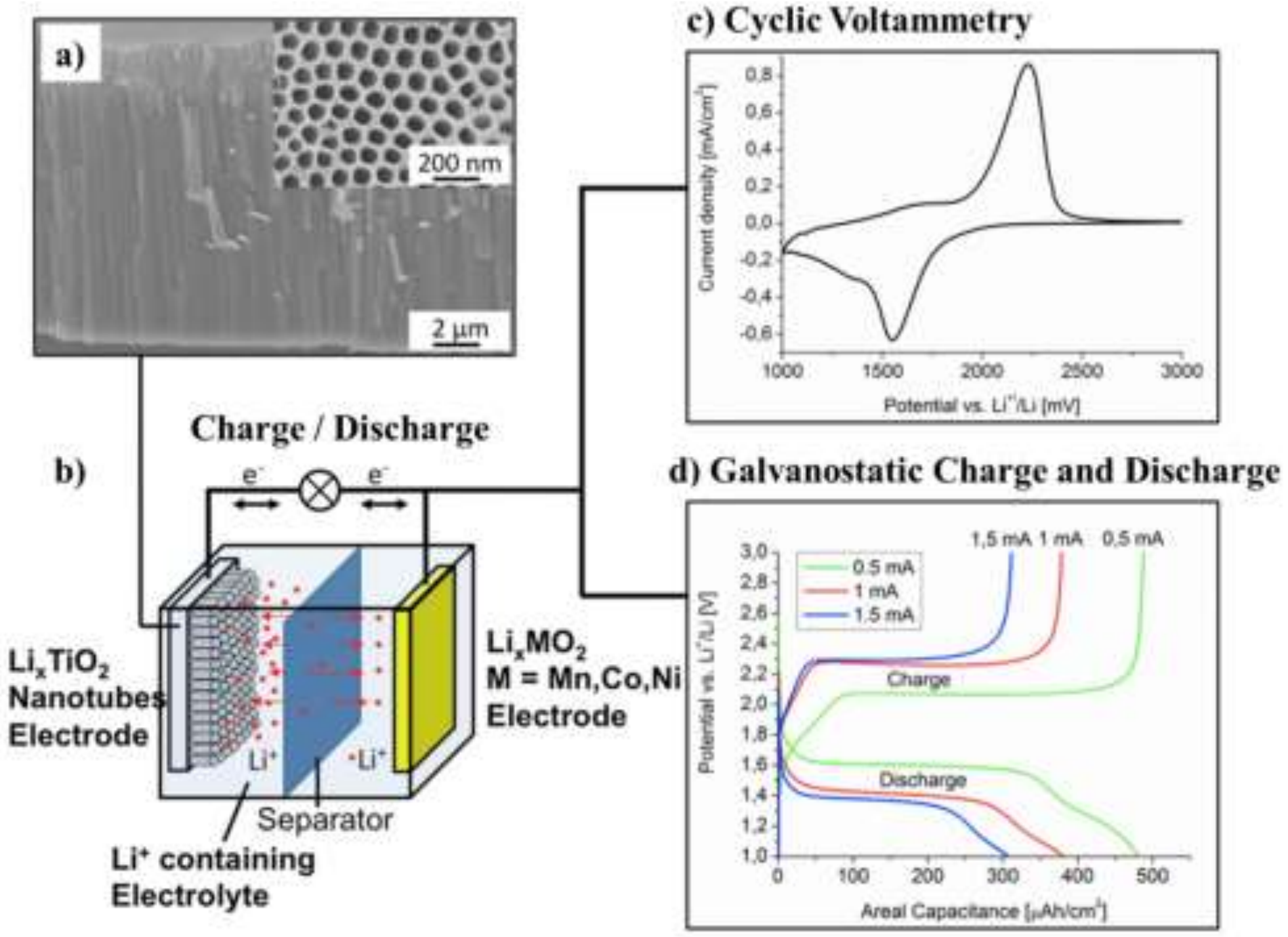

**Figure 36**

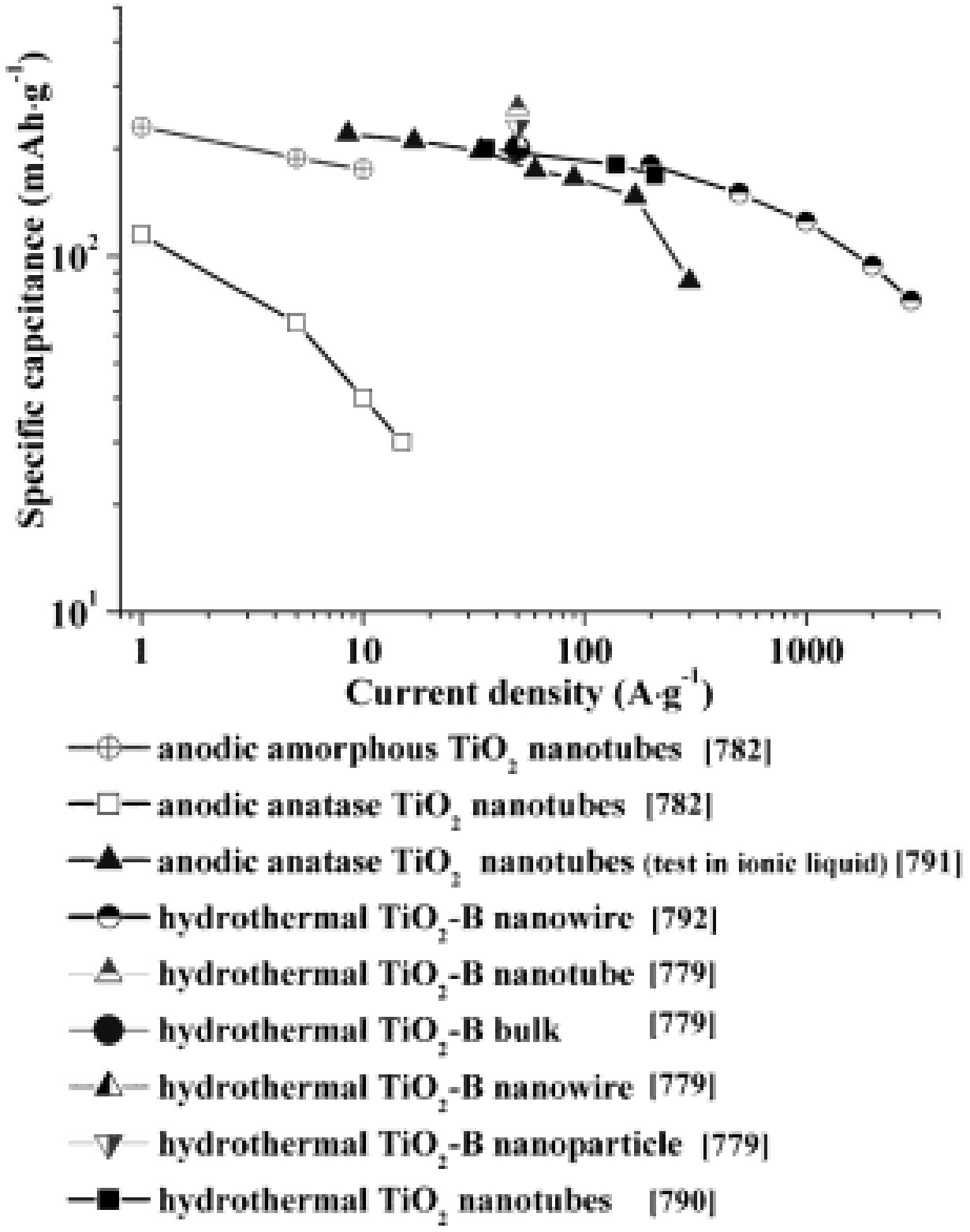
Specific capacitance (mAh·g$^{-1}$)
10$^2$
10$^1$
1
10
100
1000
Current density (A·g$^{-1}$)
anodic amorphous $TiO_2$ nanotubes [782]
anodic anatase $TiO_2$ nanotubes [782]
anodic anatase $TiO_2$ nanotubes (test in ionic liquid) [791]
hydrothermal $TiO_2$-B nanowire [792]
hydrothermal $TiO_2$-B nanotube [779]
hydrothermal $TiO_2$-B bulk [779]
hydrothermal $TiO_2$-B nanowire [779]
hydrothermal $TiO_2$-B nanoparticle [779]
hydrothermal $TiO_2$ nanotubes [790]

**Figure 37**

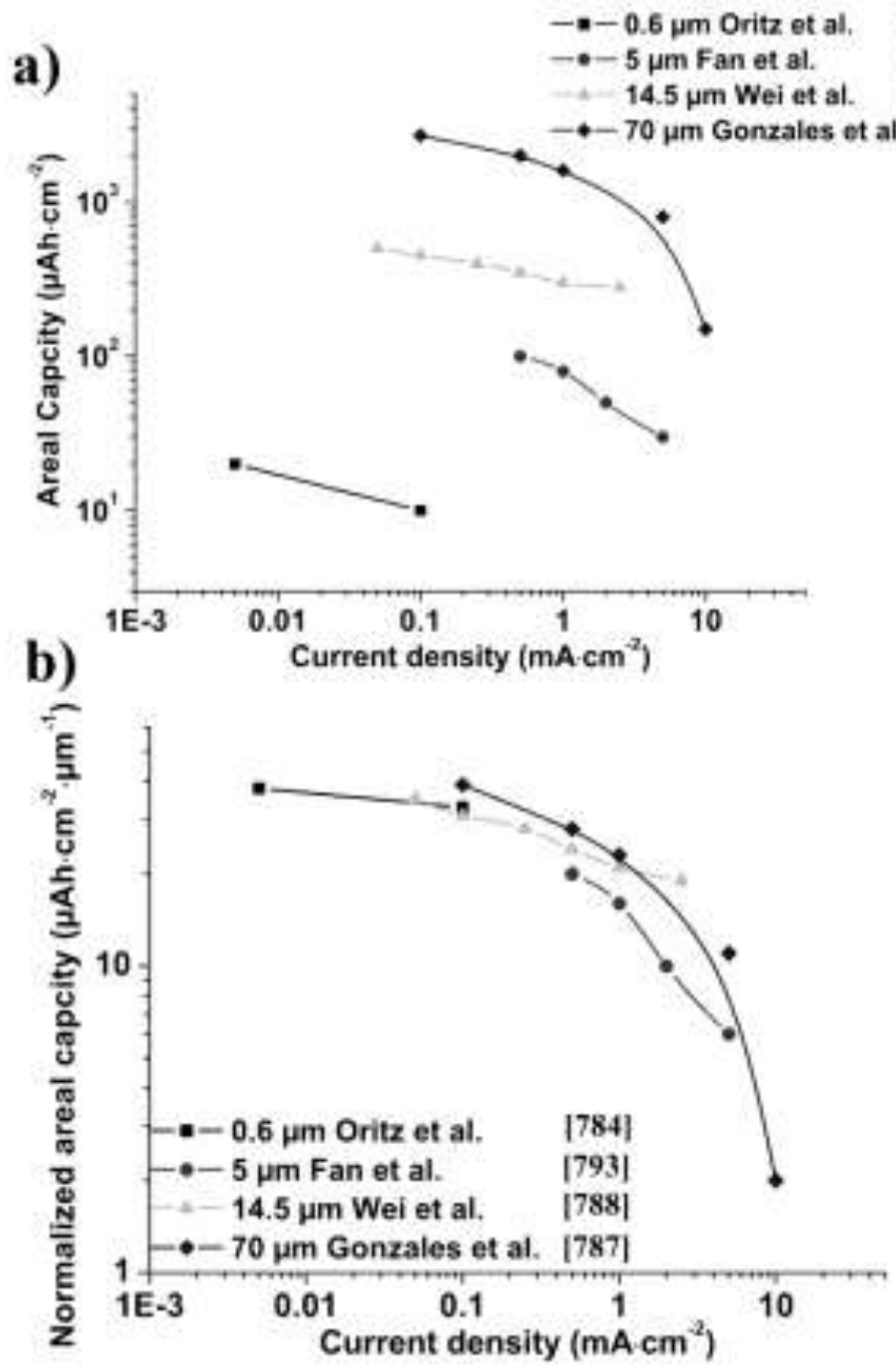
a)
0.6 µm Oritz et al. [784]
5 µm Fan et al. [793]
14.5 µm Wei et al. [788]
70 µm Gonzales et al. [787]
Areal Capcity (µAh·cm⁻²)
Current density (mA·cm⁻²)
1E-3
0.01
0.1
1
10
b)
Normalized areal capcity (µAh·cm⁻²·µm⁻¹)
0.6 µm Oritz et al. [784]
5 µm Fan et al. [793]
14.5 µm Wei et al. [788]
70 µm Gonzales et al. [787]
Current density (mA·cm⁻²)
1E-3
0.01
0.1
1
10

**Figure 38**

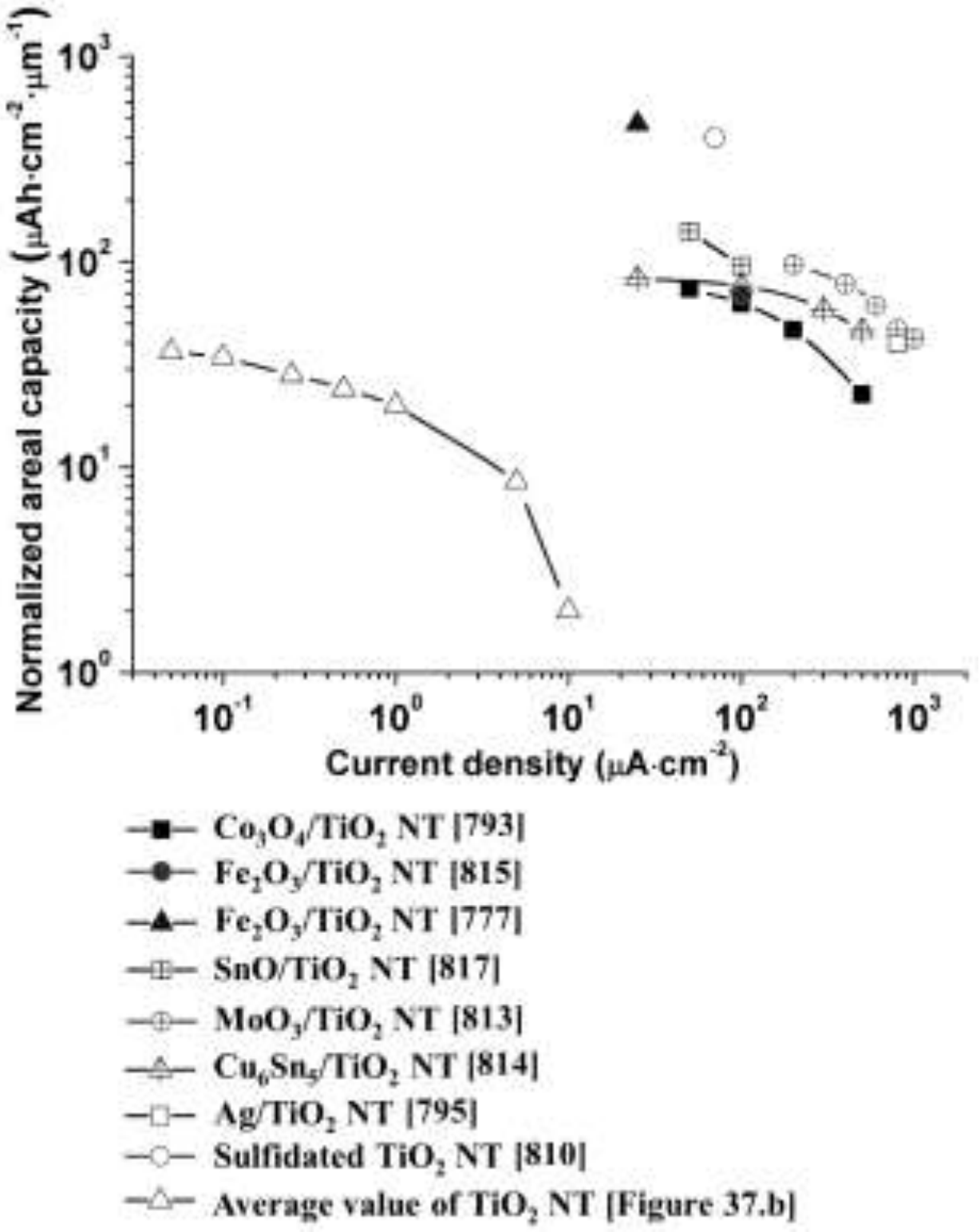

**Figure 39**

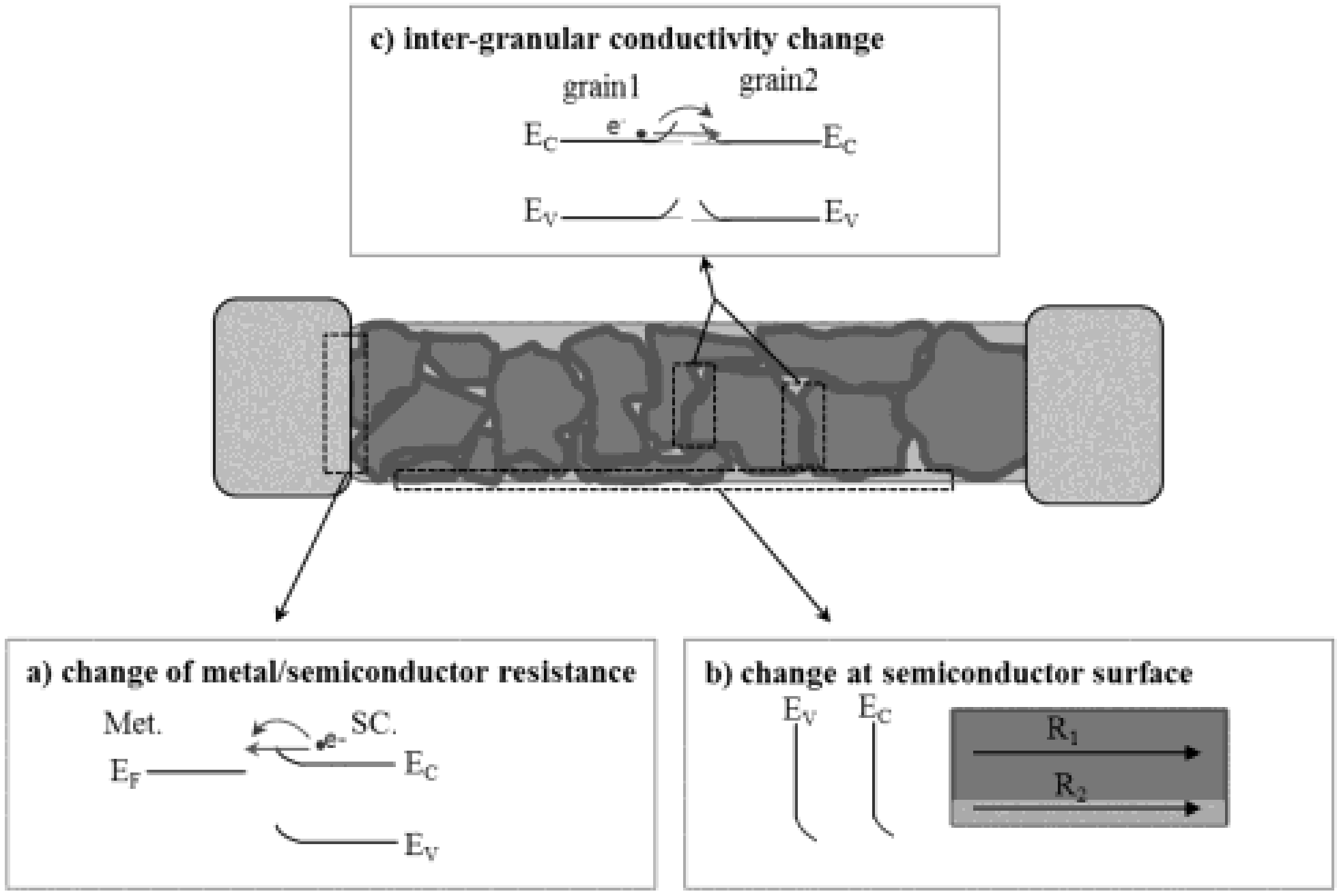

**Figure 40**

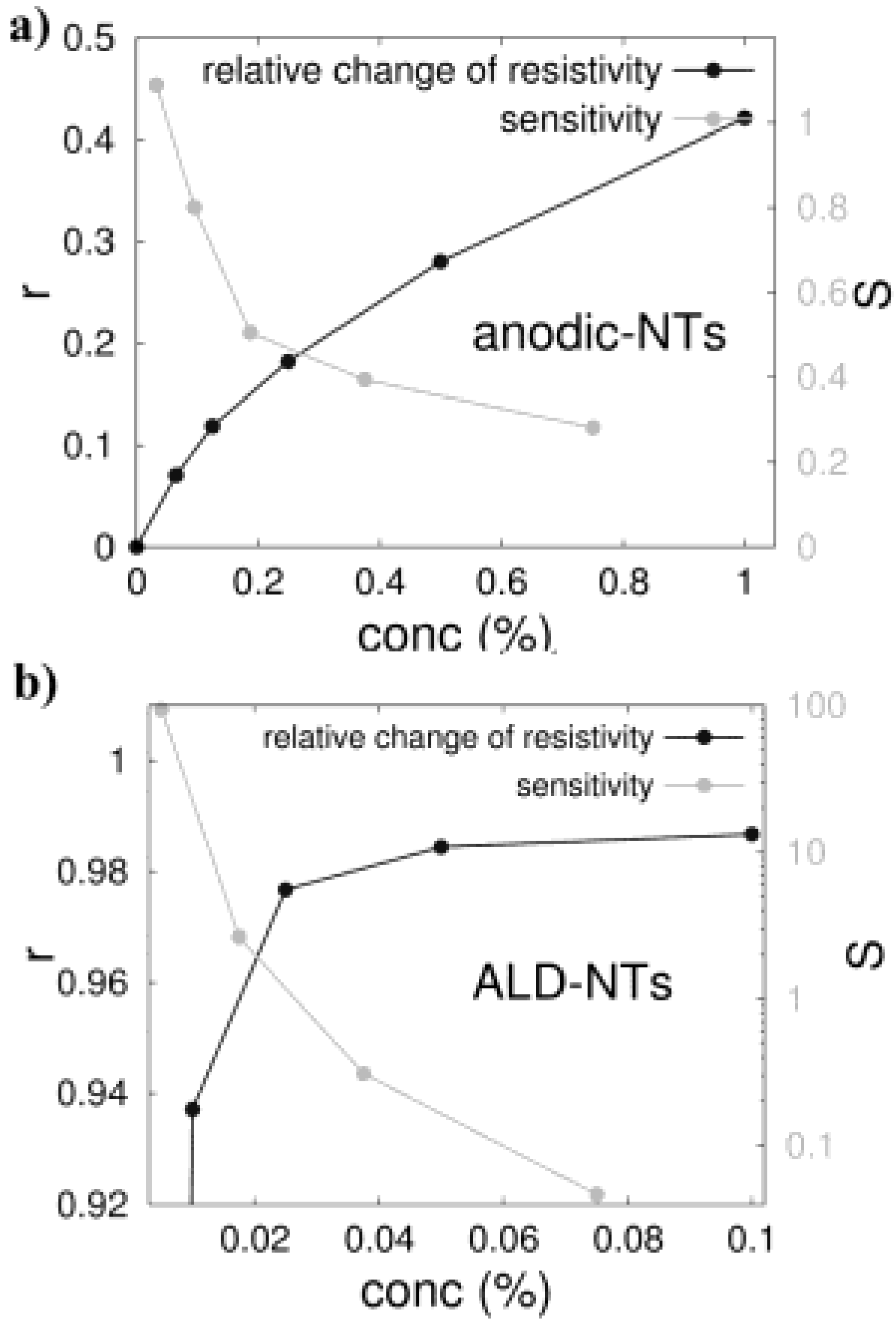

**Figure 41**

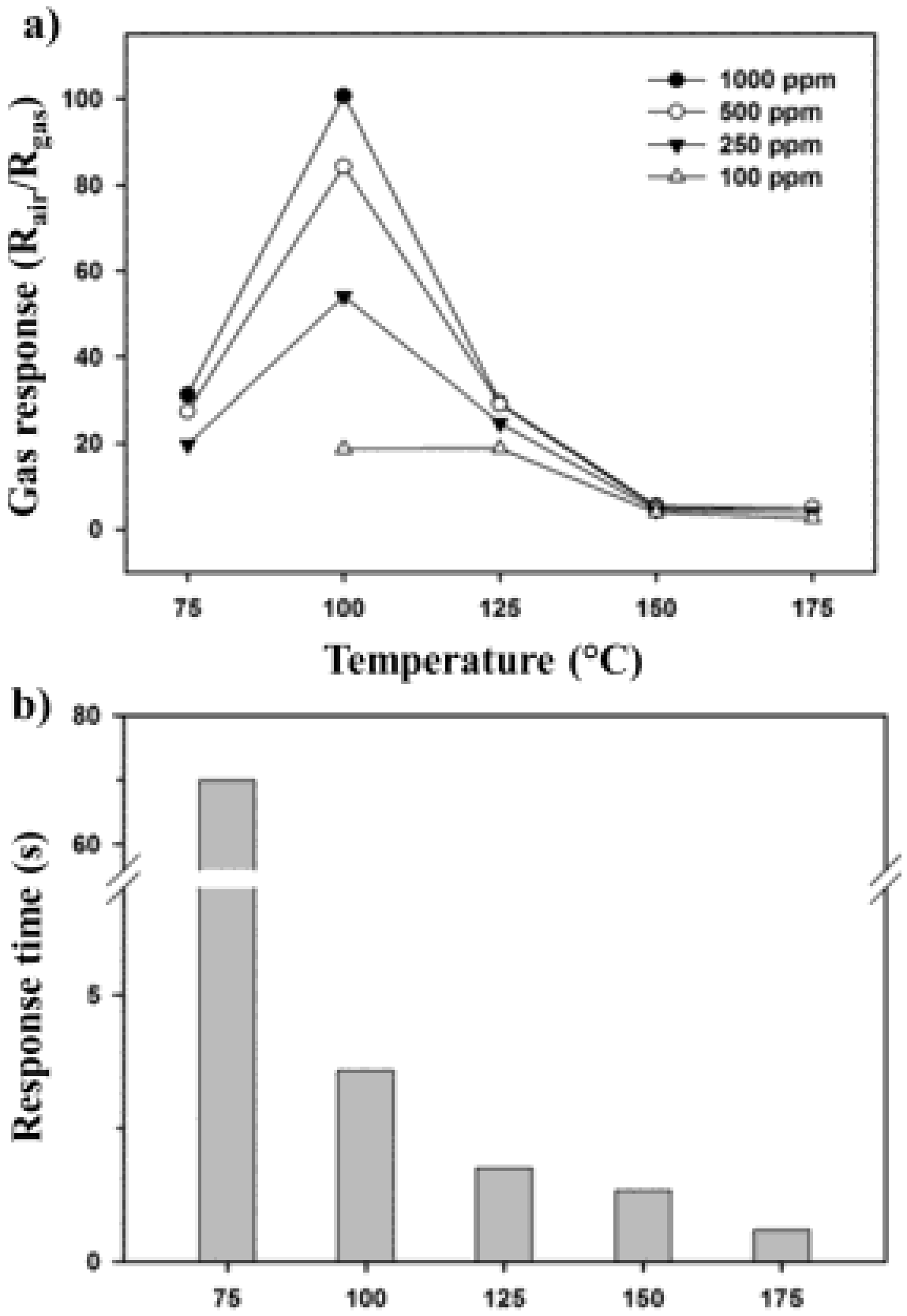

**Figure 42**

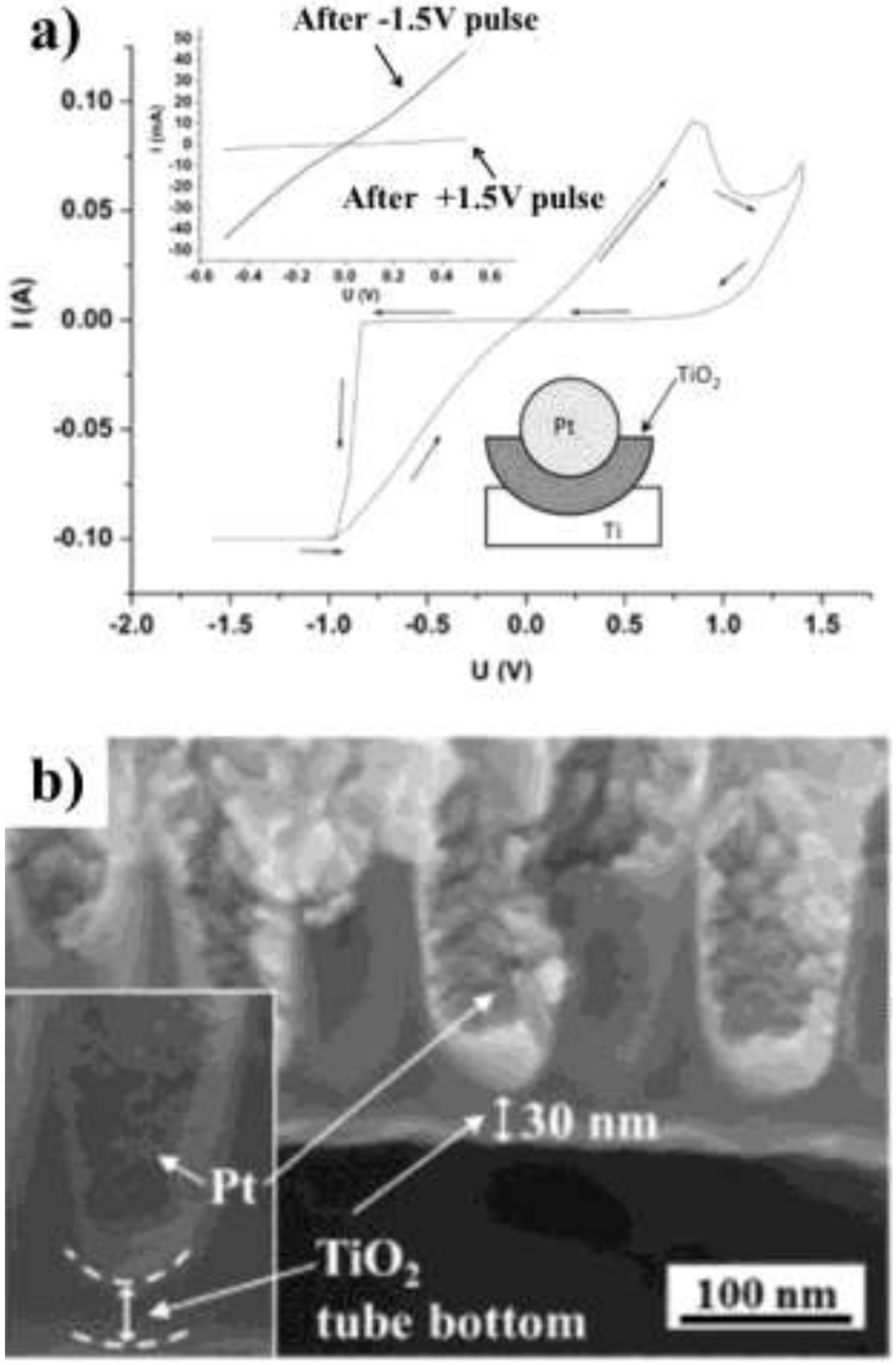

**Figure 43**

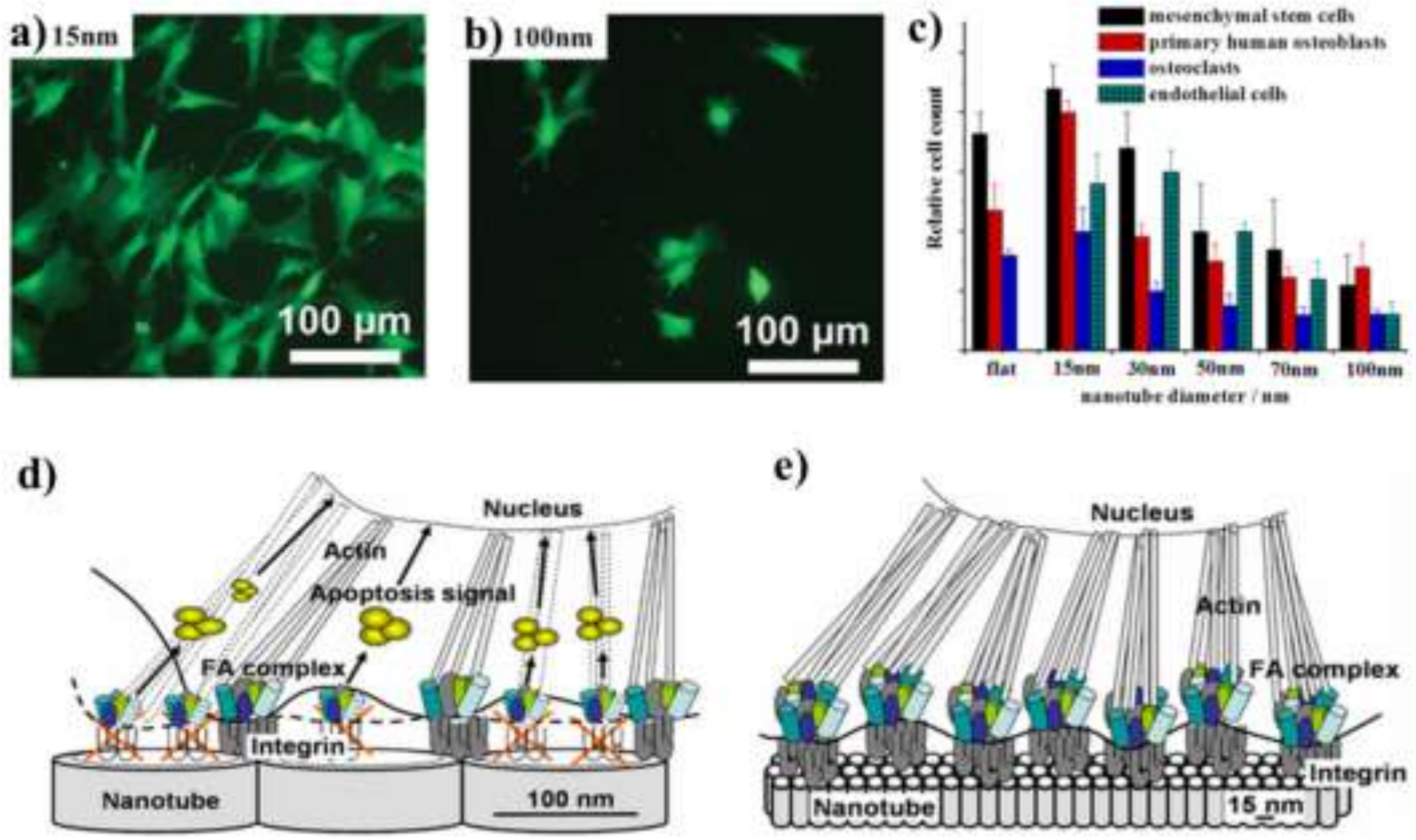

**Figure 44**

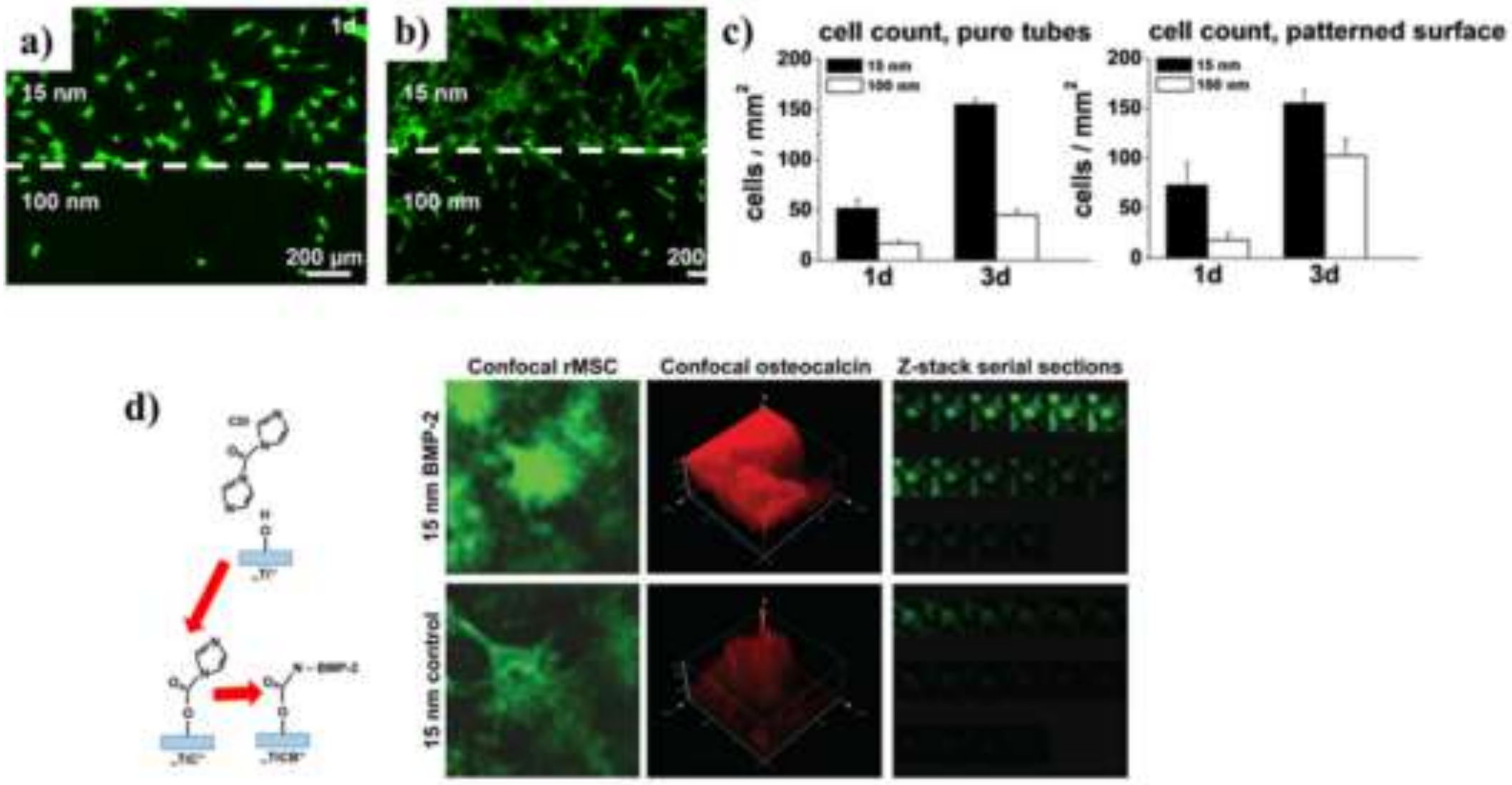

**Figure 45**

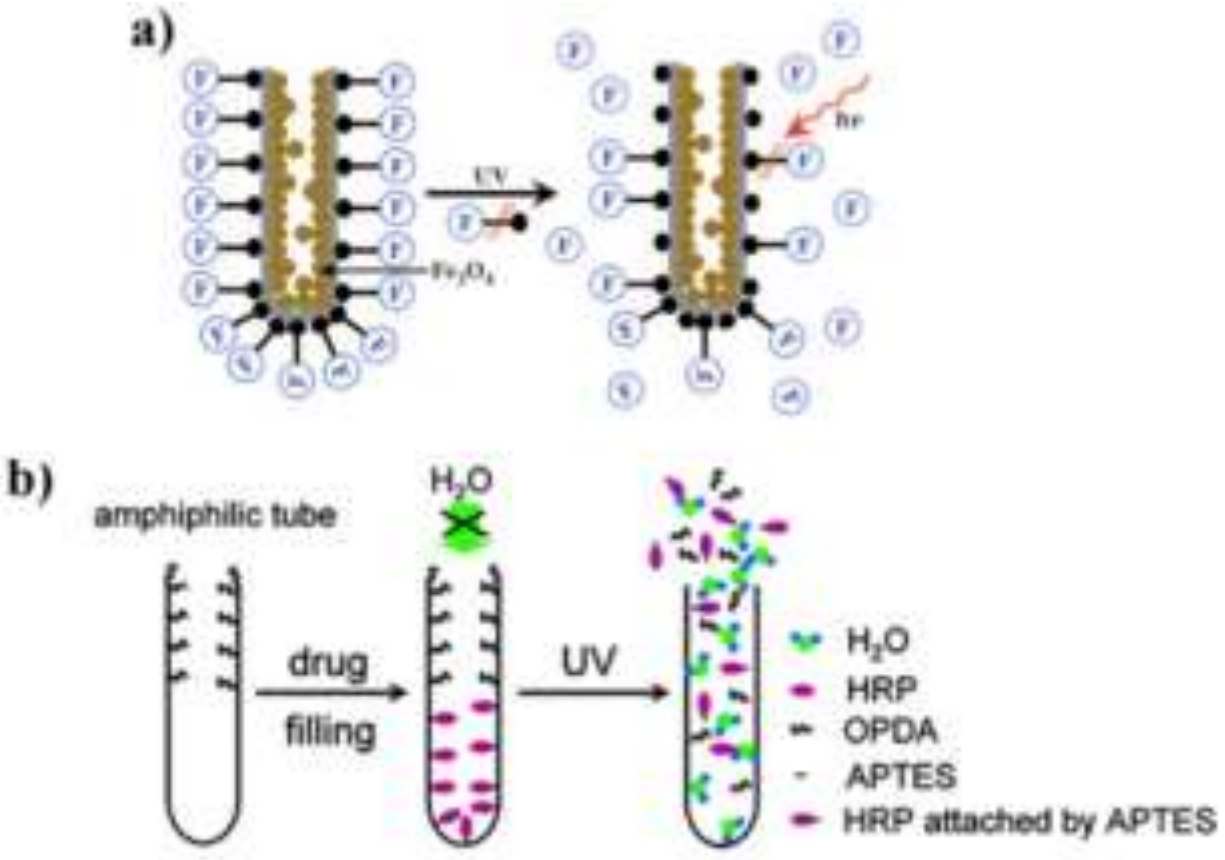